\documentclass[11pt]{article}
\usepackage{epsfig}
\usepackage{rotating}
\newcommand{\gsim}{~{}_{\textstyle\sim}^{\textstyle >}~}

\def\OEE{\Omega_{\rm IB}}

\newcommand{\RE}{{\rm Re}}
\newcommand{\IM}{{\rm Im}}
\newcommand{\vcb}{|V_{cb}|}
\newcommand{\vtd}{|V_{td}|}
\newcommand{\vub}{|V_{ub}/V_{cb}|}
\newcommand{\vts}{|V_{ts}|}

\def\R1{\varepsilon_1}
\def\E8{\varepsilon_8}

\def\r#1{(\ref{#1})}
\def\eps{\varepsilon}
\def\epe{\varepsilon'/\varepsilon}
\def\as{\alpha_s}
\newcommand{\eqn}{\ref}
\def\Heff{{\cal H}_{\rm eff}}

\newcommand{\mt}{m_{\rm t}}
\newcommand{\mtb}{\overline{m}_{\rm t}}

\newcommand{\mc}{m_{\rm c}}
\newcommand{\ms}{m_{\rm s}}
\newcommand{\md}{m_{\rm d}}
\newcommand{\mb}{m_{\rm b}}
\newcommand{\mw}{M_{\rm W}}
\newcommand{\mz}{M_{\rm Z}}
\newcommand{\gev}{\, {\rm GeV}}
\newcommand{\mev}{\, {\rm MeV}}
\newcommand{\bsi}{B_6^{(1/2)}}
\newcommand{\bei}{B_8^{(3/2)}}
\newcommand{\Lms}{\Lambda_{\overline{\rm MS}}}

\newcommand{\bea}{\begin{eqnarray}}
\newcommand{\eea}{\end{eqnarray}}
\newcommand{\bd}{\begin{displaymath}}
\newcommand{\ed}{\end{displaymath}}
\newcommand{\aem}{\alpha}

\newcommand{\beq}{\begin{equation}}
\newcommand{\eeq}{\end{equation}}
\newcommand{\be}{\begin{equation}}
\newcommand{\ee}{\end{equation}}
\newcommand{\bi}{\begin{itemize}}
\newcommand{\ei}{\end{itemize}}
\newcommand{\ord}{{\cal O}}

\def\kpnn{$K^+\rightarrow\pi^+\nu\bar\nu$}
\def\kpn{K^+\rightarrow\pi^+\nu\bar\nu}
\def\klpn{K_{\rm L}\rightarrow\pi^0\nu\bar\nu}
\def\klpnn{$K_{\rm L}\rightarrow\pi^0\nu\bar\nu$}

\newcommand{\kmm}{K_{\rm L} \to \mu^+ \mu^-}
\newcommand{\kpe}{K_{\rm L} \to \pi^0 e^+ e^-}

\def\aspi{\frac{\as}{4\pi}}

\newcommand{\imlt}{\IM\lambda_t}
\newcommand{\relt}{\RE\lambda_t}
\newcommand{\relc}{\RE\lambda_c}

\renewcommand{\baselinestretch}{1.3}

\begin{document}
\thispagestyle{empty}
\phantom{xxx}
\vskip1truecm
\begin{flushright}
 TUM-HEP-402/01 \\
January 2001
\end{flushright}
\vskip1.8truecm
\centerline{\LARGE\bf Flavour Dynamics:
 CP Violation and Rare Decays }
   \vskip1truecm
\centerline{\Large\bf Andrzej J. Buras}
\bigskip
\centerline{\sl Technische Universit{\"a}t M{\"u}nchen}
\centerline{\sl Physik Department} 
\centerline{\sl D-85748 Garching, Germany}
\vskip1truecm
\centerline{\bf Abstract}
These lectures give an up to date description of
CP violation and rare decays of K and B mesons and consist
of ten chapters:
i) Grand view of the field including CKM matrix and
the unitarity triangle, ii) General aspects of the theoretical
framework based on effective weak Hamiltonians, the operator
product expansion and the renormalization group, 
iii) Particle-antiparticle mixing and various types of CP violation,
iv) Standard analysis of the unitarity
triangle, v) The ratio $\epe$,
vi) Rare decays $K^+\to\pi^+\nu\bar\nu$ and $K_L\to\pi^0\nu\bar\nu$,
vii) Express review of other rare decays, viii) 
CP violation in B decays, ix) A brief look beyond the Standard
Model discussing in particular the models with minimal flavour
violation, x) Perspectives for the coming years.
\vskip1.52truecm

\centerline{\it Lectures given at the 38th Course of}
\centerline{\bf the Erice International School of Subnuclear Physics:}
\centerline{\bf Theory and Experiment Heading for New Physics}
\centerline{\it 27 August--5 September, 2000}
%%% end title page %%%%%%%%%%%%%

\newpage

\thispagestyle{empty}

\mbox{}

\newpage

\pagenumbering{roman}

\tableofcontents

\newpage

\pagenumbering{arabic}

\setcounter{page}{1}
\centerline{\Large\bf Flavour Dynamics:
 CP Violation and Rare Decays }
   \vskip1truecm
\centerline{\large\bf Andrzej J. Buras}
\bigskip
\centerline{\sl Technische Universit{\"a}t M{\"u}nchen}
\centerline{\sl Physik Department} 
\centerline{\sl D-85748 Garching, Germany}
\vskip1truecm
\centerline{\bf Contents:}
1. Grand View, 2.  Theoretical Framework, 
3. Particle-Antiparticle Mixing and Various Types of CP Violation,
4. Standard Analysis of the Unitarity Triangle, 5. $\epe$ in the 
Standard Model,
6. The Decays $K^+\to\pi^+\nu\bar\nu$ and $K_L\to\pi^0\nu\bar\nu$,
7. Express Review of Rare K- and B-Decays, 8. 
CP Violation in B Decays, 9. Looking Beyond the Standard
Model, 10. Perspectives.
\vskip1.52truecm

\section{Grand View}
\setcounter{equation}{0}
\subsection{Preface}
Flavour dynamics and the related origin of quarks and lepton masses  
and mixings are among the least understood topics in the
elementary particle physics. While the definite understanding of
flavour dynamics will most probably come from a fundamental
theory at very short distance scales, as the GUT scale or 
the Planck scale, it is commonly accepted that the study of
CP-violating and rare decay processes  plays an important role
in the search for this fundamental theory.

In this context an important issue is the
question whether the Standard Model (SM) of fundamental interactions is
capable of describing the violation of CP symmetry observed in
nature. Actually this question has already been answered through
the studies of a dynamical generation of the baryon asymmetry in the
universe, which is necessary for our existence. It turns out that
the size of CP violation in the SM is too small
to generate a large enough matter-antimatter asymmetry 
observed in the universe today.

On the other hand it is conceivable that the physics responsible for 
the baryon asymmetry involves only very short distance scales,
as the GUT scale or 
the Planck scale, and the related CP violation is unobservable in the
experiments performed by humans. Yet even if  such an unfortunate situation
is a real possibility, it is unlikely that the SM provides
an adequate description of CP violation at scales accessible to experiments
peformed on our planet in this millennium. On the one hand the
Kobayashi-Maskawa (KM) picture of CP violation  is so economical that it
is hard to believe that it will pass future experimental tests. On
the other hand almost any extention of the SM contains
additional sources of CP violating effects. As some kind of new physics
is required in order to understand the patterns of quark and lepton
masses and mixings and generally to understand the flavour dynamics, 
it is very likely that this physics will bring
new sources of CP violation modifying KM picture considerably.

Similarly to CP violation,  particle-antiparticle mixing and rare decays
of hadrons and leptons play an important role in the tests of the SM
and of its extentions. As  particle-antiparticle mixing
and  rare decay branching ratios depend sensitively on the masses
and couplings of particles involved, these transitions constitute
an excellent machinery to study the flavour dynamics of quarks,
leptons and other particles like sparticles in the supersymmetric
 extentions of the SM.

As of January 2001 all existing data on CP violation and rare decays can
be described by the SM within the theoretical and experimental
uncertainties. An important exception are the neutrino oscillations,
which implying  non-vanishing neutrino masses changed the SM
picture in the lepton sector considerably. 
It is exciting that in the coming years the new data on
CP violation and rare decays as well as $B_s^0-\bar B_s^0$ mixing coming
from a number of laboratories in Europe, USA and Japan may change the
SM picture in the quark sector as well.

These lectures provide a rather non-technical description of this fascinating
field. There is unavoidably an overlap with our Les Houches  
\cite{AJBLH} and Lake Louise lectures \cite{AJBLAKE} and with the 
reviews \cite{BBL} and \cite{BF97}.
On the other hand new developments since the summer 1999 have been taken
into account, as far as the space allowed for it, and all numerical
results have been  updated. Moreover the discussions of various types
of CP violation and of the physics beyond the SM have been considerably 
extended. In particular we discuss in detail the models with minimal
flavour violation, presenting an improved lower bound on the angle
$\beta$ in the unitarity triangle. 
Finally we provide the complete list of references to NLO calculations
for weak decays performed until the end of 2000.

The first decade of the new millennium began strictly speaking
one month ago. It is a common expectation that this decade will
bring important, possibly decisive, insights into the structure
of flavour dynamics that can be most efficiently studied through
rare and CP-violating decays. We hope that these lecture notes
will be helpful in following the new developments. In this respect
the recent books \cite{BULIND,Branco,Bigi}, the working group reports
\cite{BABAR,LHCB} and most recent reviews \cite{REV} are strongly
recommended.

\subsection{Some Facts about the Standard Model}
In the first eight sections of these lectures we will dominantly work 
in the context of the SM with three generations of 
quarks and leptons and
the interactions described by the gauge group 
$ SU(3)_C\otimes SU(2)_L\otimes U(1)_Y$ spontaneously broken to
$SU(3)_C\otimes U(1)_Q$.
There are excellent text books on the dynamics of the SM 
\cite{Donoghu}--\cite{Muta}.
Let us therfore collect here only those ingredients of this model which
are fundamental for the subject of these lectures.

\bi
\item
The strong interactions are mediated by eight gluons $G_a$, the
electroweak interactions by $W^{\pm}$, $Z^0$ and $\gamma$.
\item
Concerning {\it Electroweak Interactions}, the left-handed leptons and
quarks are put into $SU(2)_L$ doublets:
\begin{equation}\label{2.31}
\left(\begin{array}{c}
\nu_e \\
e^-
\end{array}\right)_L\qquad
\left(\begin{array}{c}
\nu_\mu \\
\mu^-
\end{array}\right)_L\qquad
\left(\begin{array}{c}
\nu_\tau \\
\tau^-
\end{array}\right)_L
\end{equation}
\begin{equation}\label{2.66}
\left(\begin{array}{c}
u \\
d^\prime
\end{array}\right)_L\qquad
\left(\begin{array}{c}
c \\
s^\prime
\end{array}\right)_L\qquad
\left(\begin{array}{c}
t \\
b^\prime
\end{array}\right)_L       
\end{equation}
with the corresponding right-handed fields transforming as singlets
under $ SU(2)_L $. The primes in (\ref{2.66}) will be
discussed in a moment. 
\item
The charged current processes mediated by $W^{\pm}$ are
flavour violating with the strength of violation given by
the gauge coupling $g_2$  and effectively at low energies 
by the Fermi constant 
\begin{equation}\label{2.100}
\frac{G_{\rm F}}{\sqrt{2}}=\frac{g^2_2}{8 \mw^2}
\end{equation}
and a {\it unitary} $3\times3$
{\rm CKM} matrix. 
\item
The {\rm CKM} matrix \cite{CAB,KM} connects the {\it weak
eigenstates} $(d^\prime,s^\prime,b^\prime)$ and the corresponding {\it mass 
eigenstates} $d,s,b$ through
\begin{equation}\label{2.67}
\left(\begin{array}{c}
d^\prime \\ s^\prime \\ b^\prime
\end{array}\right)=
\left(\begin{array}{ccc}
V_{ud}&V_{us}&V_{ub}\\
V_{cd}&V_{cs}&V_{cb}\\
V_{td}&V_{ts}&V_{tb}
\end{array}\right)
\left(\begin{array}{c}
d \\ s \\ b
\end{array}\right)\equiv\hat V_{\rm CKM}\left(\begin{array}{c}
d \\ s \\ b
\end{array}\right).
\end{equation}
In the leptonic sector the analogous mixing matrix is a unit matrix
due to the masslessness of neutrinos in the SM.
\item
The unitarity of the CKM matrix assures the absence of
flavour changing neutral current transitions at the tree level.
This means that the
elementary vertices involving neutral gauge bosons ($G_a$, $Z^0$,
$\gamma$) and the neutral Higgs are flavour conserving.
This property is known under the name of GIM mechanism \cite{GIM}.
\item
The fact that the $V_{ij}$'s can a priori be complex
numbers allows  CP violation in the SM \cite{KM}. 
\item
An important property of the strong interactions described by
Quantum Chromodynamics (QCD) is {\it the asymptotic freedom}
\cite{ASYM}.
This property implies that at short distance scales $\mu >\ord(1~\gev)$
the strong interaction effects in weak decays can be evaluated
by means of perturbative methods with the expansion parameter
$\alpha_{\overline{MS}}(\mu)$ \cite{BBDM}. The existing analyses
of high energy processes give  
$\alpha_{\overline{MS}}(\mz)=0.118\pm 0.003$ \cite{Bethke}.
\item
At long distances, corresponding to $\mu<\ord( 1\gev)$, 
$\alpha_{\overline{MS}}(\mu)$
becomes  large and QCD effects in weak decays relevant
to these scales can only be evaluated by means of non-perturbative methods. 
\ei

\subsection{CKM Matrix}
\subsubsection{General Remarks}
We know from the text books that the CKM matrix can be
parametrized by
three angles and a single complex phase.
This phase is  necessary to describe
CP violation within the framework of the SM.

Many parametrizations of the CKM
matrix have been proposed in the literature. The classification 
of different
parametrizations can be found in \cite{FX1}.  We will use
two parametrizations in these lectures: the standard parametrization 
\cite{CHAU} recommended by the Particle Data Group  \cite{PDG}  
and the Wolfenstein parametrization \cite{WO}. In the context of
the models for fermion masses and mixings a useful parametrization
has been proposed by Fritzsch and Xing \cite{FX2}. 
In this parametrization, in contrast
to the standard and the Wolfenstein parametrization, the complex phase
recides only in the $2\times 2$ submatrix involving  u, d, s and c
quarks.

\subsubsection{Standard Parametrization}
            \label{sec:sewm:stdparam}
With
$c_{ij}=\cos\theta_{ij}$ and $s_{ij}=\sin\theta_{ij}$ 
($i,j=1,2,3$), the standard parametrization is
given by:
\begin{equation}\label{2.72}
\hat V_{\rm CKM}=
\left(\begin{array}{ccc}
c_{12}c_{13}&s_{12}c_{13}&s_{13}e^{-i\delta}\\ -s_{12}c_{23}
-c_{12}s_{23}s_{13}e^{i\delta}&c_{12}c_{23}-s_{12}s_{23}s_{13}e^{i\delta}&
s_{23}c_{13}\\ s_{12}s_{23}-c_{12}c_{23}s_{13}e^{i\delta}&-s_{23}c_{12}
-s_{12}c_{23}s_{13}e^{i\delta}&c_{23}c_{13}
\end{array}\right)\,,
\end{equation}
where $\delta$ is the phase necessary for {\rm CP} violation.
$c_{ij}$ and
$s_{ij}$ can all be chosen to be positive
and  $\delta$ may vary in the
range $0\le\delta\le 2\pi$. However, the measurements
of CP violation in $K$ decays force $\delta$ to be in the range
 $0<\delta<\pi$. 

From phenomenological applications we know that 
$s_{13}$ and $s_{23}$ are small numbers: $\ord(10^{-3})$ and ${\cal
O}(10^{-2})$,
respectively. Consequently to an excellent accuracy $c_{13}=c_{23}=1$
and the four independent parameters are given as 
\begin{equation}\label{2.73}
s_{12}=| V_{us}|, \quad s_{13}=| V_{ub}|, \quad s_{23}=|
V_{cb}|, \quad \delta.
\end{equation}

The first three can be extracted from tree level decays mediated
by the transitions $s \to u$, $b \to u$ and $b \to c$ respectively.
The phase $\delta$ can be extracted from CP violating transitions or 
loop processes sensitive to $| V_{td}|$. The latter fact is based
on the observation that
 for $0\le\delta\le\pi$, as required by the analysis of CP violation
in the $K$ system,
there is a one--to--one correspondence between $\delta$ and $|V_{td}|$
given by
\begin{equation}\label{10}
| V_{td}|=\sqrt{a^2+b^2-2 a b \cos\delta},
\qquad
a=| V_{cd} V_{cb}|,
\qquad
b=| V_{ud} V_{ub}|\,.
\end{equation} 

The main  phenomenological advantages of (\ref{2.72}) over other
parametrizations proposed in the literature are basically these
two: 
\begin{itemize}
\item
$s_{12}$, $s_{13}$ and $s_{23}$ being related in a very simple way
to $| V_{us}|$, $| V_{ub}|$ and $|V_{cb}|$ respectively, can be
measured independently in three decays.
\item
The CP violating phase is always multiplied by the very small
$s_{13}$. This shows clearly the suppression of CP violation
independently of the actual size of $\delta$.
\end{itemize}

For numerical evaluations the use of the standard parametrization
is strongly recommended. However once the four parameters in
(\ref{2.73}) have been determined it is often useful to make
a change of basic parameters in order to expose the structure of
the results more transparently. This brings us to the Wolfenstein
parametrization \cite{WO} and its generalization given in 
\cite{BLO}.

\subsubsection{Wolfenstein Parameterization }\label{Wolf-Par}
 The Wolfenstein parametrization 
is an approximate parametrization of the CKM matrix in which
each element is expanded as a power series in the small parameter
$\lambda=| V_{us}|=0.22$,
\begin{equation}\label{2.75} 
\hat V=
\left(\begin{array}{ccc}
1-{\lambda^2\over 2}&\lambda&A\lambda^3(\varrho-i\eta)\\ -\lambda&
1-{\lambda^2\over 2}&A\lambda^2\\ A\lambda^3(1-\varrho-i\eta)&-A\lambda^2&
1\end{array}\right)
+\ord(\lambda^4)\,,
\end{equation}
and the set (\ref{2.73}) is replaced by
\begin{equation}\label{2.76}
\lambda, \qquad A, \qquad \varrho, \qquad \eta \, .
\end{equation}

Because of the
smallness of $\lambda$ and the fact that for each element 
the expansion parameter is actually
$\lambda^2$, it is sufficient to keep only the first few terms
in this expansion. 

The Wolfenstein parametrization is certainly more transparent than
the standard parametrization. However, if one requires sufficient 
level of accuracy, the higher order terms in $\lambda$ have to
be included in phenomenological applications.
This can be done in many ways.
The
point is that since (\ref{2.75}) is only an approximation the {\em exact}
definiton of the parameters in (\ref{2.76}) is not unique by terms of the 
neglected order
${\cal O}(\lambda^4)$. 
This situation is familiar from any perturbative expansion, where
different definitions of expansion parameters (coupling constants) 
are possible.
This is also the reason why in different papers in the
literature different ${\cal O}(\lambda^4)$ terms in (\ref{2.75})
 can be found. They simply
correspond to different definitions of the parameters in (\ref{2.76}).
Since the physics does not depend on a particular definition, it
is useful to make a choice for which the transparency of the original
Wolfenstein parametrization is not lost. Here we present one
way of achieving this.

\subsubsection{Wolfenstein Parametrization Beyond LO}
An efficient and systematic way of finding higher order terms in $\lambda$
is to go back to the standard parametrization (\ref{2.72}) and to
 {\it define} the parameters $(\lambda,A,\varrho,\eta)$ through 
\cite{BLO,schubert}
\begin{equation}\label{2.77} 
s_{12}=\lambda\,,
\qquad
s_{23}=A \lambda^2\,,
\qquad
s_{13} e^{-i\delta}=A \lambda^3 (\varrho-i \eta)
\end{equation}
to {\it  all orders} in $\lambda$. 
It follows  that
\begin{equation}\label{2.84} 
\varrho=\frac{s_{13}}{s_{12}s_{23}}\cos\delta,
\qquad
\eta=\frac{s_{13}}{s_{12}s_{23}}\sin\delta.
\end{equation}
(\ref{2.77}) and (\ref{2.84}) represent simply
the change of variables from (\ref{2.73}) to (\ref{2.76}).
Making this change of variables in the standard parametrization 
(\ref{2.72}) we find the CKM matrix as a function of 
$(\lambda,A,\varrho,\eta)$ which satisfies unitarity exactly.
Expanding next each element in powers of $\lambda$ we recover the
matrix in (\ref{2.75}) and in addition find explicit corrections of
$\ord(\lambda^4)$ and higher order terms:

\be
V_{ud}=1-\frac{1}{2}\lambda^2-\frac{1}{8}\lambda^4 +\ord(\lambda^6)
\ee
\be
V_{us}=\lambda+\ord(\lambda^7),\qquad 
V_{ub}=A \lambda^3 (\varrho-i \eta)
\ee
\be
V_{cd}=-\lambda+\frac{1}{2} A^2\lambda^5 [1-2 (\varrho+i \eta)]+
\ord(\lambda^7)
\ee
\be
V_{cs}= 1-\frac{1}{2}\lambda^2-\frac{1}{8}\lambda^4(1+4 A^2) +\ord(\lambda^6)
\ee
\be
V_{cb}=A\lambda^2+\ord(\lambda^8), \qquad
V_{tb}=1-\frac{1}{2} A^2\lambda^4+\ord(\lambda^6)
\ee
\be
V_{td}=A\lambda^3 \left[ 1-(\varrho+i \eta)(1-\frac{1}{2}\lambda^2)\right]
+\ord (\lambda^7)
\ee
\begin{equation}\label{2.83d}
 V_{ts}= -A\lambda^2+\frac{1}{2}A(1-2 \varrho)\lambda^4
-i\eta A \lambda^4 +\ord(\lambda^6)~.
\end{equation}

We note that by definition
$V_{ub}$ remains unchanged and the
corrections to $V_{us}$ and $V_{cb}$ appear only at $\ord(\lambda^7)$ and
$\ord(\lambda^8)$, respectively.
Consequently to an 
 an excellent accuracy we have:
\begin{equation}\label{CKM1}
V_{us}=\lambda, \qquad V_{cb}=A\lambda^2,
\end{equation}
\begin{equation}\label{CKM2}
V_{ub}=A\lambda^3(\varrho-i\eta),
\qquad
V_{td}=A\lambda^3(1-\bar\varrho-i\bar\eta)
\end{equation}
with \cite{BLO}
\begin{equation}\label{2.88d}
\bar\varrho=\varrho (1-\frac{\lambda^2}{2}),
\qquad
\bar\eta=\eta (1-\frac{\lambda^2}{2}).
\end{equation}
The advantage of this generalization of the Wolfenstein parametrization
over other generalizations found in the literature is the absence of
relevant corrections to $V_{us}$, $V_{cb}$ and $V_{ub}$ and an elegant
change in $V_{td}$ which allows a simple generalization of the 
so-called unitarity triangle beyond LO. For these reasons this
generalization of the Wolfenstein parametrization has been adopted
by most authors in the literature.

Finally let us collect useful approximate analytic expressions
for $\lambda_i=V_{id}V^*_{is}$ with $i=c,t$:
\begin{equation}\label{2.51}
 \IM\lambda_t= -\IM\lambda_c=\eta A^2\lambda^5=
\mid V_{ub}\mid \mid V_{cb} \mid \sin\delta~, 
\end{equation}
\begin{equation}\label{2.52}
 \RE\lambda_c=-\lambda (1-\frac{\lambda^2}{2})~,
\end{equation}
\begin{equation}\label{2.53}
 \RE\lambda_t= -(1-\frac{\lambda^2}{2}) A^2\lambda^5 (1-\bar\varrho) \,.
\end{equation}
Expressions (\ref{2.51}) and (\ref{2.52}) represent to an accuracy of
0.2\% the exact formulae obtained using (\ref{2.72}). The expression
(\ref{2.53}) deviates by at most 2\% from the exact formula in the
full range of parameters considered. For $\varrho$ close to zero
this deviation is below 1\%. 
After inserting the expressions (\ref{2.51})--(\ref{2.53}) in the exact
formulae for quantities of interest, a further expansion in $\lambda$
should not be made. 
\subsubsection{Unitarity Triangle}
The unitarity of the CKM-matrix implies various relations between its
elements. In particular, we have
\begin{equation}\label{2.87h}
V_{ud}^{}V_{ub}^* + V_{cd}^{}V_{cb}^* + V_{td}^{}V_{tb}^* =0.
\end{equation}
Phenomenologically this relation is very interesting as it involves
simultaneously the elements $V_{ub}$, $V_{cb}$ and $V_{td}$ which are
under extensive discussion at present.

The relation (\ref{2.87h})  can be
represented as a ``unitarity'' triangle in the complex 
$(\bar\varrho,\bar\eta)$ plane. 
The invariance of (\ref{2.87h})  under any phase-transformations
implies that the  corresponding triangle
is rotated in the $(\bar\varrho,\bar\eta)$  plane under such transformations. 
Since the angles and the sides
(given by the moduli of the elements of the
mixing matrix)  in this triangle remain unchanged, they
 are phase convention independent and are  physical observables.
Consequently they can be measured directly in suitable experiments.  
One can construct additional five unitarity triangles corresponding
to other orthogonality relations, like the one in (\ref{2.87h}).
They are discussed in \cite{Kayser}. Some of them should be useful
when LHC-B and BTeV experiments will provide data.
The areas of all unitarity triangles are equal and related to the measure of 
CP violation 
$J_{\rm CP}$ \cite{CJ,js}:
\begin{equation}
\mid J_{\rm CP} \mid = 2\cdot A_{\Delta},
\end{equation}
where $A_{\Delta}$ denotes the area of the unitarity triangle.

\begin{figure}[hbt]
\vspace{0.10in}
\centerline{
\epsfysize=2.1in
\epsffile{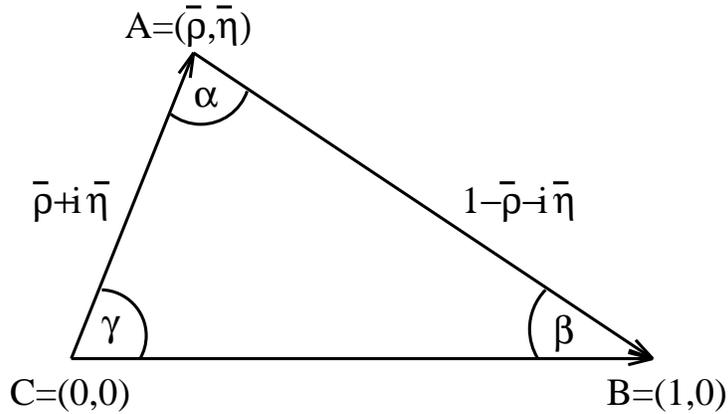}
}
\vspace{0.08in}
\caption{Unitarity Triangle.}\label{fig:utriangle}
\end{figure}

The construction of the unitarity triangle proceeds as follows:

\bi
\item
We note first that
\begin{equation}\label{2.88a}
V_{cd}^{}V_{cb}^*=-A\lambda^3+\ord(\lambda^7).
\end{equation}
Thus to an excellent accuracy $V_{cd}^{}V_{cb}^*$ is real with
$| V_{cd}^{}V_{cb}^*|=A\lambda^3$.
\item
Keeping $\ord(\lambda^5)$ corrections and rescaling all terms in
(\ref{2.87h})
by $A \lambda^3$ 
we find
\begin{equation}\label{2.88b}
 \frac{1}{A\lambda^3}V_{ud}^{}V_{ub}^*
=\bar\varrho+i\bar\eta,
\qquad
\qquad
 \frac{1}{A\lambda^3}V_{td}^{}V_{tb}^*
=1-(\bar\varrho+i\bar\eta)
\end{equation}
with $\bar\varrho$ and $\bar\eta$ defined in (\ref{2.88d}). 
\item
Thus we can represent (\ref{2.87h}) as the unitarity triangle 
in the complex $(\bar\varrho,\bar\eta)$ plane 
as shown in fig. \ref{fig:utriangle}.
\ei

Let us collect useful formulae related to this triangle:
\bi
\item
Using simple trigonometry one can express $\sin(2\phi_i$), $\phi_i=
\alpha, \beta, \gamma$, in terms of $(\bar\varrho,\bar\eta)$ as follows:
\begin{equation}\label{2.89}
\sin(2\alpha)=\frac{2\bar\eta(\bar\eta^2+\bar\varrho^2-\bar\varrho)}
  {(\bar\varrho^2+\bar\eta^2)((1-\bar\varrho)^2
  +\bar\eta^2)},  
\end{equation}
\begin{equation}\label{2.90}
\sin(2\beta)=\frac{2\bar\eta(1-\bar\varrho)}{(1-\bar\varrho)^2 + \bar\eta^2},
\end{equation}
 \begin{equation}\label{2.91}
\sin(2\gamma)=\frac{2\bar\varrho\bar\eta}{\bar\varrho^2+\bar\eta^2}=
\frac{2\varrho\eta}{\varrho^2+\eta^2}.
\end{equation}
\item
The lengths $CA$ and $BA$ in the
rescaled triangle  to be denoted by $R_b$ and $R_t$,
respectively, are given by
\begin{equation}\label{2.94}
R_b \equiv \frac{| V_{ud}^{}V^*_{ub}|}{| V_{cd}^{}V^*_{cb}|}
= \sqrt{\bar\varrho^2 +\bar\eta^2}
= (1-\frac{\lambda^2}{2})\frac{1}{\lambda}
\left| \frac{V_{ub}}{V_{cb}} \right|,
\end{equation}
\begin{equation}\label{2.95}
R_t \equiv \frac{| V_{td}^{}V^*_{tb}|}{| V_{cd}^{}V^*_{cb}|} =
 \sqrt{(1-\bar\varrho)^2 +\bar\eta^2}
=\frac{1}{\lambda} \left| \frac{V_{td}}{V_{cb}} \right|.
\end{equation}
\item
The angles $\beta$ and $\gamma$ of the unitarity triangle are related
directly to the complex phases of the CKM-elements $V_{td}$ and
$V_{ub}$, respectively, through
\beq\label{e417}
V_{td}=|V_{td}|e^{-i\beta},\quad V_{ub}=|V_{ub}|e^{-i\gamma}.
\eeq
\item
The angle $\alpha$ can be obtained through the relation
\beq\label{e419}
\alpha+\beta+\gamma=180^\circ
\eeq
expressing the unitarity of the CKM-matrix.
\ei

The triangle depicted in fig. \ref{fig:utriangle}, $|V_{us}|$ 
and $\vcb$ give the full description of the CKM matrix. 
Looking at the expressions for $R_b$ and $R_t$, we observe that within
the SM the measurements of four CP
{\it conserving } decays sensitive to $\mid V_{us}\mid$, $\mid V_{ub}\mid$,   
$\mid V_{cb}\mid $ and $\mid V_{td}\mid$ can tell us whether CP violation
($\bar\eta \not= 0$) is predicted in the SM. 
This fact is often used to determine
the angles of the unitarity triangle without the study of CP violating
quantities. 

Indeed, measuring the ratio $\vub$ in tree-level B decays and $\vtd$
through $B^0_d-\bar B^0_d$ mixing allows to determine $R_b$ and $R_t$
respectively. If so determined $R_b$ and $R_t$ satisfy
\be\label{con}
          1-R_b < R_t <1+R_b
\ee
then $\bar\eta$ is predicted to be non-zero on the basis of CP conserving
transitions in the B-system alone without any reference  
 to CP violation discovered in $K_L\to\pi^+\pi^-$
in 1964 \cite{CRONIN}. Moreover one finds
\be\label{eta}
\bar\eta=\pm\sqrt{R^2_b-\bar\varrho^2}~, \qquad
\bar\varrho=\frac{1+R^2_b-R^2_t}{2}.
\ee  

\subsection{Grand Picture}
\begin{figure}[hbt]
  \vspace{-0.10in} \centerline{
\begin{turn}{-90}
  \mbox{\epsfig{file=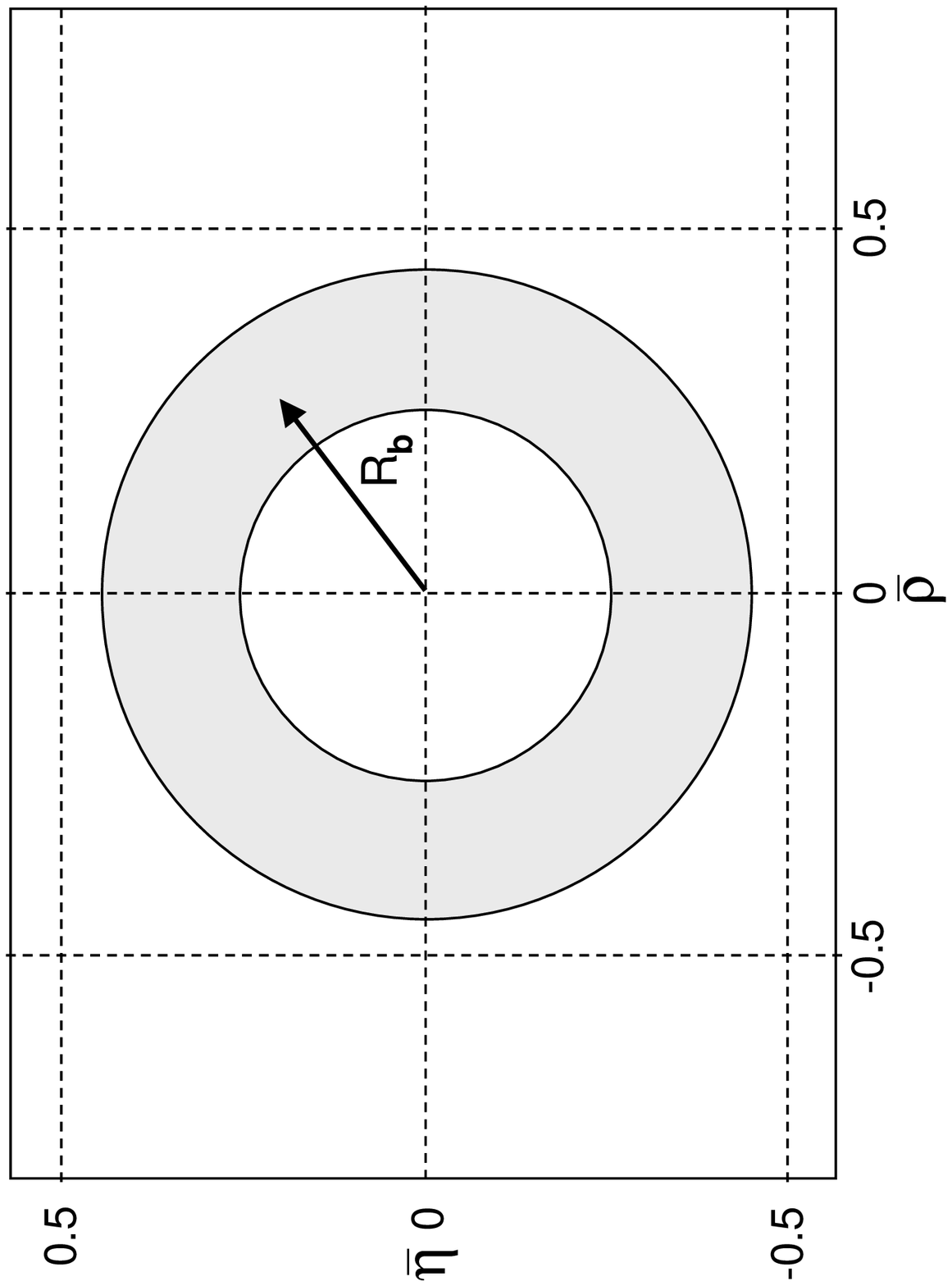,width=0.5\linewidth}}
\end{turn}
} \vspace{-0.08in}
\caption[]{``Unitarity Clock".
\label{L:2}}
 \end{figure}

What do we know about the CKM matrix and the unitarity triangle on the
basis of {\it tree level} decays? A detailed answer to this question
can be found in the reports of
the Particle Data Group \cite{PDG} as well as other reviews 
to be mentioned in Section 4,
where references to the relevant experiments and related theoretical
work can be found. Using the information given there we find in particular 

\begin{equation}\label{vcb}
|V_{us}| = \lambda =  0.2205 \pm 0.0018\,
\quad\quad
\vcb=0.041\pm0.002,
\end{equation}
\begin{equation}\label{v13}
\frac{|V_{ub}|}{\vcb}=0.085\pm0.018, \quad\quad
|V_{ub}|=(3.49\pm0.76)\cdot 10^{-3}.
\end{equation}
Using (\ref{CKM1}) and (\ref{2.94}) we find then ($\lambda=0.22$)
\be
 A=0.847\pm0.041,\qquad R_b=0.38\pm 0.08~.
\ee
This tells us only that the apex $A$ of the unitarity triangle lies
in the band shown in fig. \ \ref{L:2}. In order to answer the question where
the apex $A$ lies on this ``unitarity clock'' we have to look at different
decays. Most promising in this respect are the so-called ``loop induced''
decays and transitions and CP-violating B decays which will be discussed 
in these lectures.

These two different routes for explorations of the CKM matrix
and of the related unitarity triangle may answer the important question, 
whether
the Kobayashi-Maskawa picture
of CP violation is correct and more generally whether the Standard
Model offers a correct description of weak decays of hadrons. Indeed,
in order
to answer these important questions it is essential to calculate as
many branching ratios as possible, measure them experimentally and
check if they all can be described by the same set of the parameters
$(\lambda,A,\varrho,\eta)$.
In the language of the unitarity triangle
the question is whether the various curves in the $(\bar\varrho,\bar\eta)$ 
plane extracted from different decays and transitions will cross each other 
at a single point as shown in fig.~3 and whether
the angles $(\alpha,\beta,\gamma)$ in the
resulting triangle will agree with those extracted from
CP-asymmetries in B decays and CP-conserving B decays. 
It is truely exciting that during the present decade we should be
able to  answer  all these questions and in the case of the
inconsistencies in the $(\bar\varrho,\bar\eta)$ plane get some hints
about the physics beyond the SM. One obvious inconsistency
would be the violation of the constraint (\ref{con}).

\begin{figure}[hbt]
  \vspace{0.10in} \centerline{
\begin{turn}{-90}
  \mbox{\epsfig{file=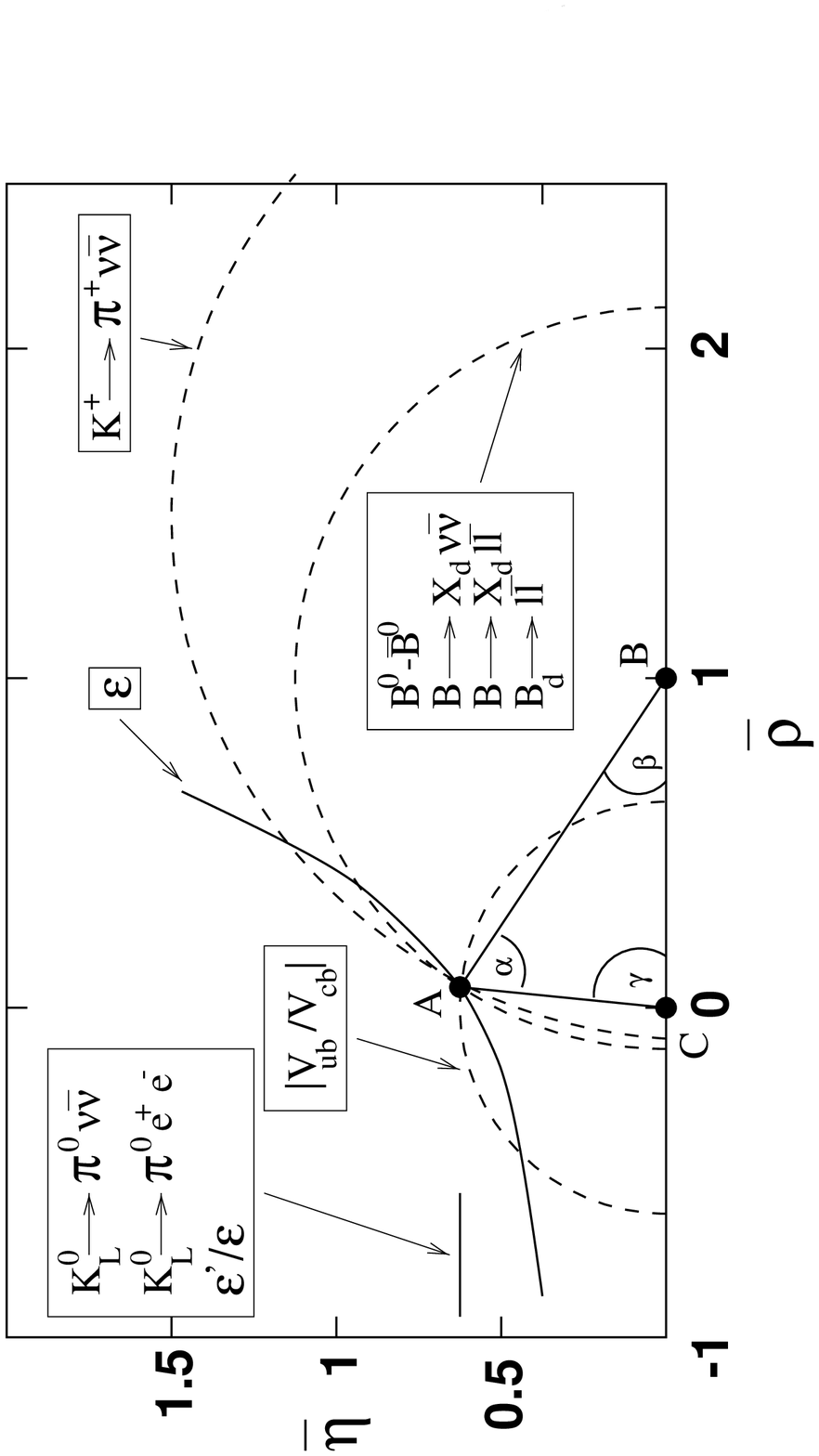,width=0.4\linewidth}}
\end{turn}
} \vspace{0.08in}
\caption[]{
  The ideal Unitarity Triangle.
\label{fig:2011}}
\end{figure}

Clearly the plot in fig.~\ref{fig:2011} is highly idealized because in order
to extract such nice curves from various decays one needs perfect
experiments and perfect theory. 
One of the goals of these lectures is 
to identify those decays
for which at least the theory is under control. For such decays,
if they can be measured with a sufficient precision, the curves
in fig.~\ref{fig:2011} are  not fully unrealistic.
Let us then briefly discuss the theoretical framework for weak
decays.

\section{Theoretical Framework}
\setcounter{equation}{0}
\subsection{OPE and Renormalization Group}
The basis for any serious phenomenology of weak decays of
hadrons is the {\it Operator Product Expansion} (OPE) \cite{OPE,ZIMM},
which allows to write
the effective weak Hamiltonian simply as follows
\be\label{b1}
{\cal H}_{eff}=\frac{G_F}{\sqrt{2}}\sum_i V^i_{\rm CKM}C_i(\mu)Q_i~.
\ee
Here $G_F$ is the Fermi constant and $Q_i$ are the relevant local
operators which govern the decays in question. 
They are built out of quark and lepton fields.
The Cabibbo-Kobayashi-Maskawa
factors $V^i_{\rm CKM}$ \cite{CAB,KM} 
and the Wilson coefficients $C_i(\mu)$ \cite{OPE} describe the 
strength with which a given operator enters the Hamiltonian.
The latter coefficients can be considered as scale dependent
``couplings'' related to ``vertices'' $Q_i$ and as discussed below
can be calculated using perturbative methods as long as $\mu$ is
not too small.

An amplitude for a decay of a given meson 
$M= K, B,..$ into a final state $F=\pi\nu\bar\nu,~\pi\pi,~DK$ is then
simply given by
\be\label{amp5}
A(M\to F)=\langle F|{\cal H}_{eff}|M\rangle
=\frac{G_F}{\sqrt{2}}\sum_i V^i_{CKM}C_i(\mu)\langle F|Q_i(\mu)|M\rangle,
\ee
where $\langle F|Q_i(\mu)|M\rangle$ 
are the matrix elements of $Q_i$ between M and F, evaluated at the
renormalization scale $\mu$. 

The essential virtue of OPE is this one. It allows to separate the problem
of calculating the amplitude
$A(M\to F)$ into two distinct parts: the {\it short distance}
(perturbative) calculation of the coefficients $C_i(\mu)$ and 
the {\it long-distance} (generally non-perturbative) calculation of 
the matrix elements $\langle Q_i(\mu)\rangle$. The scale $\mu$
separates, roughly speaking, the physics contributions into short
distance contributions contained in $C_i(\mu)$ and the long distance 
contributions
contained in $\langle Q_i(\mu)\rangle$. 
Thus $C_i$ include the top quark contributions and
contributions from other heavy particles such as W-, Z-bosons and charged
Higgs particles or supersymmetric particles in the supersymmetric extensions
of the SM. 
Consequently $C_i(\mu)$ depend generally 
on $m_t$ and also on the masses of new particles if extensions of the 
SM are considered. This dependence can be found by evaluating 
so-called {\it box} and {\it penguin} diagrams with full W-, Z-, top- and 
new particles exchanges and {\it properly} including short distance QCD 
effects. The latter govern the $\mu$-dependence of $C_i(\mu)$.

The value of $\mu$ can be chosen arbitrarily but the final result
must be $\mu$-independent.
Therefore 
the $\mu$-dependence of $C_i(\mu)$ has to cancel the 
$\mu$-dependence of $\langle Q_i(\mu)\rangle$. In other words it is a
matter of choice what exactly belongs to $C_i(\mu)$ and what to 
$\langle Q_i(\mu)\rangle$. This cancellation
of the $\mu$-dependence involves generally several terms in the expansion 
in (\ref{amp5}).
The  coefficients $C_i(\mu)$ 
depend also
on the renormalization scheme.
This scheme dependence must also be canceled
by the one of $\langle Q_i(\mu)\rangle$ so that the physical amplitudes are 
renormalization scheme independent. Again, as in the case of the 
$\mu$-dependence, the cancellation of
the renormalization scheme dependence involves generally several 
terms in the expansion (\ref{amp5}).

Although $\mu$ is in principle arbitrary,
it is customary 
to choose
$\mu$ to be of the order of the mass of the decaying hadron. 
This is $\ord (\mb)$ and $\ord(\mc)$ for B decays and
D decays respectively. In the case of K decays the typical choice is
 $\mu=\ord(1-2~\gev)$
instead of $\ord(m_K)$, which is much too low for any perturbative 
calculation of the couplings $C_i$.
Now due to the fact that $\mu\ll  M_{W,Z},~ m_t$, large logarithms 
$\ln\mw/\mu$ compensate in the evaluation of
$C_i(\mu)$ the smallness of the QCD coupling constant $\alpha_s$ and 
terms $\alpha^n_s (\ln\mw/\mu)^n$, $\alpha^n_s (\ln\mw/\mu)^{n-1}$ 
etc. have to be resummed to all orders in $\alpha_s$ before a reliable 
result for $C_i$ can be obtained.
This can be done very efficiently by means of the renormalization group
methods. 
The resulting {\it renormalization group improved} perturbative
expansion for $C_i(\mu)$ in terms of the effective coupling constant 
$\alpha_s(\mu)$ does not involve large logarithms and is more reliable.
The related technical issues are discussed in detail in \cite{AJBLH}
and \cite{BBL}.

All this looks rather formal but in fact should be familiar.
Indeed,
in the simplest case of the $\beta$-decay, ${\cal H}_{eff}$ takes 
the familiar form
\be\label{beta}
{\cal H}^{(\beta)}_{eff}=\frac{G_F}{\sqrt{2}}
\cos\theta_c[\bar u\gamma_\mu(1-\gamma_5)d \otimes
\bar e \gamma^\mu (1-\gamma_5)\nu_e]~,
\ee
where $V_{ud}$ has been expressed in terms of the Cabibbo angle. In this
particular case the Wilson coefficient is equal unity and the local
operator, the object between the square brackets, is given by a product 
of two $V-A$ currents. 
Equation (\ref{beta}) represents the Fermi theory for $\beta$-decays 
as formulated by Sudarshan and
Marshak \cite{SUMA} and Feynman and Gell-Mann \cite{GF} more than 
forty years ago, 
except that in (\ref{beta})
the quark language has been used and following Cabibbo a small departure of
$V_{ud}$ from unity has been incorporated. In this context the basic 
formula (\ref{b1})
can be regarded as a generalization of the Fermi Theory to include all known
quarks and leptons as well as their strong and electroweak interactions as
summarized by the SM. 

Due to the interplay of electroweak 
and strong interactions the structure of the local operators is 
much richer than in the case of the $\beta$-decay. They can be classified 
with respect to the Dirac structure, colour structure and the type of quarks 
and leptons relevant for a given decay. Of particular interest are the 
operators involving quarks only. In the case of the $\Delta S=1$
transitions the relevant set of operators is given as follows:
 
{\bf Current--Current :}
\begin{equation}\label{OS1} 
Q_1 = (\bar s_{\alpha} u_{\beta})_{V-A}\;(\bar u_{\beta} d_{\alpha})_{V-A}
~~~~~~Q_2 = (\bar s u)_{V-A}\;(\bar u d)_{V-A} 
\end{equation}

{\bf QCD--Penguins :}
\begin{equation}\label{OS2}
Q_3 = (\bar s d)_{V-A}\sum_{q=u,d,s}(\bar qq)_{V-A}~~~~~~   
 Q_4 = (\bar s_{\alpha} d_{\beta})_{V-A}\sum_{q=u,d,s}(\bar q_{\beta} 
       q_{\alpha})_{V-A} 
\end{equation}
\begin{equation}\label{OS3}
 Q_5 = (\bar s d)_{V-A} \sum_{q=u,d,s}(\bar qq)_{V+A}~~~~~  
 Q_6 = (\bar s_{\alpha} d_{\beta})_{V-A}\sum_{q=u,d,s}
       (\bar q_{\beta} q_{\alpha})_{V+A} 
\end{equation}

{\bf Electroweak--Penguins :}
\begin{equation}\label{OS4} 
Q_7 = {3\over 2}\;(\bar s d)_{V-A}\sum_{q=u,d,s}e_q\;(\bar qq)_{V+A} 
~~~~~ Q_8 = {3\over2}\;(\bar s_{\alpha} d_{\beta})_{V-A}\sum_{q=u,d,s}e_q
        (\bar q_{\beta} q_{\alpha})_{V+A}
\end{equation}
\begin{equation}\label{OS5} 
 Q_9 = {3\over 2}\;(\bar s d)_{V-A}\sum_{q=u,d,s}e_q(\bar q q)_{V-A}
~~~~~Q_{10} ={3\over 2}\;
(\bar s_{\alpha} d_{\beta})_{V-A}\sum_{q=u,d,s}e_q\;
       (\bar q_{\beta}q_{\alpha})_{V-A} \,.
\end{equation}
Here, $\alpha,\beta$ denote colours and $e_q$ denotes the electric quark charges reflecting the
electroweak origin of $Q_7,\ldots,Q_{10}$. Finally,
$(\bar s u)_{V-A}\equiv \bar s_\alpha\gamma_\mu(1-\gamma_5) u_\alpha$. 

Clearly, in order to calculate the amplitude $A(M\to F)$ the matrix 
elements $\langle Q_i(\mu)\rangle$ have to be evaluated. 
Since they involve long distance contributions one is forced in
this case to use non-perturbative methods such as lattice calculations, the
1/N expansion (N is the number of colours), QCD sum rules, hadronic sum rules,
chiral perturbation theory and so on. In the case of certain B-meson decays,
the {\it Heavy Quark Effective Theory} (HQET) also turns out to be a 
useful tool.
Needless to say, all these non-perturbative methods have some limitations.
Consequently the dominant theoretical uncertainties in the decay amplitudes
reside in the matrix elements $\langle Q_i(\mu)\rangle$.

The fact that in many cases the matrix elements $\langle Q_i(\mu)\rangle$
 cannot be reliably
calculated at present, is very unfortunate. One of the main goals of the
experimental studies of weak decays is the determination of the CKM factors 
$V_{\rm CKM}$
and the search for the physics beyond the SM. Without a reliable
estimate of $\langle Q_i(\mu)\rangle$ this goal cannot be achieved unless 
these matrix elements can be determined experimentally or removed from the 
final measurable quantities
by taking suitable ratios and combinations of decay amplitudes or branching
ratios. Flavour symmetries like $SU(2)_{\rm F}$ and 
$SU(3)_{\rm F}$ relating various
matrix elements can be useful in this respect, provided flavour
breaking effects can be reliably calculated. 
However, this can be achieved rarely  and often
 one has to face directly the calculation of 
$\langle Q_i(\mu)\rangle$. We will discuss these problems later on.

One of the outstanding issues in the calculation of 
$\langle Q_i(\mu)\rangle$ is the compatibility (``matching'') of 
$\langle Q_i(\mu)\rangle$
with $C_i(\mu)$.  $\langle Q_i(\mu)\rangle$ have to carry
the correct $\mu$ and renormalization scheme dependence 
in order to ensure the $\mu$
and scheme independence of physical amplitudes. Most of the non-perturbative
methods struggle still with this problem. 
Moreover, it has been emphasised recently in \cite{Donoghue}
that the presence of higher dimensional operators can in the case
of low matching scales complicate further this issue.
It appears to me that in the
future lattice methods have the best chance to get the matching in
question under control. On the other hand, analytic solutions would
certainly be preferable.

\subsection{Wilson Coefficients at NLO}
In order to achieve
sufficient precision for the theoretical predictions it is desirable to have
accurate values of $C_i(\mu)$ . Indeed it has been realized at the end of
the 1980's
that the leading term (LO) in the renormalization group improved perturbation
theory, in which the terms $\alpha^n_s (\ln\mw/\mu)^n$ are summed, is 
generally insufficient and the
inclusion of next-to-leading corrections  (NLO) corresponding to summing
the terms $\alpha^n_s (\ln\mw/\mu)^{n-1}$ is necessary. 
In particular,
the proper matching of $C_i(\mu)$ and
$\langle Q_i(\mu)\rangle$ discussed above can only be done
meaningfully after NLO corrections have been taken into account.
One finds then that unphysical  $\mu$- and renormalization scheme dependences
in the decay amplitudes and branching ratios resulting from the truncation of
the perturbative series are considerably reduced by including NLO
corrections.
It is then instructive to discuss briefly the general formulae for
$C_i(\mu)$ at the NLO level. Detailed exposition can be found in
\cite{BBL} and \cite{AJBLH}.

The general expression for $ C_i(\mu) $ is given by:
\begin{equation}\label{CV}
 \vec C(\mu) = \hat U(\mu,M_W) \vec C(M_W)   
\end{equation}
where $ \vec C $ is a column vector built out of $ C_i $'s.
$\vec C(M_W)$ are the initial conditions which depend on the 
short distance physics at high energy scales.
In particular they depend on $\mt$. We set the high energy scale
at $\mw$, but other choices are clearly possible.
$ \hat U(\mu,M_W) $, the evolution matrix,
is given as follows:
\begin{equation}\label{UM}
 \hat U(\mu,M_W) = T_g \exp \left[ 
   \int_{g(M_W)}^{g(\mu)}{dg' \frac{\hat\gamma^T(g')}{\beta(g')}}\right] 
\end{equation}
with $g$ denoting the QCD effective coupling constant and $T_g$ an
ordering operation defined in \cite{AJBLH}. $ \beta(g) $
governs the evolution of $g$ and $ \hat\gamma $ is the anomalous dimension
matrix of the operators involved. The structure of this equation
makes it clear that the renormalization group approach goes
 beyond the usual perturbation theory.
Indeed $ \hat  U(\mu,M_W) $ sums automatically large logarithms
$ \log M_W/\mu $ which appear for $ \mu\ll M_W $. In the so-called leading
logarithmic approximation (LO) terms $ (g^2\log M_W/\mu)^n $ are summed.
The next-to-leading logarithmic correction (NLO) to this result involves
summation of terms $ (g^2)^n (\log M_W/\mu)^{n-1} $ and so on.
This hierarchic structure gives the renormalization group improved
perturbation theory.

As an example let us consider only QCD effects and the case of a single
operator so that (\ref{CV}) reduces to
\begin{equation}\label{CC}
  C(\mu) =  U(\mu,M_W)  C(M_W)   
\end{equation}
with $C(\mu)$ denoting the coefficient of the operator in question.
 Keeping the first two terms in the expansions of
 $\gamma(g)$ and $\beta(g)$ in powers of $g$:
\begin{equation}
\gamma (g) = \gamma^{(0)} \frac{\alpha_s}{4\pi} + \gamma^{(1)}
\left(\frac{\alpha_s}{4\pi}\right)^2, 
\quad\quad
 \beta (g) = - \beta_0 \frac{g^3}{16\pi^2} - \beta_1 
\frac{g^5}{(16\pi^2)^2}
\end{equation}
and inserting these expansions into (\ref{UM}) gives:
\begin{equation}\label{UMNLO}
 U (\mu, M_W) = \Biggl\lbrack 1 + {{\alpha_s (\mu)}\over{4\pi}} J
\Biggl\rbrack \Biggl\lbrack {{\alpha_s (M_W)}\over{\alpha_s (\mu)}}
\Biggl\rbrack^P \Biggl\lbrack 1 - {{\alpha_s (M_W)}\over{4\pi}} J
\Biggl\rbrack 
\end{equation}
where
\begin{equation}
P = {{\gamma^{(0)}}\over{2\beta_0}},~~~~~~~~~ J = {{P}\over{\beta_0}}
\beta_1 - {{\gamma^{(1)}}\over{2\beta_0}}. 
\end{equation} 
General
formulae for $ \hat U (\mu, M_W) $ in the case of operator mixing and
valid also for electroweak effects can be found in \cite{BBL}. 
The leading
logarithmic approximation corresponds to setting $ J = 0 $ in (\ref{UMNLO}).

At NLO, $C(\mw)$ is given by
\be\label{init}
C(\mw)=C_0+\frac{\as(\mw)}{4\pi} C_1
\ee
where $C_0$ and $C_1$ depend generally on $\mt$, $\mw$ and the masses
of the new particles in the extentions of the SM. It should be 
stressed that the renormalization scheme dependence of $C_1$ is canceled
by the one of $J$ in the last square bracket in (\ref{UMNLO}). The
scheme dependence of J in the first square bracket in (\ref{UMNLO})
is canceled by the scheme dependence of $\langle Q(\mu)\rangle$. The
power $P$ is scheme independent.
The methods for the calculation of $ \hat U (\mu, M_W) $ and the
discussion of the cancellation of the $\mu$- and scheme dependence 
are presented in detail in \cite{AJBLH}.

As an example consider the case of the operator 
$(\bar b d)_{V-A}(\bar b d)_{V-A}$ relevant for $B^0_d-\bar B^0_d$
mixing. In this case using the so-called NDR renormalization scheme
one has \cite{WEISZ,BJW90}
\begin{equation}\label{UMNLOb}
 U (\mb, \mw) = \Biggl\lbrack 1 + 1.63 {{\alpha_s (\mb)}\over{4\pi}} 
\Biggl\rbrack \Biggl\lbrack {{\alpha_s (\mw)}\over{\alpha_s (\mb)}}
\Biggl\rbrack^{6/25} \Biggl\lbrack 1 - 1.63{{\alpha_s (\mw)}\over{4\pi}} 
\Biggl\rbrack=0.86~.
\end{equation}
where we have used $f=5$, $\as(\mz)=0.118$ and $\as(\mb)=0.222$. 
The departure of
$ U (\mb, \mw)$ from unity is rather small in this example, due
to the small value of the power $P$. In the case of
$(V-A)\otimes(V+A)$ operators the renormalization group effects
are larger.
\subsection{Status of the NLO Calculations}
\subsubsection{General Comments}
During the last decade the NLO corrections to $C_i(\mu)$ have
been calculated within the SM
for the most important and
interesting decays. They will be taken into account in these lectures. 
In table~\ref{TAB1}  we give
references to all NLO calculations within the SM done until the end of 2000.
While these calculations improved considerably the precision of
theoretical predictions in weak decays and can be considered as an
important progress in this field, the pioneering LO calculations
for current-current operators \cite{LOCC}, penguin operators \cite{LOP}
and $\Delta S=2$ operators \cite{LODF2} should not be forgotten.
\subsubsection{NNLO Calculations}
In the case of the CP violating ratio $\epe$ and the rare decays
$B \to X_s l^+l^-$ and $K_L\to\pi^0e^+e^-$, the NLO matching conditions
for electroweak operators  do not involve QCD corrections to box and
penguin diagrams and  consequently the renormalization scale 
dependence in the top quark mass in these processes
is not negligible. In order to reduce this unphysical dependence,
QCD corrections to the relevant box and penguin diagrams have
to be computed. In the renormalization group improved perturbation
theory these corrections are a part of next-next-to-leading (NNLO)
corrections. In \cite{BGH} and \cite{BMU574} such corrections
have been computed for $\epe$ and the rare decays in question,
respectively. 
\subsubsection{Two-Loop Anomalous Dimensions Beyond the SM}
In the extentions of the SM new operators are present. The
two loop anomalous dimensions for all $\Delta F=2$ four-quark
dimension-six operators have been computed in \cite{CET0,BMU}.
In \cite{BMU} also the corresponding results for $\Delta F=1$ can be found.
The applications of these results to
$(\Delta M_K,\varepsilon_K)$ and $\Delta B=1$ decays in the MSSM can
be found in \cite{CET} and \cite{China}, respectively.

\subsubsection{Two-Loop Electroweak Corrections}
In order to reduce scheme and scale dependences related to the
definition of electroweak parameters like $\sin^2\theta_W$,
and $\alpha_{QED}$,  two-loop electroweak contributions to rare decays
have to be computed. For $K^0_L \rightarrow \pi^0\nu\bar{\nu}$, 
$B \rightarrow l^+l^-$ and $B \rightarrow X_{\rm s}\nu\bar{\nu}$ they
can be found in \cite{BB97}, for $B^0-\bar B^0$ mixing in \cite{GKP}
and for $B\rightarrow X_s \gamma$ in \cite{CZMA,STRUMIA,KN98,GH00}.
\subsubsection{NLO Calculations Beyond the SM}
There exist also a number of partial or complete NLO QCD calculations 
within the Two-Higgs-Doublet Model and the MSSM. In the case
of the Two-Higgs-Doublet Model such calculations for $B^0-\bar B^0$
mixing and $B\to X_s\gamma$ can be found in \cite{UKJS} and
\cite{GAMB,BG98,strum} respectively. The corresponding calculations
for $B\to X_s\gamma$ in the MSSM can be found in \cite{GAMB2} and
\cite{BMU567}. The latter paper gives also the results for
$B\to X_s {\rm gluon}$.
Finally gluino-mediated NLO-QCD corrections to $B^0-\bar B^0$ mixing
in the MSSM have been considered in \cite{Soff}.

\begin{table}[thb]
\caption{References to NLO Calculations within the SM}
\label{TAB1}
\begin{center}
\begin{tabular}{|l|l|}
\hline
\bf \phantom{XXXXXXXX} Decay & \bf Reference \\
\hline
\hline
\multicolumn{2}{|c|}{$\Delta F=1$ Decays} \\
\hline
current-current operators     & \cite{ACMP,WEISZ} \\
QCD penguin operators         & \cite{BJLW1,BJLW,ROMA1,ROMA2,MIS1} \\
electroweak penguin operators & \cite{BJLW2,BJLW,ROMA1,ROMA2} \\
magnetic penguin operators    & \cite{MisMu:94,CZMM} \\
$Br(B)_{SL}$                  & \cite{ACMP,Buch:93,Bagan} \\
inclusive $\Delta S=1$ decays       & \cite{JP} \\
\hline
\multicolumn{2}{|c|}{Particle-Antiparticle Mixing} \\
\hline
$\eta_1$                   & \cite{HNa} \\
$\eta_2,~\eta_B$           & \cite{BJW90,UKJS} \\
$\eta_3$                   & \cite{HNb} \\
\hline
\multicolumn{2}{|c|}{Rare $K$- and $B$-Meson Decays} \\
\hline
$K^0_L \rightarrow \pi^0\nu\bar{\nu}$, $B \rightarrow l^+l^-$,
$B \rightarrow X_{\rm s}\nu\bar{\nu}$ & \cite{BB1,BB2,MU98,BB98} \\
$K^+   \rightarrow \pi^+\nu\bar{\nu}$, $K_{\rm L} \rightarrow \mu^+\mu^-$
                                      & \cite{BB3,BB98} \\
$K^+\to\pi^+\mu\bar\mu$               & \cite{BB5} \\
$K_{\rm L} \rightarrow \pi^0e^+e^-$         & \cite{BLMM} \\
$B\rightarrow X_s \mu^+\mu^-$           & \cite{Mis:94,BuMu:94} \\
$B\rightarrow X_s \gamma$      & 
\cite{AG2}-\cite{BG98} \\
$B\rightarrow X_s {\rm gluon}$      & 
\cite{GH97,BMU567,GL00} \\
$\Delta\Gamma_{B_s}$     &  \cite{BBGLN} \\
inclusive B $\to$ Charmonium & \cite{BMR} \\
$B\to D\pi$, $B\to\pi\pi$    & \cite{BBNS1}\\
\hline
\end{tabular}
\end{center}
\end{table}

\subsection{QCD Factorization for Exclusive Non-Leptonic B-Meson Decays}
A simple method for the evaluation of the hadronic matrix elements of
four quark operators relevant for B decays is the factorization approach 
in which the matrix
elements are expressed in terms of products of meson decay constants
and formfactors \cite{NeuStech}. 
In its naive formulation, this approach gives the
$\mu$-independent hadronic matrix elements and consequently 
$\mu$-dependent decay amplitudes. Moreover final state interactions are
not taken into account. Various generalizations of this method have
been proposed in the literature \cite{GFACT} with the hope to include 
non-factorizable contributions and to remove the $\mu$-dependence from 
the decay
amplitudes. Critical reviews of these attempts can be found in 
\cite{BSF,BBNS1}.
Parallel to these efforts general parametrizations of decay amplitudes
by means of flavour flow diagrams \cite{DIAG} and Wick contractions 
\cite{IWICK,BSWICK} supplemented
by dynamical assumptions have been proposed. These parametrizations
may turn out to be useful when more data will be available.

Recently factorization for a large class of non-leptonic two-body
B-meson decays has been shown by Beneke, Buchalla, Neubert and
 Sachrajda \cite{BBNS1} to follow from QCD in the
heavy-quark limit. The resulting factorization formula incorporates
elements of the naive factorization approach but allows to
compute systematically non-factorizable corrections. In this approach
the $\mu$-dependence of hadronic matrix elements is under control.
Moreover spectator quark effects are taken into account and final
state interaction phases can be computed perturbatively. While,
in my opinion, an important progress in evaluating non-leptonic 
amplitudes has been made in \cite{BBNS1}, the usefulness of this approach
at the quantitative level has still to be demonstrated when the
data improve. In particular the role of the $1/m_b$ corrections
has to be considerably better understood. Recent lectures on this
approach can be found in \cite{NEUTASI}.

There is an alternative perturbative QCD approach to non-leptonic
decays \cite{Li} which has been developed earlier from the QCD 
hard-scattering
approach. Some elements of this approach are present in the
QCD factorization formula of \cite{BBNS1}. The main difference between these
two approaches is the treatment of soft spectator contributions
which are assumed to be negligible in the perturbative QCD approach.
While the QCD factorization approach is  more general and systematic, 
the perturbative QCD approach is an interesting possibility.
Competition is always healthy and only
time will show which of these two frameworks is more successful and
whether they have to be replaced by still more powerful approches
in the future.

Finally a new method to calculate the $B\to\pi\pi$ hadronic matrix
elements from QCD light-cone sum rules has been proposed very recently
 by Khodjamirian \cite{KOD}. This work may shed light on the
importance of $1/\mb$ and soft-gluon effects in the QCD factorization
approach. Reviews of QCD light-cone sum rules can be found in
\cite{LCQCD}.

\subsection{Inclusive Decays}
So far we have discussed only  {\it exclusive} decays. It turns out that
in the case of {\it inclusive} decays of heavy mesons, like B-mesons,
things turn out to be easier. In an inclusive decay one sums over all 
or over a special class of accessible final states.
A well known example is the decay $B\to X_s\gamma$, where $X_s$ includes
all accessible final states with the net strange quantum number $S=1.$

At first sight things look as complicated as in the case of exclusive decays.
It turns out, however, that the resulting branching ratio can be calculated
in the expansion in inverse powers of $\mb$ with the leading term 
described by the spectator model
in which the B-meson decay is modelled by the decay of the $b$-quark:
\be\label{hqe}
{\rm Br}(B\to X)={\rm Br}(b\to q) +\ord(\frac{1}{\mb^2})~. 
\ee
This formula is known under the name of the Heavy Quark Expansion (HQE)
\cite{HQE1,HQE2}.
Since the leading term in this expansion represents the decay of the quark,
it can be calculated in perturbation theory or more correctly in the
renormalization group improved perturbation theory. It should be emphasized
that also here the basic starting point is the effective Hamiltonian 
 (\ref{b1})
and that the knowledge of $C_i(\mu)$ is essential for 
the evaluation of
the leading term in (\ref{hqe}). But there is an important difference 
relative to the
exclusive case: the matrix elements of the operators $Q_i$ can be 
``effectively"
evaluated in perturbation theory. 
This means, in particular, that their $\mu$ and renormalization scheme
dependences can be evaluated and the cancellation of these dependences by
those present in $C_i(\mu)$ can be explicitly investigated.

Clearly in order to complete the evaluation of $Br(B\to X)$ also the 
remaining terms in
(\ref{hqe}) have to be considered. These terms are of a non-perturbative 
origin, but
fortunately they are suppressed by at least two powers of $m_b$. 
They have been
studied by several authors in the literature \cite{HQE2} with the result 
that they affect
various branching ratios by less then $10\%$ and often by only a few percent.
Consequently the inclusive decays give generally more precise theoretical
predictions at present than the exclusive decays. On the other hand their
measurements are harder. There are of course some important theoretical
issues related to the validity of HQE in (\ref{hqe}) which appear in the 
literature under the name of quark-hadron duality. 
Since these matters are rather involved I will not discuss them
here.

\begin{figure}[hbt]
\centerline{
\epsfysize=4.5in
\epsffile{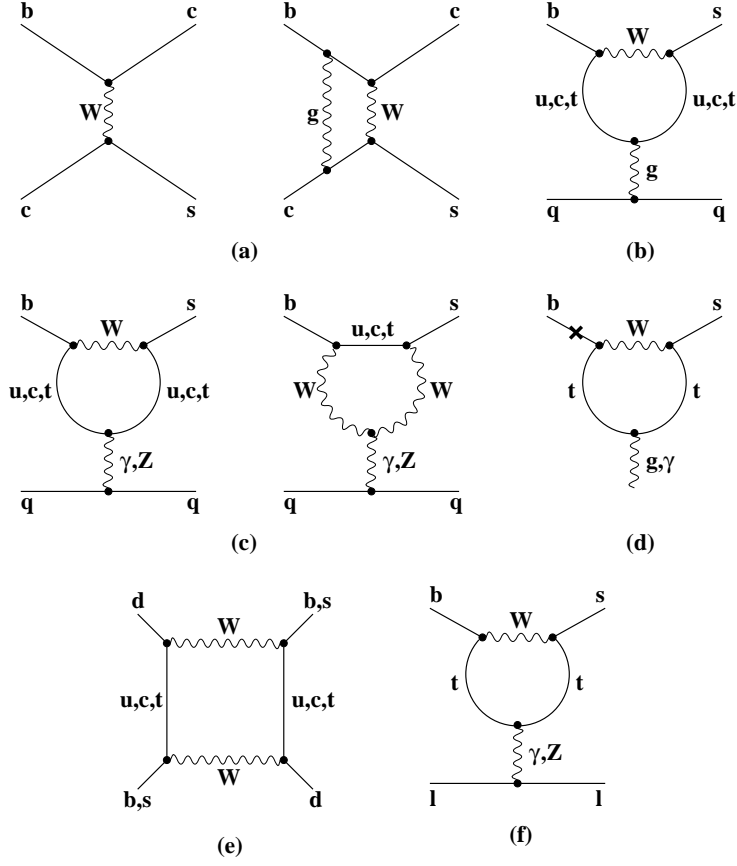}
}
\caption{Typical Penguin and Box Diagrams.}
\label{fig:fdia}
\end{figure}

\subsection{Penguin--Box Expansion}
The rare and CP violating decays of K and B mesons are governed 
by various penguin and box diagrams with internal top quark and charm quark
exchanges. Some examples are shown in fig.~\ref{fig:fdia}. 
Evaluating these diagrams one finds
a set of basic universal (process independent) 
$\mt$-dependent functions $F_r(x_t)$ \cite{IL} where $x_t=\mt^2/\mw^2$. 
Explicit expressions for these
functions will be given below. 

It is useful to express the OPE formula (\ref{amp5}) directly in terms
of the functions $F_r(x_t)$ \cite{PBE0}:
\begin{equation}
A({M\to F}) = P_0(M\to F) + \sum_r P_r(M\to F) \, F_r(x_t),
\label{generalPBE1}
\end{equation}
where the sum runs over all possible functions contributing to a given
amplitude. 
$P_0$  summarizes contributions stemming from internal quarks
other than the top, in particular the charm quark. 
In the OPE formula (\ref{amp5}), the functions $F_r(x_t)$ are
hidden in the initial conditions for $ C_i(\mu)$ represented
by $\vec C(\mw)$ in (\ref{CV}).

The coefficients $P_0$ and $P_r$ are process dependent and
include QCD corrections contained in the evolution matrix
$\hat U(\mu,\mw)$. They depend also on hadronic matrix
elements of local operators and the relevant CKM factors.
An efficient and straightforward method for finding the coefficients
$P_r$ is presented in \cite{PBE0}.
I would like to call (\ref{generalPBE1}) {\it Penguin-Box Expansion} 
(PBE).
We will encounter many
examples of PBE in the course of 
these lectures.

Originally  PBE was designed to expose the $\mt$-dependence
of FCNC processes \cite{PBE0}. After the top quark mass has been
measured precisely this role of PBE is less important.
On the other hand,
PBE is very well suited for the study of the extentions of the
SM in which new particles are exchanged in the loops.
If there are no new local operators
the mere change is to modify the functions $F_r(x_t)$ which now
acquire the dependence on the masses of new particles such as
charged Higgs particles and supersymmetric particles. The process
dependent coefficients $P_0$ and $P_r$ remain unchanged. 
The effects of new physics can be then transparently seen.
However, if
new effective operators with different Dirac and colour structures
are present, new functions multiplied by new coefficients
$P_r(M\to F)$ contribute to (\ref{generalPBE1}).

In the rest of this section we present the functions $F_r(x_t)$
within the SM.
To this end, let us denote by $B_0$, $C_0$ and $D_0$ the functions 
$F_r(x_t)$
resulting from $\Delta F=1$ ($F$ stands for flavour) box diagram,
$Z^0$-penguin and $\gamma$-penguin diagram respectively. These
diagrams are gauge dependent and it is useful to introduce
gauge independent combinations \cite{PBE0}
\be\label{XYZ}
X_0=C_0-4 B_0,\qquad Y_0=C_0-B_0,\qquad Z_0=C_0+\frac{1}{4}D_0~.
\ee
Then the set of gauge independent basic functions which govern
the FCNC processes in the SM is given to a very good
approximation as follows ($x_i=m^2_i/\mw^2$):
\begin{equation}\label{S0}
 S_0(x_t)=2.46~\left(\frac{\mt}{170\gev}\right)^{1.52},
\quad\quad S_0(x_c)=x_c,
\ee
\begin{equation}\label{BFF}
S_0(x_c, x_t)=x_c\left[\ln\frac{x_t}{x_c}-\frac{3x_t}{4(1-x_t)}-
 \frac{3 x^2_t\ln x_t}{4(1-x_t)^2}\right],
\end{equation}
\be\label{XA0}
X_0(x_t)=1.57~\left(\frac{\mt}{170\gev}\right)^{1.15},
\quad\quad
Y_0(x_t)=1.02~\left(\frac{\mt}{170\gev}\right)^{1.56},
\end{equation}
\begin{equation}
 Z_0(x_t)=0.71~\left(\frac{\mt}{170\gev}\right)^{1.86},\quad\quad
   E_0(x_t)= 0.26~\left(\frac{\mt}{170\gev}\right)^{-1.02},
\end{equation}
\begin{equation}
 D'_0(x_t)=0.38~\left(\frac{\mt}{170\gev}\right)^{0.60}, \quad\quad 
E'_0(x_t)=0.19~\left(\frac{\mt}{170\gev}\right)^{0.38}.
\end{equation}
The first three functions correspond to
$\Delta F=2$ box diagrams with $(t,t)$, $(c,c)$ and $(t,c)$ exchanges.
$E_0$ results from QCD penguin diagram with off-shell gluon, 
$D'_0$ and $E'_0$ from $\gamma$ and QCD penguins with on-shell
photons and gluons respectively. 
The subscript ``$0$'' indicates that 
these functions
do not include QCD corrections to the relevant penguin and box diagrams.

In the range $150\gev \le \mt \le 200\gev$ 
these approximations reproduce the
exact expressions to an accuracy better than 1\%. 
These formulae will allow us to exhibit elegantly the $\mt$ dependence
of various branching ratios in the phenomenological sections of
these lectures.
Exact expressions for all functions can be found in \cite{AJBLH}.
 
Generally, several basic functions contribute to a given decay,
although decays exist which depend only on a single function.
We have the following correspondence between the most interesting FCNC
processes and the basic functions:
\begin{center}
\begin{tabular}{lcl}
$K^0-\bar K^0$-mixing &\qquad\qquad& $S_0(x_t)$, $S_0(x_c,x_t)$, 
$S_0(x_c)$ \\
$B_{d,s}^0-\bar B_{d,s}^0$-mixing &\qquad\qquad& $S_0(x_t)$ \\
$K \to \pi \nu \bar\nu$, $B \to X_{d,s} \nu \bar\nu$ 
&\qquad\qquad& $X_0(x_t)$ \\
$K_{\rm L}\to \mu \bar\mu$, $B \to l\bar l$ &\qquad\qquad& $Y_0(x_t)$ \\
$K_{\rm L} \to \pi^0 e^+ e^-$ &\qquad\qquad& $Y_0(x_t)$, $Z_0(x_t)$, 
$E_0(x_t)$ \\
$\varepsilon'$ &\qquad\qquad& $X_0(x_t)$, $Y_0(x_t)$, $Z_0(x_t)$,
$E_0(x_t)$ \\
$B \to X_s \gamma$ &\qquad\qquad& $D'_0(x_t)$, $E'_0(x_t)$ \\
$B \to X_s \mu^+ \mu^-$ &\qquad\qquad&
$Y_0(x_t)$, $Z_0(x_t)$, $E_0(x_t)$, $D'_0(x_t)$, $E'_0(x_t)$
\end{tabular}
\end{center}

The supersymmetric contributions to the functions $S_0$, $X_0$,
$Y_0$, $Z_0$ and $E_0$ within the  MSSM with minimal
flavour violation (see Section 9) have been recently compiled
in \cite{EP00}. See also \cite{BERTOL}-\cite{AAA}. QCD corrections
to these functions can be extracted from papers in table~\ref{TAB1}
and from the section on NLO calculations beyond the SM.
In the SM it is convenient in most cases to include these corrections
in the coefficients $P_r$. Beyond the SM it is better to retain them
in $F_r$ as these corrections depend on the new parameters present
in the extentions of the SM.
 
\section{Particle-Antiparticle Mixing and Various Types\\ of CP
Violation}
        \label{sec:epsBBUT}
\setcounter{equation}{0}
\subsection{Preliminaries}
Let us next discuss the formalism of particle--antiparticle mixing
and CP violation. Much more elaborate discussion can be found in
two recent books \cite{Branco,Bigi}. We will concentrate here on
$K^0-\bar K^0$ mixing, $B_{d,s}^0-\bar B^0_{d,s}$ mixings and
CP violation in K-meson and B-meson decays. As this section is
rather long it is useful to specify our goals. These are:
\begin{itemize}
\item
Presentation of basic concepts of particle--antiparticle mixing
and CP violation.
\item
Introduction of the CP violating parameters $\varepsilon$ and
$\varepsilon'$ that describe the so--called {\it indirect} a
{\it direct} CP violation in $K_L \to \pi\pi$, respectively.
\item
Presentation of a different and more useful classification of
different types of CP violation that distinguishes between:
CP violation in mixing, CP violation in decay and CP violation
in the interference between mixing and decay.
\item
Derivation of a number of formulae that will turn out to be useful
in subsequent more phenomenological sections.
\end{itemize}

It is important to emphasize at this moment that particle--antiparticle
 mixing and CP violation have been  of fundamental
importance for the construction and testing of the SM.
They have also proven often to be 
undefeatable challenges for suggested extensions of this model.
In this context the seminal papers of Glashow, Iliopoulos and Maiani
\cite{GIM} and of Kobayashi and Maskawa \cite{KM} should be mentioned.
From the calculation of the
$K_{\rm L}-K_{\rm S}$ mass difference, Gaillard and Lee \cite{GALE} 
were able to estimate the
value of the charm quark mass before charm discovery. On the
other hand $B_d^0-\bar B_d^0$ mixing \cite{ARGUS} gave the first 
indication of a large top quark mass. 
Next CP violation in the $K^0-\bar K^0$ mixing
offers within the SM a plausible description of
CP violation in $K_L\to\pi\pi$ discovered in 1964 \cite{CRONIN}. 
Finally the very small values of the measured
$K_{\rm L}-K_{\rm S}$ mass difference and of the CP violating parameter
$\varepsilon$ put severe restrictions on the flavour structure and
the pattern of complex phases in the extentions of the SM.
 This is in particular the case of general supersymmetric
extentions of the SM, in which the mismatch in the alignment of
the quark mass matrices and the squark mass matrices is very restricted
by the $K_{\rm L}-K_{\rm S}$ mass difference and $\varepsilon$.

It is important to stress that
in the SM the phenomena discussed in this section
appear first at the one--loop level and as such they are
sensitive measures of the top quark couplings $V_{ti}(i=d,s,b)$ and 
in particular of the phase $\delta=\gamma$.
They allow then to construct the unitarity triangle as explicitly
demonstrated in Section 4.

Let us next enter some details. The following subsection borrows a lot 
from \cite{CHAU83,BSSII}. The discussion of different types of 
CP violation benefited from several very nice lectures by Nir 
 \cite{NIRSLAC}, although the presentation of this topic below differs
occassionally from his.

\begin{figure}[hbt]
\vspace{0.10in}
\centerline{
\epsfysize=1.5in
%\rotate[r]{
\epsffile{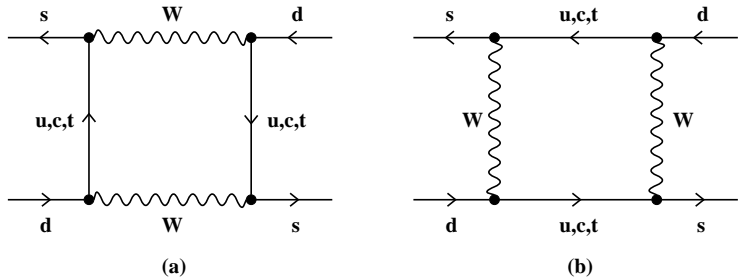}
}%}
\vspace{0.08in}
\caption[]{Box diagrams contributing to $K^0-\bar K^0$ mixing
in the SM.
\label{L:9}}
\end{figure}

\subsection{Express Review of $K^0-\bar K^0$ Mixing}
$K^0=(\bar s d)$ and $\bar K^0=(s\bar d)$ are flavour eigenstates which 
in the SM
may mix via weak interactions through the box diagrams in fig.
\ref{L:9}.
We will choose the phase conventions so that 
\be
CP|K^0\rangle=-|\bar K^0\rangle, \qquad   CP|\bar K^0\rangle=-|K^0\rangle.
\ee

In the absence of mixing the time evolution of $|K^0(t)\rangle$ is
given by
\be
|K^0(t)\rangle=|K^0(0)\rangle \exp(-iHt)~, 
\qquad H=M-i\frac{\Gamma}{2}~,
\ee
where $M$ is the mass and $\Gamma$ the width of $K^0$. Similar formula
exists for $\bar K^0$.

On the other hand, in the presence of flavour mixing the time evolution 
of the $K^0-\bar K^0$ system is described by
\be\label{SCH}
i\frac{d\psi(t)}{dt}=\hat H \psi(t) \qquad  
\psi(t)=
\left(\begin{array}{c}
|K^0(t)\rangle\\
|\bar K^0(t)\rangle
\end{array}\right)
\ee
where
\be
\hat H=\hat M-i\frac{\hat\Gamma}{2}
= \left(\begin{array}{cc} 
M_{11}-i\frac{\Gamma_{11}}{2} & M_{12}-i\frac{\Gamma_{12}}{2} \\
M_{21}-i\frac{\Gamma_{21}}{2}  & M_{22}-i\frac{\Gamma_{22}}{2}
    \end{array}\right)
\ee
with $\hat M$ and $\hat\Gamma$ being hermitian matrices having positive
(real) eigenvalues in analogy with $M$ and $\Gamma$. $M_{ij}$ and
$\Gamma_{ij}$ are the transition matrix elements from virtual and physical
intermediate states respectively.
Using
\be
M_{21}=M^*_{12}~, \qquad 
\Gamma_{21}=\Gamma_{12}^*~,\quad\quad {\rm (hermiticity)}
\ee
\be
M_{11}=M_{22}\equiv M~, \qquad \Gamma_{11}=\Gamma_{22}\equiv\Gamma~,
\quad {\rm (CPT)}
\ee
we have
\be\label{MM12}
\hat H=
 \left(\begin{array}{cc} 
M-i\frac{\Gamma}{2} & M_{12}-i\frac{\Gamma_{12}}{2} \\
M^*_{12}-i\frac{\Gamma^*_{12}}{2}  & M-i\frac{\Gamma}{2}
    \end{array}\right)~.
\ee

Diagonalizing (\ref{SCH}) we find:

{\bf Eigenstates:}
\be\label{KLS}
K_{L,S}=\frac{(1+\bar\varepsilon)K^0\pm (1-\bar\varepsilon)\bar K^0}
        {\sqrt{2(1+\mid\bar\varepsilon\mid^2)}}
\ee
where $\bar\varepsilon$ is a small complex parameter given by
\be\label{bare3}
\frac{1-\bar\varepsilon}{1+\bar\varepsilon}=
\sqrt{\frac{M^*_{12}-i\frac{1}{2}\Gamma^*_{12}}
{M_{12}-i\frac{1}{2}\Gamma_{12}}}=
\frac{\Delta M-i\frac{1}{2}\Delta\Gamma}
{2 M_{12}-i\Gamma_{12}}
=\frac{2 M^*_{12}-i\Gamma^*_{12}}{\Delta M-i\frac{1}{2}\Delta\Gamma}
\equiv r\exp(i\kappa)~.
\ee
with $\Delta\Gamma$ and $\Delta M$ given below.

{\bf Eigenvalues:}
\be
M_{L,S}=M\pm \RE Q  \qquad \Gamma_{L,S}=\Gamma\mp 2 \IM Q
\ee
where
\be
Q=\sqrt{(M_{12}-i\frac{1}{2}\Gamma_{12})(M^*_{12}-i\frac{1}{2}\Gamma^*_{12})}.
\ee
Consequently we have
\be\label{deltak}
\Delta M= M_L-M_S = 2\RE Q~,
\quad\quad
\Delta\Gamma=\Gamma_L-\Gamma_S=-4 \IM Q.
\ee

It should be noted that the mass eigenstates $K_S$ and $K_L$ differ from 
CP eigenstates
\begin{equation}
K_1={1\over{\sqrt 2}}(K^0-\bar K^0),
  \qquad\qquad CP|K_1\rangle=|K_1\rangle~,
\end{equation}
\begin{equation}
K_2={1\over{\sqrt 2}}(K^0+\bar K^0),
  \qquad\qquad CP|K_2\rangle=-|K_2\rangle~,
\end{equation}
by 
a small admixture of the
other CP eigenstate:
\begin{equation}
K_{\rm S}={{K_1+\bar\varepsilon K_2}
\over{\sqrt{1+\mid\bar\varepsilon\mid^2}}},
\qquad
K_{\rm L}={{K_2+\bar\varepsilon K_1}
\over{\sqrt{1+\mid\bar\varepsilon\mid^2}}}\,.
\end{equation}

Since $\bar\varepsilon$ is $\ord(10^{-3})$, one has
 to a very good approximation:
\be\label{deltak1}
\Delta M_K = 2 \RE M_{12}, \qquad \Delta\Gamma_K=2 \RE \Gamma_{12}~,
\ee
where we have introduced the subscript K to stress that these formulae apply
only to the $K^0-\bar K^0$ system.

The 
$K_{\rm L}-K_{\rm S}$
mass difference is experimentally measured to be \cite{PDG}
\begin{equation}\label{DMEXP}
\Delta M_K=M(K_{\rm L})-M(K_{\rm S}) = 
(3.489\pm 0.008) \cdot 10^{-15} \gev\,.
\end{equation}
In the SM roughly $70\%$ of the measured $\Delta M_K$
is described by the real parts of the box diagrams with charm quark
and top quark exchanges, whereby the contribution of the charm exchanges
is by far dominant. This is related to the smallness of the real parts
of the CKM top quark couplings compared with the corresponding charm
quark couplings. 
Some non-negligible contribution comes from the box diagrams with
simultaneous charm and top exchanges.
The remaining $20 \%$ of the measured $\Delta M_K$ is attributed to long 
distance contributions which are difficult to estimate \cite{GERAR}.
Further information with the relevant references can be found in 
\cite{HNa}.

The situation with $\Delta \Gamma_K$ is rather different.
It is fully dominated by long distance effects. Experimentally
one has $\Delta\Gamma_K\approx-2 \Delta M_K$. We will use
this relation in what follows.

Generally to observe CP violation one needs an interference between
various amplitudes that carry complex phases. As these phases are
obviously convention dependent, the CP-violating effects depend only
on the differences of these phases. 
In this context it should be stressed that
the small parameter $\bar\varepsilon$  depends on the 
phase convention
chosen for $K^0$ and $\bar K^0$. Therefore it may not 
be taken as a physical measure of CP violation.
On the other hand $\RE~\bar\varepsilon$ and $r$, defined in
(\ref{bare3})  are independent of
phase conventions. In particular the departure of $r$ from 1
measures CP violation in the $K^0-\bar K^0$ mixing:
\be
r=1+\frac{2 |\Gamma_{12}|^2}{4 |M_{12}|^2+|\Gamma_{12}|^2}
    \IM\left(\frac{M_{12}}{\Gamma_{12}}\right)
\approx 1+\IM\left(\frac{M_{12}}{\Gamma_{12}}\right)~.
\ee

This type of CP violation can be best isolated in semi-leptonic
decays of the $ K_L$ meson. The non-vanishing
asymmetry
\be\label{ASLK}
a_{\rm SL}(K_L)=\frac{\Gamma(K_L\to \pi^-e^+\nu_e )-
                 \Gamma( K_L\to \pi^+e^-\bar\nu_e )}
{\Gamma(K_L\to \pi^-e^+\nu_e )+
                 \Gamma( K_L\to \pi^+e^-\bar\nu_e )}
          = \left(\IM\frac{\Gamma_{12}}{M_{12}}\right)_K
\ee
signals this type of CP violation.
Equivalently
\be\label{ASLK1}
a_{SL}(\rm K_L)=\frac{1-r^2}{1+r^2}=2 \RE \bar\varepsilon
\ee
Note that $a_{SL}(\rm K_L)$ is determined purely by the quantities
related to  $K^0-\bar K^0$ mixing. 
Specifically, it measures
the difference between the phases of $\Gamma_{12}$ and
$M_{12}$.

That a non--vanishing $a_{\rm SL}(K_L)$ is indeed a signal
of CP violation can also be understood in the following
manner. $K_L$, that should be a CP eigenstate $K_2$ in the case
of CP conservation, decays into CP conjugate final states with
different rates. As $\RE \bar\varepsilon>0$, $K_L$ prefers slightly
to decay into $\pi^-e^+\nu_e$ than $\pi^+e^-\bar\nu_e$.
This would not be possible in a CP-conserving world.

\subsection{The First Look at $\varepsilon$ and $\varepsilon'$}
Since a two pion final state is CP even while a three pion final state is CP
odd, $K_{\rm S}$ and $K_{\rm L}$ preferably decay to $2\pi$ and $3\pi$, 
respectively
via the following CP-conserving decay modes \cite{Zichichi}:
\begin{equation}
K_{\rm L}\to 3\pi {\rm ~~(via~K_2),}\qquad K_{\rm S}\to 2 
\pi {\rm ~~(via~K_1).}
\end{equation}
This difference is responsible for the large disparity in their
life-times. A factor of 579. 
However, $K_{\rm L}$ and $K_{\rm S}$ are not CP eigenstates and 
may decay with small branching fractions as follows:
\begin{equation}
K_{\rm L}\to 2\pi {\rm ~~(via~K_1),}\qquad K_{\rm S}\to 3 
\pi {\rm ~~(via~K_2).}
\end{equation}
This violation of CP is called {\it indirect} as it
proceeds not via explicit breaking of the CP symmetry in 
the decay itself but via the admixture of the CP state with opposite 
CP parity to the dominant one.
 The measure for this
indirect CP violation is defined as (I=isospin)
\begin{equation}\label{ek}
\varepsilon
={{A(K_{\rm L}\rightarrow(\pi\pi)_{I=0}})\over{A(K_{\rm 
S}\rightarrow(\pi\pi)_{I=0})}}.
\end{equation}

Following the derivation in \cite{CHAU83} one finds
\begin{equation}
\eps = \bar\varepsilon+i\xi= \frac{\exp(i \pi/4)}{\sqrt{2} \Delta M_K} \,
\left( \IM M_{12} + 2 \xi \RE M_{12} \right),
\quad\quad
\xi = \frac{\IM A_0}{\RE A_0}.
\label{eq:epsdef}
\end{equation}
The phase convention dependence of the term involving $\xi$ cancells
the convention dependence of $\bar\varepsilon$ so that $\varepsilon$
is free from this dependence. The isospin amplitude $A_0$ is defined
below.

The important point in the definition (\ref{ek}) is that only the
transition to $(\pi\pi)_{I=0}$ enters. The transition to
$(\pi\pi)_{I=2}$ is absent. This allows to remove a certain type
of CP violation that originates in decays only. Yet as 
$\varepsilon\not=\bar\varepsilon$ and only $\RE\varepsilon=
\RE\bar\varepsilon$, it is clear that $\varepsilon$ includes
a type of CP violation represented by $\IM\varepsilon$ which is
absent in the semileptonic asymmetry (\ref{ASLK}). We will
identify this type of CP violation in Section 3.7, where a more
systematic classification of different types of CP violation
will be given.

\begin{figure}[hbt]
\centerline{
\epsfysize=1.5in
\epsffile{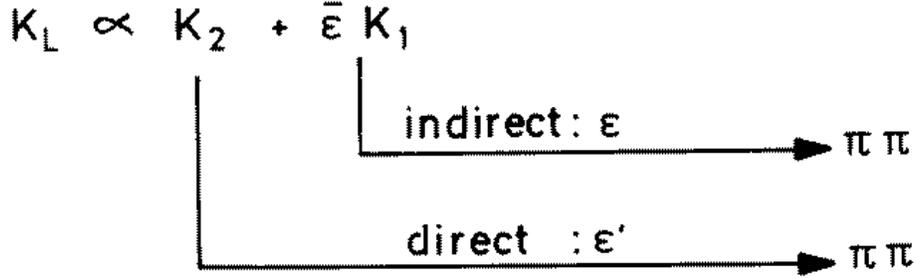}
}
\caption[]{
Indirect versus direct CP violation in $K_L \to \pi\pi$.
\label{fig:14}}
\end{figure}

While {\it indirect} CP violation reflects the fact that the mass
eigenstates are not CP eigenstates, so-called {\it direct}
CP violation is realized via a 
direct transition of a CP odd to a CP even state or vice versa (see
fig.~\ref{fig:14}). 
A measure of such a direct CP violation in $K_L\to \pi\pi$ is characterized
by a complex parameter $\varepsilon'$  defined as
\be\label{eprime0}
\varepsilon'=\frac{1}{\sqrt{2}}\left(\frac{A_{2,L}}{A_{0,S}}-
\frac{A_{2,S}}{A_{0,S}}\frac{A_{0,L}}{A_{0,S}}\right)
\ee
where short hand notation $A_{I,L}\equiv A(K_L\to (\pi\pi)_I)$
and $A_{I,S}\equiv A(K_S\to (\pi\pi)_I)$ has been used.

This time the transitions to $(\pi\pi)_{I=0}$ and
$(\pi\pi)_{I=2}$ are included which allows to study CP violation in
the decay itself. We will discuss this issue in general terms
in Section 3.7. For the time being it is useful to cast (\ref{eprime0})
into a more transparent formula
\be\label{eprime}
\varepsilon'=\frac{1}{\sqrt{2}}\IM\left(\frac{A_2}{A_0}\right)
              \exp(i\Phi_{\varepsilon'}), 
   \qquad \Phi_{\varepsilon'}=\frac{\pi}{2}+\delta_2-\delta_0, 
\ee
where
the isospin amplitudes $A_I$ in $K\to\pi\pi$
decays are introduced through
\begin{equation}\label{ISO1} 
A(K^+\rightarrow\pi^+\pi^0)=\sqrt{3\over 2} A_2 e^{i\delta_2}~,
\end{equation}
\begin{equation}\label{ISO2}
A(K^0\rightarrow\pi^+\pi^-)=\sqrt{2\over 3} A_0 e^{i\delta_0}+ \sqrt{1\over
3} A_2 e^{i\delta_2}~,
\end{equation}
\begin{equation}\label{ISO3}
A(K^0\rightarrow\pi^0\pi^0)=\sqrt{2\over 3} A_0 e^{i\delta_0}-2\sqrt{1\over
3} A_2 e^{i\delta_2}\,.
\end{equation} 
Here the subscript $I=0,2$ denotes states with isospin $0,2$
equivalent to $\Delta I=1/2$ and $\Delta I = 3/2$ transitions,
respectively, and $\delta_{0,2}$ are the corresponding strong phases. 
The weak CKM phases are contained in $A_0$ and $A_2$.
The isospin amplitudes $A_I$ are complex quantities which depend on
phase conventions. On the other hand, $\varepsilon'$ measures the 
difference between the phases of $A_2$ and $A_0$ and is a physical
quantity.
The strong phases $\delta_{0,2}$ can be extracted from $\pi\pi$ scattering. 
Then $\Phi_{\varepsilon'}\approx \pi/4$.

Experimentally $\varepsilon$ and $\varepsilon'$
can be found by measuring the ratios
\begin{equation}
\eta_{00}={{A(K_{\rm L}\to\pi^0\pi^0)}\over{A(K_{\rm S}\to\pi^0\pi^0)}},
            \qquad
  \eta_{+-}={{A(K_{\rm L}\to\pi^+\pi^-)}\over{A(K_{\rm S}\to\pi^+\pi^-)}}.
\end{equation}

Indeed, assuming $\varepsilon$ and $\varepsilon'$ to be small numbers one
finds
\be
\eta_{00}=\varepsilon-{{2\varepsilon'}\over{1-\sqrt{2}\omega}}
            ,~~~~
  \eta_{+-}=\varepsilon+{{\varepsilon'}\over{1+\omega/\sqrt{2}}}
\end{equation}
where $\omega=\RE A_2/\RE A_0=0.045$.

In the absence of direct CP violation $\eta_{00}=\eta_{+-}$.
The ratio ${\varepsilon'}/{\varepsilon}$  can then be measured through
\begin{equation}\label{BASE}
\RE(\epe)=\frac{1}{6(1+\omega/\sqrt{2})}
\left(1-\left|{{\eta_{00}}\over{\eta_{+-}}}\right|^2\right)~.
\end{equation}
To my knowledge \cite{KKHW} the experimental groups in giving 
their
results for $\RE(\epe)$ omitt the term $\omega/\sqrt{2}$ in (\ref{BASE}).
Yet in order to be consistent with the definitions 
(\ref{ek}) and (\ref{eprime0}) used by theorists
this term should be kept. 
  Consequently the existing experimental
results for $\RE(\epe)$ quoted below should be rescaled down by $3.2\%$.
Clearly at present this rescaling is academic as the experimental
error in $\RE(\epe)$ is roughly $\pm 20\%$ and the theoretical one at 
least $\pm 50\%$. We will therefore omitt this rescaling in what
follows.

\subsection{Basic Formula for $\eps$}
            \label{subsec:epsformula}
With all this information at hand let us derive a formula for $\varepsilon$
which can be efficiently used in pheneomenological applications.
The off-diagonal 
element $M_{12}$ in
the neutral $K$-meson mass matrix representing $K^0$-$\bar K^0$
mixing is given by
\begin{equation}
2 m_K M^*_{12} = \langle \bar K^0| \Heff(\Delta S=2) |K^0\rangle\,,
\label{eq:M12Kdef}
\end{equation}
where $\Heff(\Delta S=2)$ is the effective Hamiltonian for the 
$\Delta S=2$ transitions.
That $ M^*_{12}$ and not $ M_{12}$ stands on the l.h.s of this formula,
is evident from (\ref{MM12}). The factor $2 m_K$ reflects our normalization
of external states.

To lowest order in electroweak interactions $\Delta S=2$ transitions 
are induced
through the box diagrams of fig. \ref{L:9}. Including
leading and next-to-leading QCD corrections in the renormalization
group improved perturbation theory one has for $\mu<\mu_c=\ord(m_c)$
\begin{eqnarray}\label{hds2}
{\cal H}^{\Delta S=2}_{\rm eff}&=&\frac{G^2_{\rm F}}{16\pi^2}M^2_W
 \left[\lambda^2_c\eta_1 S_0(x_c)+\lambda^2_t \eta_2 S_0(x_t)+
 2\lambda_c\lambda_t \eta_3 S_0(x_c, x_t)\right] \times
\nonumber\\
& & \times \left[\as^{(3)}(\mu)\right]^{-2/9}\left[
  1 + \frac{\as^{(3)}(\mu)}{4\pi} J_3\right]  Q(\Delta S=2) + h. c.
\end{eqnarray}
where
$\lambda_i = V_{is}^* V_{id}^{}$,
  $\as^{(3)}$ is the strong coupling constant
in an effective three flavour theory and $J_3=1.895$ in the NDR scheme 
 \cite{BJW90}.
In (\ref{hds2}),
the relevant operator
\begin{equation}\label{qsdsd}
Q(\Delta S=2)=(\bar sd)_{V-A}(\bar sd)_{V-A},
\end{equation}
is multiplied by the corresponding Wilson coefficient function.
This function is decomposed into a
charm-, a top- and a mixed charm-top contribution.
The functions $S_0$  are given in (\ref{S0}) and (\ref{BFF}).

Short-distance QCD effects are described through the correction
factors $\eta_1$, $\eta_2$, $\eta_3$ and the explicitly
$\alpha_s$-dependent terms in (\ref{hds2}). 
The NLO values of $\eta_i$ are given as follows \cite{HNa,BJW90,HNb}:
\begin{equation}
\eta_1=1.38\pm 0.20,\qquad
\eta_2=0.57\pm 0.01,\qquad
  \eta_3=0.47\pm0.04~.
\end{equation}
The quoted errors reflect the remaining theoretical uncertainties due to
leftover $\mu$-dependences at $\ord(\as^2)$ and $\Lambda_{\overline{MS}}$,
the scale in the QCD running coupling.

Defining  the renormalization group 
invariant parameter $\hat B_K$ by
\begin{equation}
\hat B_K = B_K(\mu) \left[ \alpha_s^{(3)}(\mu) \right]^{-2/9} \,
\left[ 1 + \frac{\alpha_s^{(3)}(\mu)}{4\pi} J_3 \right]~,
\label{eq:BKrenorm}
\end{equation}
\begin{equation}
\langle \bar K^0| (\bar s d)_{V-A} (\bar s d)_{V-A} |K^0\rangle
\equiv \frac{8}{3} B_K(\mu) F_K^2 m_K^2
\label{eq:KbarK}
\end{equation}
and using (\eqn{hds2}) one finds
\begin{equation}
M_{12} = \frac{G_{\rm F}^2}{12 \pi^2} F_K^2 \hat B_K m_K \mw^2
\left[ {\lambda_c^*}^2 \eta_1 S_0(x_c) + {\lambda_t^*}^2 \eta_2 S_0(x_t) +
2 {\lambda_c^*} {\lambda_t^*} \eta_3 S_0(x_c, x_t) \right],
\label{eq:M12K}
\end{equation}
where $F_K=160~\mev$ is the $K$-meson decay constant and $m_K$
the $K$-meson mass. 
It should be mentioned that in the normalization of external
states in which the factor $2 m_K$ in (\ref{eq:M12Kdef}) is
absent, the r.h.s of (\ref{eq:KbarK}) is divided by $2 m_K$ so
that (\ref{eq:M12K}) remains unchanged.

To proceed further we neglect the last term in (\eqn{eq:epsdef}) as it
 constitutes at most a 2\,\% correction to $\eps$. This is justified
in view of other uncertainties, in particular those connected with
$\hat B_K$.
Inserting (\eqn{eq:M12K}) into (\eqn{eq:epsdef}) we find
\begin{equation}
\eps=C_{\eps} \hat B_K \IM\lambda_t \left\{
\RE\lambda_c \left[ \eta_1 S_0(x_c) - \eta_3 S_0(x_c, x_t) \right] -
\RE\lambda_t \eta_2 S_0(x_t) \right\} \exp(i \pi/4)\,,
\label{eq:epsformula}
\end{equation}
where we have used the unitarity relation $\IM\lambda_c^* = {\rm
Im}\lambda_t$ and  have neglected $\RE\lambda_t/\RE\lambda_c
 = \ord(\lambda^4)$ in evaluating $\IM(\lambda_c^* \lambda_t^*)$.
The numerical constant $C_\eps$ is given by
\begin{equation}
C_\eps = \frac{G_{\rm F}^2 F_K^2 m_K \mw^2}{6 \sqrt{2} \pi^2 \Delta M_K}
       = 3.837 \cdot 10^4 \, .
\label{eq:Ceps}
\end{equation}
To this end we have used the experimental value of $\Delta M_K$
in (\ref{DMEXP}) and $\mw=80.4~\gev$.

Using the standard parametrization of (\eqn{2.72}) to evaluate ${\rm
Im}\lambda_i$ and $\RE\lambda_i$, setting the values for $s_{12}$,
$s_{13}$, $s_{23}$ and $\mt$ in accordance with experiment
 and taking a value for $\hat B_K$ (see below), one can
determine the phase $\delta$ by comparing (\eqn{eq:epsformula}) with the
experimental value for $\eps$
\begin{equation}\label{eexp}
\varepsilon_{exp}
=(2.280\pm0.013)\cdot10^{-3}\;\exp{i\Phi_{\varepsilon}},
\qquad \Phi_{\varepsilon}={\pi\over 4}.
\end{equation}

Once $\delta$ has been determined in this manner one can find the
apex $(\bar\varrho,\bar\eta)$ of the unitarity triangle
in fig. \ref{fig:utriangle}   by using 
\begin{equation}\label{2.84a} 
\varrho=\frac{s_{13}}{s_{12}s_{23}}\cos\delta,
\qquad
\eta=\frac{s_{13}}{s_{12}s_{23}}\sin\delta
\end{equation}
and
\begin{equation}\label{2.88da}
\bar\varrho=\varrho (1-\frac{\lambda^2}{2}),
\qquad
\bar\eta=\eta (1-\frac{\lambda^2}{2}).
\end{equation}

For a given set ($s_{12}$, $s_{13}$, $s_{23}$,
$\mt$, $\hat B_K$) there are two solutions for $\delta$ and consequently two
solutions for $(\bar\varrho,\bar\eta)$. 
This will be evident from the analysis of the unitarity triangle discussed
in detail below.

Finally we have to say a few words about the non-perturbative
parameter $\hat B_K$, the main uncertainty in this analysis. 
References to older estimates can be found in \cite{AJBLAKE}. 
In our numerical analysis presented 
below we will use 
\begin{equation}\label{BKT}
\hat B_K=0.85\pm 0.15 \,
\end{equation}
which is very close to most recent lattice estimates, reviewed recently
in \cite{Lellouch,Football}, slightly
higher that the large-N estimates \cite{BBG0,Bijnens,DORT99}
and  somewhat lower than chiral quark model estimates \cite{BERT97}.

Recently an interesting large-N calculation of $\hat B_K$ in the
chiral limit and including next-to-leading 1/N corrections
has been performed by Peris and de Rafael \cite{PerRaf}. The nice
feature of this calculation is the explicit cancellation of the
$\mu$-dependence and the renormalization scheme dependence between the
Wilson coefficient and the matrix element of the operator $Q(\Delta S=2)$.
The resulting $\hat B_K=0.41\pm0.09$ is by a factor of two lower than
the value in (\ref{BKT}) and the lattice results. However, before a 
meaningful comparision with the lattice values can be made, higher 
order corrections in the chiral expansion have to be added to 
the result in \cite{PerRaf}.

\subsection{Express Review of $B_{d,s}^0$-$\bar B_{d,s}^0$ Mixing}
The flavour eigenstates in this case are
\be\label{fl}
B^0_d=(\bar bd),\qquad
\bar B^0_d=(b \bar d),\qquad
B^0_s=(\bar bs),\qquad
\bar B^0_s=( b \bar s)~.
\ee
They mix via the box diagrams in fig.~\ref{L:9} with $s$ replaced
by $b$ in the case of $B_{d}^0$-$\bar B_{d}^0$ mixing.
In the case $B_{s}^0$-$\bar B_{s}^0$ mixing also $d$ has to be replaced
by $s$.

Dropping the subscripts $(d,s)$ for a moment, it is customary to
denote the mass eigenstates by
\be\label{HL}
B_H=p B^0+q \bar B^0, \qquad B_L=p B^0-q \bar B^0
\ee
where
\be\label{pq}
p=\frac{1+\bar\varepsilon_B}{\sqrt{2(1+|\bar\varepsilon_B|^2)}},
\qquad
q=\frac{1-\bar\varepsilon_B}{\sqrt{2(1+|\bar\varepsilon_B|^2)}},
\ee
with $\bar\varepsilon_B$ corresponding to $\bar\varepsilon$ in
the $K^0-\bar K^0$ system. Here ``H'' and ``L'' denote 
{\it Heavy} and {\it Light} respectively. As in the $B^0-\bar B^0$ 
system one has $\Delta\Gamma\ll\Delta M$, 
 it is more suitable to distinguish the mass eigenstates by their 
masses than the corresponding life-times.

The strength of the $B^0_{d,s}-\bar B^0_{d,s}$ mixings
is described by the mass differences
\begin{equation}
\Delta M_{d,s}= M_H^{d,s}-M_L^{d,s}~.
\end{equation}
In contrast to $\Delta M_K$ , in this case the long distance contributions
are estimated to be very small and $\Delta M_{d,s}$ is very well
approximated by the relevant box diagrams. 
Moreover, due $m_{u,c}\ll m_t$ 
only the top sector can contribute significantly to 
$B_{d,s}^0-\bar B_{d,s}^0$ mixings.
The charm sector and the mixed top-charm contributions are
entirely negligible.

 $\Delta M_{d,s}$ can be expressed
in terms of the off-diagonal element in the neutral B-meson mass matrix
by using the formulae developed previously for the K-meson system.
One finds
\begin{equation}
\Delta M_q= 2 |M_{12}^{(q)}|, \qquad q=d,s.
\label{eq:xdsdef}
\end{equation}
This formula differs from $\Delta M_K=2 \RE M_{12}$ because in the
B-system $\Gamma_{12}\ll M_{12}$.

We also have
\be\label{delgamma}
\Delta\Gamma=\Gamma(B_H)-\Gamma(B_L)=
2 \frac{\RE(M_{12}\Gamma_{12}^*)}{|M_{12}|}
\ee
and
\be\label{q/p}
\frac{q}{p}=
=\frac{2 M^*_{12}-i\Gamma^*_{12}}{\Delta M-i\frac{1}{2}\Delta\Gamma}
=\frac{M_{12}^*}{|M_{12}|}
\left[1-\frac{1}{2}\IM\left(\frac{\Gamma_{12}}{M_{12}}\right)\right]
\ee
where higher order terms in the small quantity $\Gamma_{12}/M_{12}$
have been neglected.

The smallness of $\IM(\Gamma_{12}/M_{12})< \ord(10^{-3})$ has
two important consequences:
\bi
\item
The semileptonic asymmetry $a_{\rm SL}(B)$ discussed a few
pages below is even smaller than $a_{\rm SL}(K_L)$. Typically 
$\ord(10^{-4})$. These are bad news.
\item
The ratio $q/p$ is a pure phase to an excellent approximation.
These are very good news as we will see below.
\ei
Inspecting the relevant box diagrams we find
\be
(M_{12}^*)_d \propto (V_{td}V_{tb}^*)^2~,
\qquad
(M_{12}^*)_s \propto (V_{ts}V_{tb}^*)^2~.
\ee
Now, from Section 1 we know that
\be
V_{td}=\vtd e^{-i\beta}, \qquad
V_{ts}=-\vts e^{-i\beta_s}
\ee
with $\beta_s=\ord(10^{-2})$. Consequently to an excellent approximation 
\be\label{pureph}
\left(\frac{q}{p}\right)_{d,s}= e^{i2\phi_M^{d,s}},
\qquad
\phi^d_M=-\beta, \qquad \phi^s_M=-\beta_s,
\ee
with $\phi_M^{d,s}$ given entirely by the weak phases in the
CKM matrix.

\subsection{Basic Formulae for $\Delta M_{d,s}$}
            \label{subsec:BBformula}
Let us next find $\Delta M_{d,s}$.
The off-diagonal
term $M_{12}$ in the neutral $B$-meson mass matrix is given by
a formula analogous to (\ref{eq:M12Kdef})
\begin{equation}
2 m_{B_q} |M_{12}^{(q)}| = 
|\langle \bar B^0_q| \Heff(\Delta B=2) |B^0_q\rangle|,
\label{eq:M12Bdef}
\end{equation}
where 
in the case of $B_d^0-\bar B_d^0$
mixing 
\begin{eqnarray}\label{hdb2}
{\cal H}^{\Delta B=2}_{\rm eff}&=&\frac{G^2_{\rm F}}{16\pi^2}M^2_W
 \left(V^\ast_{tb}V_{td}\right)^2 \eta_{B}
 S_0(x_t)\times
\nonumber\\
& &\times \left[\alpha^{(5)}_s(\mu_b)\right]^{-6/23}\left[
  1 + \frac{\alpha^{(5)}_s(\mu_b)}{4\pi} J_5\right]  Q(\Delta B=2) + h. c.
\end{eqnarray}
Here $\mu_b=\ord(m_b)$, $J_5=1.627$,
\begin{equation}\label{qbdbd}
Q(\Delta B=2)=(\bar bd)_{V-A}(\bar bd)_{V-A}
\end{equation}
and \cite{BJW90,UKJS}
\begin{equation}
\eta_B=0.55\pm0.01.
\end{equation}
In the case of  $B_s^0-\bar B_s^0$ mixing one should simply replace
$d\to s$ in (\ref{hdb2}) and (\ref{qbdbd}) with all other quantities
 and numerical values unchanged.

Defining the renormalization group invariant parameters $\hat B_q$
in analogy to (\ref{eq:BKrenorm}) and (\ref{eq:KbarK})
one finds
 using (\ref{hdb2}) 
\begin{equation}
\Delta M_q = \frac{G_{\rm F}^2}{6 \pi^2} \eta_B m_{B_q} 
(\hat B_{B_q} F_{B_q}^2 ) \mw^2 S_0(x_t) |V_{tq}|^2,
\label{eq:xds}
\end{equation}
where $F_{B_q}$ is the $B_q$-meson decay constant.
This implies two useful formulae
\begin{equation}\label{DMD}
\Delta M_d=
0.50/{\rm ps}\cdot\left[ 
\frac{\sqrt{\hat B_{B_d}}F_{B_d}}{200\mev}\right]^2
\left[\frac{\mtb(\mt)}{170\gev}\right]^{1.52} 
\left[\frac{\vtd}{8.8\cdot10^{-3}} \right]^2 
\left[\frac{\eta_B}{0.55}\right]  
\end{equation}
and
\begin{equation}\label{DMS}
\Delta M_{s}=
15.1/{\rm ps}\cdot\left[ 
\frac{\sqrt{\hat B_{B_s}}F_{B_s}}{240\mev}\right]^2
\left[\frac{\mtb(\mt)}{170\gev}\right]^{1.52} 
\left[\frac{\vts}{0.040} \right]^2
\left[\frac{\eta_B}{0.55}\right] \,.
\end{equation}

There is a vast literature on the calculations of $F_{B_{d,s}}$ and
$\hat B_{d,s}$. The most recent lattice results are summarized in 
\cite{LL,Football}. 
They are compatible with the results obtained 
with the help of QCD sum rules   \cite{QCDSF}.
Guided by \cite{LL,Football} we will use in our numerical analysis 
 the value for
$F_{B_d}\sqrt{\hat B_{B_d}}$ given in table~\ref{tab:inputparams}.
The experimental situation on
$\Delta M_{d,s}$  is also given there.

\subsection{Classification of CP Violation}
\subsubsection{Preliminaries}
We have mentioned in Section 2 that due to the presence of hadronic
matrix elements, various decay amplitudes contain large theoretical
uncertainties. It is of interest to investigate which measurements
of CP-violating effects do not suffer from hadronic uncertainties.
To this end it is useful to make a classification of CP-violating
effects that is more transparent than the division into the
{\it indirect} and {\it direct} CP violation considered so far.
To my knowledge this
classification has been developed first for B decays but it can
also be useful for K decays. A nice detailed presentation can be
found in \cite{NIR99}.

Generally complex phases may enter particle--antiparticle mixing
and the decay process itself. It is then natural to consider
three types of CP violation:
\bi
\item
CP Violation in Mixing
\item
CP Violation in Decay
\item
CP Violation in the Interference of Mixing and Decay
\ei

As the phases in mixing and decay are convention dependent,
the CP-violating effects depend only
on the differences of these phases. This is clearly seen in
the classification given below.

\subsubsection{CP Violation in Mixing}
This type of CP violation can be best isolated in semi-leptonic
decays of neutral B and K mesons. We have discussed the asymmetry
$a_{SL}(K_L)$ before. In the case of B decays the non-vanishing
asymmetry (we suppress the indices $(d,s)$)
\be\label{ASLB}
a_{SL}(B)=\frac{\Gamma(\bar B^0(t)\to l^+\nu X)-
                 \Gamma( B^0(t)\to l^-\bar\nu X)}
                {\Gamma(\bar B^0(t)\to l^+\nu X)+
                 \Gamma( B^0(t)\to l^-\bar\nu X)}
            =\frac{1-|q/p|^4}{1+|q/p|^4}
          = \left(\IM\frac{\Gamma_{12}}{M_{12}}\right)_B
\ee
signals this type of CP violation. Here $\bar B^0(0)=\bar B^0$, 
$B^0(0)= B^0$. For $t\not=0$ the formulae analogous to 
(\ref{SCH}) should be used.
Note that the final states in (\ref{ASLB}) contain ``wrong charge''
leptons and can only be reached in the presence of $B^0-\bar B^0$
mixing. That is one studies effectively the difference between the
rates for $\bar B^0\to B^0\to l^+\nu X$
and $ B^0 \to \bar B^0 \to l^-\bar\nu X$. 
As the phases in
the transitions $B^0 \to \bar B^0$ and $\bar B^0 \to B^0$ 
differ from each other, a non-vanishing CP asymmetry follows.
Specifically $a_{\rm SL}(B)$ measures
the difference between the phases of $\Gamma_{12}$ and
$M_{12}$.

As $M_{12}$ and 
in particular $\Gamma_{12}$ suffer from large hadronic uncertainties
no precise extraction of CP-violating phases from this type of CP
violation can be expected.  
Moreover as $q/p$ is almost a pure phase, see (\ref{q/p}) and 
(\ref{pureph}), the
asymmetry is very small and outside the reach of experiments
performed in the coming years.

\subsubsection{CP Violation in Decay}
This type of CP violation is best isolated in charged B and charged K
decays as mixing effects do not enter here. However, it can also
be measured in neutral B and K decays. The relevant asymmetry is
given by
\be\label{ADECAY}
a^{\rm decay}_{f^\pm}=\frac{\Gamma(B^+\to f^+)-\Gamma(B^-\to f^-)}
                          {\Gamma(B^+\to f^+)+\Gamma(B^-\to f^-)}
=\frac{1-|\bar A_{f^-}/A_{f^+}|^2}{1+| \bar A_{f^-}/A_{f^+}|^2}
\ee
where
\be\label{AH}
A_{f^+}=\langle f^+|{\cal H}^{\rm weak}| B^+\rangle,
\qquad
\bar A_{f^-}=\langle f^-|{\cal H}^{\rm weak}| B^-\rangle~.
\ee
For this asymmetry to be non-zero one needs at least two different 
contributions with different {\it weak} ($\phi_i$) and {\it strong}
($\delta_i$) phases. These could be for instance two tree diagrams,
two penguin diagrams or one tree and one penguin. Indeed writing the
decay amplitude $A_{f^+}$ and its CP conjugate $\bar A_{f^-}$ as
\be\label{AMPL}
A_{f^+}=\sum_{i=1,2}=A_i e^{i(\delta_i+\phi_i)},
\qquad
\bar A_{f^-}=\sum_{i=1,2}=A_i e^{i(\delta_i-\phi_i)},
\ee
with $A_i$ being real, one finds 
\be\label{BDECAY}
a^{\rm decay}_{f^\pm}=\frac{ -2 A_1 A_2 \sin(\delta_1-\delta_2)
\sin(\phi_1-\phi_2)}{A_1^2+A_2^2+2 A_1 A_2 \cos(\delta_1-\delta_2)
\cos(\phi_1-\phi_2)}~.
\ee
The sign of strong phases $\delta_i$ is the same for $A_{f^+}$
and $\bar A_{f^-}$ because CP is conserved by strong interactions.
The corresponding weak phases have opposite sign. 

The presence of hadronic uncertainties in $A_i$ and the presence
of strong phases $\delta_i$ complicates the extraction of the
weak phases $\phi_i$ from data. An example of this type of
CP violation in K decays is $\varepsilon'$. We will demonstrate
this below.
\subsubsection{CP Violation in the Interference of Mixing and Decay}
This type of CP violation is only possible in neutral B and K
decays. We will use B decays for illustration suppressing the
subscripts $d$ and $s$. Moreover, we set $\Delta\Gamma=0$. Formulae
with $\Delta\Gamma\not =0$ can be found in \cite{BF97}.

 Most interesting are the decays into final states which
are CP-eigenstates. Then a time dependent asymmetry defined by
\be\label{TASY}
a_{CP}(t,f)=\frac{\Gamma(B^0(t)\to f)-
                 \Gamma( \bar B^0(t)\to f)}
                {\Gamma(B^0(t)\to f)+
                 \Gamma( \bar B^0(t)\to f)}
\ee
is given by
\begin{equation}\label{e8}
a_{CP}(t,f)=
{\cal A}^{\rm decay}_{CP}(B\to f)\cos(\Delta M
t)+{\cal A}^{\rm int}_{CP}(B\to f)\sin(\Delta M t)
\end{equation}
where we have separated the {\it decay} CP-violating contributions 
from those describing CP violation in the interference of
mixing and decay:
\begin{equation}\label{e9}
{\cal A}^{\rm decay}_{CP}(B\to f)\equiv\frac{1-\left\vert\xi_f\right\vert^2}
{1+\left\vert\xi_f\right\vert^2},
\qquad
{\cal A}^{\rm int}_{CP}(B\to f)\equiv\frac{2\mbox{Im}\xi_f}{1+
\left\vert\xi_f\right\vert^2}.
\end{equation}
The later type of CP violation is sometimes called the
{\it mixing-induced} CP violation \cite{Clarification}. 
The quantity $\xi_f$ containing essentially all the information
needed to evaluate the asymmetries (\ref{e9}) 
is given by
\begin{equation}\label{e11}
\xi_f=\frac{q}{p}\frac{A(\bar B^0\to f)}{A(B^0 \to f)}=
\exp(i2\phi_M)\frac{A(\bar B^0\to f)}{A(B^0 \to f)}
\end{equation}
with $\phi_M$, introduced in (\ref{pureph}), 
denoting the weak phase in the $B^0-\bar B^0$ mixing.
 $A(B^0 \to f)$ and $A(\bar B^0 \to f)$ are  decay amplitudes. 
The time dependence of $a_{CP}(t,f)$ allows to extract
${\cal A}^{\rm decay}_{CP}$ and ${\cal A}^{\rm int}_{CP}$ as
coefficients of $\cos(\Delta M t)$ and $\sin(\Delta M t)$
respectively.

Generally several decay mechanisms with different weak and
strong phases can contribute to $A(B^0 \to f)$. These are
tree diagram (current-current) contributions, QCD penguin
contributions and electroweak penguin contributions. If they
contribute with similar strength to a given decay amplitude
the resulting CP asymmetries suffer from hadronic uncertainies
related to matrix elements of the relevant operators $Q_i$.
The situation is then analogous to the class just discussed.
Indeed
\be\label{ratiocp}
\frac{A(\bar B^0\to f)}{A(B^0 \to f)}=-\eta_f
\left[\frac{A_T e^{i(\delta_T-\phi_T)}+A_P e^{i(\delta_P-\phi_P)}}
{A_T e^{i(\delta_T+\phi_T)}+A_P e^{i(\delta_P+\phi_P)}}\right]
\ee
with $\eta_f=\pm 1$ being the CP-parity of the final state,
depends on strong phases $\delta_{T,P}$ and hadronic matrix
elements present in $A_{T,P}$. Thus the measurement of the
asymmetry does not allow a clean determination of the weak
phases $\phi_{T,P}$. The minus sign in (\ref{ratiocp}) follows
from our CP phase convention $ CP |B^0\rangle= -|\bar B^0\rangle$,
 that has also been used in writing the phase factor in (\ref{e11}).
Only $\xi$ is phase convention independent. Explicit derivation
can be found in section 8.4.1 of \cite{BF97}.

An interesting case arises when a single mechanism dominates the 
decay amplitude or the contributing mechanisms have the same weak 
phases. Then the hadronic matrix elements and strong phases drop out and
\be\label{cp}
\frac{A(\bar B^0\to f)}{A(B^0 \to f)}=-\eta_f e^{-i2\phi_D}
\ee
is a pure phase with $\phi_D$ being the weak phase in the decay amplitude.
Consequently
\begin{equation}\label{e111}
\xi_f=-\eta_f\exp(i2\phi_M) \exp(-i 2 \phi_D),
\qquad
\mid \xi_f \mid^2=1~.
\end{equation}
In this particular case 
${\cal A}^{\rm decay}_{CP}(B\to f)$
vanishes and the CP asymmetry is given entirely
in terms of the weak phases $\phi_M$ and $\phi_D$:
\begin{equation}\label{simple}
a_{CP}(t,f)= \IM\xi_f \sin(\Delta Mt) \qquad
\IM\xi_f=\eta_f \sin(2\phi_D-2\phi_M)~.
\end{equation}
Thus the corresponding measurement of weak phases is free from
hadronic uncertainties. A well known example is the decay
$B_d\to \psi K_S$. Here $\phi_M=-\beta$ and $\phi_D=0$. As
in this case $\eta_f=-1$,  we find
\begin{equation}
a_{CP}(t,f)= -\sin(2\beta) \sin(\Delta Mt) 
\end{equation}
which allows a very clean measurement of the angle $\beta$ in the
unitarity triangle. We will return to this decay and other
decays in which this type of CP violation can be tested.

We observe that the asymmetry $a_{CP}(t,f)$ measures directly the
difference between the phases of $B^0-\bar B^0$-mixing $(2\phi_M)$
and of the decay amplitude $(2\phi_D)$. This tells us immediately 
that we are dealing with the interference of mixing and decay.
As $\phi_M$ and $\phi_D$
are obviously phase convention dependent quantities, only their
difference is physical, it is
impossible to state on the basis of a single asymmetry 
whether CP violation takes place in the decay or in the mixing.
To this end at least two asymmetries for $B^0_d (\bar B^0_d)$
decays to different final states $f_i$ have to be measured.
As $\phi_M$ does not depend on the final state, 
$\IM\xi_{f_1}\not=\IM\xi_{f_2}$ is a signal of CP violation
in the decay. The same applies to $B^0_s (\bar B^0_s)$ decays.

In the case of K decays, this type of CP violation can be
cleanly measured in the rare decay $K_L\to\pi^0\nu\bar\nu$.
Here the difference between the weak phase in the $K^0-\bar K^0$
mixing and in the decay $\bar s \to \bar d \nu\bar\nu$ matters.

We can now compare the two classifications of different types
of CP violation. CP violation in mixing is a manifestation
of indirect CP violation. CP violation in decay is a manifestation
of direct CP violation. CP violation in interference of mixing
and decay contains elements of both the indirect and direct CP
violation.

It is clear from this discussion that only in the case of the
third type of CP violation there are possibilities to measure weak
phases without hadronic uncertainties. This takes place provided
a single mechanism (diagram) is responsible for the decay or the
contributing decay mechanisms have the same weak phases.

\subsubsection{Another Look at $\varepsilon$ and $\varepsilon'$}
Let us finally investigate what type of CP violation is
represented by $\varepsilon$ and $\varepsilon'$.
Here instead of different mechanism it is sufficient to talk
about different isospin amplitudes.

In the case of $\varepsilon$, CP violation in decay is not
possible as only the isospin amplitude $A_0$ is involved.
See (\ref{ek}). We know also that only $\RE\varepsilon=
\RE\bar\varepsilon$ is related to CP violation in mixing.
Consequently:
\bi
\item 
$\RE\varepsilon$ represents CP violation in mixing,
\item
$\IM\varepsilon$ represents CP violation in the
interfernce of mixing and decay.
\ei

In order to analyze the case of $\varepsilon'$ we use the formula
(\ref{eprime}) to find
\be\label{ree}
\RE\varepsilon'=-\frac{1}{\sqrt{2}}\left\vert\frac{A_2}{A_0}\right\vert
\sin(\phi_2-\phi_0)\sin(\delta_2-\delta_0)
\ee
\be\label{iee}
\IM\varepsilon'=\frac{1}{\sqrt{2}}\left\vert\frac{A_2}{A_0}\right\vert
\sin(\phi_2-\phi_0)\cos(\delta_2-\delta_0)~.
\ee  
Consequently:
\bi
\item 
$\RE~\varepsilon'$ represents CP violation in decay as it is only
non zero provided simultaneously $\phi_2\not=\phi_0$ and 
$\delta_2\not=\delta_0$.
\item
$\IM~\varepsilon'$ exists even for $\delta_2=\delta_0$ but as
it requires $\phi_2\not=\phi_0$ it represents CP violation in 
decay as well.
\ei
Experimentally  $\delta_2\not=\delta_0$.
Within the SM, $\phi_2$ and $\phi_0$ are connected with
electroweak penguins and QCD penguins respectively.
Do these phases differ from
each other so that a nonvanishing $\varepsilon'$ is obtained?
 We will
return to this question in Section 5.

\section{Standard Analysis of the Unitarity Triangle}\label{UT-Det}
\subsection{Basic Procedure}
With all these formulae at hand we can now summarize the standard
analysis of the unitarity triangle in fig. \ref{fig:utriangle}. 
It proceeds in five steps.

{\bf Step 1:}

{}From  $b\to c$ transition in inclusive and exclusive 
leading B-meson decays
one finds $\vcb$ and consequently the scale of the unitarity triangle:
\begin{equation}
\vcb\quad \Longrightarrow\quad\lambda \vcb= \lambda^3 A~.
\end{equation}

{\bf Step 2:}

{}From  $b\to u$ transition in inclusive and exclusive $B$ meson decays
one finds $\vub$ and consequently using (\ref{2.94}) 
the side $CA=R_b$ of the unitarity triangle:
\begin{equation}\label{rb}
\left| \frac{V_{ub}}{V_{cb}} \right|
 \quad\Longrightarrow \quad R_b=\sqrt{\bar\varrho^2+\bar\eta^2}=
4.44 \cdot \left| \frac{V_{ub}}{V_{cb}} \right|~.
\end{equation}

{\bf Step 3:}

\begin{figure}[hbt]
  \vspace{0.10in} \centerline{
\begin{turn}{-90}
  \mbox{\epsfig{file=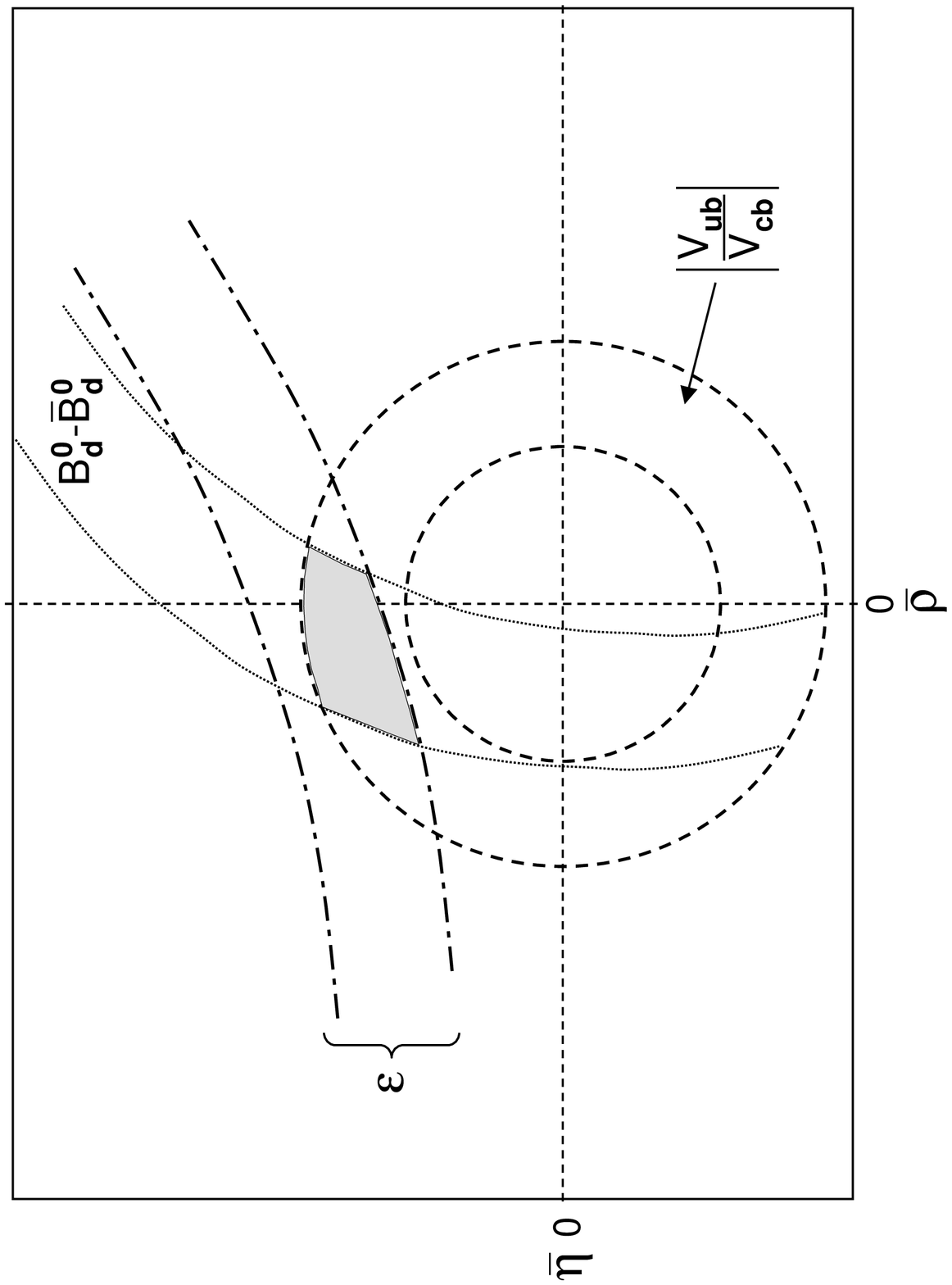,width=0.5\linewidth}}
\end{turn}
} \vspace{-0.18in}
\caption[]{Schematic determination of the Unitarity Triangle.
\label{L:10}}
 \end{figure}

{}From the experimental value of $\varepsilon$ (\ref{eexp}) 
and the formula (\ref{eq:epsformula}) one 
derives, using the approximations (\ref{2.51})--(\ref{2.53}), 
the constraint
\begin{equation}\label{100}
\bar\eta \left[(1-\bar\varrho) A^2 \eta_2 S_0(x_t)
+ P_c(\varepsilon) \right] A^2 \hat B_K = 0.226,
\end{equation}
where
\begin{equation}\label{102}
P_c(\varepsilon) = 
\left[ \eta_3 S_0(x_c,x_t) - \eta_1 x_c \right] \frac{1}{\lambda^4},
\qquad
x_t=\frac{\mt^2}{\mw^2}.
\end{equation}
 $P_c(\varepsilon)=0.31\pm0.05$ summarizes the contributions
of box diagrams with two charm quark exchanges and the mixed 
charm-top exchanges. 
The main uncertainties in the constraint (\ref{100}) reside in
$\hat B_K$ and to some extent in $A^4$ which multiplies the leading term.
Equation (\ref{100}) specifies 
a hyperbola in the $(\bar \varrho, \bar\eta)$
plane.
This hyperbola intersects the circle found in step 2
in two points which correspond to the two solutions for
$\delta$ mentioned earlier. This is illustrated in fig. \ref{L:10}.
The position of the hyperbola (\ref{100}) in the
$(\bar\varrho,\bar\eta)$ plane depends on $\mt$, $|V_{cb}|=A \lambda^2$
and $\hat B_K$. With decreasing $\mt$, $|V_{cb}|$ and $\hat B_K$ the
$\eps$-hyperbola moves away from the origin of the
$(\bar\varrho,\bar\eta)$ plane. 

{\bf Step 4:}

{}From the observed $B^0_d-\bar B^0_d$ mixing parametrized by $\Delta M_d$ 
the side $BA=R_t$ of the unitarity triangle can be determined:
\begin{equation}\label{106}
 R_t= \frac{1}{\lambda}\frac{|V_{td}|}{\vcb} = 1.0 \cdot
\left[\frac{|V_{td}|}{8.8\cdot 10^{-3}} \right] 
\left[ \frac{0.040}{\vcb} \right]
\end{equation}
with
\begin{equation}\label{VT}
\vtd=
8.8\cdot 10^{-3}\left[ 
\frac{200\mev}{\sqrt{\hat B_{B_d}}F_{B_d}}\right]
\left[\frac{170~GeV}{\mtb(\mt)} \right]^{0.76} 
\left[\frac{\Delta M_d}{0.50/{\rm ps}} \right ]^{0.5} 
\sqrt{\frac{0.55}{\eta_B}}.
\end{equation}

Since $\mt$, $\Delta M_d$ and $\eta_B$ are already rather precisely
known, the main uncertainty in the determination of $\vtd$ from
$B_d^0-\bar B_d^0$ mixing comes from $F_{B_d}\sqrt{\hat B_{B_d}}$.
Note that $R_t$ suffers from additional uncertainty in $\vcb$,
which is absent in the determination of $\vtd$ this way. 
The constraint in the $(\bar\varrho,\bar\eta)$ plane coming from
this step is illustrated in fig.~\ref{L:10}.

{\bf Step 5:}

{}The measurement of $B^0_s-\bar B^0_s$ mixing parametrized by $\Delta M_s$
together with $\Delta M_d$  allows to determine $R_t$ in a different
manner. Using (\ref{eq:xds}) one finds 
\begin{equation}\label{107b}
\frac{\vtd}{|V_{ts}|}= 
\xi\sqrt{\frac{m_{B_s}}{m_{B_d}}}
\sqrt{\frac{\Delta M_d}{\Delta M_s}},
\qquad
\xi = 
\frac{F_{B_s} \sqrt{\hat B_{B_s}}}{F_{B_d} \sqrt{\hat B_{B_d}}}.
\end{equation}

Now to an excellent accuracy \cite{BLO}:
\be\label{vts1}
\vtd=\vcb\lambda R_t, \qquad 
\vts=\vcb(1-\frac{1}{2}\lambda^2+\bar\varrho\lambda^2)~.
\ee
We note next that through the unitarity of the CKM
matrix, the present experimental upper bound on 
$\Delta M_d/\Delta M_s$ (see table~\ref{tab:inputparams}) 
and the value of $\vub$
 one has $0\le\bar\varrho\le 0.5$, where
$\xi=1.15\pm 0.06$ \cite{Football,LL} has been used. 
Consequently
$\vts$ deviates from $\vcb$ by at most $2\%$. This means
that to a very good accuracy we can set $\vts=\vcb$.
Consequently (\ref{107b}) and the first formula in (\ref{vts1})
imply
\be\label{Rt}
R_t=0.84~ \xi_{\rm eff}\sqrt{\frac{\Delta M_d}{0.50/ps}}, \qquad
\xi_{\rm eff}=\xi \sqrt{\frac{15.0/ps}{\Delta M_s}}~.
\ee
Using next $\Delta M^{{\rm max}}_d= 0.50/
\mbox{ps}$  one finds a useful  formula
\begin{equation}\label{107bu}
(R_t)_{\rm max} = 0.84~ \xi 
\sqrt{\frac{15.0/ps}{(\Delta M_s)_{\rm min}}}~.
\end{equation}
If necessary the $\ord(\lambda^2)$ corrections in (\ref{vts1}) can
be incorporated in (\ref{Rt}). This will be only required when the
error on $\xi$ will be decreased below $2\%$, which is clearly
a very difficult task.

One should 
note that $\mt$ and $|V_{cb}|$ dependences have been eliminated this way
 and that $\xi$ should in principle 
contain much smaller theoretical
uncertainties than the hadronic matrix elements in $\Delta M_d$ and 
$\Delta M_s$ separately.  
The most recent values relevant for (\ref{107bu}) are summarized
in table~\ref{tab:inputparams}.

\begin{table}[thb]
\caption[]{The ranges of the input parameters.
\label{tab:inputparams}}
\vspace{0.4cm}
\begin{center}
\begin{tabular}{|c|c|c|}
\hline
{\bf Quantity} & {\bf Central} & {\bf Error}  \\
\hline
$|V_{cb}|$ & 0.041 & $\pm 0.002$      \\
$\vub$ & $0.085$ & $\pm 0.018 $   \\
$|V_{ub}|$ & $0.00349$ & $\pm 0.00076$ \\ 
$\hat B_K$ & 0.85 & $\pm 0.15$   \\
$\sqrt{\hat B_d} F_{B_{d}}$ & $230\mev$ & $\pm 40\mev$  \\
$\mt$ & $166\gev$ & $\pm 5\gev$   \\
$(\Delta M)_d$ & $0.487/\mbox{ps}$ & $\pm 0.014/\mbox{ps}$ 
  \\
$(\Delta M)_s$  & $>15.0/\mbox{ps}$ & \\
$\xi$ & $1.15$ & $\pm 0.06$  \\
\hline
\end{tabular}
\end{center}
\end{table}

\subsection{Numerical Results}\label{sec:standard}
\subsubsection{Input Parameters}

The input parameters needed to perform the
standard analysis of the unitarity triangle
are given in table~\ref{tab:inputparams}.
In constructing this table I was guided to a large extent by the reviews
\cite{Lellouch,Football,LL,STOCCHI,STONE}. 
I am aware of the fact that other authors
would possibly use slightly different ranges for input parameters.
Still table \ref{tab:inputparams} is representative for the
present situation. 
Please note, however, that the error on $|V_{ub}|$ in
table~\ref{tab:inputparams} is by a factor of two larger than in
\cite{Football}. I do not think that our present understanding of theoretical
uncertainties in the determination of $|V_{ub}|$ is sufficiently
good that an error of $\pm 10\%$ on this element can be defended.

The great progress during the last year has been the improved lower
limit on $\Delta M_s$ from LEP and SLD as reviewed in \cite{STOCCHI}.
The value of $\mt$ refers
to the running current top quark mass defined at $\mu=\mt^{Pole}$.
It corresponds to 
$\mt^{Pole}=174.3\pm 5.1\gev$ measured by CDF and D0 \cite{CDFD0}.

\begin{figure}[hbt]
\vspace{0.10in}
\centerline{
\epsfysize=2.4in
%\rotate[r]{
\epsffile{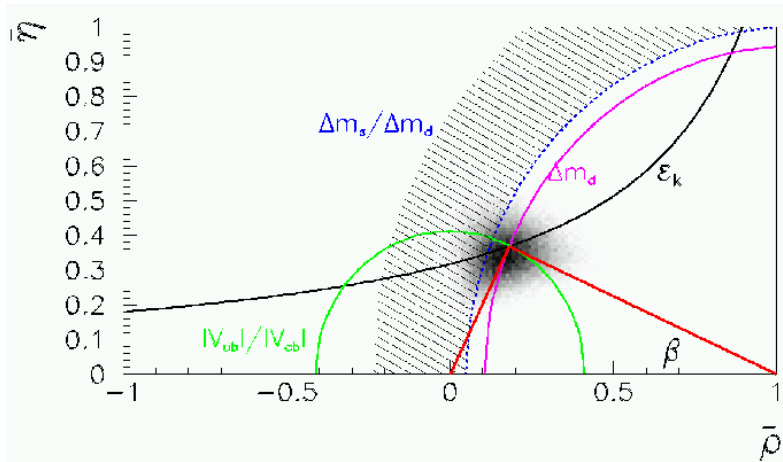}
}%}
\vspace{0.08in}
\caption[]{
The Unitarity Triangle as of January 2001.
\label{fig:2000}}
\end{figure}

\subsubsection{Various Error Analyses}
Having set the input parameters and their uncertainties there
is the question how to treat the theoretical uncertainties in
a quantitative analysis. There is a hot discussion on this
issue, that can be traced by reading 
\cite{REV,Football,STOCCHI,STONE,FRENCH,ALI00,SCHUNE}.
Basically three different approaches can be found in the
literature:
\begin{itemize}
\item
Gaussian Method: The experimentally measured numbers  
 and the theoretical input parameters are used with Gaussian errors.
Examples are the analyses in \cite{Football,ALI00,Parodi}.
There are some small differences in the actual treatment of errors
in these papers. We refer to \cite{Football} for a detailed
presentation.
\item
BaBar $95\%~ C.L.$ Scanning Method: This method has been developed
in \cite{FRENCH} and is the official method of the BaBar collaboration
\cite{BABAR}. In this method one sets the theoretical input
parameters at some fixed values and finds the allowed ($95\%~C.L.$)
region for $(\bar\varrho,\bar\eta)$ by using gaussian errors for the
experimental input parameters. Repeating this procedure for
different sets of the values of the theoretical input parameters one
obtains an envelope of $95\%~ C.L.$ regions. The latest application
of this method can be found in \cite{SCHUNE}.
\item
Simple Scanning: Both the experimentally measured numbers and the 
theoretical input parameters are scanned independently within  
ranges, given for instance in table~\ref{tab:inputparams}.
This is the method used for instance by Rosner \cite{REV} and
by myself below.
\end{itemize}

In my opinion the use of Gaussian errors for theoretical input parameters
is questionable but I do not want to enter this discussion here.
A recent attempt to justify this method can be found in 
\cite{Football,STOCCHI}. Yet, when the lattice calculations
improve dramatically I could imagine that one could defend this approach.
On the other hand the simple scanning method appears to be too conservative.
It should be stressed that only the gaussian method pretends to
give standard deviations for the output quantities. The scanning
methods can only give ranges for the output quantities.
The BaBar scanning gives
generally the ranges for quantities of interest which are comparable
to the ones found by the more naive scanning used here. The $95\%
C.L.$ ranges from the gaussian method are not so different from
the ones obtained by the scanning methods and consequently global
pictures of the unitarity triangle obtained by these methods are
compatible with each other. In order to get the full story the interested 
reader should have a look at \cite{Football,ALI00,SCHUNE}. In particular
the first paper contains very useful material.

In this section I will present the results of my simple scanning
analysis and of a Gaussian analysis by Stefan Schael who used
the same input parameters. For the rest of these lectures I will
only use simple scanning, except for $\epe$ where also the
results of the Gaussian method will be presented. The quoted results
of the scanning method for a quantity $Q$ given below should be understood 
as follows:
\be
\left\{ Q=A\pm B\right\}\equiv \left\{ A-B \le Q \le A+B\right\} 
\ee
This means that the central value $A$ does not generally correspond
to central values of the input parameters.

\subsubsection{Output of the Standard Analysis}

Using simultaneously the five steps discussed above one finds
the allowed region of $(\bar\varrho,\bar\eta)$.
In fig.~\ref{fig:2000} we show the result of an analysis by
Stefan Schael which uses the input parameters of 
table~\ref{tab:inputparams}.
Only  the dark region is allowed. From this
figure one extracts
\be\label{ABC}
\alpha=93^\circ\pm 11^\circ, \qquad 
\beta={23.6^\circ}^{+ 4.9^\circ}_{-4.3^\circ},
 \qquad \gamma=64^\circ\pm 11^\circ,
\ee
\be\label{vtdexp}
\sin 2\beta=0.73^{+0.07}_{-0.14}, \qquad \vtd=(8.0\pm 0.7)\cdot 10^{-3}~.
\ee
In this analysis Gaussian errors for all input parameters have been used.
My own, more conservative analysis that uses scanning for all input
parameters gives
\be\label{ABC2}
78.8^\circ\le\alpha\le 120^\circ, \qquad 
15.1^\circ\le\beta\le 28.6^\circ,\qquad 
37.9^\circ\le\gamma\le 76.5^\circ
\ee
\be\label{vtdexp2}
\sin 2\beta=0.67\pm0.17, \qquad \vtd=(8.0\pm 1.3)\cdot 10^{-3}~.
\ee
The "true" errors are probably between these two estimates.
The results of both analyses are summarized in table \ref{TAB2}.

\begin{figure}[thb]
\vspace{0.10in}
\centerline{
\epsfysize=2.8in
%\rotate[r]{
\epsffile{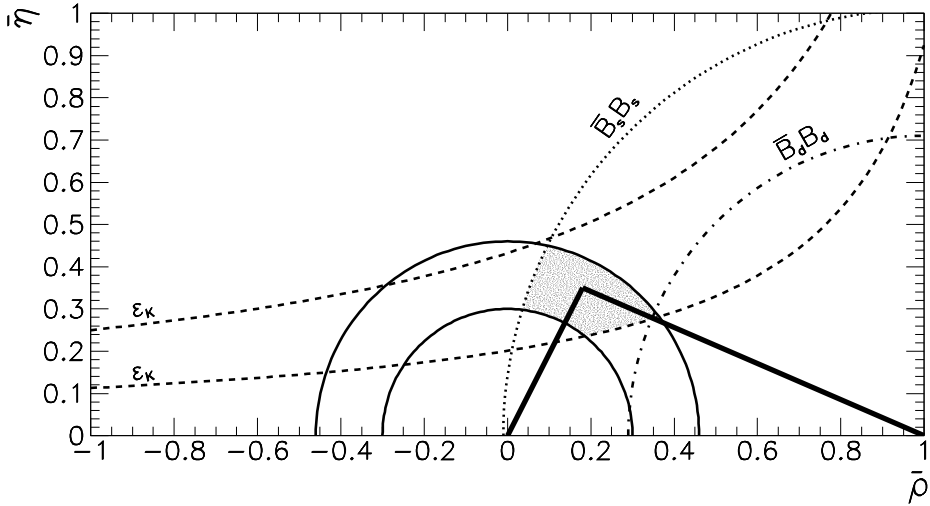}
}%}
\vspace{0.08in}
\caption[]{
Conservative Unitarity Triangle as of January 2001.
\label{fig:utdata}}
\end{figure}

The allowed region for $(\bar\varrho,\bar\eta)$ resulting from
the scanning method is presented in fig. 
\ref{fig:utdata}.
It is the shaded area on the right hand side of the  circle
representing  
the lower bound for $(\Delta M)_s$, that is $(\Delta M)_s>15/ps$.
The hyperbolas in fig.~\ref{fig:utdata}
give the constraint from $\varepsilon$ and the two circles centered
at $(0,0)$ the constraint from $\vub$.
The circle on the right comes from $B^0_{d}-\bar B^0_{d}$ mixing
and excludes the region to its right. We observe that
$B^0_{d}-\bar B^0_{d}$ mixing is almost ineffective for the chosen ranges 
of the input parameters within the SM and
the allowed region is governed by $\vub$, $\Delta M_s$ and $\varepsilon$.
We also observe that the region
$\bar\varrho<0$ is practically excluded by the lower bound on
$\Delta M_s$.
It is clear
from this figure that $(\Delta M)_s$ is a very important
ingredient in this analysis and that the measurement of $(\Delta M)_s$
giving also  lower bound on $R_t$ will have a large impact
on the plots in figs.~\ref{fig:2000} and \ref{fig:utdata}. 
Finally we find that whereas the angle $\beta$ is rather
constrained, the uncertainties in 
$\alpha$ and $\gamma$ are  substantially larger.

\begin{table}[thb]
\caption[]{Output of the Standard Analysis. 
 $\lambda_t=V^*_{ts} V_{td}$.\label{TAB2}}
\vspace{0.4cm}
\begin{center}
\begin{tabular}{|c||c||c|}\hline
{\bf Quantity} & {\bf Scanning} & {\bf Gaussian} \\ \hline
$\mid V_{td}\mid/10^{-3}$ &$6.7 - 9.3$ &$8.0\pm0.7$ \\ \hline
$\mid V_{ts}/V_{cb}\mid$ &$0.979 - 0.993$ &$0.984\pm0.004$  \\ \hline
$\mid V_{td}/V_{ts}\mid$ &$0.16 - 0.22$ &$0.20\pm0.02$  \\ \hline
$\sin(2\beta)$ &$0.50 - 0.84$ &$ 0.73^{+0.07}_{-0.14} $ \\ \hline
$\sin(2\alpha)$ &$-0.87 - 0.38$ &$-0.1\pm0.3$ \\ \hline
$\sin(\gamma)$ &$0.61 - 0.97 $ &$ 0.9\pm 0.1 $ \\ \hline
$\IM \lambda_t/10^{-4}$ &$0.94 - 1.60 $ &$1.2\pm 0.2 $ \\ \hline
$\bar\eta$ & $0.22-0.46$ &$ 0.36\pm 0.07$ \\ \hline
$\bar\varrho$ & $ 0.06 - 0.34 $ & $0.18\pm 0.09$ \\ \hline
\end{tabular}
\end{center}
\end{table}

\subsubsection{An Upper Bound on \boldmath{$\Delta M_s$}}
In view of the expected measurement of $\Delta M_s$ in 2001 at
Tevatron it is of interest
to find its upper bound within the SM. The most straightforward
manner to obtain this bound is to set $\vts_{max}=0.043$,
$\overline{\mt}(\mt)_{max}=171\gev$ and 
$(\sqrt{\hat B_{B_s}}F_{B_s})_{max}=300~\mev$ \cite{Football,LL} in
(\ref{DMS}). We find
\be\label{DMSU}
\Delta M_s< 27.5/ps
\ee
with 16.0/ps obtained for central values of the
input parameters.

\subsection{First Conclusions}
In this section we have completed the determination of the CKM matrix.
It is given by the values of $|V_{us}|$, $\vcb$ and $|V_{ub}|$ in
(\ref{vcb}) and (\ref{v13}), the results in table~\ref{TAB2} and
the unitarity triangle shown in figs.~\ref{fig:2000} and \ref{fig:utdata}. 
 We should
stress, that  soon this analysis will be improved through
the new information on the angle $\beta$ coming from B factories and
Tevatron
and on $\Delta M_s$ from Tevatron.
In particular the latter measurement 
should have an important impact on the allowed area in the 
$(\bar\varrho,\bar\eta)$ plane. 

We conclude that the SM is capable of describing
the observed indirect CP violation in $K_L$ decays, taking into
account the data on $B^0_{d,s}-\bar B^0_{d,s}$ mixings, $\vcb$ and 
$\vub$. We also observe that $R_b$ and $R_t$ from $\vub$ and
$B^0_{d,s}-\bar B^0_{d,s}$ mixings alone satisfy the condition 
(\ref{con}). Taking $(R_b)_{min}=0.30$, this condition reads
$0.70<R_t<1.30$. From (\ref{107bu}) and the lower bound on
$\Delta M_s$ one has
$R_t<1.01$. On the other hand $\vtd_{min}$ is governed by
$B_d^0-\bar B_d^0$ mixing. From (\ref{106}) and (\ref{vtdexp2})
one has then $R_t>0.71$. Consequently CP violation in B-decays
is predicted on the basis of $\vub$ and $B^0_{d,s}-\bar B^0_{d,s}$ 
mixings alone, even if our conservative analysis does not show
this so clearly.
Indeed the first result for the CP asymmetry in $B_d\to\psi K_S$ with
$\sin 2\beta=0.79\pm 0.42$ from CDF \cite{CDFB}, presented in 1999,
 established
 CP violation in B decays at $93\%$ C.L. and moreover was consistent with 
(\ref{vtdexp}) and (\ref{vtdexp2}). Yet the reality could turn out
to be  different as we will see in a moment. 

\subsection{First Results on \boldmath{$\sin 2\beta$} from B Factories}
Since the summer 2000 we have the first results on the time dependent 
CP asymmetry, $a_{\psi K_S}$, from
 BaBar \cite{BaBar} and Belle \cite{Belle}.
These results indicate that the value of the angle $\beta$
in the unitarity triangle could turn out to be substantially smaller than
expected on the basis of the standard analysis of the unitarity
triangle within the SM presented above.
Indeed the three measurements of this asymmetry
\begin{equation}\label{sinexp}
(\sin 2\beta)_{\psi K_S} =\left\{ \begin{array}{ll}
0.79\pm 0.42 &{\rm (CDF)}~\cite{CDFB}\\
0.12\pm0.37\pm0.09 & {\rm (BaBar)}~\cite{BaBar} \\
0.45\pm 0.44 \pm 0.08 &{\rm (Belle)}~\cite{Belle}
\end{array} \right.
\end{equation}
imply the grand average
\be\label{grand}
 (\sin 2\beta)_{\psi K_S}= 0.42\pm0.24~.
\label{gas}
\ee
This should be compared with the results of the standard analysis of the
unitarity triangle within the SM presented above which gave
$\sin 2\beta =0.73^{+0.07}_{-0.14}$ and $\sin 2\beta =0.67\pm 0.17$
for the gaussian and scanning analyses respectively. Similar
values can be found in the literature
\begin{equation}\label{sinth}
(\sin 2\beta)_{SM} =\left\{ \begin{array}{ll}
0.69\pm0.07 & ~\cite{Football} \\
0.70\pm 0.24 &~\cite{ALI00}\\
0.63\pm 0.12 & ~\cite{SCHUNE}
\end{array} \right.
\end{equation}
where the last two results represent $95\%$ C.L. ranges.
Clearly in view of the large spread of experimental results and
large statistical errors in (\ref{sinexp}), the SM estimates in
(\ref{vtdexp}), (\ref{vtdexp2}) and (\ref{sinth}) are compatible 
with the experimental value
of $(\sin 2\beta)_{\psi K_S}$ in (\ref{grand}). Yet the small
values of $\sin 2\beta$ found by BaBar and Belle may give some hints for
new physics contributions to $B^0_d-\bar B^0_d$ and $K^0-\bar K^0$
mixings. In particular as discussed recently in several papers
\cite{NIR00}-\cite{XING} new CP violating phases in $B^0_d-\bar B^0_d$ 
and $K^0-\bar K^0$ mixing could be
responsible for small values of $\sin 2\beta$ in (\ref{grand}).
In this context an absolut lower bound on $\sin 2\beta$ in models
with minimal flavour violation has been derived in \cite{ABRB}.
We will discuss this bound and its update in detail in Section 9. 
Finally a 
general model independent analysis of new physics in 
$B^\pm\to \psi K^\pm$ and $B^0_d\to \psi K_S$ has been presented
very recently in \cite{FM01}.

On the other hand as stressed in \cite{NIR00} the SM estimates of
$\sin 2\beta$ are sensitive to the assumed ranges for the parameters
\begin{equation}\label{par}
\vcb,\quad \vub, \quad \hat B_K, \quad \sqrt{\hat B_d} F_{B_d}, 
\quad \xi~,
\end{equation}
that enter the standard analysis of the unitarity triangle.
While for ``reasonable ranges" (see table~\ref{tab:inputparams}) 
of these parameters,
values of $\sin 2\beta \le 0.5$ are essentially excluded, such low
values within the SM could still be possible if some of the
parameters in (\ref{par}) were chosen outside these ranges.  
In particular for $\vub\le 0.06$ or
$\hat B_K\ge 1.3$ or $\xi\ge 1.4$ the value for
$\sin 2\beta$ lower than 0.5 could be obtained within the SM. We agree
with these findings.

In what follows we will assume the "reasonable ranges'' for
the parameters as given in table~\ref{tab:inputparams}
and we will use the values of the CKM parameters determined in this
section to predict
various branching ratios for rare and CP-violating decays.
This means we will not take into account in our numerical
analyses the recent results from BaBar and Belle. We
will return to them in Section 9 where some aspects of
the physics beyond the SM will be discussed. Thus the
next four sections apply only to the SM
and could be affected considerably if the improved
measurements of  the asymmetry $a_{\psi K_S}$ 
by BaBar, Belle, CDF and D0 
will confirm with higher precision 
the present low values of $\sin 2\beta$ from 
BaBar and Belle.
This would be truely exciting!

%\clearpage
\section{$\epe$ in the Standard Model}\label{EpsilonPrime}
\setcounter{equation}{0}
\subsection{Preliminaries}
Direct CP violation remains one of the important targets 
of contemporary particle physics \cite{WW}. 
In the case of $K\to\pi\pi$,
a non-vanishing value of the ratio Re($\epe$)  
would give the first
signal for direct CP violation ruling out superweak models
\cite{wolfenstein:64} in which $\varepsilon'/\varepsilon = 0$.
Until February 1999 the experimental situation on $\epe$ was rather unclear:
\begin{equation}\label{eprime2}
\RE(\varepsilon'/\varepsilon) =\left\{ \begin{array}{ll}
(23 \pm 7)\cdot 10^{-4} & {\rm (NA31)}~\cite{barr:93}~, \\
(7.4 \pm 5.9)\cdot 10^{-4} &{\rm (E731)}~ \cite{gibbons:93}.
\end{array} \right.
\end{equation}
While the result of the NA31 collaboration at CERN 
\cite{barr:93}
clearly indicated direct CP violation, the value of E731 at Fermilab
\cite{gibbons:93}, was compatible with superweak theories.
This controversy is now settled in favour of NA31.
The most recent experimental results for the ratio $\epe$ from
Fermilab and CERN presented during 1999 and 2000 read
\begin{equation}\label{eprime1}
\RE(\varepsilon'/\varepsilon) =\left\{ \begin{array}{ll}
(28.0 \pm 4.1)\cdot 10^{-4} & {\rm (KTeV)}~\cite{KTEV}~, \\
(14.0 \pm 4.3)\cdot 10^{-4} &{\rm (NA48)}~ \cite{NA48}~.
\end{array} \right.
\end{equation}
Together with the  NA31 measurement these data confidently establish direct 
CP violation in nature and taking also the
E731 result into account one finds
the grand average \cite{NA48} 
\be
\RE(\epe) = (19.2\pm 2.4)\cdot 10^{-4}~,
\label{ga}
\ee
close to the NA31 result.

While the experimentalists should be congratulated for these
results, the situation is clearly unsatisfactory.
The KTeV result is by a factor of four larger than
the E731 result. In addition the substantial difference
between KTeV and NA48 results is disturbing. 
Let us hope that these issues will be clarified soon.
In this context an independent measurement of
$\epe$ by KLOE at Frascati, 
which uses a different experimental technique than KTeV and NA48
will be very important.

There is a long history of calculations of $\epe$ in the SM
beginning with the pioneering calculations in
\cite{EGN} and  \cite{GW79} and  
the first extensive phenomenological analyses in \cite{BSS}.
Over the 1980's  these calculations
were refined through the inclusion of QED penguin effects 
for $\mt \ll \mw$ \cite{BW84,donoghueetal:86,burasgerard:87},
the inclusion of isospin breaking in the
quark masses \cite{donoghueetal:86,burasgerard:87,lusignoli:89},
through improved estimates of hadronic matrix elements in
the framework of the $1/N$ approach \cite{bardeen:87}
and the inclusion of QCD penguins, electroweak penguins 
($\gamma$ and $Z^0$ penguins) and the relevant box diagrams for arbitrary
top quark mass \cite{flynn:89,buchallaetal:90}. The strong cancellation 
between QCD penguins and electroweak penguins for $m_t > 150~\gev$ found 
in these two papers was confirmed by other authors \cite{PW91}.

During the 1990's considerable progress has been made by
calculating complete NLO corrections to $\varepsilon'$
\cite{BJLW1}--\cite{ROMA2}. Together with the NLO
corrections to $\varepsilon$ and $B^0-\bar B^0$ mixing
\cite{HNa,BJW90,HNb}, this allowed
a complete NLO analysis of $\varepsilon'/\varepsilon$ including
constraints from the observed indirect CP violation ($\varepsilon$)
and  $B_{d,s}^0-\bar B_{d,s}^0$ mixings ($\Delta M_{d,s}$). The improved
determination of the $V_{ub}$ and $V_{cb}$ elements of the CKM matrix
and in particular the determination of the top quark mass
$\mt$ had of course also an important impact on
$\varepsilon'/\varepsilon$. 

While until 1999 only a few groups were involved in detailed analyses
of $\epe$ \cite{BJLW},\cite{flynn:89}-\cite{BERT98}, 
the new results from KTeV and NA48 prompted a number
of theorists to join the efforts to calculate this important ratio.
By the end of 2000 the theoretical $\epe$-club has more than
30 members.

The developments of the last two years include new estimates of
hadronic matrix elements using lattice simulations \cite{Lellouch}, 
various
versions of the large--N approach, chiral quark model, QCD sum
rules and other approches mentioned below. Simultaneously
extensive studies of final state interactions (FSI), isospin
breaking effects and electromagnetic effects in $\epe$ have
been pursued. Finally the impact of new physics on $\epe$
has been analyzed in various extentions of the SM.

We will briefly review these topics below.
To this end we have to recall a number of useful formulae for 
$\epe$. Other reviews of $\epe$ can be found
in \cite{BERT98}-\cite{REVOTH}
\subsection{Basic Formulae}
           \label{subsec:epeformulae}
The parameter $\varepsilon'$ is given in terms of the isospin amplitudes
$A_I$ in (\ref{eprime}). Applying OPE to these amplitudes one finds
\cite{BJLW,BJL96a}
\begin{equation}
\frac{\varepsilon'}{\varepsilon} = 
\IM \lambda_t\cdot F_{\varepsilon'},
\label{eq:epe1}
\end{equation}
where
\begin{equation}
F_{\varepsilon'} = 
\left[ P^{(1/2)} - P^{(3/2)} \right] \exp(i\Phi),
\label{eq:epe2}
\end{equation}
with
\begin{eqnarray}
P^{(1/2)} & = & r \sum y_i \langle Q_i\rangle_0
(1-\OEE)~,
\label{eq:P12} \\
P^{(3/2)} & = &\frac{r}{\omega}
\sum y_i \langle Q_i\rangle_2~.~~~~~~
\label{eq:P32}
\end{eqnarray}
Here
\begin{equation}
r = \frac{G_{\rm F} \omega}{2 |\eps| \RE A_0}~, 
\qquad
\langle Q_i\rangle_I \equiv \langle (\pi\pi)_I | Q_i | K \rangle~,
\qquad
\omega = \frac{\RE A_2}{\RE A_0}.
\label{eq:repe}
\end{equation}
Since
\begin{equation}
\Phi=\Phi_{\varepsilon'}-\Phi_\varepsilon \approx 0,
\label{Phi}
\end{equation}
$F_{\varepsilon'}$ and $\epe$
are real  to an excellent approximation.
The operators $Q_i$ have been given already in (\ref{OS1})-(\ref{OS5}).
The Wilson coefficient functions $ y_i(\mu)$
were calculated including
the complete next-to-leading order (NLO) corrections in
\cite{BJLW1}--\cite{ROMA2}. The details
of these calculations can be found there and in the review
\cite{BBL}. 
Their numerical values for $\Lms^{(4)}$ corresponding to
$\alpha_{\overline{MS}}^{(5)}(\mz)=0.119\pm 0.003$
and two renormalization schemes (NDR and HV)
are given in table 
\ref{tab:wc10smu13} \cite{EP99}.

\begin{table}[htb]
\caption[]{$\Delta S=1 $ Wilson coefficients at $\mu=\mc=1.3\gev$ for
$\mt=165\gev$ and $f=3$ effective flavours.
$y_1 = y_2 \equiv 0$.
\label{tab:wc10smu13}}
\begin{center}
\begin{tabular}{|c|c|c||c|c||c|c|}
\hline
& \multicolumn{2}{c||}{$\Lms^{(4)}=290\mev$} &
  \multicolumn{2}{c||}{$\Lms^{(4)}=340\mev$} &
  \multicolumn{2}{c| }{$\Lms^{(4)}=390\mev$} \\
\hline
Scheme & NDR & HV & NDR & HV & NDR & HV \\
\hline
$y_3$ & 0.027 & 0.030 & 0.030 & 0.034 & 0.033 & 0.038 \\
$y_4$ & --0.054 & --0.056 & --0.059 & --0.061 & --0.064 & --0.067 \\
$y_5$ & 0.006 & 0.015 & 0.005 & 0.016 & 0.003 & 0.017 \\
$y_6$ & --0.082 & --0.074 & --0.092 & --0.083 & --0.105 & --0.093 \\
\hline
$y_7/\aem$ & --0.038 & --0.037 & --0.037 & --0.036 & --0.037 & --0.034 \\
$y_8/\aem$ & 0.118 & 0.127 & 0.134 & 0.143 & 0.152 & 0.161 \\
$y_9/\aem$ & --1.410 & --1.410 & --1.437 & --1.437 & --1.466 & --1.466 \\
$y_{10}/\aem$ & 0.496 & 0.502 & 0.539 & 0.546 & 0.585 & 0.593 \\
\hline
\end{tabular}
\end{center}
\end{table}

It is customary in phenomenological
applications to take $\RE A_0$ and $\omega$ from
experiment, i.e.
\begin{equation}
\RE A_0 = 3.33 \cdot 10^{-7}\gev,
\qquad
\omega = \frac{\RE A_2}{\RE A_0}= 0.045,
\label{eq:ReA0data}
\end{equation}
where the last relation reflects the so-called $\Delta I=1/2$ rule.
This strategy avoids to a large extent the hadronic uncertainties 
in the real parts of the isospin amplitudes $A_I$.

The sum in (\ref{eq:P12}) and (\ref{eq:P32}) runs over all contributing
operators. $P^{(3/2)}$ is fully dominated by electroweak penguin
contributions. $P^{(1/2)}$ on the other hand is governed by QCD penguin
contributions which are suppressed by isospin breaking in the quark
masses ($m_u \not= m_d$). The latter effect is described by
\begin{equation}
\Omega_{\rm IB} = \frac{1}{\omega} \frac{(\IM A_2)_{\rm
IB}}{\IM A_0}\,.
\label{eq:Omegaeta}
\end{equation}

The main source of uncertainty in the calculation of
$\epe$ are the hadronic matrix elements $\langle Q_i \rangle_I$ and
$\Omega_{\rm IB}$ as we will see later on.
 $\langle Q_i \rangle_I$ generally depend
on the renormalization scale $\mu$ and on the scheme used to
renormalize the operators $Q_i$. These two dependences are canceled by
those present in the Wilson coefficients $y_i(\mu)$ so that the
resulting physical $\epe$ does not (in principle) depend on $\mu$ and on the
renormalization scheme for the operators.  Unfortunately, the accuracy of
the present non-perturbative methods used to evalutate $\langle Q_i
\rangle_I$  is not
sufficient to have the $\mu$ and scheme dependences of
$\langle Q_i \rangle_I$ fully under control. 
We believe that this situation will change once the lattice calculations
improve.
A brief review of the existing methods 
including most recent developments will be given below.

\subsection{An Analytic Formula for $\epe$}
           \label{subsec:epeanalytic}
The basic formulae for $\epe$ presented above are not very transparent
and we would like to improve on this. To this end let us first
determine as many matrix elements $\langle Q_i \rangle_I$ as possible
from the leading CP conserving $K \to \pi\pi$ decays, for which the
experimental data is summarized in (\ref{eq:ReA0data}). 
This determination is particularly easy if one sets $\mu
= m_c$ \cite{BJLW}. The details of this approach will not be 
discussed here.
It sufficies to say that
this method allows to determine only the matrix
elements of the $(V-A)\otimes(V-A)$ operators. Explicit formulae
can be found in \cite{BJLW}.
For the central value of $\IM\lambda_t$
these operators give a negative contribution to $\epe$ 
of about $-2.5\cdot 10^{-4}$. This shows that they
are only relevant if  $\epe$ is below $1 \cdot 10^{-3}$.

Unfortunately the method in \cite{BJLW} does not provide
 the matrix elements of the dominant $(V-A)\otimes(V+A)$
operators and
one has to use  non-perturbative methods to estimate them.
In this context the calculations of Bertolini and collaborators
\cite{BERT98} within the chiral quark model should be mentioned.
These authors have determined the parameters of this model
by taking the constraints (\ref{eq:ReA0data}) into account.
In this manner they were able to determine not only the
matrix elements of the $(V-A)\otimes(V-A)$ operators,
as in \cite{BJLW}, but also the matrix elements of the 
$(V-A)\otimes(V+A)$ operators. This is clearly interesting.
On the other hand, it is not clear how well the chiral
quark model approximates QCD and consequently whether the
extracted values of the relevant matrix elements have
anything to do with the true QCD values. For this reason
we will not use them here.

In order to exhibit the matrix elements of the dominant
$(V-A)\otimes(V+A)$ operators
 in an analytic formula
for $\epe$, we first express them
in terms of non-perturbative parameters
$B_i^{(1/2)}$ and $B_i^{(3/2)}$ as follows:
\begin{equation}
\langle Q_i \rangle_0 \equiv B_i^{(1/2)} \, \langle Q_i
\rangle_0^{\rm (vac)}\,,
\qquad
\langle Q_i\rangle_2 \equiv B_i^{(3/2)} \, \langle Q_i
\rangle_2^{\rm (vac)} \,.
\label{eq:1}
\end{equation}
The label ``vac'' stands for the vacuum
insertion estimate of the hadronic matrix elements in question 
for
which $B_i^{(1/2)}=B_i^{(3/2)}=1$.

The Wilson coefficients $y_5$, $y_7$ and $y_8$ are much smaller
than $y_6$. On the other hand the contributions of 
$\langle Q_{7,8}\rangle_2$ are
enhanced by the factor $1/\omega\approx 22$ as seen in
(\ref{eq:P32}). But $y_7$ is substantially smaller than $y_8$
and $\langle Q_7\rangle_{0,2}$ are colour suppressed.
It is then not surprising \cite{BJLW} that $\epe$ is only
weakly sensitive to the values of the parameters
 $B_5^{(1/2)}$, $B_7^{(1/2)}$, $B_8^{(1/2)}$
and $B_7^{(3/2)}$ as long as their absolute values are not
substantially larger than 1. On the basis of existing non-perturbative
approaches we can assume that this is indeed the case.

Following \cite{BJLW} we set then
\begin{equation}
B_{7,8}^{(1/2)}(\mc) = 1,
\qquad
B_5^{(1/2)}(\mc) = B_6^{(1/2)}(\mc)\equiv \bsi,
\qquad
B_7^{(3/2)}(\mc) = B_8^{(3/2)}(\mc)\equiv \bei~.
\label{eq:B1278mc}
\end{equation}

This strategy \cite{BJLW} allows then 
to express $\epe$ or equivalently $F_{\varepsilon'}$ in terms of
of the following parameters
\be\label{pare}
\bsi,\qquad\bei,\qquad\OEE,\qquad\ms,\qquad\mt,\qquad
\Lms^{(4)}~.
\ee 
The appearence of $\ms$ originates in
\begin{equation}
 \langle Q_6\rangle_0^{\rm (vac)}\sim 1/\ms^2(\mu),
\qquad
\langle Q_8\rangle_2^{\rm (vac)}\sim 1/\ms^2(\mu)~.
\label{eq:5}
\end{equation}
$\Lms^{(4)}$ and $\mt$ enter through the coefficients $y_i$.

With all this information at hand,
it is possible   \cite{BJL96a,buraslauten:93} to cast the formal
expressions for $\epe$ in (\ref{eq:epe1})--(\ref{eq:P32})
into an analytic formula which exhibits the  dependence
on the parameters in (\ref{pare}). We set $\OEE=0.16$
for the moment. We will return to it below.
The analytic formula given below, while being rather accurate, 
exhibits
various features which are not transparent in a pure numerical
analysis. It can be used in phenomenological applications if
one is satisfied with a few percent accuracy. Needless to say, 
in the numerical analysis \cite{EP99,EP00} presented below
 we have used exact expressions.

In this formulation
the function $F_{\varepsilon'}$
is given simply as follows:
\begin{equation}
F_{\varepsilon'} =
P_0 + P_X \, X_0(x_t) + P_Y \, Y_0(x_t) + P_Z \, Z_0(x_t) 
+ P_E \, E_0(x_t)
\label{eq:3b}
\end{equation}
with the $\mt$-dependent functions given in subsection 2.6.  

The coefficients $P_i$ are given in terms of $B_6^{(1/2)}$,
 $B_8^{(3/2)}$ and $\ms(\mc)$
as follows:
\begin{equation}
P_i = r_i^{(0)} + 
r_i^{(6)} R_6 + r_i^{(8)} R_8 \, 
\label{eq:pbePi}
\end{equation}
where
\be\label{RS}
R_6\equiv \bsi\left[ \frac{137\mev}{\ms(\mc)+\md(\mc)} \right]^2,
\qquad
R_8\equiv \bei\left[ \frac{137\mev}{\ms(\mc)+\md(\mc)} \right]^2.
\ee
The $P_i$ are renormalization scale and scheme independent. They depend,
however, on $\Lms^{(4)}$. In table~\ref{tab:pbendr} we give the numerical
values of $r_i^{(0)}$, $r_i^{(6)}$ and $r_i^{(8)}$ for different values
of $\Lms^{(4)}$ at $\mu=\mc$ in the NDR renormalization scheme
\cite{EP00}. Actually at NLO only $r_0$ coefficients are renormalization
scheme dependent. The last row gives them in the HV scheme.
The inspection of table~\ref{tab:pbendr} shows
that the terms involving $r_0^{(6)}$ and $r_Z^{(8)}$ dominate the ratio
$\epe$. Moreover, the function $Z_0(x_t)$ representing a gauge invariant
combination of $Z^0$- and $\gamma$-penguins grows rapidly with $\mt$
and due to $r_Z^{(8)} < 0$ these contributions suppress $\epe$ strongly
for large $\mt$ \cite{flynn:89,buchallaetal:90}.

\begin{table}[thb]
\caption[]{Coefficients in the formula (\ref{eq:pbePi})
 for various $\Lms^{(4)}$ in 
the NDR scheme.
The last row gives the $r_0$ coefficients in the HV scheme.
\label{tab:pbendr}}
\begin{center}
\begin{tabular}{|c||c|c|c||c|c|c||c|c|c|}
\hline
& \multicolumn{3}{c||}{$\Lms^{(4)}=290\mev$} &
  \multicolumn{3}{c||}{$\Lms^{(4)}=340\mev$} &
  \multicolumn{3}{c| }{$\Lms^{(4)}=390\mev$} \\
\hline
$i$ & $r_i^{(0)}$ & $r_i^{(6)}$ & $r_i^{(8)}$ &
      $r_i^{(0)}$ & $r_i^{(6)}$ & $r_i^{(8)}$ &
      $r_i^{(0)}$ & $r_i^{(6)}$ & $r_i^{(8)}$ \\
\hline
0 &
  --3.122  &   10.905  &    1.423  &
  --3.167  &   12.409  &    1.262  & 
  --3.210  &   14.152  &    1.076  \\
$X$ &
    0.556  &    0.019  &    0      &  
    0.540  &    0.023  &    0      &  
    0.526  &    0.027  &    0      \\
$Y$ &
    0.404  &    0.080  &    0      & 
    0.387  &    0.088  &    0      & 
    0.371  &    0.097  &    0      \\
$Z$ &
    0.412  &  --0.015  &  --9.363  & 
    0.474  &  --0.017  & --10.186  & 
    0.542  &  --0.019  & --11.115  \\
$E$ &
    0.204  &  --1.276  &    0.409  & 
    0.188  &  --1.399  &    0.459  &  
    0.172  &  --1.533  &    0.515  \\
\hline
0 &
  --3.097  &    9.586  &    1.045  &
  --3.141  &   10.748  &    0.867  &
  --3.183  &   12.058  &    0.666  \\
\hline
\end{tabular}
\end{center}
\end{table}

Finally by investigating numerically the formula (\ref{eq:3b})
it is possible to find a crude approximation for
$F_{\varepsilon'}$:
\be\label{crude}
F_{\varepsilon'}\approx 13\cdot 
\left[\frac{110\mev}{\ms(2~\gev)}\right]^2
\left[\bsi(1-\OEE)-0.4\cdot \bei\left(\frac{\mt}{165\gev}\right)^{2.5}\right]
\left(\frac{\Lms^{(4)}}{340~\mev}\right)~.
\ee
This formula while exhibiting very transparently the dependence
of $\epe$ on the parameters (\ref{pare})
should not be used for any serious numerical analysis.

We observe the known features:
\bi
\item
$F_{\varepsilon'}$ increases with increasing $\Lms^{(4)}$. In fact
the renormalization group effects play an important role here.
If one did not include them $\epe$ would be typically $\ord(10^{-5})$.
On the other hand the formula (\ref{crude}) is only valid for
$\Lms^{(4)}> 200 \mev$. $F_{\varepsilon'}$ does not vanish for
$\Lms^{(4)}=0$.
\item
$F_{\varepsilon'}$ increases with increasing $\bsi$ that represents
QCD penguins. It decreases with increasing $\bei$ that represents
electroweak penguins. The latter contribution increases with
increasing $\mt$.
\item
The partial 
cancellation between QCD penguin ($\bsi$) and electroweak 
penguin ($\bei$) contributions requires 
accurate 
values of $\bsi$ and $\bei$ for an acceptable estimate of $\epe$. 
Because of the accurate value $\mt(\mt)=166\pm 5~\gev$, the uncertainty 
in $\epe$ due to the top quark mass amounts only to a few percent. 
\item
The $1/\ms^2$ dependence is an artifact of the decomposition
of the matrix elements into $B_i$--factors and the vacuum insertion
matrix elements as given in (\ref{eq:1}). We will return to this
point below.
\end{itemize}

\subsection{The Status of $\ms$, $\bsi$, $\bei$, $\OEE$ and $\Lms^{(4)}$}
\subsubsection{$\ms$}
The most recent values for $\ms(2\gev)$ extracted from lattice calculations,
QCD sum rules and $\tau$--decays are summarized in table~\ref{tabms}.
The lattice result is the most recent ``world average" of Lubicz
\cite{Lubicz}. See also \cite{GUPTA01}. 
It is higher than the older result from Gupta \cite{GUPTA98}
but could decrease a bit when the calculations with dynamical fermions
are completed. 
Significant progress has been made in the last two
years by the introduction of improved actions and non-perturbative
renormalization techniques. Quenched calculations give 
$\ms(2\gev)=(110\pm15)~\mev$ and the increased error in table~\ref{tabms}
gives the estimate of the quenching error based on the ca"culations
with two flavours. Details with references to calculations of
different groups can be found in \cite{Lubicz}.
The recent QCD sum rule values
are considerably smaller than the older ones \cite{QCDS}. The
same comment applies to $\ms$ from $\tau$--decays.

While the situation regarding $\ms$ improved considerably during the
last two years and the values obtained using different methods
are coming closer to each other, the error on $\ms$ is still large.
In our numerical analysis of $\epe$ we will use
\begin{equation}\label{ms}
\ms(\mu) =\left\{ \begin{array}{ll}
(110\pm20)\;\mev & \mu=2\gev \\
(130\pm25)\;\mev & \mu=\mc 
\end{array} \right.
\end{equation}
which is in the ball park of the results given in 
table~\ref{tabms}. Here $\mc=1.3\gev$.

Five years ago values like $\ms(2\gev)=130-150\mev$ could be found
in the literature. Beginning with the work of Gupta \cite{GUPTA98}
the values of $\ms$ decreased considerably, which makes the values
of $\epe$ larger. On the other hand QCD sum rules  allow to
derive lower bounds on the strange quark mass. It is found that generally
$\ms(2\gev)\gsim 100\,\mev$ \cite{MSBOUND}. We observe that
lattice results are very close to this bound.

\begin{table}[thb]
\caption[]{$\ms(2\gev)$ obtained using various methods. 
 \label{tabms}}
\vspace{0.4cm}
\begin{center}
\begin{tabular}{|c||c||c|}\hline
{\bf Method} & {\bf $\ms(2\gev)~[\mev]$} & {\bf Reference} \\ \hline
Lattice &$110\pm 25$ & Lubicz \cite{Lubicz} \\ \hline
QCDS & $119\pm12$ & Narison\cite{NARM} \\ \hline
QCDS & $116\pm22 $ & Jamin \cite{JAM} \\ \hline
QCDS & $115\pm8 $ & Maltman \cite{MAL} \\ \hline
$\tau$--Decays &$114\pm 23$ & Pich-Prades \cite{PiPr} \\ \hline
\end{tabular}
\end{center}
\end{table}

The large sensitivity of $\epe$ to $\ms$ has been known 
since the analyses in the 1980's. In the context of
the KTeV result this issue has been analyzed in \cite{Nierste}.
It has been found that provided $2\bsi-\bei\le 2$ the
consistency of the SM with the KTeV result
requires the $2\sigma$ bound $\ms(2\gev)\le 110\mev$.
Our analysis below is compatible with these findings.

On the other hand, it has been emphasized by 
Guido Martinelli, Eduardo de Rafael and other researchers 
in the field that the
study of $\epe$ as a function of $\ms$, independently of $\bsi$
and $\bei$, could be misleading. Indeed a correlation could exist
between $\ms$ and the latter parameters. In the large--N approach
of \cite{bardeen:87,DORT98} such correlation is not observed.
However, I can imagine that in an approach, like lattice
simulations, that calculates the hadronic matrix elements
directly, without parametrizing them in terms of $\ms$ and
$B_i$--factors, some correlations in values of these parameters
could certainly be present. It will be interesting to study this
issue once the lattice calculations improve.

From my point of view, the strange quark mass  
 must be somehow related 
to the matrix elements $\langle Q_6\rangle_0$ and 
$\langle Q_8\rangle_2$. Afterall $Q_6$ and $Q_8$ are density-density
operators and in
the large--N limit the anomalous dimensions of these operators
are equal to $-2\gamma_m$, where $\gamma_m$ is the anomalous
dimension of the mass operator \cite{burasgerard:87}. 
Consequently the $\mu$-dependence
of $\langle Q_6\rangle_0$ and 
$\langle Q_8\rangle_2$ is governed to a very good approximation
by the $\mu$-dependence of $\ms(\mu)$ as shown in (\ref{eq:5}).
This has been verified numerically beyond the large--N limit
in \cite{BJLW}. This implies that $\bsi$ and $\bei$ are
practically $\mu$-independent and from the point of view of
the $\mu$-dependence $\ms$ is indeed uncorrelated with $\bsi$
and $\bei$. On the other hand I can imagine that at fixed $\mu$,
there is some correlation between the actual values of $\ms$,
$\bsi$ and $\bei$.

In this context it should be recalled \cite{burasgerard:87}
that the $\mu$-dependence of $\langle Q_6\rangle_0$ and 
$\langle Q_8\rangle_2$,
just discussed, is to a very good approximation canceled by
the $\mu$-dependence of $y_6(\mu)$ and $y_8(\mu)$ respectively.
This means that the unphysical $\mu$-dependence in $\epe$
is very small already in the present calculations. This is
a nice feature, which can be seen explicitly within the
lare--N approach.

\subsubsection{$\bsi$ and $\bei$}
The values for $\bsi$ and $\bei$ obtained in various approaches
are collected in table~\ref{tab:317}. 
The lattice results have been obtained at $\mu=2\gev$.
The results in the large--N approach and the chiral quark model
correspond to scales below $1\gev$.
However, as explained above,
$\bsi$ and $\bei$ are only weakly dependent on $\mu$.
Consequently
the  comparison
of these parameters
obtained in different approaches at different $\mu$ is meaningful
as long as $\mu> 0.8~\gev$.

Next, the values coming from lattice and
chiral quark model are given in the NDR renormalization 
scheme. The
corresponding values in the HV scheme can be found
using approximate relations \cite{EP99} 
\be\label{NDRHV}
(\bsi)_{\rm HV}\approx 1.2 (\bsi)_{\rm NDR},
\qquad
 (\bei)_{\rm HV}\approx 1.2 (\bei)_{\rm NDR}.
\ee
The scheme dependence of these parameters in the large-N approach are 
to my knowledge not yet under full control but some progress is
being made \cite{PerRaf,Prades,Bardeen}.

Concerning the
lattice results for $B^{(1/2)}_{6}$,
the old results read
$B^{(1/2)}_{5,6}(2~\gev)=1.0 \pm 0.2$ \cite{kilcup:91,sharpe:91}.
However, over the last years it has been realized beginning
with \cite{kilcup:99} that lattice calculations of $\bsi$ are 
very uncertain. One has
to conclude that there are no solid predictions for
$B^{(1/2)}_{6}$ from the lattice at present.
There is a hope that soon results using domain wall fermions not
only for $\bsi$ and $\bei$ but for all hadronic matrix elements
entering $\epe$ will be available. A very informative and 
critical review of the present situation has been given recently
by Laurent Lellouch \cite{Lellouch}.

The result for $\bei$ from the dispersive approach \cite{Golowich} 
corresponds to $\ms(2\gev)\approx 100\mev$. In this approach
$\bei$ increases with $\ms$. The results for $\bei$ from the
$1/N$ approach of de Rafael and collaborators are not yet available. Their
analysis of the electroweak operator $Q_7$ can be found in \cite{Knecht}.

\begin{table}[thb]
\caption[]{ Results for $\bsi$ and $\bei$ obtained
in various approaches. 
\label{tab:317}}
\begin{center}
\begin{tabular}{|c|c|c|}\hline
  { Method}& $\bsi$& $B^{(3/2)}_8$  \\
 \hline
Lattice\cite{GKS,G67,APE}&$-$ &$0.69-1.06$  \\
Large$-$N\cite{DORT98,DORT99}& $0.72-1.10$ &$0.42-0.64$ \\
ChQM\cite{BERT98}& $1.07-1.58$ &$0.75-0.79$  \\
Dispersive Approach \cite{Golowich}&$-$ & $1.11\pm0.16\pm0.23$\\
Large$-$N (NJL)\cite{Prades}& $2.9\pm0.7$ &$1.3\pm0.2$  \\
Sum Rules\cite{Narison}& $2.8\pm0.5$ &$1.7\pm0.4$  \\
\hline
\end{tabular}
\end{center}
\end{table}

Table~\ref{tab:317} demonstrates very clearly that there is no
consensus on the values of $\bsi$ and $\bei$. In particular
the last two groups find both parameters to be substantially
higher than  obtained by the other groups. It should
be remarked that the Dortmund group \cite{DORT98,DORT99}
generalized the previous leading order large--N calculations
\cite{BBG0,burasgerard:87,bardeen:87} to include higher order 
$1/N$ corrections.
They find that in the chiral limit $\ord(p^2/N)$ correction
not included in the result in table~\ref{tab:317} enhance
$\bsi$ to roughly 1.5. As the chiral logs have not been taken
into account it is probably premature to use this result
in phenomenological applications.

Biased to some
extent by the results from the large-N approach and lattice
calculations, we will use
in our numerical analysis below $\bsi$ and $\bei$ in
the ranges:
\be\label{bbb}
\bsi=1.0\pm0.3,
\qquad
\bei=0.8\pm 0.2
\ee
keeping always $\bsi\ge \bei$.

\subsubsection{$\OEE$ and $\Lms^{(4)}$}
The older estimates of $\OEE$ in the $1/N$ approach
\cite{burasgerard:87} and in chiral perturbation theory
\cite{donoghueetal:86,lusignoli:89} gave the value $0.25\pm0.10$.
The most recent refined calculation in \cite{ECKER99} 
gives 
\begin{equation}
\OEE = 0.16\pm 0.03~.
\label{eq:Omegaetadata}
\end{equation}
We will adopt this value here.
On the other hand  the story of $\OEE$
is not finished yet. We will discuss it below.

The dependence of $\epe$ on $\OEE$ can be studied numerically by 
using the formula (\ref{eq:P12}) or incorporated approximately
into the analytic formula (\ref{eq:3b}) by simply replacing
$\bsi$ with an effective parameter
\be\label{eff}
(\bsi)_{\rm eff}=\bsi\frac{(1-0.9~\OEE)}{0.856}
\ee
A numerical analysis shows that using $(1-\OEE)$
overestimates the role of $\OEE$.

In table~\ref{tab:inputp} we summarize the input parameters
used in the numerical analysis of $\epe$ below.
The range for $\Lms^{(4)}$ in table~\ref{tab:inputp} 
corresponds roughly to $\alpha_s(\mz)=0.119\pm 0.003$,
which is very close to the recent determination in
\cite{Bethke}.

\begin{table}[thb]
\caption[]{Collection of input parameters.
We impose $\bsi\ge\bei$.
\label{tab:inputp}}
\vspace{0.4cm}
\begin{center}
\begin{tabular}{|c|c|c|c|}
\hline
{\bf Quantity} & {\bf Central} & {\bf Error} & {\bf Reference} \\
\hline
$\Lms^{(4)}$ & $340 \mev$ & $\pm 50\mev$ & \cite{PDG,Bethke} \\
$\ms(\mc)$ & $130\mev$    & $\pm 25\mev$ & See Text\\
$\bsi $ & 1.0 & $\pm 0.3$ & See Text\\
$\bei $ & 0.8 & $\pm 0.2$ & See Text\\
$\IM\lambda_t$ & 1.33 & $\pm 0.14$ (G)& \cite{EP99} \\
$\IM\lambda_t$ & 1.33 & $\pm 0.30$ (G)& \cite{EP99}\\
\hline
\end{tabular}
\end{center}
\end{table}

\subsection{Numerical Results for $\epe$}
 We list the results from various groups in
table \ref{tab:31738}.
The labels (G) and (S) in the second column
stand for two error estimates: ``Gaussian'' and ``Scanning'' 
respectively.
The result from the Munich group given here is an update of
the analysis in \cite{EP99} done in \cite{EP00}. 
It uses the input parameters of table~\ref{tab:inputp} where
the value of $\IM\lambda_t$ from \cite{EP00,EP99}  is slightly
higher than the one found in the previous section. 
Similarly the
result of the Rome group \cite{ROMA99} is the most recent estimate given 
in \cite{CM00}.

In \cite{EP99,EP00,ROMA99,CM00} $\epe$ has 
been found to be typically by a factor of 2-3 below the data and 
the KTeV result in (\ref{eprime1})
could only be accomodated if all relevant parameters were chosen
simultaneously close to their extreme values. 
On the other hand the NA48 result is essentially compatible with
\cite{EP99,EP00,ROMA99,CM00} within experimental and theoretical errors.
Higher values of $\epe$ than in \cite{EP99,EP00,ROMA99,CM00},
in the ballpark of (\ref{ga}), have
been found in \cite{BERT98,Dortmund,Narison,Prades,Tajpei,Beijing}. 
The result in \cite{BEL} corresponds to $\bsi=\bei=1$ 
as the Dubna-DESY group has no estimate of these non-perturbative
parameters.
Recent reviews can be found in \cite{BERT98}-\cite{REVOTH}.
Furthermore it has also been
suggested that the final state interactions (FSI) could enhance $\epe$ 
by a factor of two \cite{BERT98,PAPI99,PA99}. We say a few words about
it below.

\begin{table}[thb]
\begin{center}
\begin{tabular}{|c|c|}\hline
  {\bf Reference}&  $\epe~[10^{-4}]$ \\ \hline
Munich
\cite{EP99,EP00}&  $9.2^{+6.8}_{-4.0}$ (G) \\
Munich
\cite{EP99,EP00}&   $1.4\to 32.7$ (S) \\
\hline
Rome
\cite{ROMA99,CM00}& $8.1^{+10.3}_{-9.5}$ (G) \\
Rome
\cite{ROMA99,CM00}& $-13.0\to 37.0$ (S) \\
\hline
Trieste
\cite{BERT98}&  $22\pm 8 $ (G) \\
Trieste
\cite{BERT98}&  $9\to 48 $ (S) \\
\hline
Dortmund
\cite{Dortmund}&    $6.8\to 63.9$ (S) \\
\hline
Montpellier
\cite{Narison}&  $ 24.2\pm 8.0$ \\
\hline
Granada-Lund
\cite{Prades}&   $ 34\pm 18$ \\
\hline
Dubna-DESY
\cite{BEL} 
&  $-3.2 \to 3.3$ (S) \\
\hline
Taipei
\cite{Tajpei}&  $7 \to 16$ \\
\hline
Barcelona-Valencia
\cite{PAPI99}&  $17 \pm 6$ \\
\hline
Beijing \cite{Beijing} & $16.2 \to 39.3 $ \\
\hline
\end{tabular}
\caption[]{ Results for $\epe$ in the SM in  units of $10^{-4}$.
\label{tab:31738}}
\end{center}
\end{table}

\subsection{Renormalization Scheme Dependence}
All the results presented here apply to the NDR scheme. 
They are lower by roughly $30\%$ in the
HV scheme if the same values for $(\bsi,\bei)$ are used.
On the other hand, if simultaneously $(\bsi,\bei)$ are
transfered to the HV scheme by means of (\ref{NDRHV}), the
scheme dependence is reduced. However, the game of changing
renormalization schemes is only meaningful if a given
non-perturbative method has renormalization scheme dependence
of the matrix elements fully under control. This is not really the
case in the present approaches. Future lattice calculations have
the best chance to make progress here, but it is very desirable
to have also analytic solutions to this problem.

In this context it should be emphasized that the cancellation of
the renormalization scheme dependence in the electroweak
penguin sector requires to go beyond the NLO approximation
for $y_i(\mu)$ \cite{EP99}. The reason is that the dominant
effect of the electroweak penguins related to the functions
$X_0$, $Y_0$ and $Z_0$ in (\ref{eq:3b}) enters first at the
NLO level in the renormalization group improved perturbation
theory. Consequently the issue of the scheme dependence in this
sector is shifted to the NNLO level. In particular QCD corrections
to the one--loop functions $X_0$, $Y_0$ and $Z_0$ have to be
calculated. This calculation is now available \cite{BGH}.
However, in order to complete the NNLO analysis, corresponding
corrections to gluon penguin diagrams and the three loop
anomalous dimension  matrices must be calculated. A very
difficult task. The general structure for Wilson coefficients
at NNLO can be found in \cite{BGH}.

\subsection{Final State Interactions}
What is the role of final state interactions in $\epe$?
This question has been addressed recently by several authors 
\cite{BERT98,Dortmund,PAPI99,PA99}. Actually the issue of FSI in
$K\to\pi\pi$ decays is not new. It has been known from the
analyses in \cite{bardeen:87,Kambor} that the pion loops representing
FSI provide a sizable enhancement of $\Delta I=1/2$ transitions
and suppression of $\Delta I=3/2$ transitions helping in the
explanation of the $\Delta I=1/2$ rule. The 
analyses addressing this rule are given in \cite{DI12,Beijing,GW00}, 
where further references can be found. 

If this pattern of enhancements and suppressions also
applies to the matrix elements of the penguin operators
relevant for $\epe$, this would mean that FSI effects
enhance the QCD penguin contributions and suppress the
electroweak penguin contributions making $\epe$ larger than
in the case of neglecting these effects. In this
context an interesting proposal has been made by Pallante
and Pich \cite{PAPI99}, who following an older work by
Truong \cite{Truong}, suggested that the FSI effects in question can be
unambiguosly resummed to all orders in chiral perturbation
thory using the so-called Omnes factor. Assuming in addition
that this factor should be the dominant FSI effect relative to
large-N estimates of hadronic matrix elements, they found
that the result for $\epe$ from the Munich group \cite{EP99,EP00}
can be enhanced by roughly a factor of two.

A critical analysis
of this suggestion has been presented in \cite{AITAL}.
A nice summary of the points made in this paper can be found
in \cite{ISI00}. Personally, I think that the sign of the
effect calculated by Pallante and Pich  could be correct
but in view of the critics made in \cite{AITAL}, the
actual size of FSI in the evaluation of $\epe$ given in 
\cite{PAPI99} cannot be really trusted. 
Most probably a satisfactory solution to the problem of
FSI can only be achieved in a complete calculation of
the hadronic matrix elements within a self-consistent framework.
In this context, an interesting
recent proposal in \cite{LLs}, if realized, could put the inclusion
of FSI in the lattice calculations of hadronic matrix elements
in principle under control.

\subsection{Isospin Breaking Effects}
Isospin breaking effects are generated by the $m_u-m_d$ difference
and by electromagnetic corrections. While they are generally small
in the kaon system (roughly $1\%$) 
and can be neglected in view of other uncertainties,
they may have an impact on $\epe$. Indeed as seen in (\ref{eq:Omegaeta}),
$\OEE$ is enhanced by $1/\omega\approx 22$, implying that $\OEE$ in
the ball park of 0.20 could certainly be expected.

Let us first concentrate on the isospin breaking corrections due
to $m_u\not=m_d$. In the lowest order of chiral perturbation theory
($(m_u-m_d)p^0$) only the $\pi^0-\eta$ mixing in the strong
Lagrangian contributes and one finds $\OEE=0.13$ 
\cite{lusignoli:89,ECKER99}.
The older estimates of ($(m_u-m_d)p^2$) contributions to $\OEE$, 
which include also $\pi^0-\eta-\eta'$ mixing, within the large--N
approach \cite{burasgerard:87} and in chiral perturbation theory
\cite{donoghueetal:86,lusignoli:89} gave the total value $0.25\pm0.10$.
The most recent refined calculation in \cite{ECKER99} 
gives somewhat smaller value $\OEE=0.16\pm0.03$ that we have 
adopted in our numerical analysis.

However, as stressed in \cite{Gardner,Wolfe}, at this level also
$\ord(p^4)$ weak counterterms have to be taken into account.
At present these counterterms can only be estimated by making
some specific dynamical assumptions. As demonstrated in
\cite{Gardner,Wolfe} these additional contributions can compete
with the ones included sofar. Interestingly, they have the
tendency of decreasing $\OEE$ and even to reverse its sign
making $\epe$ larger. Unfortunately a quantitative estimate of 
these contributions is plaqued by very large uncertainties.
Consequently, if one wants to be conservative, an error of $100\%$
should be assigned to $\OEE$ with a central value arround 0.1.

Concerning the electromagnetic corrections, a recent analysis
indicates that they have only a small impact on $\OEE$ 
\cite{CIRI1,CIRI2}.
On the other hand, as pointed out in \cite{CIRI2,CIRI3}, the
electromagnatic corrections modify the amplitude decompositions
in (\ref{ISO1})--(\ref{ISO3}). 
In particular the $\Delta I=5/2$ contributions to all
three decays have to be included. These contributions could
have, in principle, an impact on $\epe$, but their size
is not understood at present \cite{CIRI2}-\cite{Gardner2}.
\subsection{Summary}
As of the beginning of the third millennium, we know confidently that
there is a direct CP violation in $K_L\to \pi\pi$ decays.
The ratio $\epe$ is measured to be around $2\cdot 10^{-3}$
but values as low as $1\cdot 10^{-3}$ are in view of the most
recent NA48 results and the older E731 results not yet excluded.
We should know the truth in the next few years.

While the present theoretical status of $\epe$ is rather
unsatisfactory, with various estimates ranging from
$5\cdot 10^{-4}$ to $4\cdot 10^{-3}$, $\epe$ within the SM
agrees within theoretical and experimental uncertainties
with the data. The sign and the order of magnitude have
been correctly predicted by theorists.
Unfortunately, in view of very large hadronic and substantial parametric 
uncertainties, it is impossible at present  
to see what is precisely the impact of 
the $\epe$-data on the CKM matrix. 

The short distance contributions to $\epe$ within the SM are fully
under control. On the other hand considerable progress has to be made in
the evaluation of the matrix elements $\langle Q_i\rangle_{0,2}$. 
The calculations
should not be confined to $\langle Q_6\rangle_0$ and 
$\langle Q_8\rangle_2$ but should include
matrix elements of all contributing operators. Personally, I believe
that lattice calculations have eventually the best chance to achieve
this goal, but this may still take several years of extensive work 
\cite{Lellouch}.
On the other, it is very important to continue analytic efforts
in order to confront the future lattice results. Finally isospin
breaking corrections should also be taken into account. Here
I expect analytic tools to be more powerful than lattice simulations
for some time.

In any case in view of very large theoretical uncertainties 
and sizable experimental errors there is still a lot of room for 
non-standard contributions to $\epe$.
Indeed  results from NA31, KTeV and NA48
prompted several  analyses
of $\epe$ within various extensions of the SM
like general supersymmetric models 
\cite{Nierste,MM99,BS99,KN,SIL99},
models with anomalous gauge couplings \cite{HE}, four-generation models
\cite{Huang}, left-right symmetric models \cite{Jang} and 
models with additional
fermions and gauge bosons \cite{Frampton}.
Unfortunately several of these extensions have
many free parameters and are not very conclusive at present.
One should also remember that the hadronic uncertainties in the
SM are also present in its extensions.
The situation may change in the future when the calculations
of hadronic matrix elements will improve and new data from high
energy colliders will restrict the possible ranges of parameters
involved. For instance a recent analysis of the bounds
on anomalous gauge couplings from LEP2 data indicates that
substantial enhancements of $\epe$ from this sector are very
unlikely \cite{TERR}.

On the other hand from the point of view of Munich and Rome groups 
 the $\epe$ data puts models 
in which there are new positive contributions to $\eps$ and 
negative contibutions to 
$\varepsilon'$ in  difficulties. In particular 
as analyzed in \cite{EP99} the two Higgs Doublet Model II
can either be ruled out  with improved 
hadronic matrix elements or a  
powerful lower bound on $\tan\beta$ can 
be obtained from $\epe$.
In the Minimal Supersymmetric SM (MSSM), in addition to charged
Higgs exchanges in loop diagrams, also charginos contribute.
For suitable  choice of the
supersymmetric parameters, the chargino contribution
can enhance $\epe$  with respect to the SM expectations \cite{GG}.
Yet, as found in \cite{GG,EP00},
the most conspicuous effect of
minimal supersymmetry is a depletion of $\epe$. We will quantify this
in section 9.
The situation can be different in more general
models in which there are more parameters than
in the two Higgs doublet model II and in the MSSM, in particular
new CP violating phases.
As an example, in general supersymmetric models
$\epe$ can be considerably enhanced
through the contributions of the chromomagnetic
penguins \cite{GMS,Nierste,MM99,BS99}, $Z^0$-penguins
with the opposite sign to the one in the SM
\cite{ISI,BS98,BS99,CHANOWITZ} and isospin breaking effects 
in the squark sector \cite{KN}.

The future of $\epe$ in the SM and in its extensions depends on 
the progress in the reduction of parametric and hadronic uncertainties. 
In any case $\epe$ already played a decisive role in establishing direct 
CP violation in nature and its quite large value gives additional strong 
motivation for searching for this phenomenon in 
cleaner K decays like 
$K_L\to\pi^0\nu\bar\nu$ and 
$K_L\to\pi^0 e^+ e^-$, in B decays, in D decays and elsewhere.
We now turn to discuss some of these topics.

\section{ The Decays $K^+\to\pi^+\nu\bar\nu$ and
$K_{\rm L}\to\pi^0\nu\bar\nu$}
         \label{sec:HeffRareKB}
\setcounter{equation}{0}
\subsection{General Remarks}
            \label{sec:HeffRareKB:overview}
We will now move to discuss
the semileptonic rare FCNC
transitions $\kpn$ and $K_{\rm L}\to\pi^0\nu\bar\nu$.
Within the SM these decays are loop-induced
semileptonic FCNC processes determined only 
by $Z^0$-penguin and box diagrams
and  are governed by the single
function $X_0(x_t)$ given in (\ref{XA0}).

A particular and very important virtue of $K\to\pi\nu\bar\nu$
is their clean theoretical character.
This is related to the fact that
the low energy hadronic
matrix elements required are just the matrix elements of quark currents
between hadron states, which can be extracted from the leading
(non-rare) semileptonic decays. Other long-distance contributions
are negligibly small \cite{RS}.
The contributions of higher dimensional operators are found to be
negligibly small in the case of $K_{\rm L}\to\pi^0\nu\bar\nu$ 
\cite{GBGI} and below $5\%$ of the charm contribution in the
case of $K^+\to\pi^+\nu\bar\nu$ \cite{FalkLP}.
 As a consequence of these features,
the scale ambiguities, inherent to perturbative QCD, 
constitute  the dominant theoretical uncertainties 
present in the analysis of these decays.
These theoretical uncertainties have been considerably reduced
through the inclusion of
the next-to-leading QCD corrections 
 \cite{BB1}--\cite{BB3}. 

The investigation of these low energy rare decay processes in
conjunction with their theoretical cleanliness, allows to probe,
albeit indirectly, high energy scales of the theory and in particular
to measure $V_{td}$ and $\IM\lambda_t= \IM V^*_{ts} V_{td}$
from $K^+\to\pi^+\nu\bar\nu$ and $K_{\rm L}\to\pi^0\nu\bar\nu$
respectively.
Moreover, the combination of these two decays offers one of the
cleanest measurements of $\sin 2\beta$ \cite{BB4}.
However, the very fact
that these processes are based on higher order electroweak effects
implies
that their branching ratios are expected to be very small and not easy to
access experimentally.

\subsection{The Decay \kpnn}
            \label{sec:HeffRareKB:kpnn}
\subsubsection{The effective Hamiltonian}
The effective Hamiltonian for $\kpn$  can
be written as
\begin{equation}\label{hkpn} 
{\cal H}_{\rm eff}={G_{\rm F} \over{\sqrt 2}}{\alpha\over 2\pi 
\sin^2\Theta_{\rm W}}
 \sum_{l=e,\mu,\tau}\left( V^{\ast}_{cs}V_{cd} X^l_{\rm NL}+
V^{\ast}_{ts}V_{td} X(x_t)\right)
 (\bar sd)_{V-A}(\bar\nu_l\nu_l)_{V-A} \, .
\end{equation}
The index $l$=$e$, $\mu$, $\tau$ denotes the lepton flavour.
The dependence on the charged lepton mass resulting from the box-graph
is negligible for the top contribution. In the charm sector this is the
case only for the electron and the muon but not for the $\tau$-lepton.

The function $X(x_t)$ relevant for the top part is given by
\begin{equation}\label{xx9} 
X(x_t)=X_0(x_t)+\aspi X_1(x_t) 
=\eta_X\cdot X_0(x_t), \qquad\quad \eta_X=0.994,
\end{equation}
with the QCD correction \cite{BB1}--\cite{BB98}.
\begin{equation}\label{xx1}
X_1(x_t)=\tilde X_1(x_t)+
8x_t{\partial X_0(x_t)\over\partial x_t}\ln x_\mu\,.
\end{equation}
Here $x_\mu=\mu_t^2/M^2_W$ with $\mu_t=\ord(m_t)$ and
$\tilde X_1(x_t)$ is a complicated function given in 
\cite{BB1}--\cite{BB98}.
The $\mu_t$-dependence of the last term in (\ref{xx1}) cancels to the
considered order the $\mu_t$-dependence of the leading term 
$X_0(x_t(\mu))$.
The leftover $\mu_t$-dependence in $X(x_t)$ is below $1\%$.
The factor $\eta_X$ summarizes the NLO 
corrections represented by the second
term in (\ref{xx9}).
With $\mt\equiv \mtb(\mt)$ the QCD factor $\eta_X$
is practically independent of $m_t$ and $\Lambda_{\overline{MS}}$
and is very close to unity.

The expression corresponding to $X(x_t)$ in the charm sector is the function
$X^l_{\rm NL}$. It results from the NLO calculation \cite{BB3} and is given
explicitly in \cite{BB98}.
The inclusion of NLO corrections reduced considerably the large
$\mu_c$ dependence
(with $\mu_c={\cal O}(m_c)$) present in the leading order expressions
for the charm contribution
 \cite{novikovetal:77}.
Varying $\mu_c$ in the range $1\gev\le\mu_c\le 3\gev$ changes $X_{\rm NL}$
by roughly $24\%$ after the inclusion of NLO corrections to be compared
with $56\%$ in the leading order. Further details can be found in
\cite{BBL,BB3}. The impact of the $\mu_c$ uncertainties on the
resulting branching ratio $Br(\kpn)$ is discussed below.

The
numerical values for $X^l_{\rm NL}$ for $\mu = \mc$ and several values of
$\Lms^{(4)}$ and $\mc(\mc)$ can be found in \cite{BB98}. 
The net effect of QCD corrections is to suppress the charm contribution
by roughly $30\%$. For our purposes we need only
\begin{equation}\label{p0k}
P_0(X)=\frac{1}{\lambda^4}\left[\frac{2}{3} X^e_{\rm NL}+\frac{1}{3}
 X^\tau_{\rm NL}\right]=0.42\pm0.06
\end{equation}
where the error results from the variation of $\Lms^{(4)}$ and $\mc(\mc)$.
The contribution of dimension eight operators, not included here,
is estimated to be below the error in (\ref{p0k}) \cite{FalkLP}.

\subsubsection{Deriving the Branching Ratio}
The relevant hadronic
matrix element of the weak current $(\bar sd)_{V-A}$ in (\ref{hkpn}) 
can be extracted
with the help of isospin symmetry from
the leading decay $K^+\to\pi^0e^+\nu$.
Consequently the resulting theoretical
expression for  the branching fraction $Br(K^+\to\pi^+\nu\bar\nu)$ can
be related to the experimentally well known quantity
$Br(K^+\to\pi^0e^+\nu)$. Let us demonstrate this.

The effective Hamiltonian for the tree level decay $K^+\to\pi^0 e^+\nu$
is given by
\begin{equation}\label{kp0} 
{\cal H}_{\rm eff}(K^+\to\pi^0 e^+\nu)
={G_{\rm F} \over{\sqrt 2}}
 V^{\ast}_{us}
 (\bar su)_{V-A}(\bar\nu_e e)_{V-A} \, .
\end{equation}
Using isospin symmetry we have
\be\label{iso1}
\langle \pi^+|(\bar sd)_{V-A}|K^+\rangle=\sqrt{2}
\langle \pi^0|(\bar su)_{V-A}|K^+\rangle.
\ee
Consequently neglecting differences in the phase space of these two decays,
due to $m_{\pi^+}\not=m_{\pi^0}$ and $m_e\not=0$, we find 
\be\label{br1}
\frac{Br(\kpn)}{Br(K^+\to\pi^0 e^+\nu)}=
{\alpha^2\over |V_{us}|^2 2\pi^2 
\sin^4\Theta_{\rm W}}
 \sum_{l=e,\mu,\tau}\left| V^{\ast}_{cs}V_{cd} X^l_{\rm NL}+
V^{\ast}_{ts}V_{td} X(x_t)\right|^2~.
\end{equation}
\subsubsection{Basic Phenomenology}
Using (\ref{br1}) 
and including isospin breaking corrections one finds
\begin{equation}\label{bkpn}
Br(\kpn)=\kappa_+\cdot\left[\left({\imlt\over\lambda^5}X(x_t)\right)^2+
\left({\relc\over\lambda}P_0(X)+{\relt\over\lambda^5}X(x_t)\right)^2
\right]~,
\end{equation}
\begin{equation}\label{kapp}
\kappa_+=r_{K^+}{3\alpha^2 Br(K^+\to\pi^0e^+\nu)\over 2\pi^2
\sin^4\Theta_{\rm W}}
 \lambda^8=4.11\cdot 10^{-11}\,,
\end{equation}
where we have used
\begin{equation}\label{alsinbr}
\alpha=\frac{1}{129},\qquad \sin^2\Theta_{\rm W}=0.23, \qquad
Br(K^+\to\pi^0e^+\nu)=4.82\cdot 10^{-2}\,.
\end{equation}
Here $\lambda_i=V^\ast_{is}V_{id}$ with $\lambda_c$ being
real to a very high accuracy. $r_{K^+}=0.901$ summarizes isospin
breaking corrections in relating $\kpn$ to $K^+\to\pi^0e^+\nu$.
These isospin breaking corrections are due to quark mass effects and 
electroweak radiative corrections and have been calculated in
\cite{MP}. Finally $P_0(X)$ is given in (\ref{p0k}).

Using the improved Wolfenstein parametrization and the approximate
formulae (\ref{2.51}) -- (\ref{2.53}) we can next put 
(\ref{bkpn}) into a more transparent form \cite{BLO}:
\begin{equation}\label{108}
Br(K^{+} \to \pi^{+} \nu \bar\nu) = 4.11 \cdot 10^{-11} A^4 X^2(x_t)
\frac{1}{\sigma} \left[ (\sigma \bar\eta)^2 +
\left(\varrho_0 - \bar\varrho \right)^2 \right]\,,
\end{equation}
where
\begin{equation}\label{109}
\sigma = \left( \frac{1}{1- \frac{\lambda^2}{2}} \right)^2\,.
\end{equation}

The measured value of $Br(K^{+} \to \pi^{+} \nu \bar\nu)$ then
determines  an ellipse in the $(\bar\varrho,\bar\eta)$ plane  centered at
$(\varrho_0,0)$ with 
\begin{equation}\label{110}
\varrho_0 = 1 + \frac{P_0(X)}{A^2 X(x_t)}
\end{equation}
and having the squared axes
\begin{equation}\label{110a}
\bar\varrho_1^2 = r^2_0, \qquad \bar\eta_1^2 = \left( \frac{r_0}{\sigma}
\right)^2
\end{equation}
where
\begin{equation}\label{111}
r^2_0 = \frac{1}{A^4 X^2(x_t)} \left[
\frac{\sigma \cdot Br(K^{+} \to \pi^{+} \nu \bar\nu)}
{4.11 \cdot 10^{-11}} \right]\,.
\end{equation}
Note that $r_0$ depends only on the top contribution.
The departure of $\varrho_0$ from unity measures the relative importance
of the internal charm contributions. $\varrho_0\approx 1.35$.

The ellipse defined by $r_0$, $\varrho_0$ and $\sigma$ given above
intersects with the circle (\ref{2.94}).  This allows to determine
$\bar\varrho$ and $\bar\eta$  with 
\begin{equation}\label{113}
\bar\varrho = \frac{1}{1-\sigma^2} \left( \varrho_0 - \sqrt{\sigma^2
\varrho_0^2 +(1-\sigma^2)(r_0^2-\sigma^2 R_b^2)} \right), \qquad
\bar\eta = \sqrt{R_b^2 -\bar\varrho^2}
\end{equation}
and consequently
\begin{equation}\label{113aa}
R_t^2 = 1+R_b^2 - 2 \bar\varrho,
\end{equation}
where $\bar\eta$ is assumed to be positive.
Given $\bar\varrho$ and $\bar\eta$ one can determine $V_{td}$:
\begin{equation}\label{vtdrhoeta}
V_{td}=A \lambda^3(1-\bar\varrho-i\bar\eta),\qquad
|V_{td}|=A \lambda^3 R_t.
\end{equation}
The determination of $|V_{td}|$ and of the unitarity triangle requires
the knowledge of $V_{cb}$ (or $A$) and of $|V_{ub}/V_{cb}|$. Both
values are subject to theoretical uncertainties present in the existing
analyses of tree level decays. Whereas the dependence on
$|V_{ub}/V_{cb}|$ is rather weak, the very strong dependence of
$Br(\kpn)$ on $A$ or $V_{cb}$ makes a precise prediction for this
branching ratio difficult at present. We will return to this below.
The dependence of $Br(\kpn)$ on $\mt$ is also strong. However $\mt$
is known already  within $\pm 4\%$ and
consequently the related uncertainty in 
$Br(\kpn)$ is substantialy smaller than the corresponding uncertainty 
due to $V_{cb}$.

\subsubsection{Numerical Analysis of \kpnn}
\label{sec:Kpnn:NumericalKp}
The uncertainties 
in the prediction for $Br(\kpn)$ and in the determination of  $|V_{td}|$
related to the choice of the renormalization scales $\mu_t$
and $\mu_c$ in the top part and the charm part, respectively
have been investigated in \cite{BBL}.
To this end the scales $\mu_c$ and $\mu_t$ entering $m_c(\mu_c)$
and $m_t(\mu_t)$, respectively, have been varied in the ranges
$1\gev\leq\mu_c\leq 3\gev$ and $100\gev\leq\mu_t\leq 300\gev$.
It has been found that including
the full next-to-leading corrections reduces the uncertainty in the
determination of $|V_{td}|$ from $\pm 14\%$ (LO) to $\pm 4.6\%$ (NLO).
The main bulk of this theoretical error stems
from the charm sector. 
In the case of $Br(\kpn)$,
the theoretical uncertainty
due to $\mu_{c,t}$ is reduced from $\pm 22\%$ (LO) to $\pm 7\%$ (NLO).

Scanning the input parameters of table \ref{tab:inputparams}
we find 
\begin{equation}\label{kpnr}
Br(\kpn)=
(7.5 \pm 2.9)\cdot 10^{-11} 
\end{equation}
where the error comes dominantly from the uncertainties in the CKM
parameters. 

The predicted branching ratio decreased during the
last years due to the improved lower bound on $\Delta M_s$.
Indeed,
it is possible to derive an upper bound on $Br(\kpn)$ \cite{BB98}:
\be\label{boundk}
Br(\kpn)_{\rm max}=\frac{\kappa_+}{\sigma}
\left[ P_0(X)+A^2 X(x_t)\frac{r_{sd}}{\lambda}
\sqrt{\frac{\Delta M_d}{\Delta M_s}}\right]^2
\ee
where $r_{ds}=\xi\sqrt{m_{B_s}/m_{B_d}}$ with $\xi$ defined in 
(\ref{107b}). This equation translates
 a lower bound on $\Delta M_s$ into an upper bound on $Br(\kpn)$.
This bound is very clean and does not involve theoretical
hadronic uncertainties except for $r_{sd}$. Using
\be
\sqrt{\frac{\Delta M_d}{\Delta M_s}}<0.18~,~\quad
A<0.89~,~\quad P_0(X)<0.48~,~\quad X(x_t)<1.56~,~\quad
r_{sd}<1.22
\ee
we find
\be
Br(\kpn)_{\rm max}=11.5 \cdot 10^{-11}~.
\ee
This limit could be further strengthened with improved input.
However, this bound is strong enough to indicate a clear
conflict with the SM if $Br(\kpn)$ should
be measured at $1.5\cdot 10^{-10}$.

\subsubsection{$\vtd$ from $K^+\to\pi^+\nu\bar\nu$}
Once $Br(K^+\to\pi^+\nu\bar\nu)\equiv Br(K^+)$ is measured, $\vtd$ can be
extracted subject to various uncertainties:
\be\label{vtda}
\frac{\sigma(\vtd)}{\vtd}=\pm 0.04_{scale}\pm \frac{\sigma(\vcb)}{\vcb}
\pm 0.7 \frac{\sigma(\bar\mc)}{\bar\mc}
\pm 0.65 \frac{\sigma( Br(K^+))}{Br(K^+)}~.
\ee
Taking $\sigma(\vcb)=0.002$, $\sigma(\bar\mc)=100\mev$ and
$\sigma( Br(K^+))=10\%$ and adding the errors in quadrature we find that
$\vtd$ can be determined with an accuracy of $\pm 10\%$.
This number
is increased to $\pm 11\%$ once the uncertainties due to $\mt$,
$\alpha_s$ and $|V_{ub}|/\vcb$ are taken into account. Clearly this
determination can be improved although a determination of $\vtd$ with
an accuracy better than $\pm 5\%$ seems rather unrealistic.

\subsubsection{Summary and Outlook}
The accuracy of the SM prediction for $Br(\kpn)$ has
improved considerably during the last decade. 
This progress can be traced back to the
improved values of $\mt$ and $\vcb$ and to the inclusion of NLO
QCD corrections which considerably reduced the scale uncertainties
in the charm sector. 

Now, what about the experimental status of this decay ?
One of the high-lights of 1997 was the observation by AGS E787
collaboration at Brookhaven \cite{Adler97} 
of one event consistent with the signature expected for this decay.
Recently an analysis of a larger data set has been published 
\cite{Adler00}. No further events were seen, leading to 
the branching ratio:
\be\label{kp97}
Br(K^+ \rightarrow \pi^+ \nu \bar{\nu})=
(1.5^{+3.4}_{-1.2})\cdot 10^{-10}~.
\end{equation}
The central value is  by a factor of 2 above the SM
expectation but in view of large errors the result is compatible with the
SM. The analysis of additional
data on $K^+\to \pi^+\nu\bar\nu$ present on tape at AGS E787 should improve
this result in the near future considerably.
In view of the clean character of this decay a measurement of its
branching ratio at the level of $ 1.5 \cdot 10^{-10}$ with a small error 
would signal the presence of physics beyond the SM.

The experimental outlook for this decay has been recently reviewed by
Littenberg \cite{LITT00}. A new experiment, AGS E949
\cite{AGS2} is scheduled to run in 2001-3. It is expected to
reach a sensitivity of $\sim 10^{-11}/{\rm event}$. In the long term,
the CKM experiment at Fermilab \cite{Coleman} should be able to
reach $\sim 10^{-12}/{\rm event}$ sensitivity. At this level an
accurate measurement of the branching ratio should be possible.
\subsection{The Decay $K_{\rm L}\to\pi^0\nu\bar\nu$}
            \label{sec:HeffRareKB:klpinn1}
\subsubsection{The effective Hamiltonian}
The effective
Hamiltonian for $K_{\rm L}\to\pi^0\nu\bar\nu$
is given as follows:
\begin{equation}\label{hxnu}
{\cal H}_{\rm eff} = {G_{\rm F}\over \sqrt 2} {\alpha \over
2\pi \sin^2 \Theta_{\rm W}} V^\ast_{ts} V_{td}
X (x_t) (\bar sd)_{V-A} (\bar\nu\nu)_{V-A} + h.c.\,,   
\end{equation}
where the function $X(x_t)$, present already in $\kpn$,
includes NLO corrections and is given in (\ref{xx9}). 

As we will demonstrate shortly, $\klpn$  proceeds in the SM 
almost
entirely through direct CP violation \cite{littenberg:89}.
The indirectly CP violating contribution and the CP conserving
contribution analyzed in \cite{GBGI} are fully negligible.
Within the SM $\klpn$ is completely dominated by short-distance loop diagrams 
with top quark
exchanges. The charm contribution can be fully
neglected and the theoretical uncertainties present in $\kpn$ due to
$m_c$, $\mu_c$ and $\Lambda_{\overline{MS}}$ are absent here. 
Consequently the rare decay $\klpn$ is even cleaner than $\kpn$
and is very well suited for the determination of 
the Wolfenstein parameter $\eta$ and in particular $\imlt$.

We have stated that the
decay $\klpn$ is dominated by {\it direct} CP violation. Now
as discussed in Section 3
the standard definition of the direct CP violation 
requires the presence of strong phases which are
absent in $\klpn$. Consequently the violation of
CP symmetry in $\klpn$ can only arise through the interference between
$K^0-\bar K^0$ mixing and the decay amplitude. This type of CP
violation has been discussed already in Section 3.
However, as already pointed out by Littenberg \cite{littenberg:89}
and demonstrated explictly in a moment,
the contribution of CP violation to $\klpn$ via $K^0-\bar K^0$ mixing 
alone is tiny. It gives $Br(\klpn) \approx 2\cdot 10^{-15}$ 
and its interference with the directly CP-violating contribution is
$\ord (10^{-13})$.
Consequently, in this sence,  CP violation in $\klpn$ with
$Br(\klpn) = {\cal O}(10^{-11})$ is a manifestation of CP violation
in the decay and as such deserves the name of {\it direct} CP violation.
In other words the difference in the magnitude of CP violation in
$K_{\rm L}\to\pi\pi~(\varepsilon)$ and $\klpn$ is a signal of direct
CP violation and measuring $\klpn$ at the expected level would
be another signal of this phenomenon. More details on this
issue can be found in \cite{NIR96,BUCH96,BB96}.

\subsubsection{Deriving the Branching Ratio}
Let us derive the basic formula for $Br(\klpn)$ in a manner analogous
to the one for  $Br(K^+ \to \pi^+ \nu \bar\nu)$. To this end we
consider one neutrino flavour and define the complex function
\begin{equation}\label{hxnu1}
F = {G_{\rm F}\over \sqrt 2} {\alpha \over
2\pi \sin^2 \Theta_{\rm W}} V^\ast_{ts} V_{td}
X (x_t).   
\end{equation}
Then the effective Hamiltonian in (\ref{hxnu}) can be written as
\begin{equation}\label{hxnu2}
{\cal H}_{\rm eff} =  F (\bar sd)_{V-A} (\bar\nu\nu)_{V-A}+
F^\ast (\bar ds)_{V-A} (\bar\nu\nu)_{V-A}~.
\end{equation}
Now, from (\ref{KLS}) we have
\be\label{KLS1}
K_L=\frac{1}{\sqrt{2}}
[(1+\bar\varepsilon)K^0+ (1-\bar\varepsilon)\bar K^0]
\ee
where we have neglected
$\mid\bar\varepsilon\mid^2\ll 1$. Thus the amplitude
for $K_L\to\pi^0\nu\bar\nu$ is given by
\be\label{ampkl0}
A(K_L\to\pi^0\nu\bar\nu)=
\frac{1}{\sqrt{2}}
\left[F(1+\bar\varepsilon) \langle \pi^0|(\bar sd)_{V-A}|K^0\rangle
+ 
F^\ast (1-\bar\varepsilon) \langle \pi^0|(\bar ds)_{V-A}|\bar K^0\rangle
 \right] (\bar\nu\nu)_{V-A}.
\ee
Recalling
\be\label{DEF}
CP|K^0\rangle = - |\bar K^0\rangle, \quad\quad
C|K^0\rangle =  |\bar K^0\rangle
\ee
we have
\be
\langle \pi^0|(\bar ds)_{V-A}|\bar K^0\rangle=-
\langle \pi^0|(\bar sd)_{V-A}|K^0\rangle,
\ee
where the minus sign is crucial for the subsequent steps.

Thus we can write
\be\label{bmpkl0}
A(K_L\to\pi^0\nu\bar\nu)=
\frac{1}{\sqrt{2}}
\left[F(1+\bar\varepsilon) -F^\ast (1-\bar\varepsilon)\right]
 \langle \pi^0|(\bar sd)_{V-A}| K^0\rangle
 (\bar\nu\nu)_{V-A}.
\ee
Now the terms $\bar\varepsilon$ can be safely neglected in comparision
with unity, which implies that the indirect CP violation
(CP violation in the $K^0-\bar K^0$ mixing) is negligible in this decay.
We have then
\be
F(1+\bar\varepsilon) -F^\ast (1-\bar\varepsilon)=
{G_{\rm F}\over \sqrt 2} {\alpha \over
\pi \sin^2 \Theta_{\rm W}} \IM (V^\ast_{ts} V_{td})
\cdot X(x_t).   
\end{equation}
Consequently using isospin relation
\be
\langle \pi^0|(\bar ds)_{V-A}|\bar K^0\rangle=
\langle \pi^0|(\bar su)_{V-A}|K^+\rangle
\ee
together with (\ref{kp0}) and taking into account the difference
in the lifetimes of $K_L$ and $K^+$ we have after summation over three
neutrino flavours
\be\label{br2}
\frac{Br(K_L\to\pi^0\nu\bar\nu)}{Br(K^+\to\pi^0 e^+\nu)}=
3\frac{\tau(K_L)}{\tau(K^+)}
{\alpha^2\over |V_{us}|^2 2\pi^2 
\sin^4\Theta_{\rm W}}
 \left[\IM \lambda_t \cdot X(x_t)\right]^2
\end{equation}
where $\lambda_t=V^{\ast}_{ts}V_{td}$.
\subsubsection{Master Formulae for $Br(\klpn)$}
\label{sec:Kpnn:MasterKL}
Using (\ref{br2}) we can write $Br(\klpn)$ simply as
follows
\begin{equation}\label{bklpn}
Br(K_{\rm L}\to\pi^0\nu\bar\nu)=\kappa_{\rm L}\cdot
\left({\imlt\over\lambda^5}X(x_t)\right)^2
\end{equation}
\begin{equation}\label{kapl}
\kappa_{\rm L}=\frac{r_{K_{\rm L}}}{r_{K^+}}
 {\tau(K_{\rm L})\over\tau(K^+)}\kappa_+ =1.80\cdot 10^{-10}
\end{equation}
with $\kappa_+$ given in (\ref{kapp}) and
$r_{K_{\rm L}}=0.944$ summarizing isospin
breaking corrections in relating $\klpn$ to $K^+\to\pi^0e^+\nu$
\cite{MP}.

Using the Wolfenstein
parametrization and (\ref{xx9}) we can rewrite (\ref{bklpn}) as
\begin{equation}
Br(K_{\rm L}\to\pi^0\nu\bar\nu)=
3.0\cdot 10^{-11}
\left [ \frac{\eta}{0.39}\right ]^2
\left [\frac{\mtb(\mt)}{170~GeV} \right ]^{2.3} 
\left [\frac{\mid V_{cb}\mid}{0.040} \right ]^4 \,.
\label{bklpn1}
\end{equation}

The determination of $\eta$ using $Br(\klpn)$ requires the knowledge
of $V_{cb}$ and $\mt$. The very strong dependence on $V_{cb}$ or $A$
makes a precise prediction for this branching ratio difficult at
present.

On the other hand inverting (\ref{bklpn}) and using (\ref{xx9})
 one finds \cite{BB96}:
\begin{equation}\label{imlta}
\IM\lambda_t=1.36\cdot 10^{-4} 
\left[\frac{170\gev}{\mtb(\mt)}\right]^{1.15}
\left[\frac{Br(\klpn)}{3\cdot 10^{-11}}\right]^{1/2}\,.
\end{equation}
without any uncertainty in $\vcb$.
(\ref{imlta}) offers
 the cleanest method to measure $\IM\lambda_t$;
even better than the CP asymmetries
in $B$ decays discussed in section 8 that require the knowledge
of $\vcb$ to determine $\IM\lambda_t$. Measuring $Br(\klpn)$
with $10\%$ accuraccy allows to determine $\IM\lambda_t$
with an error of $5\%$ \cite{AJBLH,BB96}.
\subsubsection{Numerical Analysis of \klpnn}
\label{sec:Kpnn:NumericalKL}
The $\mu_t$-uncertainties present in the function $X(x_t)$ have 
already been
discussed in connection with $\kpn$. After the inclusion of NLO
corrections they are so small that they can be neglected for all
practical purposes. 
Scanning the input parameters of table \ref{tab:inputparams}
we find 
\begin{equation}\label{klpnr4}
Br(\klpn)=
(2.6 \pm 1.2)\cdot 10^{-11} 
\end{equation}
where the error comes dominantly from the uncertainties in the CKM
parameters. 
\subsubsection{Summary and Outlook}
The accuracy of the SM prediction for $Br(\klpn)$ has
improved considerably during the last decade. 
This progress can be traced back mainly to the
improved values of $\mt$ and $\vcb$ and to some extent to 
the inclusion of NLO QCD corrections.

The present upper bound on $Br(K_{\rm L}\to \pi^0\nu\bar\nu)$ from
the KTeV  experiment at Fermilab \cite{KTeV00X} reads 
\begin{equation}\label{KLD}
Br(\klpn)<5.9 \cdot 10^{-7}\,.
\end{equation}
This is about four orders of magnitude above the SM expectation
(\ref{klpnr4}).
Moreover this bound is substantially weaker than the 
{\it model independent} bound \cite{NIR96}
from isospin symmetry:
\begin{equation}
Br(\klpn) < 4.4 \cdot Br(\kpn)
\end{equation}
which through (\ref{kp97})  gives
\begin{equation}\label{B108}
Br(\klpn) < 2.6 \cdot 10^{-9} ~(90\% C.L.)
\end{equation}

The experimental outlook for this decay has been recently reviewed by
Littenberg \cite{LITT00}.
The KEK E391a experiment \cite{KEKKL} should reach sensitivity
of $\sim 10^{-10}/{\rm event}$ which would give some events
only in the presence of new physics contributions (see Section 9).
KAMI at Fermilab \cite{FNALKL} should be able to reach a
sensitivity of $< 10^{-12}/{\rm event}$.
Finally a very interesting new experiment KOPIO
at Brookhaven (BNL E926) \cite{KOPIO} 
expects to reach the single event sensitivity of $2\cdot 10^{-12}$.
Both KAMI and KOPIO should provide useful measurements of this
gold-plated decay.

\begin{figure}[hbt]
\vspace{0.10in}
\centerline{
\epsfysize=2.7in
\epsffile{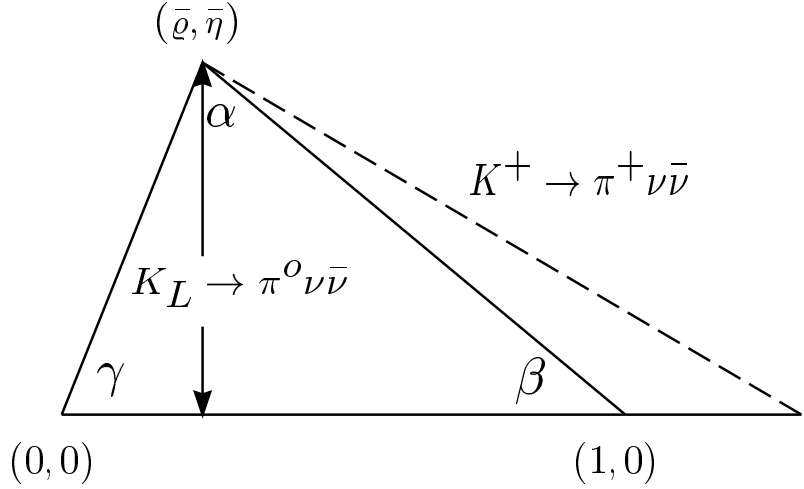}
}
\vspace{0.08in}
\caption{Unitarity triangle from $K\to\pi\nu\bar\nu$.}\label{fig:KPKL}
\end{figure}

\subsection{Unitarity Triangle and $\sin 2\beta$ from $K\to\pi\nu\bar\nu$}
\label{sec:Kpnn:Triangle}
The measurement of $Br(\kpn)$ and $Br(\klpn)$ can determine the
unitarity triangle completely, (see fig.~\ref{fig:KPKL}), 
provided $\mt$ and $V_{cb}$ are known \cite{BB4}.
Using these two branching ratios simultaneously allows to eliminate
$|V_{ub}/V_{cb}|$ from the analysis which removes a considerable
uncertainty. Indeed it is evident from (\ref{bkpn}) and
(\ref{bklpn}) that, given $Br(\kpn)$ and $Br(\klpn)$, one can extract
both $\imlt$ and $\relt$. One finds \cite{BB4,BBL}
\begin{equation}\label{imre}
\imlt=\lambda^5{\sqrt{B_2}\over X(x_t)}\qquad
\relt=-\lambda^5{{\relc\over\lambda}P_0(X)+\sqrt{B_1-B_2}\over X(x_t)}\,,
\end{equation}
where we have defined the ``reduced'' branching ratios
\begin{equation}\label{b1b2}
B_1={Br(\kpn)\over 4.11\cdot 10^{-11}}\qquad
B_2={Br(\klpn)\over 1.80\cdot 10^{-10}}\,.
\end{equation}
Using next the expressions for $\imlt$, $\relt$ and $\relc$ given
in (\ref{2.51})--(\ref{2.53}) we find
\begin{equation}\label{rhetb}
\bar\varrho=1+{P_0(X)-\sqrt{\sigma(B_1-B_2)}\over A^2 X(x_t)}\,,\qquad
\bar\eta={\sqrt{B_2}\over\sqrt{\sigma} A^2 X(x_t)}
\end{equation}
with $\sigma$ defined in (\ref{109}). An exact treatment of the CKM
matrix shows that the formulae (\ref{rhetb}) are rather precise
\cite{BB4}. 

Using (\ref{rhetb}) one finds subsequently \cite{BB4}
\begin{equation}\label{sin}
r_s=r_s(B_1, B_2)\equiv{1-\bar\varrho\over\bar\eta}=\cot\beta\,, \qquad
\sin 2\beta=\frac{2 r_s}{1+r^2_s}
\end{equation}
with
\begin{equation}\label{cbb}
r_s(B_1, B_2)=\sqrt{\sigma}{\sqrt{\sigma(B_1-B_2)}-P_0(X)\over\sqrt{B_2}}\,.
\end{equation}
Thus within the approximation of (\ref{rhetb}) $\sin 2\beta$ is
independent of $V_{cb}$ (or $A$) and $m_t$.

It should be stressed that $\sin 2\beta$ determined this way depends
only on two measurable branching ratios and on the function
$P_0(X)$ which is completely calculable in perturbation theory.
Consequently this determination is free from any hadronic
uncertainties and its accuracy can be estimated with a high degree
of confidence. 

An extensive numerical analysis of the formulae above has been presented
in \cite{BB96}. 
Assuming that the branching ratios are known to within $\pm 10\%$
and $\mt$ within $\pm 3~\gev$ one finds the results in table~\ref{tabkb2}
\cite{BB96}. 
We observe that respectable determinations of all considered 
quantities except for 
$\bar\varrho$ can be obtained.
Of particular interest are the accurate determinations of
$\sin 2\beta$ and of ${\rm Im}\lambda_t$.
The latter quantity as seen in (\ref{imlta}) 
can be obtained from
$K_{\rm L}\to\pi^0\nu\bar\nu$ alone and does not require knowledge
of $V_{cb}$.
The importance of measuring accurately  ${\rm Im}\lambda_t$ is evident.
It plays a central role in the phenomenology of CP violation
in $K$ decays and is furthermore equivalent to the 
Jarlskog parameter $J_{\rm CP}$ \cite{CJ}, 
the invariant measure of CP violation in the SM, 
$J_{\rm CP}=\lambda(1-\lambda^2/2){\rm Im}\lambda_t$.

\begin{table}
\caption[]{Illustrative example of the determination of CKM
parameters from $K\to\pi\nu\bar\nu$.
\label{tabkb2}}
\vspace{0.4cm}
\begin{center}
\begin{tabular}{|c|c|c|}\hline
&$\sigma(|V_{cb}|)=\pm 0.002$ & $\sigma(|V_{cb}|)=\pm 0.001$
\\ 
\hline
$\sigma(|V_{td}|) $& $\pm 10\% $ & $ \pm 9\% $
 \\ 
\hline 
$\sigma(\bar\varrho) $ & $\pm 0.16$ &$\pm 0.11$
  \\
\hline
$\sigma(\bar\eta)$ & $\pm 0.04$&$\pm 0.03$
 \\
\hline
$\sigma(\sin 2\beta)$ & $\pm 0.05$&$\pm 0.05$
 \\
\hline
$\sigma({\rm Im}\lambda_t)$&$\pm 5\%$ &$\pm 5\%$ 
 \\
\hline
\end{tabular}
\end{center}
\end{table}

The accuracy to which $\sin 2\beta$ can be obtained from
$K\to\pi\nu\bar\nu$ is, in the  example discussed above, 
comparable to the one expected
in determining $\sin 2\beta$ from CP asymmetries in $B$ decays prior to
LHCB and BTeV experiments.  In this case $\sin 2\beta$ is determined best by
measuring CP violation in $B_d\to J/\psi K_{\rm S}$.
Using the formula  for the corresponding time-integrated 
CP asymmetry one finds an
interesting connection between rare $K$ decays and $B$ physics \cite{BB4}
\begin{equation}\label{kbcon}
{2 r_s(B_1,B_2)\over 1+r^2_s(B_1,B_2)}=
-a_{\mbox{{\scriptsize CP}}}(B_d\to J/\psi K_{\mbox{{\scriptsize S}}})
{1+x^2_d\over x_d}
\end{equation}
which must be satisfied in the SM. 
Here $x_d=\Delta M_d/\Gamma(B^0_d)$ is a $B_d^0-\bar B_d^0$ mixing 
parameter.
We stress that except
for $P_0(X)$  all quantities in
(\ref{kbcon}) can be directly measured in experiment and that this
relationship is essentially independent of $m_t$ and $V_{cb}$.
Due to very small theoretical uncertainties in (\ref{kbcon}), this
relation is particularly suited for tests of CP violation in the
SM and offers a powerful tool to probe the physics
beyond it. We will return to this topic in Section 9.

The relation (\ref{kbcon}) is one of several correlations
between $K\to\pi\nu\bar\nu$ and observables in B physics.
Another example is the upper bound in (\ref{boundk}).
A recent numerical analysis of such correlations can be found
in \cite{Perez}. Finally analyses of $K\to \pi\nu\bar\nu$ in
non-supersymmetric extensions of the SM can be found in \cite{KLBSM}.
$K\to \pi\nu\bar\nu$ in supersymmetric extensions is discussed in 
section 9.

\section{ Express Review of Rare $K$ and $B$ Decays}
         \label{sec:RareKB}
\setcounter{equation}{0}
\subsection{Rare $K$ Decays}
\subsubsection{$K_L \to \pi^0 e^+e^-$}
There are three contributions to this decay: CP-conserving,
indirectly CP-violating and directly CP-violating.
Unfortunately out of these three contributions only the
directly CP-violating can be calculated reliably.
This contribution is dominated by $Z^0$-penguin diagrams.
The enhancement of these diagrams for large $\mt$, pointed out
in the context of this decay in \cite{KLMT}, gave in fact the
motivation to study  $Z^0$-penguin effects in $\epe$
\cite{flynn:89,buchallaetal:90}.
Including NLO corrections \cite{BLMM} and scanning the input parameters 
of table \ref{tab:inputparams}
we find
\be
  Br(\kpe)_{\rm dir}=(4.3 \pm 2.1) \cdot 10^{-12}\,, 
  \label{eq:brkpep}
\ee
where the errors come dominantly from the uncertainties in the CKM
parameters. 
The calculations of indirectly CP-violating contribution
are plagued by theoretical uncertainties \cite{KL}.
On the other hand, this contribution taken alone is given
simply by
\be
  Br(\kpe)_{\rm indir}=|\varepsilon|^2 \frac{\tau(K_L)}{\tau(K_S)}
Br(K_S\to \pi^0e^+e^-)=
3.0\cdot 10^{-3}Br(K_S\to \pi^0e^+e^-) \,, 
  \label{eq:brpep}
\ee
with $\tau(K_{L,S})$ denoting the $K_{L,S}$ life-times and 
$Br(K_S\to \pi^0e^+e^-)$ hopefully measured in the next years
by KLOE at Frascati. The two CP violating contributions will
in general interfer with each other. Given the present uncertainty
on $Br(K_S\to \pi^0e^+e^-)$ the total CP-violating contribution
could be as high as ${\rm few}\times 10^{-11}$ \cite{AEIP} but
taking into account the theoretical estimates of the indirectly
CP-violating contribution, one should expect it below $10^{-11}$
within the SM.

The upper bound on the CP-conserving contribution governed
by $K_L\to\pi^0\gamma\gamma\to\pi^0e^+e^-$  can be
obtained with the help of chiral perturbation theory \cite{KL}
and the data on $K_L\to\pi^0\gamma\gamma$. The recent results
on the latter decay \cite{KGG} imply that this contribution is
smaller than $2\cdot 10^{-12}$. Consequently it is smaller
than expected by some authors in the past. As this contribution
does not interfere with the remaining larger CP-violating 
contributions, it does
not present a significant problem but in order to be able to extract
CKM parameters from this decay its better estimate is clearly needed.

The most recent experimental bound from KTeV \cite{ke}
reads 
\begin{equation}
  \label{eq:brklexp}
  Br(\kpe)<5.1 \cdot 10^{-10}~(90\% C.L.). 
\end{equation}
Considerable improvements are expected in the coming
years.

\subsubsection{$\kmm$}
The $\kmm$ branching ratio can be decomposed generally as follows:
\begin{equation}
  \label{eq:deckmm}
  BR(\kmm)=\vert \RE A \vert^2 + \vert \IM A \vert^2\,,
\end{equation}
where $\RE A$ denotes the dispersive contribution and $\IM A$ the
absorptive one. The latter contribution can be determined in a model
independent way from the $K_L \to \gamma \gamma$ branching ratio. The
resulting $\vert \IM A \vert^2=(7.07 \pm 0.18) \cdot 10^{-9}$ 
is very close to the experimental
branching ratio \cite{mpmm}
\be
Br(\kmm)=(7.18 \pm 0.17) \cdot 10^{-9}~, 
\ee
so that $\vert \RE A \vert^2$ is substantially smaller and extracted to 
be \cite{mpmm}
\begin{equation}
  \label{eq:reaexp}
  \vert \RE A_{\rm exp} \vert^2 < 3.7 \cdot 10^{-10} \qquad {\rm 
  (90\% \, \,C.L.).}
\end{equation}
Now $\RE A$ can be decomposed as 
\begin{equation}
  \label{eq:decomp}
  \RE A = \RE A_{\rm LD} + \RE A_{\rm SD}\,,  
\end{equation}
with 
\begin{equation}
  \label{eq:reasdbr}
  \vert \RE A_{\rm SD} \vert^2 \equiv Br(\kmm)_{\rm SD}
\end{equation}
representing the short-distance contribution 
which can be calculated reliably.
An
improved estimate of the long-distance contribution $\RE A_{\rm LD}$
has been  presented in
\cite{dambrosio}
\begin{equation}
  \label{eq:reald}
  \vert \RE A_{LD} \vert < 2.9 \cdot 10^{-5} \qquad {\rm (90\% \,\,C.L.).} 
\end{equation}
Together with \r{eq:reaexp} this gives 
\begin{equation}
  \label{eq:sdlimitbr}
  Br(\kmm)_{\rm SD} < 2.3 \cdot 10^{-9}.
\end{equation}
This result is  close to the one presented 
by Gomez Dumm and Pich \cite{pich}. 
More pesimistic view on the extraction of the short
distance part from $Br(\kmm)$  can be found in \cite{GVAL}.

The bound in (\ref{eq:sdlimitbr})
should be compared with the short distance contribution within
the SM for which we find
\begin{equation}
\label{kmusm}
 Br(\kmm)_{\rm SD}=(8.7\pm 3.7)\cdot 10^{-10}.
\end{equation}
This implies that there is a considerable room for new physics
contributions. 
Reviews of rare K decays are given in \cite{AJBLH,BF97,LITT00,CPRARE}.

\subsection{Rare B Decays}
\subsubsection{Preliminaries}
The most interesting here are the decays $B\to X_{s,d}\gamma$,
$B\to X_{s,d} l^+l^-$, $B\to X_{s,d}\nu\bar\nu$, $B_{s,d}\to l^+l^-$
and the exclusive counterparts of the first three decays with
$X_{s,d}$ replaced by $K^*(\varrho)$ or $K(\pi)$. Within the SM the dominant
operators are
\be\label{OP1}
Q_7=\bar s_L\sigma_{\mu\nu}m_b b_R F^{\mu\nu},
\qquad Q_9=\bar s_L\gamma^\mu b_L \bar l\gamma_\mu l~,
\ee
\be\label{OP2}
Q_{10}=\bar s_L\gamma^\mu b_L \bar l\gamma_\mu\gamma_5 l,
\qquad Q_\nu=\bar s_L\gamma^\mu b_L \bar \nu\gamma_\mu(1-\gamma_5)\nu~.
\ee
$B\to X_{s,d}\gamma$ is governed by $Q_7$ and
$B\to X_{s,d} l^+l^-$ by $Q_7$, $Q_9$ and $Q_{10}$. 
 $B\to X_{s,d}\nu\bar\nu$ and $B_{s,d}\to l^+l^-$
involve only $Q_\nu$ and $Q_{10}$ respectively.

There is a vast literature on these decays within the SM and its
extensions. I can recommend in particular the most recent analyses and 
reviews in \cite{GREUB00,Mikolaj,AH1,AH2,BHI,Lunghi,KrLu,Hewett} 
where further references can be found.
Theoretically cleanest are the decays $B\to X_{s,d}\nu\bar\nu$
 and  $B_{s,d}\to l^+l^-$ as they involve only $Z^0$-penguin and box
diagram contributions. $B\to X_{s,d}\gamma$ and $B\to X_{s,d} l^+l^-$
involving magnetic $\gamma$-penguins and the standard photon-penguins
are subject to long distance contributions which have to be taken
into account. In the case of exclusive channels there are additional
uncertainties in the relevant hadronic formfactors. On the experimental
side the exclusive channels are easier to measure. Moreover, knowing
them experimentally will help to distinguish between
$B\to X_{s} l^+l^-$ and $B\to X_{d} l^+l^-$.
 
\subsubsection{$B \to X_s \gamma$ }
A lot of efforts have been put into predicting the branching
ratio for the inclusive radiative decay $B \to X_s \gamma$ 
including NLO-QCD
 corrections and higher order electroweak corrections. The
relevant references can be found in table~\ref{TAB1} and 
\cite{AJBLH,GREUB00,Mikolaj},
where theoretical details are given. The final result of these efforts
can be summarized by \cite{GH00}
\be\label{bsg}
Br(B \to X_s \gamma)_{\rm th}
=( 3.29 \pm 0.21\pm0.21) \cdot 10^{-4}~.
\ee
The two errors correspond to parametric and scale uncertainties.
The main theoretical achievement was the reduction of the scale dependence
through NLO calculations, in particular those given in 
\cite{GREUB} and \cite{CZMM}. In
the leading order this uncertainty alone would induce an error of
$\pm 0.6$ \cite{AGMU}.

The theoretical result in (\ref{bsg})
 should be compared with experimental data from CLEO \cite{CLEOBS}, 
ALEPH \cite{ALEPHBS} and BELLE \cite{BELLEBS}, whose combined branching
ratio reads
\begin{equation}\label{bsgex}
 Br(B \to X_s \gamma)_{\rm exp}=
(3.21 \pm 0.40)\cdot 10^{-4}~.
\end{equation}
Clearly, the SM result agrees well with the data. In
order to see whether any new physics can be seen in this decay,
the theoretical and in particular experimental errors should
be reduced. This is certainly a difficult task.
On the other hand already the available data put powerful constraints
on the parameter space of the supersymmetric extensions of the SM.
Most recent analyses can be found in \cite{EP00,DGG00,CGNW00}.

It is easier to measure the exclusive branching ratios 
$Br(B^\pm\to K^{*\pm}\gamma)$ and $Br(B_d^0\to K^{*0}\gamma)$.
 The most recent data from CLEO, BaBar and Belle can be found
in \cite{CBB}. They are found in the ball park of $3\cdot 10^{-5}$ and
$ 5\cdot 10^{-5}$ respectively. Theoretical calculations of these
branching ratios are plagued by uncertainties in the relevant
formfactors. Most recent theoretical reviews are given in \cite{Pirjol}.

\subsubsection{$B\to X_s l^+l^-$ and $B\to K^*l^+l^-$}
These rare decays have been the subject of many theoretical studies.
The NLO-QCD corrections have been calculated in \cite{Mis:94,BuMu:94}.
Recently dominant NNLO corrections to the relevant Wilson coefficients
have been calculated in \cite{BMU574}. The remaining scale dependence
in the branching ratios is roughly $13\%$. In order to remove this 
dependence two-loop matrix elements of the relevant four-quark operators
have to be calculated.
 
The main interest in these decays is their large sensitivity to
new physics contributions that are encoded in the Wilson coefficients
$C_7$, $C_9$ and $C_{10}$ of the operators in (\ref{OP1}) and 
(\ref{OP2}). These contributions can be studied through
the dilepton mass distribution, the leptonic forward-backward 
asymmetry and CP-asymmetries. Most recent analyses can be found in
\cite{AH1,AH2,BHI,Lunghi,KrLu,Hewett} . While $B\to X_s\gamma$ is only 
sensitive to the absolute value
of $C_7$, the decays $B\to X_s l^+l^-$ and $B\to K^*l^+l^-$ 
depend also on the sign of this coefficient that in more complicated
models can be opposite to the one in the SM. In addition these
decays involve the coefficients $C_9$ and $C_{10}$ that in
the extensions of the SM can take rather different values than found
in the SM. Of particular interest is the leptonic forward-backward
 asymmetry (FBA) in
$B\to K^*\mu^+\mu^-$ and $B\to K^*e^+e^-$ which vanishes for a certain 
value of the dilepton mass. The position of this zero is a sensitive
function of the ratio $C_7/\RE(C_9)$ with small uncertainties from
the formfactors \cite{Burdman,Charles,AH2}. Moreover FBA is proportional
to $C_{10}$ and consequently sensitive to the magnitude and the
sign of this coefficient. 
It is clear that this asymmetry will offer useful tests of the
SM and of its extensions. 
Similar comments apply to CP asymmetries that are very small in the SM 
but can be substantial in  general supersymmetric models. The calculations
of $1/\mb^2$ and $1/\mc^2$ can be found in \cite{Overmb,AH3} and 
\cite{BUC97,CRS},
respectively.
In the process of including
non-perturbative corrections induced by intermediate $c\bar c$ states
one has to avoid double counting. The most efficient method here at present
is the dispersion approach of Kr\"uger and Sehgal \cite{KSMETHOD},
which makes use of experimental data.

Within the SM we have \cite{AH2,AH3}
\be\label{TH}
Br(B\to X_s \mu^+\mu^-)=(5.7\pm1.2)\cdot 10^{-6}, 
\qquad  Br(B\to K^*\mu^+\mu^-)=(2.0\pm0.7)\cdot 10^{-6}, 
\ee
to be compared with the experimental  upper bounds
\be\label{EXP}
Br(B\to X_s\mu^+\mu^-)< 4.2\cdot 10^{-5}~\cite{Glenn},\qquad
Br(B\to K^*\mu^+\mu^-)< 4.0\cdot 10^{-6}~\cite{Affolder1}~.
\ee
The first events for these decays are expected  from 
B factories and Tevatron already this year.
\subsubsection{$B\to X_{s,d}\nu\bar\nu$}
            \label{sec:HeffRareKB:klpinn2}
The decays $B\to X_{s,d}\nu\bar\nu$ are the theoretically
cleanest decays in the field of rare $B$-decays.
They are dominated by the same $Z^0$-penguin and box diagrams
involving top quark exchanges which we encountered already
in the case of $\kpn$ and $\klpn$, except for the appropriate
change of the external quark flavours. Since the change of external
quark flavours has no impact on the $m_t$ dependence,
the latter is fully described by the function $X(x_t)$ in
(\ref{xx9}) which includes
the NLO corrections. The charm contribution 
is fully negligible
here and the resulting effective Hamiltonian is very similar to
the one for $\klpn$ given in (\ref{hxnu}). 
For the decay $B\to X_s\nu\bar\nu$ it reads
\begin{equation}\label{bxnu}
{\cal H}_{\rm eff} = {G_{\rm F}\over \sqrt 2} {\alpha \over
2\pi \sin^2 \Theta_{\rm W}} V^\ast_{tb} V_{ts}
X (x_t) (\bar bs)_{V-A} (\bar\nu\nu)_{V-A} + h.c.   
\end{equation}
with $s$ replaced by $d$ in the
case of $B\to X_d\nu\bar\nu$.
 
The theoretical uncertainties related to the renormalization
scale dependence are as in $\klpn$ and 
can be essentially neglected. The same applies to long distance
contributions considered in \cite{BUC97}.
The calculation of the branching fractions for $B\to X_{s,d}\nu\bar\nu$ 
can be done in the spectator model corrected for short distance QCD effects.
Normalizing to $Br(B\to X_c e\bar\nu)$ and summing over three neutrino 
flavours one finds

\begin{equation}\label{bbxnn}
\frac{Br(B\to X_s\nu\bar\nu)}{Br(B\to X_c e\bar\nu)}=
\frac{3 \alpha^2}{4\pi^2\sin^4\Theta_{\rm W}}
\frac{|V_{ts}|^2}{|V_{cb}|^2}\frac{X^2(x_t)}{f(z)}
\frac{\kappa(0)}{\kappa(z)}\,.
\end{equation}
Here $f(z)$ is the phase-space factor for $B\to X_c
e\bar\nu$ with $z=\mc^2/\mb^2$  and $\kappa(z)=0.88$ 
\cite{CM78,KIMM} is the
corresponding QCD correction. The
factor $\kappa(0)=0.83$ represents the QCD correction to the matrix element
of the $b\to s\nu\bar\nu$ transition due to virtual and bremsstrahlung
contributions.
In the case of $B\to X_d\nu\bar\nu$ one has to replace $V_{ts}$ by
$V_{td}$ which results in a decrease of the branching ratio by
roughly an order of magnitude.

Setting $Br(B\to X_ce\bar\nu)=10.4\%$, $f(z)=0.54$,
$\kappa(z)=0.88$ and using the values in (\ref{alsinbr})
 we have
\begin{equation}
Br(B \to X_s \nu\bar\nu) = 3.7 \cdot 10^{-5} \,
\frac{|V_{ts}|^2}{|V_{cb}|^2} \,
\left[ \frac{\mtb(\mt)}{170\gev} \right]^{2.30} \, .
\label{eq:bxsnnnum}
\end{equation}
Taking next, 
$f(z)=0.54\pm 0.04$ and 
$Br(B\to X_ce\bar\nu)=(10.4\pm 0.4)\%$
and scanning the input parameters of table \ref{tab:inputparams}
we find
\begin{equation}\label{klpnr3}
Br(B \to X_s \nu\bar\nu)=
(3.5 \pm 0.7)\cdot 10^{-5}
\end{equation}
to be compared with the experimental upper bound:
\begin{equation}\label{124}
Br(B\to X_s \nu\bar\nu) < 7.7\cdot 10^{-4} 
\quad
(90\%\,\,\mbox{C.L.})
\end{equation}
obtained for the first time by ALEPH \cite{Aleph96}.
This is only a factor of 20 above the SM expectation.
Even if the actual measurement of this decay is very difficult,
all efforts should be made to reach this goal. One should also 
make attempts to measure $Br(B\to X_d \nu\bar\nu)$. Indeed 
\begin{equation}\label{bratio}
\frac{Br(B\to X_d\nu\bar\nu)}{Br(B\to X_s\nu\bar\nu)}=
\frac{|V_{td}|^2}{|V_{ts}|^2}
\end{equation} 
offers the
cleanest direct determination of $\vtd/\vts$ as all uncertainties related
to $\mt$, $f(z)$ and $Br(B\to X_ce\bar\nu)$ cancel out.

It might be much easier to measure the exclusive mode $B\to K^*\nu\bar\nu$.
 Most recent discussions can be found in \cite{NUNUK}.
\subsubsection{$B_{s,d}\to l^+l^-$}
The decays $B_{s,d}\to l^+l^-$ are after $B\to X_{s,d}\nu\bar\nu$ 
the theoretically cleanest decays in the field of rare $B$-decays.
They are dominated by the $Z^0$-penguin and box diagrams
involving top quark exchanges which we encountered already
in the case of $B\to X_{s,d}\nu\bar\nu$   except that due to
charged leptons in the final state the charge flow in the
internal lepton line present in the box diagram is reversed.
This results in a different $\mt$ dependence summarized
by the function  $Y(x_t)$, the NLO generalization \cite{BB2,MU98,BB98}
of the function $Y_0(x_t)$ given in (\ref{XA0}).
The charm contributions are fully negligible
here and the resulting effective Hamiltonian is given 
for $B_s\to l^+l^-$ as follows:
\begin{equation}\label{hyll}
{\cal H}_{\rm eff} = -{G_{\rm F}\over \sqrt 2} {\alpha \over
2\pi \sin^2 \Theta_{\rm W}} V^\ast_{tb} V_{ts}
Y (x_t) (\bar bs)_{V-A} (\bar ll)_{V-A} + h.c.   \end{equation}
with $s$ replaced by $d$ in the
case of $B_d\to l^+l^-$.

The function $Y(x_t)$ is given by
\begin{equation}\label{yyx}
Y(x_t) = Y_0(x_t) + \aspi Y_1(x_t) \equiv \eta_Y Y_0(x_t),
\qquad \eta_Y=1.012
\end{equation}
where $Y_1(x_t)$ can be found in \cite{BB2,MU98,BB98}.
The leftover $\mu_t$-dependence in $Y(x_t)$ is tiny and amounts to
an uncertainty of $\pm 1\%$ at the level of the branching ratio.
With $\mt\equiv \mtb(\mt)$ the QCD factor $\eta_Y$
depends only very weakly on $m_t$. 
The dependence on
$\Lambda_{\overline{MS}}$ can be neglected. 

The branching ratio for $B_s\to l^+l^-$ is given by \cite{BB2}
\begin{equation}\label{bbll}
Br(B_s\to l^+l^-)=\tau(B_s)\frac{G^2_{\rm F}}{\pi}
\left(\frac{\alpha}{4\pi\sin^2\Theta_{\rm W}}\right)^2 F^2_{B_s}m^2_l m_{B_s}
\sqrt{1-4\frac{m^2_l}{m^2_{B_s}}} |V^\ast_{tb}V_{ts}|^2 Y^2(x_t)
\end{equation}
where $B_s$ denotes the flavour eigenstate $(\bar bs)$ and $F_{B_s}$ is
the corresponding decay constant. Using
(\ref{alsinbr}) and (\ref{yyx})  we find in the
case of $B_s\to\mu^+\mu^-$
\begin{equation}\label{bbmmnum}
Br(B_s\to\mu^+\mu^-)=3.3\cdot 10^{-9}\left[\frac{\tau(B_s)}{1.5~
\mbox{ps}}\right]
\left[\frac{F_{B_s}}{210\mev}\right]^2 
\left[\frac{|V_{ts}|}{0.040}\right]^2 
\left[\frac{\mtb(\mt)}{170\gev}\right]^{3.12}.
\end{equation}

The main uncertainty in this branching ratio results from
the uncertainty in $F_{B_s}$.
Scanning the input parameters of table \ref{tab:inputparams}
together with $\tau(B_s)=1.5$ ps and $F_{B_s}=(210\pm 30)\mev$ 
we find
\begin{equation}\label{klpnr1}
Br(B_s\to\mu^+\mu^-)=
(3.2 \pm 1.5)\cdot 10^{-9}~.
\end{equation}

For $B_d\to\mu^+\mu^-$ a similar formula holds with obvious
replacements of labels $(s\to d)$. Provided the decay constants
$F_{B_s}$ and $F_{B_d}$ will have been calculated reliably by
non-perturbative methods or measured in leading leptonic decays one
day, the rare processes $B_{s}\to\mu^+\mu^-$ and $B_{d}\to\mu^+\mu^-$
should offer clean determinations of $|V_{ts}|$ and $|V_{td}|$. 
In particular the ratio
\begin{equation}
\frac{Br(B_d\to\mu^+\mu^-)}{Br(B_s\to\mu^+\mu^-)}
=\frac{\tau(B_d)}{\tau(B_s)}
\frac{m_{B_d}}{m_{B_s}}
\frac{F^2_{B_d}}{F^2_{B_s}}
\frac{|V_{td}|^2}{|V_{ts}|^2}
\end{equation}
having smaller theoretical uncertainties than the separate
branching ratios should offer a useful measurement of
$\vtd/\vts$. Since $Br(B_d\to\mu^+\mu^-)= {\cal O}(10^{-10})$
this is, however, a very difficult task. For $B_s \to \tau^+\tau^-$
and $B_s\to e^+e^-$ one expects branching ratios ${\cal O}(10^{-6})$
and ${\cal O}(10^{-13})$, respectively, with the corresponding branching 
ratios for $B_d$-decays by one order of magnitude smaller.

The bounds on $B_{s,d}\to \mu\bar \mu$ are still
 many orders of magnitude away from SM expectations.
The $90\% C.L.$  bounds from CDF read
\begin{equation}\label{MUBOUND}
Br(B_s\to\mu^+\mu^-)\le 
2.0\cdot 10^{-6}~~~~~(90\% C.L.)~\cite{CDFMU}
\end{equation}
and $Br(B_d\to\mu^+\mu^-)\le 6.8\cdot 10^{-7}$. CLEO \cite{CLEOMU}
provided the bound $Br(B_d\to\mu^+\mu^-)\le 6.1\cdot 10^{-7}$.
CDF should reach in Run II the
sensitivity of $1\cdot 10^{-8}$ and $4\cdot 10^{-8}$ for
$B_d\to \mu\bar\mu$ and $B_s\to \mu\bar\mu$, respectively.
Thus if the SM is the whole story one will have to
wait until LHC-B and BTeV to see any events. On the other hand
in a Two-Higgs-Doublet-Model and in particular in the MSSM
one can find substantially larger branching ratios provided
$\tan\beta$ is large \cite{LARGET}. This means that either
this decay will be measured in Run II at Fermilab or the allowed
parameter space in these models will be considerably reduced.
The usefulness of this decay and of $B\to\tau^+\tau^-$ 
 in tests of the physics beyond the SM
is discussed in these papers and in \cite{GLN96}.

\section{ CP Violation in B Decays}
\setcounter{equation}{0}
\subsection{Preliminaries}
CP violation in B decays is certainly one of the most important 
targets of B-factories and of dedicated B-experiments at hadron 
facilities. It is well known that CP-violating effects are expected
to occur in a large number of channels at a level attainable 
experimentally in the near future.
 Moreover there exist channels which
offer the determination of CKM phases essentially without any hadronic
uncertainties. 
Therefore, it is expected that
during the coming years very important progress in our understanding
of CP violation will come through the measurements of CP asymmetries
in B decays as well as through various strategies for extracting the angles
$\alpha,~\beta$ and $\gamma$ from two-body B decays.
They are extensively discussed in the books \cite{Branco,Bigi},
 in the working group reports in
\cite{BABAR,LHCB} and in \cite{BF97,NIR99,NQ,NIR08,RF97,B95}.

Here after recalling some useful formulae, I will confine the
discussion to three topics:
\bi
\item
An express review of the classic strategies for the determination
of the angles of the unitarity triangle,
\item
Short description of recent strategies for the determination of the
angle $\gamma$ from the decays $B\to\pi K$,
\item
Strategies involving U--spin symmetry
\ei

\subsection{A Few Useful Formulae and Examples}
Let us begin our discussion with neutral $B$ decays to CP Eigenstates.
A time dependent asymmetry $a_{CP}(t,f)$ in the decay 
$B^0\to f$ with $f$ being a CP eigenstate is given in (\ref{e8})
where we have separated the {\it direct} CP-violating contributions 
from those describing CP violation in the interference of mixing
and decay. As demonstrated in Section 3.7,
an interesting case arises when a single mechanism dominates the 
decay amplitude or the contributing mechanisms have the same weak 
phases. Then $a_{CP}(t,f)$ is given simply by
\begin{equation}\label{e111s}
a_{CP}(t,f)=\eta_f \sin(2\phi_D-2\phi_M) \sin(\Delta Mt)
\end{equation}
where $\phi_D$ is the weak phase in the decay amplitude and $\phi_M$
the weak phase in the $B^0_{d,s}-\bar B^0_{d,s}$ mixing. $\eta_f=\pm 1$
is the CP parity of the final state.
In this particular case the hadronic matrix elements drop out,
 the direct CP-violating contribution
vanishes and the CP asymmetry is given entirely
in terms of the weak phases $\phi_D$ and $\phi_M$. 

If a single tree diagram dominates, the factor $\sin(2\phi_D-2\phi_M)$
can be calculated by using
\begin{equation}\label{dtree}
\phi_D =\left\{ \begin{array}{ll}
\gamma & b\to u \\
0 & b\to c \end{array} \right.
\qquad
\phi_M =\left\{ \begin{array}{ll}
-\beta & B^0_d \\
-\beta_s & B^0_s \end{array} \right.
\end{equation}
where we have indicated the basic transition of the b-quark into
a lighter quark. $\beta_s=\ord(10^{-2})$.
On the other hand if the penguin diagram with internal
top exchange dominates one has
\begin{equation}\label{pen}
\phi_D =\left\{ \begin{array}{ll}
-\beta & b\to d \\
0 & b\to s \end{array} \right.
\qquad
\phi_M =\left\{ \begin{array}{ll}
-\beta & B^0_d \\
-\beta_s & B^0_s~. \end{array} \right.
\end{equation}
Let us practice these formulae.
Assuming that 
$B_d\to \psi K_S$ and $B_d\to \pi^+\pi^-$ are dominated by
tree diagrams with $b\to c$ and $b\to u$ transitions respectively
we readly find
\begin{equation}\label{113c}
 a_{CP}(t,\psi K_S)=-\sin(2\beta) \sin(\Delta M_d t),
\ee
\be\label{113d}
   a_{CP}(\pi^+\pi^-)=-\sin(2\alpha) \sin(\Delta M_d t).
\end{equation}
Now in the case of $B_d\to \psi K_S$ the penguin diagrams have
to a very good approximation the same phase ($\phi_D=0)$ as
the tree contribution and are moreover Zweig suppressed.
Consequently (\ref{113c}) is very accurate. This is not
the case for $B_d\to \pi^+\pi^-$ where the penguin contribution
could be substantial. Heaving weak phase $\phi_D=-\beta$,
which differs from the tree phase $\phi_D=\gamma$, this penguin
contribution changes effectively $\sin (2\alpha)$  to
$\sin (2\alpha+\theta_P)$
where $\theta_P$ is a function of $\beta$ and hadronic parameters.
Strategies to determine $\theta_P$ and consequently $\alpha$ are
discussed below.

Similarly the pure penguin dominated decay $B_d\to \phi K_S$
is governed by the $b\to s$ penguin  with internal top exchange
which implies that in this decay the angle $\beta$ is measured.
The accuracy of this measurement is a bit lower than using 
$B_d\to \psi K_S$ as penguins with internal u and c exchanges
may introduce a small pollution.

Finally we can consider the asymmetry in $B_s\to\psi\phi$, 
an analog of $B_d \to \psi K_s$. In the leading order of the
Wolfenstein parametrization the asymmetry $a_{CP}(t,\psi\phi)$ vanishes.
Including higher order terms in $\lambda$ one finds \cite{B95}
\begin{equation}\label{DU}
a_{CP}(t,\psi\phi)=2\lambda^2\eta \sin(\Delta M_s t)
\end{equation}
where $\lambda$ and $\eta$ are the Wolfenstein parameters.

\subsection{Classic Strategies}
\subsubsection{The Angle \boldmath{$\alpha$}}
The classic determination of $\alpha$ by means of the
time dependent CP  asymmetry in the decay
$B_d^0 \rightarrow \pi^+ \pi^-$ as given by (\ref{113d})
is affected by the ``QCD penguin pollution" which has to be
taken care of in order to extract $\alpha$. We have just mentioned
this problem.
The  CLEO results for penguin dominated $B \to\pi K$ decays 
indicated \cite{CLEO99} that
this pollution could be substantial as stressed in particular
in \cite{ITAL}.
The well known strategy to deal with this "penguin problem''
is the isospin analysis of Gronau and London \cite{CPASYM}. It
requires however the measurement of $Br(B^0\to \pi^0\pi^0)$ which is
expected to be below $10^{-6}$: a very difficult experimental task.
For this reason several, rather involved, strategies \cite{SNYD} 
have been proposed which
avoid the use of $B_d \to \pi^0\pi^0$ in conjunction with
$a_{CP}(\pi^+\pi^-,t)$. They are reviewed in 
\cite{BF97,BABAR,LHCB}. 
 It is to be seen which of these methods
will eventually allow us to measure $\alpha$ with a respectable precision.

The most promising at present appears the method of Quinn and Snyder  
\cite{SNYD}.
It uses the Dalitz plot for $B^0_d\to \varrho\pi\to \pi^+\pi^-\pi^0$.
This method allows to determine both $\sin 2\alpha$ and $\cos 2\alpha$,
reducing possible discrete ambiguities. The remaining discrete
ambiguity can be removed with the help of CP asymmetry in
$B^0_d\to\pi^+\pi^-$ and a theoretical assumption \cite{HRQ}.

While the determination of $\alpha$ remains a
challenge for both theorists and experimentalists, the situation looks
more promising now than in 1999. First the new measurements of
$Br(B^0_d\to\pi^+\pi^-)$ from BaBar and Belle
\begin{equation}\label{pipi}
 Br(B^0_d\to\pi^+\pi^-)=\left\{ \begin{array}{ll}
(9.3 \pm 2.4\pm 1.3)\cdot 10^{-6} & {\rm (BaBar)}~\cite{BrBaBar} \\
(6.3 \pm 4.0)\cdot 10^{-6} &{\rm (Belle)}~ \cite{Bellepi},
\end{array} \right.
\end{equation}
show that this decay mode may be less suppressed relative to
$B^0_d\to K^+\pi^-$ than was suggested by CLEO, implying smaller
penguin pollution. In addition the QCD factorization approach
\cite{BBNS1,NEUB} could help in calculating the penguin pollution 
from first principles.

\subsubsection{The Angle \boldmath{$\beta$}}
The CP-asymmetry in the decay $B_d \rightarrow \psi K_S$ allows
 in the SM
a direct measurement of the angle $\beta$ in the unitarity triangle
without any theoretical uncertainties \cite {BSANDA}. The
relevant formula is given in (\ref{113c}).
Of considerable interest \cite{RF97,PHI} is also the pure penguin decay
$B_d \rightarrow \phi K_S$, which is expected to be sensitive
to physics beyond the SM. Comparision of $\beta$
extracted from $B_d \rightarrow \phi K_S$ with the one from
$B_d \rightarrow \psi K_S$ should be important in this
respect. An analogue of $B_d \rightarrow \psi K_S$ in $B_s$-decays
is $B_s \rightarrow \psi \phi$. As shown in (\ref{DU}), the 
corresponding CP asymmetry measures here
$\eta$ \cite{B95} in the Wolfenstein parametrization. It is very
small, however, and this fact makes it a good place to look for the 
physics beyond the SM. In particular the CP violation
in $B^0_s-\bar B^0_s$ mixing from new sources beyond the Standard
Model should be probed in this decay.
Another useful channel for $\beta$ is $B_d\to D^+ D^-$.

\subsubsection{The Angle \boldmath{$\gamma$}}
The two theoretically cleanest methods for the determination of $\gamma$
are: i) the full time dependent analysis of 
$B_s\to D^+_s K^{-}$ and $\bar B_s\to D^-_s K^{+}$  \cite{adk}
and ii) the well known triangle construction due to Gronau and Wyler 
\cite{Wyler}
which uses six decay rates $B^{\pm}\to D^0_{CP} K^{\pm}$,
$B^+ \to D^0 K^+,~ \bar D^0 K^+$ and  $B^- \to D^0 K^-,~ \bar D^0 K^-$.
Both methods are  unaffected by penguin contributions. 
The first method is experimentally very
challenging because of the
expected large $B^0_s-\bar B^0_s$ mixing. The second method is problematic
because of the small
branching ratios of the colour supressed channel $B^{+}\to D^0 K^{+}$
and its charge conjugate,
giving a rather squashed triangle and thereby
making
the extraction of $\gamma$ very difficult. Variants of the latter method
which could be more promising have been proposed in \cite{DUN2,V97}.
Very recently the usefulness of $B_c\to D D_s$ for the extraction of
$\gamma$ was stressed in \cite{FW01}. 
It appears that these methods will give useful results at later stages
of CP-B investigations. In particular the first and the last method 
will be feasible
only at LHC-B. Other recent strategies for $\gamma$ will be discussed 
below.

\subsection{Constraints for \boldmath{$\gamma$} from $B\to\pi K$}
The most recent developments are related to the extraction of
the angle $\gamma$ from the decays $B\to PP$ (P=pseudoscalar).
Several of these modes have been observed 
by the CLEO collaboration \cite{CLEO99}. In the summer of 2000
BaBar \cite{BrBaBar} and Belle \cite{BrBelle} announced their first 
results for $B\to K\pi$  branching ratios. In the future they should 
allow us to obtain direct 
information on $\gamma$.
At present, there are only experimental results 
available for the combined branching ratios of these modes, i.e.\ averaged 
over decay and its charge conjugate, suffering from large 
uncertainties. They are collected in table~\ref{Kpi}.

\begin{table}[thb]
\caption[]{ Branching ratios for $B\to K\pi$ Values in units of $10^{-6}$.
\label{Kpi}}
\begin{center}
\begin{tabular}{|c|c|c|c|}\hline
Decay & CLEO & BaBar & Belle  
\\ \hline
$B_d^0\to\pi^\mp K^\pm$ &$ 17.2^{+2.5}_{-2.4}\pm1.2$ &
$12.5^{+3.0+1.3}_{-2.6-1.7}$ &$ 17.4^{+5.1}_{-4.6}\pm 3.4$ \\
$B^\pm\to\pi^0 K^\pm$ &$ 11.6^{+3.0+1.4}_{-2.7-1.3}$ &
 & $18.8^{+5.5}_{-4.9}\pm 2.3$ \\
$B^\pm\to \pi^\pm K$& $ 18.2^{+4.6}_{-4.0}\pm 1.6$ &
 &  \\
$B^0_d\to \pi^0 K$ & $14.6^{+5.9+2.4}_{-5.1-3.3}$ &
 & $21.0^{+9.3+2.5}_{-7.8-2.3}$ \\
\hline
 \end{tabular}
\end{center}
\end{table}

There has been a large activity in this field during the last three years.
The main issues here are the final state interactions (FSI) \cite{FSI}, 
SU(3) symmetry
breaking effects and the importance of electroweak penguin
contributions. Several interesting ideas have been put forward
to extract the angle $\gamma$ in spite of large hadronic
uncertainties in $B\to \pi K$ decays 
\cite{FM,GRRO,GPAR1,GPAR3,GPAR2,NRBOUND}.
Reviews can be found in \cite{RFA,GPAR3,NEU99}.

Three strategies for bounding and determining $\gamma$ have been 
proposed. The ``mixed" strategy \cite{FM} uses 
$B^0_d\to \pi^0 K^\pm$ and $B^\pm\to\pi^\pm K$. The ``charged" strategy
\cite{NRBOUND} involves $B^\pm\to\pi^0 K^\pm,~\pi^\pm K$ and
the ``neutral" strategy \cite{GPAR3} the modes 
$B_d^0\to \pi^\mp K^\pm,~\pi^0K^0$. 
General parametrizations for the 
study of the FSI, SU(3) symmetry
breaking effects and of the electroweak penguin
contributions in these channels have been presented 
in \cite{GPAR1,GPAR3,GPAR2}.

As demonstrated in \cite{FM,GPAR1,NRBOUND,GPAR3,GPAR2}, 
already CP-averaged $B\to\pi K$ branching ratios
may imply interesting bounds on $\gamma$ 
that may remove a large portion of the allowed range from the analysis
of the unitarity triangle. In particular combining the neutral and
charged strategies \cite{GPAR3} one finds that the most recent
CLEO data favour $\gamma$  in the second quadrant, which is in 
conflict with the standard analysis of the unitarity triangle as
we have seen in section 4. Other arguments for $\cos\gamma<0$ using
$B\to PP,~PV$ and $VV$ decays were given in \cite{CLEO99,HHY}.
Simultaneously to $\gamma$, the CLEO data provide some information
on strong phases. The present pattern of these phases indicates either
new-physics contributions to the electroweak penguin sector, or a
manifestation of large non-factorizable $SU(3)$-breaking effects
 \cite{GPAR3}. There is no doubt that these
strategies will be useful in the future.

Finally ratios of CP-averaged $B\to \pi K$ and $B\to\pi\pi$
rates as functions of $\gamma$ have been studied within the
QCD factorization approach in \cite{BBNS2}. Interestingly, the
ratio $R_\pi$ of the $B^0_d\to\pi^+\pi^-$ and $B^0_d\to \pi^\mp K^\pm$
decay rates is in disagreement with the CLEO experimental
value $0.25\pm 0.10$, unless the weak phase $\gamma$ were 
significantly larger
than $90^\circ$. Similar results using QCD factorization approach
have been found in \cite{Muta2}. 
This is in accordance with the findings in
\cite{GPAR3,CLEO99,HHY}. On the other hand the most recent value
$R_\pi=0.74\pm0.29$ from BaBar \cite{BrBaBar} does not necessarily 
require 
$\gamma>90^\circ$ and is in accordance with the expectations.
 With improved data, also from Belle, the situation should become
clearer already  this year.
\subsection{Employing U-Spin Symmetry}
 New strategies for $\gamma$ using the U-spin symmetry have
been proposed in \cite{RF99,RF991}. The first strategy involves
the decays $B^0_{d,s}\to \psi K_S$ and $B^0_{d,s}\to D^+_{d,s} D^-_{d,s}$
 \cite{RF99}.
The second strategy involves $B^0_s\to K^+ K^-$ and $B^0_d\to\pi^+\pi^-$
\cite{RF991}. These strategies are unaffected by FSI and are only limited
by U-spin breaking effects. They are promising for
Run II at FNAL and in particular for LHC-B. 

A method of determining $\gamma$, using $B^+\to K^0\pi^+$ and the
U-spin related processes $B_d^0\to K^+\pi^-$ and $B^0_s\to \pi^+K^-$,
was presented in \cite{GRCW}. A general discussion of U-spin symmetry 
in charmless B decays and more references to this topic can be
found in \cite{G00}.

\subsection{Summary}
This review together with the general discussion of CP asymmetries in
Section 3, was only a short excursion into the reach field of CP
violation in B decays. Clearly this field being dominated by
the interplay of flavour and QCD dynamics will give us an insight
into sofar poorly explored sector of particle physics. It is
exciting that in the coming years the new experimental data will
uncover various patterns of CP asymmetries and branching ratios bringing
some order into the structure of non-leptonic B decays and hopefully
giving some clear signals of new physics.

\section{ Looking Beyond the Standard Model}
\setcounter{equation}{0}
\subsection{General Remarks}
We begin the discussion of the physics beyond the Standard
Model with a few general remarks. As the new particles
in the extensions of the SM
are generally substantally heavier than $W^\pm$, the
impact of new physics on charged current tree level decays should
be marginal. On the other hand these new contributions
could have an important impact on
loop induced decays. From these two observations we
conclude:

\bi
\item
New physics should have only marginal impact on the determination
of $|V_{us}|$, $\vcb$ and $|V_{ub}|$.
\item
There is no impact on the calculations of the low energy non-perturbative
parameters $B_i$ except that new physics can bring new local
operators implying new parameters $B_i$.
\item
New physics could have substantial impact on rare and CP violating
decays and consequently on the determination of the unitarity triangle.
\ei

\subsection{CP Violation Beyond the SM}
The pattern of CP violation in the extensions of the SM
deviates generally from the KM picture of CP violation.
Let us discuss it briefly by comparing the special features
of CP violation in the SM with those which appear
in its extentions. A more detailed discussion can be found in
\cite{NIR99}.

\begin{itemize}
\item
In the SM CP is explicitly broken through complex phases in the
Yukawa couplings. In the extensions of the SM CP can also be 
spontaneously broken. The latter case takes place if the
scalar vacuum expectation values contain phases which cannot be
removed.
\item
In the SM there is a single complex phase ($\delta$ or $\eta$).
In the extensions of the SM new complex phases can be present.
Examples are models with extended Higgs sector and supersymmetric
models.
\item
In the SM CP violation occurs only in charged current weak 
interactions of quarks, which are necessarly flavour
violating. As a consequence CP violation is strongly suppressed
in neutral current transitions ($Z^0$, $\gamma$, $H^0$) as it
can appear there only as a loop effect. Similarly CP violation 
is very
strongly suppressed in flavour diagonal transitions, which
implies for instance unmeasurable electric dipole moments.
\item
In the extensions of the SM CP violation may occur not only in
charged current transitions but also in neutral current transitions
at the tree level. Moreover it can occur in flavour diagonal
interactions. Such interactions can be found in  supersymmetric
models and generally in models with an extended Higgs sector.
Moreover flavour violating processes mediated by gluinos
$(\ord(\alpha_s))$ can be CP--violating. Consequently in contrast
to the SM large CP--violating effects in neutral
and flavour diagonal transitions are possible. Good examples
are substantial, possibly in the near feature measurable,
electric dipole moments in some supersymmetric models and
models with an extended Higgs sector.
\end{itemize}

\subsection{Classification of New Physics}
Classification of new physics contributions can be done in various ways. 
We find it useful to classify these contributions from the point
of view of the operator structure of the effective weak Hamiltonian,
the complex phases present in the Wilson coefficients of the
relevant operators and the distinction whether the flavour changing
transitions are governed by the CKM matrix or by new sources of
flavour violation.

Let us then group the extensions of the SM in five
classes. For the first four classes we assume that there are only three
generations of quarks and leptons. The last class allows for
more generations.

{\bf Class A}

\bi
\item
 There are no new complex phases and flavour changing transitions are
 governed  by the CKM matrix.
\item
 There are no new operators beyond those present in the SM.
\item
 The Wilson coefficients of the SM operators receive new contributions
  through diagrams involving new internal particles.
\ei
 These new contributions will modify the SM expressions
for rare and CP violating decays but these modifications can
be formulated in a very transparent manner as we will see
below. The presence of new contributions will have 
generally impact on the determination
 of $\alpha$, $\beta$, $\gamma$, $\vtd$ and $\IM\lambda_t$.
 In an analysis of the unitarity triangle that uses
 the SM formulae of Section 4 these new contributions will be signaled by
\bi
\item
Inconsistencies in the determination of $(\bar\varrho,\bar\eta)$
through $\varepsilon$, $B^0_{s,d}-\bar B^0_{s,d}$ mixing and
rare decays.
\item
Disagreement of $(\bar\varrho,\bar\eta)$ extracted 
from loop induced decays
with $(\bar\varrho,\bar\eta)$ extracted using
CP asymmetries.
\ei
Examples are the Two Higgs Doublet Model II and the constrained MSSM
if $\tan\beta$ is not too large. This class of models, to be named
MFV--models \cite{CDGG,UUT} (MFV= Minimal Flavour Violation) will be 
discussed in some detail below. 

{\bf Class B}

This class of models differs from class A through the contributions
of new operators not present in the SM. It is assumed, however,
that no new complex phases
beyond the CKM phase are present.
Examples are again the two Higgs doublet model II and the constrained MSSM
with large $\tan\beta$ and all new phases set to zero.

{\bf Class C}

This class of models differs from class A through the presence of
new complex phases in the Wilson coefficients of the usual SM
operators. Contributions of new operators can
be, however, neglected. In these models new flavour changing transitions
appear that are not governed by the CKM matrix. An example is the
MSSM with  not too a large $\tan\beta$ and with non-diagonal elements
in the squark mass matrices.

 This kind of new physics will also be signaled by inconsistencies
 in the $(\bar\varrho,\bar\eta)$ plane. However, new complications
 arise. Because of new phases CP violating asymmetries measure
 generally different quantities than $\alpha$, $\beta$ and $\gamma$.
 For instance the CP asymmetry in $B\to \psi K_S$ will no longer
 measure $\beta$ but $\beta+\theta_{NP}$ where $\theta_{NP}$
 is a new phase. Strategies for dealing with such situations
 have been developed. See for instance \cite{NIR96,BNEW,MPR,BRS} 
 and references  therein.

 {\bf Class D}

 Here we group models with new complex phases, new operators
 and new flavour changing contributions which are not governed
 by the CKM matrix. The phenomenology in this class of models
 is more involved than in the classes B and C \cite{MPR,GGMS}.

 Examples of models in classes C and D are multi-Higgs models 
 with complex phases in the  Higgs sector, general SUSY models, 
 models with spontaneous
 CP violation and left-right symmetric  models.

 {\bf Class E}

 Here we group the models in which
the unitarity of the three generation CKM matrix does not
hold. 
Examples are four generation models and models with tree
level FCNC transitions. If this type of physics is present,
the unitarity triangle does not close or some inconsistencies
in the $(\bar\varrho,\bar\eta)$ plane take place.

Clearly in order to sort out which type of new physics is
responsible for deviations from the SM
expectations one has to study many loop induced decays
and many CP asymmeteries. Some ideas in this direction
can be found in \cite{NIR96,BNEW}.

\subsection{Models with Minimal Flavour Violation}

\subsubsection{General Remarks}
We will assume in accordance with the experimental findings that
all new particles have masses higher than $\mw$. As in this
class of models  the effective local operators are the same as in
the SM, the impact of new contributions in a
renormalization group analysis is then only felt in the initial conditions
for the Wilson coefficients taken usually at $\mu={\cal O}(\mw)$.  The
renormalization group transformation from $\mu={\cal O}(\mw)$ down to
$\mu={\cal O}(1~\gev)$ is on the other hand the same as in the SM. 
 This simplifies the inclusion of QCD corrections in the new
contributions considerably.
The only thing one has to do is to calculate QCD corrections to the
new contributions in the full theory. The remaining QCD corrections
present in the effective theory are the same as in the SM. 
Similarly all hadronic matrix elements and the related
non--perturbative parameters are as in the SM.

In view of the special manner in which the  new contributions affect
the decay amplitudes, it is useful to use the penguin--box expansion
 discussed in section 2.6 which suppressing CKM parameters reads
\begin{equation}        \label{PENEW}
        A= P_0+\sum_r P_r F_r~.
\end{equation}
The coefficients $P_i$ depend on non-perturbative parameters
and include NLO-QCD and QED corrections related to effective theory. 
They are common to the SM and all MFV models discussed here.
The functions $F_r$ resulting from various box and penguin diagrams
contain both the SM and new physics contributions involving
new exchanges such as charged Higgs particles, charginos, squarks etc.
They depend, in addition to $\mt$, on the masses of new particles
such as
charged Higgs particles, charginos, squarks and sleptons as well as on
a number of new physics parameters. 
They include also QCD corrections calculated in the full theory.

The explicit expressions
for $F_r$ in the SM without QCD corrections have been listed in 
section 2.6. The corresponding
generalization valid for the Two--Higgs--Doublet Model II and the MSSM
can be found in \cite{EP00}.
Clearly the calculations of QCD corrections in the full and effective
theories  must be compatible with each other. This
is, however, well understood by now \cite{AJBLH,BBL}.

Before presenting some specific results in a restricted MSSM with
minimal flavour violation, we would like
to discuss two specific features of this class of models.

\subsubsection{Universal Unitarity Triangle}
Even if this class of extentions of the SM is very simple,
the extraction of the CKM parameters in these 
models appears at first sight to be more complicated than in the
SM because new physics contributions depend on unknown parameters, 
like the masses and couplings of new particles, that pollute the 
extraction of the CKM parameters.

Yet as pointed out recently \cite{UUT}, in this class of extensions of 
the SM it is possible to
construct measurable quantities that depend on the CKM parameters
but are not polluted by new physics contributions. This means that
these quantities allow a direct determination of the ``true'' values
of the CKM parameters which are common to the SM and this particular class
of its extensions. Correspondingly there exists a {\it universal unitarity
triangle} common to all these models. 

The determination of this universal unitarity triangle and of the 
corresponding CKM
parameters has four virtues:
\begin{itemize}
\item
The CKM parameters can be determined without the knowledge of new
unknown parameters present in these particular extensions of the SM.
\item
Because the extracted values of the CKM parameters are also valid in these 
models, the dependence of various quantities on the new parameters 
becomes more transparent. In short: the determination of the CKM 
parameters and of the new parameters 
can be separated from each other.
\item
The comparison of the predictions for a given observable in the SM
and in this kind of extensions can then be done keeping the CKM
parameters fixed.
\item
Interestingly,
the extraction of the universal CKM parameters
is essentially free from hadronic uncertainties.
\end{itemize}

In what follows we will list the set of quantities which allow a
determination of the universal unitarity triangle. Subsequently we
will indicate how the models in this class can be distinguished
from each other and from more complicated models which
bring in new complex phases and new operators. We refer to \cite{UUT}
 for details.

The strategy is very simple. As the new physics contributions
enter only through the functions $F_r$, one has to look
for certain ratios in which these functions cancel out.

The determination of $R_t$, one of the sides of the universal unitarity
triangle, can be best achieved by using the ratio
$\Delta M_d/\Delta M_s$. Indeed as discussed in section 4.1
one has to a very good accuraccy
\begin{equation}\label{vtdts}
\left|\frac{V_{td}}{V_{ts}}\right|^2=\lambda^2
\frac{(1-\bar\varrho)^2+\bar\eta^2}{1+\lambda^2(2\bar\varrho-1)}
\approx \lambda^2 R_t^2~.
\end{equation}
Consequently using
\begin{equation}\label{107x}
\frac{\vtd}{|V_{ts}|}= 
\xi\sqrt{\frac{m_{B_s}}{m_{B_d}}}
\sqrt{\frac{\Delta M_d}{\Delta M_s}},
\qquad
\xi = 
\frac{F_{B_s} \sqrt{\hat B_{B_s}}}{F_{B_d} \sqrt{\hat B_{B_d}}}
\end{equation}
one can determine $R_t$ independently of new parameters characteristic 
for a given model and of $\mt$.
If necessary the $\ord(\lambda^2)$ corrections in (\ref{vtdts}) can
be incorporated. This will be only required when the
error on $\xi$ will be decreased below $2\%$, which is clearly
a very difficult task.

While the ratio $\Delta M_d/\Delta M_s$ will be the first one to serve
our purposes, there are at least two other quantities which allow
a clean measurement of $R_t$ within the class of the extensions of the
SM considered.
These are the ratios
\begin{equation}\label{bxnn}
\frac{Br(B\to X_d\nu\bar\nu)}{Br(B\to X_s\nu\bar\nu)}=
\left|\frac{V_{td}}{V_{ts}}\right|^2
\end{equation}
\begin{equation}\label{bmumu}
\frac{Br(B_d\to\mu^+\mu^-)}{Br(B_s\to\mu^+\mu^-)}=
\frac{\tau_{B_d}}{\tau_{B_s}}\frac{m_{B_d}}{m_{B_s}}
\frac{F^2_{B_d}}{F^2_{B_s}}
\left|\frac{V_{td}}{V_{ts}}\right|^2
\end{equation}
which similarly to $\Delta M_d/\Delta M_s$ measure 
$\vtd/\vts$ 
As discussed in section 7, out of these three ratios the cleanest 
is (\ref{bxnn}), which is essentially free
of hadronic uncertainties. Next comes (\ref{bmumu}), involving
$SU(3)$ breaking effects in the ratio of $B$-meson decay constants.
Finally, $SU(3)$ breaking in the ratio of bag parameters
$\hat B_{B_d}/\hat B_{B_s}$ enters in addition in (\ref{107x}). These 
$SU(3)$ breaking effects should eventually be calculable with
high precision from lattice QCD.

Note that the branching ratio for the rare decay 
$K^+\to\pi^+\nu\bar\nu$ provides a clean measurement of $V_{td}$ and
consequently of $R_t$ as discussed in section 6.
However, this branching ratio alone
cannot serve our purposes because it is sensitive to new physics 
contributions through the function X.

In order to complete the determination of $\bar\varrho$ and $\bar\eta$
in the universal unitarity triangle 
one can use $\sin2\beta$ extracted either from the CP asymmetry
in $B_d\to\psi K_S$  or from $K\to\pi\nu\bar\nu$ decays. We
have discussed the relevant formulae in sections 8 and 6 respectively. 
They remain valid in the models in question as there are no
new complex phases involved. As these formulae do not depend on
the functions $F_r$,
both extractions of $\sin 2\beta$  are  independent of
the new parameters characteristic for a given model. 

Concerning the determination of the angle $\gamma$,
any  method for the determination of $\gamma$
in which new physics of the type considered here can be eliminated
or neglected can be used here, in particular the methods 
in \cite{adk,Wyler,DUN2,V97} that involve only tree diagrams.

Once $R_t$ and $\sin 2\beta$
have been determined as discussed above, $\bar\varrho$ and $\bar\eta$
can be found through \cite{AJB94,B95}
\begin{equation}\label{5u}
\bar\eta=a\frac{R_t}{\sqrt{2}}\sqrt{\sin 2\beta \cdot r_{-b}(\sin 2\beta)}\,,
\quad\quad
\bar\varrho = 1-\bar\eta r_{b}(\sin 2\beta)\,
\end{equation}
where 
\be
r_b(z)=(1+b\sqrt{1-z^2})/z, \qquad a,b=\pm~.
\ee
Thus for given values of $(R_t,\sin 2\beta)$ there are four solutions
for  $(\bar\varrho,\bar\eta)$ corresponding to $(a,b)=(+,+),~(+,-),
~(-,+),~(-,-)$.
As described in \cite{AJB94}
three of these solutions can be eliminated by using further information, 
for instance coming from 
$|V_{ub}/V_{cb}|$ and $\varepsilon$, so that eventually the solution 
corresponding to $(a,b)=(+,+)$ is singled out,
\begin{equation}\label{5a}
\bar\eta=\frac{R_t}{\sqrt{2}}\sqrt{\sin 2\beta \cdot r_{-}(\sin 2\beta)}\,,
\quad\quad
\bar\varrho = 1-\bar\eta r_{+}(\sin 2\beta)\,~.
\end{equation}
A numerical analysis can be found in \cite{UUT}.

On the other hand $\bar\varrho$ and $\bar\eta$ following from
$R_t$ and $\gamma$ are simply given by
\begin{equation}\label{6a}
\bar\eta=R_b \sin\gamma,
\quad\quad
\bar\varrho = R_b \cos\gamma
\end{equation}
with
\begin{equation}\label{7a}
R_b=\cos\gamma\pm\sqrt{R^2_t-\sin^2\gamma}.
\end{equation}
Comparing the resulting $R_b$ with the one extracted from $\vub$
 (see (\ref{v13})) one of the two solutions can be eliminated.
Similarly using $\vub$ and $\gamma$ one can construct the
universal unitarity triangle by means of (\ref{6a}).

 Alternatively one could use the measurement
of $R_b$ by means of $\vub$ together with $R_t$ from 
$\Delta M_d/\Delta M_s$ to find
\begin{equation}\label{16a}
\bar\eta=\sqrt{R_b^2-\bar\varrho^2}~,
\quad\quad
\bar\varrho = \frac{1}{2} (1+R^2_b-R^2_t).
\end{equation}
This determination
suffers from hadronic uncertainties in the extraction
of $\vub$ but could turn out to be useful once the
uncertainties in $\vub$ have been reduced. We will give
numerical examples below. 

We observe that all these different methods
determine the ``true" values of 
$\bar\eta$ and $\bar\varrho$ independently of 
new physics contributions in
the class of models considered.
Since $\lambda$ and $\vcb=A\lambda^2$ are determined from tree level
K and B decays they are insensitive to new physics as well. Thus
the full CKM matrix can be determined in this manner. The corresponding
universal unitarity triangle common to all the models considered
can be found directly from  formulae like (\ref{5u}), (\ref{5a})
(\ref{6a}) and (\ref{16a}).

Having determined the CKM parameters by one of these methods
 one can calculate $\varepsilon$,
$\epe$, $\Delta M_d$, $\Delta M_s$ and branching ratios for
rare decays. As these quantities depend on the parameters characteristic
for a given model the results for the SM, the MSSM and other models
of this class will generally differ from
each other. Consequently by comparing these predictions with the
data one will be able to find out which
of these  models is singled out by experiment.
Equivalently, $\varepsilon$,
$\epe$, $\Delta M_d$, $\Delta M_s$ and branching ratios for
rare decays allow to determine non-universal unitarity triangles
that depend on the model considered. Only those unitarity
triangles which are the same as the universal triangle survive
the test. 

It is of course possible that  new physics is more complicated than
discussed here and that new complex phases and new operators
beyond those present in the SM have to be taken into account.
These types of effects would be signaled by:
\begin{itemize}
\item
Inconsistencies between different constructions of the universal
triangle,
\item
Disagreements of the data with the $\Delta M_{d,s}$ and the branching
ratios for rare K and B decays predicted on the basis of the
universal unitarity triangle for all models of the class considered
here.
\end{itemize}

In our opinion the universal unitarity triangle provides 
a transparent strategy to distinguish between the MFV models 
 and to  search for physics beyond the SM. 
Presently we do not know this triangle as all the
available measurements used in section 4 for the construction of the unitarity
triangle are sensitive to  physics beyond the SM. 
It is exciting, however, that in the coming years this triangle will be
known once $\Delta M_s$ has been measured and $\sin 2\beta$ extracted 
from the CP asymmetry in $B_d^0\to \psi K_S$. At later stages $\vub$,  
$K\to\pi\nu\bar\nu$,
$B\to X_{d,s}\nu\bar\nu$, $B_{d,s}\to \mu^+\mu^-$
 and future determinations of $\gamma$ through CP asymmetries in
B decays will also be very useful in this respect.

\subsubsection{\boldmath{$\sin 2\beta$} from \boldmath{$\Delta M_s$}
and \boldmath{$\vub$}}
In view of the forthcoming precise measurements of 
$(\sin 2\beta)_{\psi K_S}$ it is of interest to investigate its
dependence on $\Delta M_s$ and $\vub$ in the MFV models. We show
this in table~\ref{sinbeta}. To this end we have set $\xi=1.16$. 
The interesting feature is a very weak dependence on $\Delta M_s$
for high values of $\vub$. For low values of $\vub$ this
dependence is rather strong.
We observe that  
$\sin 2\beta$ has to be above 0.5 when  $\Delta M_s\le 18/ps$. 
On the other hand for
$\Delta M_s= 26/ps$ it can be as low as 0.17. The question then
arises whether such low  values are consistent with other known 
measurements.

\begin{table}[thb]
\caption[]{ Values of $\sin 2\beta$ in the
MFV models for specific values of $\Delta M_s$ and $\vub$.
\label{sinbeta}}
\begin{center}
\begin{tabular}{|c|c|c|c|c|c|}\hline
$\Delta M_s/\vub $& $ 0.065 $ & $0.075$ & $0.085$ & $0.095$ & $0.105$ 
\\ \hline
$17.0/ps$      & $0.528$ & $0.608$ & $0.682$  & $0.749$ & $0.809$ \\
$18.0/ps$      & $0.514$ & $0.599$ & $0.677$  & $0.747$ & $0.809$ \\
$19.0/ps$      & $0.496$ & $0.587$ & $0.669$  & $0.742$ & $0.807$ \\
$20.0/ps$      & $0.473$ & $0.572$ & $0.659$  & $0.736$ & $0.803$ \\
$22.0/ps$      & $0.412$ & $0.531$ & $0.632$  & $0.718$ & $0.793$ \\
$24.0/ps$      & $0.322$ & $0.476$ & $0.595$  & $0.694$ & $0.777$ \\
$26.0/ps$      & $0.165$ & $0.403$ & $0.549$  & $0.664$ & $0.757$ \\
\hline
 \end{tabular}
\end{center}
\end{table}

\subsubsection{An Absolut Lower Bound on \boldmath{$\sin 2\beta$}}
 It turns out that in the MFV models there exists an {\it absolute} 
 lower bound on   $\sin 2\beta$ that follows from the interplay
of $\Delta M_d$ and $\varepsilon$ and   depends only on $\vcb$, $\vub$ 
 and the nonperturbative  parameters $\hat B_K$, $F_{B_d}\sqrt{\hat B_d}$ 
 and $\xi$ \cite{ABRB}. 
 The derivation of this bound is straightforward. We first
generalize the SM expressions for the $\varepsilon$--hyperbola
in (\ref{100}) and $R_t$ resulting from $\Delta M_d$ in (\ref{106}) and
 (\ref{VT}) to the MFV
models simply as follows
\begin{equation}\label{100a}
\bar\eta \left[(1-\bar\varrho) A^2 \eta_2 F_{tt}
+ P_c(\varepsilon) \right] A^2 \hat B_K = 0.226~.
\end{equation}
\begin{equation}\label{RT}
R_t= 1.26~ \frac{ R_0}{A}\frac{1}{\sqrt{F_{tt}}},
\qquad
 R_0= 1.03~\sqrt{\frac{(\Delta M)_d}{0.50/{\rm ps}}}
          \left[\frac{200~\mev}{\sqrt{\hat B_d} F_{B_d}}\right]
          \sqrt{\frac{0.55}{\eta_B}}~.
\ee
That is $S_0(x_t)$ in the SM formulae and given in (\ref{S0}) is just replaced
by $F_{tt}$, that now in addition to the contributions of the
box diagrams with top quark exchanges includes all possible
new physics contributions within a given MFV model.

The most important feature of the formulae (\ref{100a}) and
(\ref{RT}) relevant for the discussion below is that in the context
of the standard analysis of the unitarity triangle the different 
MFV models can be characterized by the value of the function $F_{tt}$.
This has been stressed recently in particular by Ali and London 
\cite{ALI00}.
 Moreover, as shown in \cite{ABRB}, the new physics effects 
cancel in the ratio $\eta_2/\eta_B$.
In view of this it is convenient in what follows
to use the SM values  $\eta_2=0.57$,
$\eta_B=0.55$ and absorb all QCD corrections
related to new physics contributions into $F_{tt}$.

Noting that
\be\label{ss}
\sin 2\beta=\frac{2\bar\eta(1-\bar\varrho)}{R^2_t}
\ee
and combining (\ref{100a}) and (\ref{RT}) we find \cite{BLO,ABRB}
\be\label{main}
\sin 2\beta=\frac{1.26}{ R^2_0\eta_2}
\left[\frac{0.226}{A^2 B_K}-\bar\eta P_0(\varepsilon)\right]
\ee
wherby the first term in the parenthesis is typically by a factor of
2--3 larger than the second term. This dominant term is independent
of $F_{tt}$ and involves the QCD corrections only in 
the ratio $\eta_2/\eta_B$.  Consequently it is independent of $\mt$ and 
the new parameters in the extensions of the SM. The dependence on 
new physics is only present in  $\bar\eta$ entering the second term
that would be absent if charm contribution to $\varepsilon$ was
negligible.
In particular for $\bar\varrho>0$, the value of $\bar\eta$ decreases
with increasing $F_{tt}$. 

In spite of the sensitivity of the second term in (\ref{main}) to
new physics contributions, there exists an absolut lower bound on 
$\sin 2\beta$ in the MFV models, simply because for $\hat B_K>0$  
the unitarity of the CKM matrix implies
\be\label{RB}
0\le \bar\eta \le R_b~.
\ee 
At first sight one would think that the lower bound for $\sin 2\beta$
is attained for $\bar\eta = R_b$, but this is clearly not the 
case as $\bar\eta$ depends on the values of the  
parameters in (\ref{par}).
Consequently there is a correlation between the values of the
two terms in (\ref{main}).  

In fig.~\ref{lowerbound} we show $(\sin 2\beta)_{\rm min}$ as
a function of $F_{tt}$ obtained in \cite{ABRB} by means of the
scanning method for slightly different set of input parameters than
given in table~\ref{tab:inputparams}. To this end the standard
analysis of section 4 has been used except that $S_0(x_t)$ has
been replaced by $F_{tt}$. While with increasing $F_{tt}<6.2$,
$(\sin 2\beta)_{\rm min}$ decreases, it increases for larger
values of $F_{tt}$, so that indeed an absolut minimum for
$\sin 2\beta$ in MFV models is found. In the case of the
future ranges of the input parameters considered in \cite{ABRB},
the minimum is found for $F_{tt}\approx 5.2$.
For $F_{tt}\ge 13.5~(7.8)$ in the case
of the present (future) input parameters no solutions for
$\sin 2\beta$ are found.

\begin{figure}
\unitlength1mm
\begin{picture}(121,85)
  \put(25,5){\psfig{file=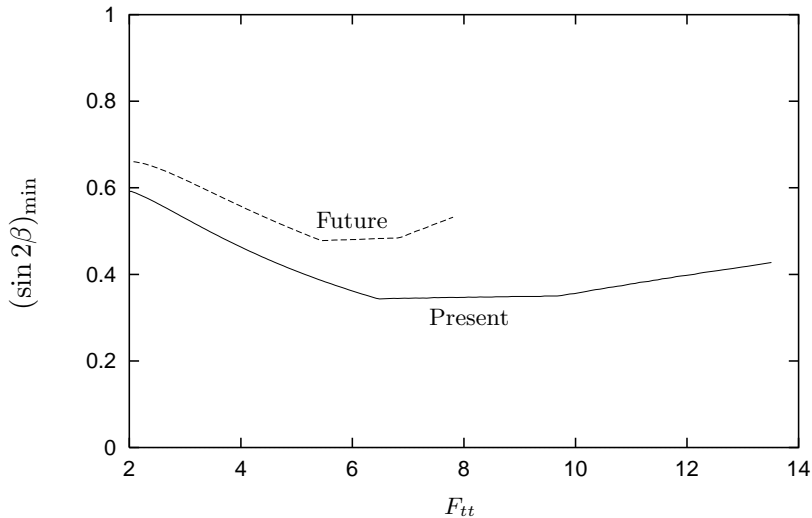,width=10.4cm}} 
  \put(60,40){\footnotesize Future}
  \put(75,27){\footnotesize Present}
  \put(77,2){\footnotesize $F_{tt}$}

\put(23,38){%
%
\makebox(0,0)[b]{\shortstack{$(\sin 2 \beta)_{\rm min}$}}%
%
}%

\end{picture}
  \caption{Lower bound for $\sin 2\beta$ as a function of
 $F_{tt}$ for present and future ranges of the input 
parameters \cite{ABRB}.}
\label{lowerbound}
\end{figure}

 A number of supersymmetric
MFV models has been reviewed by Ali and London \cite{ALI00},
where references to the original literature can be found. 
Using the results of \cite{ALI00}
we find as characteristic values  
$F_{tt}=3.0$, $F_{tt}=3.4$, $F_{tt}=4.3$ for  minimal SUGRA models,
non-minimal SUGRA models and non-SUGRA models with EDM constraints
respectively.
Moreover, $F_{tt}=2.46$ and $F_{tt}=5.2$ for the SM
and the MSSM version of \cite{EP00} respectively.
We observe a rather weak dependence of $(\sin 2\beta)_{\rm min}$
on $F_{tt}$. This is in agreement with the analysis of \cite{ALI00},
where $\sin 2\beta$ has been studied in the range $2.5\le F_{tt}\le4.3$.
Consequently the measurement of $(\sin 2\beta)_{\psi K_S}$ will
not be able to distinguish easily different MFV models. 

On the other hand, the existence of an absolut bound on $\sin 2\beta$
in the MFV models allows in principle to rule out this class of models
if $(\sin 2\beta)_{\psi K_S}$ is found below $(\sin 2\beta)_{\rm min}$.
The implications of such a possibility would be rather profound.
With the measurement of $(\sin 2\beta)_{\psi K_S}$ alone one would be
able to conclude  that new CP violating phases and/or new local
operators in the weak effective Hamiltonians for $K^0-\bar K^0$
and $B^0_{d,s}-\bar B^0_{d,s}$ mixings are necessary to describe
the data.

Repeating the analysis of \cite{ABRB} we find using the input
parameters of  table~\ref{tab:inputparams}
\begin{equation}\label{absolut}
(\sin 2\beta)_{\rm min} = 0.42
\end{equation}
to be compared with 0.34 in \cite{ABRB}.
We would like to emphasize that this bound should be considered
as conservative. Afterall it has been obtained by scanning 
independently all parameters in question. 
The main reason for the improved bound is the higher value of
$F_{B_d}\sqrt{\hat B_d}$ used in the new analysis.
The  plot of $(\sin 2\beta)_{\rm min}$ versus $F_{tt}$
in the case of the parameters of  table~\ref{tab:inputparams}
is given in fig.~\ref{fignew} where this time the "future"
represents simply the parameters of  table~\ref{tab:inputparams}
with the constraint $F_{B_d}\sqrt{\hat B_d}\ge 210~\mev$.

\begin{figure}
\unitlength1mm
\begin{picture}(121,85)
  \put(25,5){\psfig{file=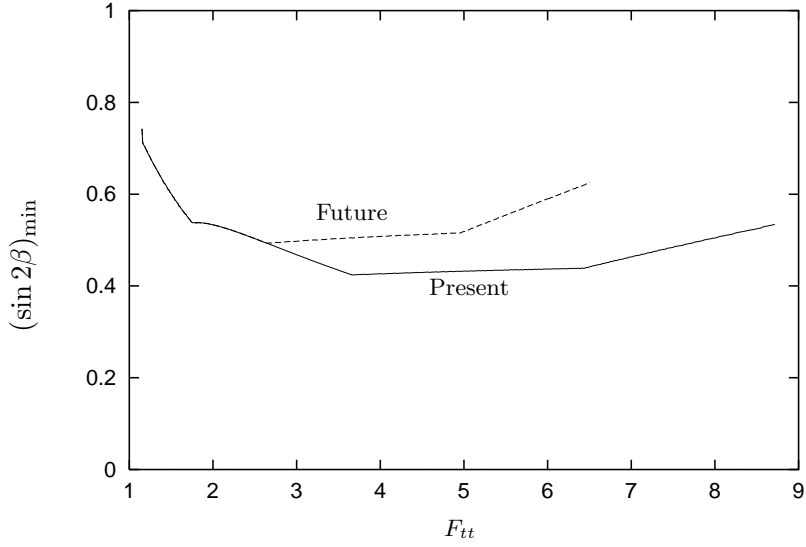,width=10.4cm}} 
  \put(60,44){\footnotesize Future}
  \put(75,34){\footnotesize Present}
  \put(77,2){\footnotesize $F_{tt}$}

\put(23,38){%
%
\makebox(0,0)[b]{\shortstack{$(\sin 2 \beta)_{\rm min}$}}%
%
}%

\end{picture}
  \caption{Updated lower bound for $\sin 2\beta$ as a function of
 $F_{tt}$ for present and future ranges of the input parameters.}
\label{fignew}
\end{figure}

The anatomy of the bound is given in table~\ref{ANA}. Here we show
$(\sin 2\beta)_{\rm min}$ as a function of $\hat B_K$ and $\vcb$
with all the remaining parameters scanned within the 
ranges of  table~\ref{tab:inputparams}.
In the fourth column we show the impact of the reduced uncertainty
in $F_{B_d}\sqrt{\hat B_d}$. In the fifth column the impact of
the future measurement of $\Delta M_s$ is shown. 
This measurement giving an upper bound on $\Delta M_s$
will  provide the lower bound on $R_t$ by means of (\ref{Rt}) 
in addition to the known upper bound. This in turn
will exclude high values of $F_{tt}$, see (\ref{RT}),
implying an improved lower bound on $\sin 2\beta$. 
The chosen value  $\xi_{\rm eff}\ge 1.05$ with
$\xi_{\rm eff}$ defined in (\ref{Rt}) corresponds for
instance to $(\Delta M)_s\le 18.0/{\rm ps}$ for $\xi=1.15$.
The numbers in the parentheses are explained below.

On the basis of this table and additional numerical analysis
we find the following features in accordance with
(\ref{main}):

\begin{itemize}
\item
$(\sin 2\beta)_{\rm min}$ increases with decreasing $\hat B_K$
and $\vcb$ and increasing $\vub$ and $F_{B_d}\sqrt{\hat B_d}$.
\item
In the ranges considered, the dependence of $(\sin 2\beta)_{\rm min}$
on $\hat B_K$, $\vcb$ and $F_{B_d}\sqrt{\hat B_d}$ is 
stronger than on $\vub$. This is evident from (\ref{main}) in
which $\vub$ is not explicitly present but affects the bound only 
through the value of $\bar\eta$ in the subleading term and 
indirectly through its impact on the allowed ranges of the
remaining parameters.
\item
On the other hand when the upper bound on $\Delta M_s$ will be known,
the lower bound on $\vub$ will have an important on 
$(\sin 2\beta)_{\rm min}$. This is seen in table~\ref{sinbeta} and in the
fifth column of table~\ref{ANA} where the values in the parentheses show
the impact of the lower bound $\vub\ge 0.085$. In this case 
the dependence of the bound on
$\hat B_K$ and $\vcb$ essentially disappears in the ranges considered. 
\item
One can check that $(\sin 2\beta)_{\rm min}$ is roughly proportional
to $F^2_{B_d}{\hat B_d}$ and its dependence on $\hat B_K$ and $A$ can be
approximately given by a single variable $\tau=A^2 \hat B_K$. These
features are clear from (\ref{main}), table~\ref{ANA} and 
fig.~\ref{fignew}. 
The observed small departures
from these regularities are caused by the correlations between various
parameters as discussed below (\ref{RB}).
\item
With the parameters of table~\ref{tab:inputparams}
 the impact of the measurement of $\Delta M_s$ is only felt
for $\hat B_K\ge0.85$. It was larger in the case of parameters
considered in \cite{ABRB}.
With increasing  $\xi_{\rm eff}$ the impact of 
the measurement of $\Delta M_s$ becomes larger. 
\end{itemize}

\begin{table}[thb]
\caption[]{ Values of $(\sin 2\beta)_{\rm min}$ in the
MFV models for specific values of $\hat B_K$ and $\vcb$
and different scenarios as explained in the text. 
\label{ANA}}
\begin{center}
\begin{tabular}{|c|c|c|c|c|}\hline
 $\hat B_K $& $ \vcb $ & Present & $F_{B_d}\sqrt{\hat B_d}>210\mev$ &
  $\xi_{eff}\ge 1.05$ \\ \hline
       & $0.039$ & 0.71  & 0.82  & 0.71~(0.71)\\
 $0.70$& $0.041$ & 0.65  & 0.75  & 0.65~(0.70)\\
       & $0.043$ & 0.59  & 0.69  &  0.59~(0.70)\\
 \hline
      & $0.039$ & 0.59  &  0.68  &  0.59~(0.70)\\
 $0.85$& $0.041$ & 0.54 & 0.63   &  0.54~(0.70)\\
       & $0.043$ & 0.50 & 0.57   &  0.50~(0.70)\\
 \hline
      & $0.039$ & 0.51  &  0.59  &   0.56~(0.70)\\
 $1.00$& $0.041$ & 0.46 & 0.54  &   0.52~(0.70)\\
       & $0.043$ & 0.42 & 0.49   &   0.49~(0.70)\\
 \hline
 \end{tabular}
\end{center}
\end{table}

At present the lower bound on $\sin 2\beta$ is fully consistent
with the experimental findings. On the other hand this
bound could  become stronger
when our knowledge of the input parameters in question improves and
when the {\it upper bound} on 
$\Delta M_s$ will be experimentally known.
In particular if the upper bounds on $\hat B_K$ and $\vcb$ and
lower bounds on $\vub$, $F_{B_d}\sqrt{\hat B_d}$ and $\xi_{\rm eff}$
will be improved,  $(\sin 2\beta)_{\rm min}$ could be 
shifted above 0.5. 
If simultaneously  the future accurate measurements of $a_{\psi K_S}$
will confirm the low values reported by BaBar and Belle,  
all MFV models will be excluded.

Clearly other measurements, in particular those of the
rare decay branching ratios and various CP asymmetries, 
will have an additional
impact on $(\sin 2\beta)_{\rm min}$, but this is
a different story. 

Finally as pointed out in \cite{ABRB}, the absolute lower bound
on $\sin 2\beta$ implies within the MFV models an absolute lower
bound on the angle $\gamma$. We find
\begin{equation}\label{absgam}
(\sin \gamma)_{\rm min} = 0.31
\end{equation}
with $\gamma$ in the first quadrant. The second quadrant in the
MFV models is excluded through the lower bound on $\Delta M_s$.

It will be exciting to watch the experimental progress in
the values of $a_{\psi K_S}$ and $\Delta M_s$ and the
theoretical progress on $\hat B_K$, $\vub$, $F_{B_d}\sqrt{\hat B_d}$
and $\xi$. Possibly we will know already next summer that new
CP violating phases and/or new operators in the effective
weak Hamiltonians are mandatory.

On the other hand if $(\sin 2\beta)_{\psi K_S}$ will be found
above 0.5 the MFV models will remain as a vital alternative to
more complicated models unless one can exclude them by different
measurements as discussed in the section on the universal
unitarity triangle. In this context and in the distinction 
between different
MFV models the measured value of $\Delta M_s$
will play an important role as $\vts\approx\vcb$ and $\Delta M_s$
is directly proportional to $F_{tt}$. To this end 
$F_{B_s}\sqrt{\hat B_s}$ has to be precisely known.

\subsubsection{Results in the MSSM}
There are many new contributions in MSSM
such as charged Higgs, chargino, neutralino and gluino contributions.
However, in the case of
minimal flavour and CP violation it is a good approximation to keep only
charged Higgs and chargino contributions. 

The most recent analysis of $\epe$ and of rare decays in this scenario
can be found in \cite{EP00}. In this analysis constraints on the
supersymmetric parameters from $\varepsilon$, $\Delta M_{d,s}$,
$B\to X_s\gamma$, $\Delta\varrho$ in the electroweak precision
studies and from the lower bound on the neutral Higgs mass
have been imposed. Supersymmetric contributions affect both the loop
functions $F_r$ present in the SM and the values of the extracted
CKM parameters like $\vtd$ and $\IM\lambda_t$. As the supersymmetric
contribution to the function $F_{tt}$ in (\ref{100a}) are always positive,
the extracted values of $\vtd$ and $\IM\lambda_t$ are always smaller
than in the SM. Consequently quantities sensitive to $\vtd$ and 
$\IM\lambda_t$ are generally suppressed relative to the SM expectations.
Only for special values of supersymmetric parameters, the
supersymmetric contributions to the loop functions $X_0$, $Y_0$ and
$Z_0$ can overcompensate the suppression of $\vtd$ and $\IM\lambda_t$
so that some enhancements of branching ratios are possible.

Let us define by $T(Q)$ the MSSM prediction for a given quantity Q
normalized to the SM result. Setting $|V_{ub}|$, $\vcb$ and the
non-perturbative parameters $B_i$, all unaffected by SUSY contributions,
at their central values one finds \cite{EP00}
\be
0.53\le T(\epe)\le 1.07, \qquad 0.65\le T(K^+\to \pi^+\nu\bar\nu)\le 1.02
\ee
\be
0.41\le T(K_L\to \pi^0\nu\bar\nu)\le 1.03, \qquad 
0.73\le T(K_L\to \pi^0 e^+e^-)\le 1.10~.
\ee
We observe that suppressions by a factor of 2 relative to the SM
expectations are still possible.

As the CKM element $\vts$ is practically fixed by the unitarity
of the CKM matrix and consequently unaffected by supersymmetric
contributions, the supersymmetric contributions to
$B\to X_s\mu\bar\mu$ and $B_s\to \mu\bar\mu$ enter only through
the loop functions. Consequently visible enhancements, but also
suppressions, in the corresponding branching ratios are found
\be
0.73\le T(B\to X_s\mu\bar\mu)\le 1.34, \qquad 
0.68\le T(B_s\to \mu\bar\mu)\le 1.53~.
\ee

Reference \cite{EP00} provides a compendium of phenomenologically
relevant formulae in the MSSM, that should turn out to be useful
once the non-perturbative parameters $B_i$ will be better know and
the relevant branching ratios have been measured.
The study of the unitarity triangle can be found in \cite{ALI00}.

\subsection{Going Beyond the MSSM}
In general supersymmetric models the effects of supersymmetric 
contributions to $\epe$ and to rare branching ratios can be
much larger than discussed above. In these models new CP-violating
phases and new operators are present. Moreover the structure
of flavour violating interactions is much richer than in the
MFV models. The new flavour violating interactions are present
because generally the sfermion mass matrices can be non-diagonal
in the basis in which all quark-squark-gaugino vertices and
quark and lepton mass matrices are flavour diagonal.
Instead
of diagonalizing sfermion mass matrices it is convenient
to consider their off-diagonal terms as new flavour violating
interactions. Denoting by $\Delta$ the off-diagonal terms in
the sfermion mass matrices, the sfermion propagators can be 
expanded as a series in terms of $\delta=\Delta/\tilde m^2$,
where $\tilde m$ is the average sfermion mass. As long as $\Delta$
is significantly smaller than $\tilde m^2$, we can just take the
first few terms of this expansion and compute any given process in
terms of these $\delta$'s. This is equivalent to an approximate
diagonalization of the squark mass matrices around their
diagonal part.

The method just described is the so--called mass--insertion
approximation \cite{HKR}. It has been reviewed in the classic
papers \cite{GGMS,MPR}, where further references can be found.
The basic parameters in this approach are the insertions
\be
(\delta_{ij})_{LL}, \quad (\delta_{ij})_{LR}, \quad 
(\delta_{ij})_{RL},\quad (\delta_{ij})_{RR}
\ee
with $i,j=1,2,3$ denoting flavour indices and L,R the helicities
of the fermionic partners of sfermions. These parameters being
generaly complex constitute new sources of CP violation.
Their values can be constrained through the existing data
on flavour violating ($i\not =j$) and flavour conserving
($i=j$) processes. For instance $(\delta_{12})_{XY}$ can
be constrained through $\Delta S=1$ and $\Delta S=2$ transitions,
whereas  $(\delta_{13})_{XY}$ and $(\delta_{23})_{XY}$
through analogous transitions involving $B_d^0$ and $B_s^0$
mesons, respectively.  

There is a vast literature on the phenomenological applications
of the mass insertion method to rare and CP-violating decays.
Here I would like to review very briefly my own work on this
subject in a wonderful collaboration with my Italian friends
Gilberto Colangelo, Gino Isidori, Andrea Romanino and in 
particular Luca Silvestrini \cite{BS99,BS98,BRS}.
In \cite{BRS} a model independent analysis of $K\to \pi\nu\bar\nu$
with an application to general supersymmetric models has
been presented. A similar analysis can also be found
in \cite{GN1}. In \cite{BS98} rare kaon decays in models with
an enhanced $\bar s dZ$ vertex, for instance supersymmetric
models, have been considered. The
generalization of this analysis to B decays has been presented
in \cite{BHI}. Let me  concentrate here on \cite{BS99}. 

Despite the presence of a large number of parameters within
the mass insertion approach to general supersymmetric models,
only a few of them are allowed to contribute substantially
to $\epe$. Phenomenological constraints, coming mainly from
$\Delta S=2$ transitions \cite{GGMS}, make the contribution
of most of them to $\Delta S=1$ amplitudes very small
compared to the SM one. The only parameters which survive
are the left-right mass insertions contributing to the Wilson
coefficients of Z- and magnetic-penguin operators.
The reason for this simplification is a dimensional one:
these are the only two classes of operators of dimension less
than six contributing to $\epe$. Supposing that the enhancement
of the Wilson coefficients of either of these two (or both)
type of operators is responsible for the observed value
of $\epe$, a corresponding effect in the rare kaon decays
should be observed. In \cite{BS99} the relation between these
new contributions to $\epe$ and the corresponding contributions
to the rare decays $K_L \to \pi^0 \nu\bar\nu$, 
$K^+\to\pi^+\nu\bar\nu$, $K_L\to \pi^0 e^+e^-$ and $K_L \to \mu^+ \mu^-$
has been analysed in detail. 

Before presenting the results of this analysis let me recall
the observation of Colangelo and Isidori \cite{ISI} that the branching ratios
of rare kaon decays could be considerably enhanced, in a generic
supersymmetric model, by large contributions to the effective
$\bar sdZ$ vertex due to a double left--right mass insertion.
This double mass insertions had not been included in the analyses
\cite{BRS,GN1}, where only single mass insertions were taken into
account. Consequently only modest enhancements of rare decay
branching ratios, up to factors 2--3 at most, can be found in these
papers, as opposed to the possible enhancement of more than
one order of magnitude found in \cite{ISI}. This interesting
observation has been challenged by Silvestrini and myself 
in \cite{BS98}, where we have shown that the data on $\epe$
and $K_L\to \mu \bar\mu$ may constrain considerably the
double left-right mass insertion and the corresponding enhancement
of the rare kaon branching ratios. Our model independent
analysis which went beyond supersymmetry, but assumed that
the only new effect is an enhanced $\bar s d Z$ vertex,
resulted in the following bounds:
\be\label{BS}
Br(\kpn)\le 2.3 \cdot 10^{-10}, \qquad Br(\klpn)\le 2.4 \cdot 10^{-10}
\ee
and $Br(\kpe)\le 3.6 \cdot 10^{-11}$, which are substantially
stronger than the bounds found in \cite{ISI}.

The purpose of our joint collaboration \cite{BS99} was to use
the strategy in \cite{BS98} specifically in the case of
supersymmetry and to include also the effects of chromomagnetic
and $\gamma$-magnetic penguins to $\epe$ and $\kpe$, respectively.
The latter two contributions were assumed to be small in the
models considered in \cite{BS98}.
Moreover, we have investigated whether the large double mass
insertions suggested in \cite{ISI} could be further constrained
within the specific framework of supersymmetry.

As demonstrated in \cite{BS99},  radiative effects relate the double
left-right mass insertion to the single left-left one.
The phenomenological constraints on the latter imply then a stringent
bound on the supersymmetric contribution to the Z-penguin.
Using this bound and those coming from the data on $\epe$ and
$K_L\to \mu\bar\mu$ one finds
\be\label{BSNEW}
Br(\kpn)\le 1.7 \cdot 10^{-10}, \qquad Br(\klpn)\le 1.2 \cdot 10^{-10}
\ee
and $Br(\kpe)\le 2.0 \cdot 10^{-11}$, that is slightly stronger
bounds than given in (\ref{BS}). Larger values are possible, in principle,
but rather unlikely. Thus the most probable maximal enhancements
over the SM expectations are roughly by factors 2, 4 and 4
for these three branching ratios, respectively. 
This analysis confirms the previous findings
in \cite{MM99} that the most natural enhancement of $\epe$, within
supersymmetric models, comes from chromomagnetic penguins.
In this case sizable enhancement of $Br(\kpe)$, as seen above,
can be expected.

There is of course a number of other analyses of general supersymmetric
effects in rare and CP-violating decays. Some references are collected
in \cite{SUSY}. In particular, the constraints on phases of
supersymmetric flavour conserving couplings can be best obtained
from the upper bounds on electric dipole moments \cite{POKOR}.

\subsection{Spontaneous CP Violation}
Spontaneous CP violation (SCPV) is a very interesting topic that we
cannot cover in these lectures. It requires extended Higgs sector.
A very comprehensive presentation of SCPV can be found in \cite{Branco}
where left-right symmetric models and multi-Higgs models are discussed
in detail. Some recent papers on SCPV are given in \cite{SCPV}.

\section{Perspectives}
\setcounter{equation}{0}
I hope I have convinced the students that the field of CP violation
and rare decays plays an important 
role in the deeper understanding of the SM 
and particle physics in general.
Indeed the field of weak decays and of CP violation is one of the least
understood sectors of the SM.
Even if the SM is still consistent with the existing data for
weak decay processes, the near future could change 
this picture
considerably through the advances in experiment and theory.
In particular the experimental work
done in the next ten
years at BNL, CERN, CORNELL, DA$\Phi$NE, DESY, 
FNAL, KEK, SLAC and eventually LHC will certainly 
have considerable impact on this field.

Let us then
make a list of things we could expect in the next ten years.
This list is certainly very biased by my own interests but could
be useful anyway. Here we go:

\begin{itemize}
\item
The error on the CKM elements $\vcb$ and $\vub$ will be decreased 
below 0.0015 and 0.01, respectively. This progress should come mainly from
CLEO III, $B$-factories and new theoretical efforts. It will have
considerable impact on the unitarity triangle and will improve
theoretical predictions for rare and CP-violating decays sensitive
to these elements.
\item
The error on $\mt$ should be decreased down to $\pm 3\gev$
at Tevatron in the Main Injector era and to $\pm 1\gev$ at LHC.
\item
The measurement of a non-vanishing ratio $\epe$ by NA31,
KTeV and NA48, excluding confidently the superweak models, 
has been an important achievement. 
Yet, as we have discussed in section 5, the experimental value
for $\epe$ requires considerable improvements before we
could be satisfied with it. In particular, the difference between
KTeV and NA48 results by roughly a factor of two is rather
disturbing. Let us hope that some progress will be made in this
direction in the coming years.
It should be stressed that the determination of $\epe$ with 
the accuraccy of  $\pm (1-2) \cdot 10^{-4}$ 
from NA48, KTeV and KLOE will give some insight into the 
physics of 
direct CP violation inspite of large theoretical uncertainties. 
In this respect measurements of CP-violating asymmetries in charged $B$
and $K$ decays will also play an outstanding role. 
The situation concerning hadronic uncertainties is quite similar
to $\epe$. 
Therefore one should hope 
that some definite progress in calculating relevant hadronic matrix elements 
will also be made. 
\item
One of the most exciting measurements in this year 
will be the measurement of $\Delta M_s$. LEP and SLD experiments
have done already a fantastic progress by providing
the lower bound $\Delta M_s\ge 15/ps$. The actual measurement
of $\Delta M_s$ should come first from Run II at FNAL.
With the improved calculations of $\xi$ in (\ref{107b}) this will have
an important impact on the determination of $\vtd$, on the
unitarity triangle and as discussed in section 9 on models
with minimal flavour violation.
 \item
Clearly future precise studies of CP violation by BaBar, Belle,
CDF, D0, CLEO III, LHC-B and BTeV providing 
direct measurements of $\alpha$, $\beta$ and $\gamma$ may totally
revolutionize our field. The first results from CDF, BaBar and Belle
are very encouraging.
During the recent years several, in some cases quite sophisticated and
involved, strategies have been developed to extract these angles with
small or even no hadronic uncertainties. Certainly the future will bring
additional methods to determine $\alpha$, $\beta$ and $\gamma$. 
Obviously it is very desirable to have as many such strategies as possible
available in order to overconstrain the unitarity triangle and to resolve 
certain discrete ambiguities which are a characteristic feature of these 
methods. A recent review on discrete ambiguities with the relevant 
literature can be found in \cite{Oliver}.
\item
Improved data for $K^+\to\pi^+\nu\bar\nu$ should
be reported by AGS E787 collaboration this year. In view of the theoretical 
cleanliness of this decay the measured branching ratio
 at the $1.5\cdot 10^{-10}$
level would signal physics beyond the SM.
A precise measurement of this very
important decay requires, however, new experimental ideas and
new efforts. The new efforts \cite{AGS2,Coleman} in this direction allow 
to hope that
a measurement of $Br(\kpn)$ with an accuracy of $\pm 10 \%$ should
be possible before 2005. This will provide an important test for
the SM and its extensions.
\item
The newly approved experiment at BNL \cite{KOPIO} and the planned
experiments at KEK \cite{KEKKL} and FNAL \cite{FNALKL} to
measure $Br(\klpn)$ at the $\pm 10\%$ level before 2005 may make a decisive
impact on the field of CP violation. 
$\klpn$ allows the
cleanest determination of $\imlt$ and taken together with $\kpn$
a very clean determination of $\sin 2 \beta$.
\item
The theoretical status of $K_{\rm L}\to \pi^0 e^+ e^-$ and of 
$K_{\rm L}\to \mu\bar\mu$, 
should be improved to confront future
data. Experiments at DA$\Phi$NE should be very helpful in this
respect. 
\item
The future improved inclusive measurements $B \to X_{s,d} \gamma$ 
confronted with improved SM predictions could
give the first signals of new physics. It appears that the errors
on the input parameters could be lowered further and the
theoretical error on $Br(B\to X_s\gamma)$ could be decreased
confidently down to $\pm 8 \%$ in the next years. The same
accuracy in the experimental branching ratio will hopefully
come  from BaBar and Belle. 
This may, however, be insufficient to
disentangle new physics contributions although such an accuracy
should put important constraints on the physics beyond the Standard
Model. It would also be desirable to look for $B \to X_d \gamma$,
but this is clearly a much harder task.
\item
Similar comments apply to transitions $B \to X_s l^+l^-$ and
$B \to K^* l^+l^-$
which are much reacher and  more sensitive to new physics contributions
than $ B \to X_{s,d} \gamma$. Observations of
$B \to X_s \mu\bar\mu$ and $B\to K^*\mu\bar\mu$ are expected
from Run II at FNAL and B factories already this year.
The distributions of various kind when measured should
be very useful in the tests of the SM and its extensions.
\item
The measurement of $B \to X_{s,d}\nu\bar\nu$ and 
$B_{s,d}\to \mu\bar\mu$ will take most probably longer time but
as stressed in these lectures all efforts should be made to measure
these transitions. 
\item
According to the SM, CP violation outside the K-meson and B-meson
systems is expected to be essentially unobservable. On the other
hand, new sources of CP violation present in multi-Higgs models
or models with supersymmetry could give rise to measurable effects.
Attempts to trace these effects include charm meson decays,
$D^0-\bar D^0$ mixing, hyperon decays at Fermilab (E756) \cite{Hyperon} 
and the
searches for the electron and neutron electric dipole moments.
The most interesting at present are the
results from CLEO \cite{CLEOD} and FOCUS (FNAL) \cite{FOCUSD}
 which possibly indicate 
$D^0-\bar D^0$ mixing at a much higher level than expected in
the SM and in the ball park of some supersymmetric expectations 
\cite{D0D0}.
 Top quarks and Higgs particles \cite{Higgs1,Higgs2} may also be used 
to probe CP violation once they are produced in large numbers at the
Tevatron, the LHC and future linear colliders. 
Finally   CP violation in neutrino oscillations \cite{LIND}
and in the context of baryogenesis \cite{WBUCH} as well as 
lepton flavour violation \cite{Feng} belong to the
class of phenomena outside the SM. 
Considerable progress in this area
should be made in this decade.
\item
On the theoretical side,
one should hope that the non-perturbative
methods will be considerably improved so that various $B_i$ parameters
will be calculated with sufficient precision. It is very important
that simultaneously with advances in lattice QCD, further efforts
are being made in finding efficient analytical tools for calculating
QCD effects in the long distance regime. This is, in particular very
important in the field of non-leptonic decays, where the progress
in lattice calculations is slow. 
The accumulation of data for non-leptonic $B$
and $D$ decays in the coming years  should teach us more 
about the role of non-factorizable contributions and in particular
about the final state interactions. 
In this context, in the case of K-decays, important
lessons will come from DA$\Phi$NE which is an excellent machine
for testing chiral perturbation theory and other non-perturbative
methods. 
\end{itemize}

This list of topics shows that flavour dynamics and the related CP
violation and rare processes have a great future. They will surely
play an  important role in particle physics in this decade. 
Clearly the next ten years should be very exciting in this field.

{\bf Acknowledgements}

I would like to thank the directors of the Erice school, Profs.
A. Zichichi, G. `t Hooft and G. Veneziano
for inviting me to such a wonderful place and  great hospitality.
Special thanks go to the students for interesting questions
and discussions. I would also like to thank Gerhard Buchalla, Robert 
Buras, Adam Falk, Robert Fleischer, Gino Isidori, Janusz Rosiek and 
Stefan Schael for discussions and 
Frank Kr\"uger for  critical and illuminating comments on rare B decays. 
Last but certainly not least I would like to thank all my
collaborators for such a wonderfull time we had together and my
secretery Elke Kr\"uger for a great help in preparing my transparencies.

This work has been supported in part by the German Bundesministerium
f\"ur Bildung and Forschung under the contract 05HT9WOA0.

\renewcommand{\baselinestretch}{0.95}

\vfill\eject


\small
\begin{thebibliography}{999}
\bibitem{AJBLH}
A.J. Buras, hep-ph/9806471,  in {\it Probing the Standard
Model of Particle Interactions}, eds. R. Gupta, A. Morel,
E. de Rafael and F. David  
(Elsevier Science B.V., Amsterdam, 1998), page 281.
\bibitem{AJBLAKE}
A.J. Buras, hep-ph/9905437, in {\it Electroweak Physics},
eds. A. Astbury et al, World Scientific 2000, page 1.
\bibitem{BBL}
{ G. Buchalla, A.J. Buras and M. Lautenbacher,} 
{ Rev. Mod. Phys} {\bf 68} (1996) 1125.
\bibitem{BF97}
{ A.J. Buras and R. Fleischer,} hep-ph/9704376, in \cite{BULIND} page
65. 
\bibitem{BULIND}
Heavy Flavours II, eds. A.J. Buras and M. Lindner,
World Scientific, 1998.
\bibitem{Branco}
G. Branco, L. Lavoura and J. Silva, (1999), CP Violation,
Oxford Science Publications, Clarendon Press, Oxford.
\bibitem{Bigi}
I.I. Bigi and A.I. Sanda, (2000), CP Violation, Cambridge Monographs
on Particle Physics, Nuclear Physics and Cosmology,
Cambridge University Press, Cambridge.
\bibitem{BABAR}
The BaBar Physics Book, eds. P. Harrison and H. Quinn, (1998),
SLAC report 504.
\bibitem{LHCB}
B Decays at the LHC, eds. P. Ball, R. Fleischer, G.F. Tartarelli, 
P. Vikas and G. Wilkinson, hep-ph/0003238.
\bibitem{REV}
P. Ball, hep-ph/0010024; I.I. Bigi, hep-ph/0011021;
 R. Fleischer, hep-ph/0011323; J.L. Rosner, hep-ph/0011355; M. Gronau,
 hep-ph/0011392;
J. Ellis, hep-ph/0011396;  L. Wolfenstein, hep-ph/0011400;
 D. Wyler, hep-ph/0101259.
\bibitem{Donoghu}
J.F. Donoghue, E. Golowich and B.R. Holstein, (1992), Dynamics of the
Standard Model, Cambridge Monographs
on Particle Physics, Nuclear Physics and Cosmology,
Cambridge University Press, Cambridge.
\bibitem{Peskin}
M.E. Peskin and D.V. Schroeder, (1995), An Introduction to Quantum Field
Theory, Addison-Wesley Publishing Company.
\bibitem{Weinberg}
S. Weinberg, The Quantum Theory of Fields, (1995), Cambridge University
Press.
\bibitem{Pokorski}
S. Pokorski, Gauge Field Theory, (1999), Cambridge Monographs on
Mathematical Physics.
\bibitem{Muta}
T. Muta, Foundations of Quantum Chromodynamics, (1998), World Scientific.
\bibitem{CAB}
N. Cabibbo, Phys. Rev. Lett. {\bf 10} (1963) 531.
\bibitem{KM}
{ M. Kobayashi and K. Maskawa},
 { Prog. Theor. Phys.} {\bf 49} (1973) 652.
\bibitem{GIM}
{ S.L. Glashow, J. Iliopoulos and L. Maiani}
{ Phys. Rev.} {\bf D 2} (1970) 1285.
\bibitem{ASYM}
D.J. Gross and F. Wilczek, { Phys. Rev. Lett.} {\bf 30} (1973) 1343;
 H.D. Politzer, { Phys. Rev. Lett.} {\bf 30} (1973) 1346;
G. `t Hooft, unpublished.
\bibitem{BBDM}
W.A. Bardeen, A.J. Buras, D.W. Duke and T. Muta,
{ Phys. Rev.} {\bf D 18} (1978) 3998.
\bibitem{Bethke}
S. Bethke, J. Phys. {\bf G26} (2000) R27.
\bibitem{FX1}
H. Fritzsch and Z.Z. Xing,  { Phys. Rev.} {\bf D 57} (1998) 594;
Prog. Part. Nucl. Phys. {\bf 45} (2000) 1.
\bibitem{CHAU}
{ L.L. Chau and W.-Y. Keung}, 
{ Phys. Rev. Lett.} {\bf 53} (1984) 1802.
\bibitem{PDG}
{ Particle Data Group,} { Euro. Phys. J.} {\bf C 15} (2000) 1.
\bibitem{WO}
{ L. Wolfenstein}, { Phys. Rev. Lett.} {\bf 51} (1983) 1945.
\bibitem{FX2}
H. Fritzsch and Z.Z. Xing,  Phys. Lett. {\bf B 413} (1997) 396. 
\bibitem{BLO}
{ A.J. Buras, M.E. Lautenbacher and G. Ostermaier,}
{ Phys. Rev.} {\bf D 50} (1994) 3433.
\bibitem{schubert}
{ M. Schmidtler and K.R. Schubert}, { Z. Phys.} {\bf C 53}
(1992) 347.
\bibitem{Kayser}
R. Aleksan, B. Kayser and D. London, Phys. Rev. Lett. {\bf 73} (1994) 18;
J.P. Silva and L. Wolfenstein, { Phys. Rev.} {\bf D 55} (1997) 5331;
I.I. Bigi and A.I. Sanda, hep-ph/9909479.
\bibitem{CJ}
{ C. Jarlskog,} { Phys. Rev. Lett.} {\bf 55}, (1985) 1039;
{ Z. Phys.} {\bf C29} (1985) 491.
\bibitem{js}
{ C. Jarlskog and R. Stora},
{ Phys. Lett.} {\bf B 208} (1988) 268.
\bibitem{CRONIN}
{ J.H. Christenson, J.W. Cronin, V.L. Fitch and R. Turlay},
{ Phys. Rev. Lett.} {\bf 13} (1964) 128.
\bibitem{OPE}
K.G. Wilson, { Phys. Rev.} {\bf 179} (1969) 1499;
K.G. Wilson and W. Zimmermann, { Comm. Math. Phys.} 
{\bf 24} (1972) 87.
\bibitem{ZIMM}
W. Zimmermann, in Proc. 1970 Brandeis Summer Institute in
Theor. Phys, (eds. S. Deser, M. Grisaru and H. Pendleton),
MIT Press, 1971, p.396; 
{ Ann. Phys.} {\bf 77} (1973) 570.
\bibitem{SUMA}
{ E.C.G. Sudarshan and R.E. Marshak}, Proc. Padua-Venice Conf. on 
Mesons and Recently Discovered Particles (1957).
\bibitem{GF}
{ R.P. Feynman and M. Gell-Mann,}
 { Phys. Rev.} {\bf 109} (1958) 193.
\bibitem{Donoghue}
V. Cirigliano, J.F. Donoghue and E. Golowich, JHEP {\bf 0010} (2000)
048;
E. Golowich, hep-ph/0008338; J.F. Donoghue, hep-ph/0012072.
\bibitem{ACMP}
{ G. Altarelli, G. Curci, G. Martinelli and S. Petrarca,}
{ Nucl. Phys.} {\bf B 187} (1981) 461.
\bibitem{WEISZ}
{ A.J. Buras and P.H. Weisz,}
{ Nucl. Phys.} {\bf B 333} (1990) 66.
\bibitem{BJLW1}
{ A.J. Buras, M. Jamin, M.E. Lautenbacher and P.H. Weisz,}
{ Nucl. Phys.} {\bf B 370} (1992) 69;
{ Nucl. Phys.} {\bf B 400} (1993) 37.
\bibitem{BJLW2}
{ A.J. Buras, M. Jamin and M.E. Lautenbacher,}
{ Nucl. Phys.} {\bf B 400} (1993) 75.
\bibitem{BJLW}
{ A.J. Buras, M. Jamin and M.E. Lautenbacher,}
{ Nucl. Phys.} {\bf B 408} (1993) 209.
\bibitem{ROMA1}
{ M. Ciuchini, E. Franco, G. Martinelli and L. Reina,}
{ Phys. Lett.} {\bf B 301} (1993) 263.
\bibitem{ROMA2}
{ M. Ciuchini, E. Franco, G. Martinelli and L. Reina,}
{ Nucl. Phys.} {\bf B 415} (1994) 403.
\bibitem{MIS1}
K. Chetyrkin, M. Misiak and M{\"u}nz,
 { Nucl.~Phys.} {\bf B520} (1998) 279.
\bibitem{MisMu:94}
{ M.Misiak and M. M{\"u}nz,}
{ Phys. Lett.} {\bf B344} (1995) 308.
\bibitem{CZMM}
{ K.G. Chetyrkin, M. Misiak and M. M{\"u}nz,} 
{ Phys. Lett.} {\bf B400} (1997) 206; Erratum-ibid.
 {\bf B425} (1998) 414.
\bibitem{Buch:93}
{ G. Buchalla,} { Nucl. Phys.} {\bf B 391} (1993) 501.
\bibitem{Bagan}
{ E. Bagan, P. Ball, V.M. Braun and P. Gosdzinsky,}
{ Nucl. Phys.} {\bf B 432} (1994) 3;
{ E. Bagan} { et al.,} { Phys. Lett.} {\bf B 342} (1995) 362;
{\bf B 351} (1995) 546; A. Lenz, U. Nierste and G. Ostermaier,
{Phys. Rev.} {\bf D56} (1997) 7228, {Phys. Rev.} {\bf D59} (1999) 
034008.   
\bibitem{JP}
{ M. Jamin and A. Pich,}
{ Nucl.~Phys.} {\bf B425} (1994) 15.
\bibitem{HNa}
{ S. Herrlich and U. Nierste,}
{ Nucl. Phys.} {\bf B419} (1994) 292. 
\bibitem{BJW90}
{ A.J. Buras, M. Jamin, and P.H. Weisz,}
{ Nucl. Phys.} {\bf B347} (1990) 491.
\bibitem{UKJS}
J. Urban, F. Krauss, U. Jentschura and G. Soff, 
{ Nucl. Phys.} {\bf B523} (1998) 40. 
\bibitem{HNb}
{ S.~Herrlich and  U.~Nierste},
{ Phys. Rev.} {\bf D52} (1995) 6505; 
{ Nucl. Phys.} {\bf B476} (1996) 27. 
\bibitem{BB1}
{ G. Buchalla and A.J. Buras,}
{ Nucl. Phys.} {\bf B 398} (1993) 285.
\bibitem{BB2}
{ G. Buchalla and A.J. Buras,}
{ Nucl. Phys.} {\bf B 400} (1993) 225.
\bibitem{MU98}
M. Misiak and J. Urban, { Phys. Lett.} {\bf B541} (1999) 161.
\bibitem{BB98}
G. Buchalla and A.J. Buras, { Nucl. Phys.} {\bf B 548} (1999) 309.
\bibitem{BB3}
{ G. Buchalla and A.J. Buras,}
{ Nucl. Phys.} {\bf B 412} (1994) 106.
\bibitem{BB5}
{ G. Buchalla and A.J. Buras,}
{ Phys. Lett.} {\bf B 336} (1994) 263.
\bibitem{BLMM}
{ A. J. Buras, M. E. Lautenbacher, M. Misiak and M. M{\"u}nz,}
{ Nucl.~Phys.} {\bf B423} (1994) 349.
\bibitem{Mis:94}
{ M. Misiak,}
{ Nucl.~Phys.} {\bf B393} (1993) 23;
{ Erratum}, { Nucl.~Phys.} {\bf B439} (1995) 461.
\bibitem{BuMu:94}
{ A.J. Buras and M. M{\"u}nz,}
{ Phys. Rev.} {\bf D 52} (1995) 186.
\bibitem{AG2} 
{  A.~Ali, and  C.~Greub,} { Z.Phys.} {\bf C49} (1991) 431;  
{ Phys.~Lett.} {\bf B259} (1991) 182;
{ Phys.~Lett.} {\bf B361} (1995) 146.
\bibitem{Yao1} {  K.~Adel and Y.P.~Yao,} 
{ Modern Physics Letters} {\bf A8} (1993) 1679;
{ Phys. Rev.} {\bf D 49} (1994) 4945.
\bibitem{Pott} 
{ N. Pott,} { Phys. Rev.} {\bf D 54} (1996) 938.
\bibitem{GREUB}
{ C. Greub, T. Hurth and D. Wyler,} { Phys.~Lett.} {\bf B380} 
(1996) 385; { Phys. Rev.} {\bf D 54} (1996) 3350.
\bibitem{GH97}
{ C. Greub and T. Hurth,} { Phys. Rev.} {\bf D 56} (1997) 2934;
Nucl. Phys. Proc. Suppl. {\bf 74} (1999) 247.
\bibitem{BKP2}
A.J. Buras, A. Kwiatkowski and N. Pott, 
{ Phys. Lett.} {\bf B 414} (1997) 157,
{ Nucl. Phys.} {\bf B 517} (1998) 353. 
\bibitem{GAMB}
M. Ciuchini, G. Degrassi, P. Gambino and G.F. Giudice, 
{ Nucl. Phys.} {\bf B 527} (1998) 21. 
\bibitem{BG98}
F.M. Borzumati and Ch. Greub, { Phys. Rev.} {\bf D 58} (1998) 07004; 
{ Phys. Rev.} {\bf D 59} (1999) 057501. 
\bibitem{BMU567}
Ch. Bobeth, M. Misiak and J. Urban,
{ Nucl. Phys.} {\bf B 567} (2000) 153.
\bibitem{GL00}
Ch. Greub and P. Linger, { Phys. Lett.} {\bf B 494} (2000) 237;
hep-ph/0009144.
\bibitem{BBGLN}
M. Beneke, G. Buchalla, C. Greub, A. Lenz and U. Nierste,
{ Phys. Lett.} {\bf B 459} (1999) 631; M. Beneke and A. Lenz, 
hep-ph/0012222.
\bibitem{BMR}
M. Beneke, F. Maltoni and I.Z. Rothstein, 
{ Phys. Rev.} {\bf D 59} (1999) 054003.
\bibitem{BBNS1}
M. Beneke, G. Buchalla, M. Neubert and C.T. Sachrajda,
 { Phys. Rev. Lett.}  {\bf 83} (1999) 1914; 
{ Nucl. Phys.} {\bf B 591} (2000) 313.
\bibitem{LOCC}
G. Altarelli and L. Maiani, { Phys. Lett.} {\bf B 52} (1974) 351;
M.K. Gaillard and B.W. Lee, { Phys. Rev. Lett.} {\bf 33} (1974) 108.
\bibitem{LOP}
A.I. Vainshtein, V.I. Zakharov and M.A. Shifman, JETP {\bf 45} (1977) 670;
 F.J. Gilman and M.B. Wise, { Phys. Rev.} {\bf D20} (1979) 2392.
\bibitem{LODF2}
F.J. Gilman and M.B. Wise, { Phys. Rev.} {\bf D27} (1983) 1128.
\bibitem{BGH}
A.J. Buras, P. Gambino and U.A. Haisch,
{ Nucl. Phys.} {\bf B 570} (2000) 117.
\bibitem{BMU574}
Ch. Bobeth, M. Misiak and J. Urban,
{ Nucl. Phys.} {\bf B 574} (2000) 291.
\bibitem{CET0}
M. Ciuchini, E. Franco, V. Lubicz, G. Martineli, I. Scimemi and
L. Silvestrini, { Nucl. Phys.} {\bf B 523} (1998) 501.
\bibitem{BMU}
A.J. Buras, M. Misiak and J. Urban,
{ Nucl. Phys.} {\bf B 586} (2000) 397.
\bibitem{CET}
M. Ciuchini, et al. JHEP {\bf 9810} (1998) 008.
\bibitem{China}
T.-F. Feng, X.-Q. Li and G.-L. Wang, hep-ph/0101081.
\bibitem{BB97}
{ G. Buchalla} and { A.J. Buras}, 
{ Phys. Rev.} {\bf D57} (1998) 216.
\bibitem{GKP}
P. Gambino, A. Kwiatkowski and N. Pott, 
{ Nucl. Phys.} {\bf B 544} (1999) 532. 
\bibitem{CZMA}
A. Czarnecki and W.J. Marciano, { Phys. Rev. Lett.} {\bf 81} (1998) 277.
\bibitem{STRUMIA}
A. Strumia, { Nucl. Phys.} {\bf B 532} (1998) 28. 
\bibitem{KN98}
A.L. Kagan and M. Neubert, Eur. Phys. J. {\bf C7} (1999) 5.
\bibitem{GH00}
P. Gambino and U. Haisch, JHEP {\bf 0009} (2000) 001.
\bibitem{strum} 
P. Ciafaloni, A. Romanino, and A. Strumia, 
{ Nucl. Phys.} {\bf B 524} (1998) 361.
\bibitem{GAMB2}
M. Ciuchini, G. Degrassi, P. Gambino and G.F. Giudice, 
{ Nucl. Phys.} {\bf B 534} (1998) 3. 
\bibitem{Soff}
F. Krauss and G. Soff, hep-ph/9807238; T.-F. Feng, X.-Q. Li, W.-G. Ma 
and F. Zhang, { Phys. Rev.} {\bf D 63} (2001) 015013.
\bibitem{NeuStech}
For a review with relevant references see M. Neubert and B. Stech,
hep-ph/9705292, in \cite{BULIND}, page 294.
\bibitem{GFACT}
These are reviewed in \cite{AJBLH} and \cite{NeuStech}.
\bibitem{BSF}
A.J. Buras and L. Silvestrini,
{ Nucl. Phys.} {\bf B548} (1999) 293; 
A.J. Buras, { Nucl. Phys.} {\bf B434} (1995) 606.
\bibitem{DIAG}
D. Zeppenfeld, Z. Phys. {\bf C 8} (1981) 77;
L.L. Chau, { Phys. Rev.} {\bf D 43} (1991) 2176;
M. Gronau, J.L. Rosner and D. London, Phys. Rev. Lett. {\bf 73} (1994) 21;
O.F. Hernandez, M. Gronau, J.L. Rosner and D. London,
{ Phys. Lett.} {\bf B 333} (1994) 500, { Phys. Rev.} {\bf D 50} (1994) 4529.
\bibitem{IWICK}
M. Ciuchini, E. Franco, G. Martinelli and L. Silvestrini,
{ Nucl. Phys.} {\bf B501} (1997) 271.
\bibitem{BSWICK}
 A.J. Buras and L. Silvestrini,
{ Nucl. Phys.} {\bf B569} (2000) 3.
\bibitem{NEUTASI}
M. Neubert, hep-ph/0012204.
\bibitem{Li}
C.-H.V. Chang, H.-n. Li, { Phys. Rev.} {\bf D55} (1997) 5577;
T.-W. Yeh and H.-n. Li, { Phys. Rev.} {\bf D56} (1997) 1615;
H. -Y. Cheng, H.-n. Li and K.-C. Yang
{ Phys. Rev.} {\bf D60} (1999) 094005;
Y.-Y. Keum, H.-n. Li and A.I. Sanda, hep-ph/0004004, 
hep-ph/0004173; H.-n. Li, hep-ph/0101145.
\bibitem{KOD}
A. Khodjamirian, hep-ph/0012271.
\bibitem{LCQCD}
A. Khodjamirian and R. R\"uckl, hep-ph/9801443, in \cite{BULIND},
page 345; P. Ball and V.M. Braun, { Phys. Rev.} {\bf D58} (1998) 
0944016; V.M. Braun, hep-ph/9911206; P. Colangelo and
A. Khodjamirian, hep-ph/0010175.
\bibitem{HQE1}
J. Chay, H. Georgi and B. Grinstein,
{ Phys. Lett.} {\bf B 247} (1990) 399;
I.I. Bigi, N.G. Uraltsev and A.I. Vainshtein,
{ Phys. Lett.} {\bf B 293} (1992) 430
[E: {\bf B 297} (1993) 477];
I.I. Bigi, M.A. Shifman, N.G. Uraltsev and A.I. Vainshtein,
Phys. Rev. Lett. {\bf 71} (1993) 496;
B. Blok, L. Koyrakh, M.A. Shifman and A.I. Vainshtein,
{ Phys. Rev.} {\bf D 49} (1994) 3356 [E: {\bf D 50} (1994) 3572];
A.V. Manohar and M.B. Wise,
{ Phys. Rev.} {\bf D 49} (1994) 1310;
 Th. Mannel, { Nucl. Phys.} {\bf B413} (1994) 396.
\bibitem{HQE2}
Pedagogical reviews of heavy quark symmetry, heavy quark expansions
 and their applications
can be found for instance in M. Neubert, Phys. Rep. {\bf 245}
(1994) 259; A.V. Manohar and M.B. Wise, Heavy Quark Physics
(Cambridge University Press, Cambridge, 2000); M. Shifman,
hep-ph/9510377; A.F. Falk, hep-ph/0007339.
\bibitem{IL}
{ T. Inami and C.S. Lim,}
{ Progr. Theor. Phys.} {\bf 65} (1981) 297.
\bibitem{PBE0}
{ G. Buchalla, A.J. Buras and M.K. Harlander,} { Nucl. Phys.}
 {\bf B 349} (1991) 1.
\bibitem{EP00}
A.J. Buras, P. Gambino, M. Gorbahn, S. J\"ager and L. Silvestrini,
{ Nucl. Phys.} {\bf B 592} (2000) 55. 
\bibitem{BERTOL}
{ S. Bertolini, F. Borzumati, A. Masiero and G. Ridolfi,}
 { Nucl. Phys.} {\bf B353} (1991) 591.
\bibitem{GG}
{ E. Gabrielli and G.F. Giudice}
{ Nucl. Phys.} {\bf B433} (1995) 3; Erratum {\sl Nucl. Phys.} {\bf B507} 
(1997) 549.
\bibitem{MW96}
P. Cho, M. Misiak and D. Wyler, { Phys. Rev.} {\bf D 54} (1996) 3329.
\bibitem{AAA}
A. Ali, Th. Mannel and Ch. Greub, { Zeit. Phys.} {\bf C 67} (1995) 417.
\bibitem{GALE}
{ M.K. Gaillard and B.W. Lee,} 
{ Phys. Rev.} {\bf D10} (1974) 897.
\bibitem{ARGUS}
{ H. Albrecht et al. (ARGUS)}, { Phys. Lett.} {\bf B192} (1987) 245;
{ M. Artuso et al. (CLEO)}, { Phys. Rev. Lett.} {\bf 62} (1989) 2233.
\bibitem{CHAU83}
L.L. Chau, { Physics Reports}, {\bf 95} (1983) 1.
\bibitem{BSSII}
A.J. Buras, W. Slominski and H. Steger,
{ Nucl. Phys.} {\bf B245} (1984) 369.
\bibitem{NIRSLAC}
Y. Nir, SLAC-PUB-5874 (1992); hep-ph/9904271, hep-ph/9911321.
\bibitem{GERAR}
J. Bijnens, J.-M. G{\'e}rard and G. Klein, 
{ Phys. Lett.} {\bf B257} (1991) 191.
\bibitem{Zichichi}
For historical remarks concerning CP violation prior to its discovery
see A. Zichichi, Subnuclear Physics, The First 50 Years: Highlights
from Erice to ELN, World Scientific, 2000, page 37. 
\bibitem{KKHW}
This remark originated in recent discussions with K. Kleinknecht
and H. Wahl.
\bibitem{Lellouch}
L. Lellouch, hep-lat/0011088.
\bibitem{Football}
M. Ciuchini, G. D'Agostini, E. Franco, V. Lubicz, G. Martinelli, F. Parodi,
P. Roudeau and A. Stocchi, hep-ph/0012308.  
\bibitem{BBG0}
{W.A. Bardeen, A.J. Buras and J.-M. G\'erard,}
{ Phys. Lett.} {\bf B211} (1988) 343;
 {J-M. G\'erard,} { Acta Physica Polonica} {\bf B21} (1990) 257. 
\bibitem{Bijnens}
{ J. Bijnens and J. Prades,} { Nucl. Phys.} {\bf B444} (1995) 523;
JHEP {\bf 0001} (2000) 002.
\bibitem{DORT99}
T. Hambye, G.O. K\"ohler and P.H. Soldan, 
Eur. Phys. J. {\bf C10} (1999) 271.
\bibitem{BERT97}
S. Bertolini, J.O. Eeg, M. Fabbrichesi and E.I. Lashin,
{ Nucl. Phys.} {\bf B514} (1998) 63.
\bibitem{PerRaf}
S. Peris and E. de Rafael,  { Phys.\ Lett.} {\bf B490} (2000) 213;
S. Peris, hep-ph/0010162.
\bibitem{LL}
L. Lellouch and C.-J.D. Lin, hep-ph/0011086; J. Flynn and C.-J.D. Lin,
hep-ph/0012154; C.T. Sachrajda, hep-lat/0101003.
\bibitem{QCDSF}
{ E. Bagan, P. Ball, V.M. Braun and H.G. Dosch},
{ Phys. Lett.} {\bf B278} (1992) 457;
{ M. Neubert}, { Phys. Rev.} {\bf D45} (1992) 2451;
{S. Narison,}
{ Phys. Lett.} {\bf B322} (1994) 247  and references therein.
\bibitem{NIR99}
Y. Nir, hep-ph/9911321.
\bibitem{Clarification}
Strictly speaking ${\cal A}^{\rm int}_{CP}(B\to f)$ contains also
CP violation in decay that only vanishes when $|\xi|=1$ as discussed
below. 
\bibitem{STOCCHI}
A. Stocchi, hep-ph/0010222; hep-ph/0012215.
\bibitem{STONE}
S. Stone, hep-ph/0012162.
\bibitem{CDFD0}
F. Abe et al. (CDF Collaboration), Phys. Rev. Lett. {\bf 82} (1999) 271;
B. Abbott et al. (D0 Collaboration), { Phys. Rev.} {\bf D60} (1999) 052001.
\bibitem{FRENCH}
Y. Grossman, Y. Nir, S. Plaszczynski and M.-H. Schune,
{ Nucl.~Phys.} {\bf B511} (1998) 69.
\bibitem{ALI00}
A. Ali and D. London, Eur. Phys. J. {\bf C9} (1999) 687;
Phys. Rep. {\bf 320}, (1999), 79; hep-ph/0002167; hep-ph/0012155.
\bibitem{SCHUNE}
S. Plaszczynski and M.-H. Schune, hep-ph/9911280.
\bibitem{Parodi}
S. Mele, {Phys.\ Rev.} {\bf D59} (1999) 113011;
M. Bargiotti et al., La Rivista del Nuovo Cimento, Vol. 23, N.3 (2000) 1;
 S. Schael, Phys. Rept.
{\bf 313} (1999) 293; M.~Ciuchini, E.~Franco, L.~Giusti, V.~Lubicz and 
G.~Martinelli, Nucl.\ Phys.\  {\bf B573} (2000) 201;
F. Caravaglios, F. Parodi, P. Roudeau, and A. Stocchi, hep-ph/0002171;
P. Faccioli, hep-ph/0011269.
\bibitem{CDFB}
T. Affolder et al., CDF collaboration,
{ Phys. Rev.} {\bf D61} (2000) 072005.
\bibitem{BaBar}
D. Hitlin, BaBar collaboration, plenary talk at ICHEP 
(Osaka, Japan, July 31, 2000), SLAC-PUB-8540.
\bibitem{Belle}
H. Aihara, Belle collaboration, plenary talk at ICHEP 
(Osaka, Japan, July 31, 2000).
\bibitem{NIR00}
G. Eyal, Y. Nir and G. Perez, JHEP {\bf 0008} (2000) 028.
\bibitem{SW}
J.P. Silva and L. Wolfenstein, hep-ph/0008004; A. Masiero, M. Piai and
O. Vives, hep-ph/0012096.
\bibitem{NK00}
A.L. Kagan and M. Neubert, { Phys. Lett.} {\bf B492} (2000) 115.
\bibitem{XING}
Z.Z. Xing, hep-ph/0008018.
\bibitem{ABRB}
A.J. Buras and R. Buras, hep-ph/0008273.
\bibitem{FM01}
R. Fleischer and Th. Mannel, hep-ph/0101276.
\bibitem{WW}
{ B. Winstein and L. Wolfenstein,} { Rev. Mod. Phys.} {\bf 65} (1993)
1113.
\bibitem{wolfenstein:64}
{ L.~Wolfenstein},
 { Phys. Rev. Lett.} {\bf 13} (1964) 562.
\bibitem{barr:93}
{ G.D. Barr} { et~al.},
{ Phys. Lett.} {\bf B317} (1993) 233.
\bibitem{gibbons:93}
{ L.K. Gibbons} { et~al.},
{ Phys. Rev. Lett.} {\bf 70} (1993) 1203.
\bibitem{KTEV} A. Alavi-Harati et al., 
{ Phys. Rev. Lett.} {\bf 83} (1999) 22.
\bibitem{NA48} V. Fanti et al., 
{ Phys. Lett.} {\bf B465} (1999) 335; T. Gershon (NA48), hep-ph/0101034.
\bibitem{EGN}
J. Ellis, M.K. Gaillard and D.V. Nanopoulos,
{ Nucl. Phys.} {\bf B109} (1976) 213.
\bibitem{GW79}
{ F.J. Gilman and M.B. Wise,} { Phys. Lett.} {\bf B83} (1979) 83;
{ B. Guberina and R.D. Peccei,} { Nucl. Phys.} {\bf B163} (1980) 289.
\bibitem{BSS}
{ F.J. Gilman and J.S. Hagelin}, { Phys. Lett.} {\bf B126} (1983) 111;
{ A.J. Buras, W. Slominski and H. Steger,} { Nucl. Phys.} {\bf B238} 
(1984) 529. 
\bibitem{BW84}
{ J. Bijnens and M.B. Wise,} { Phys. Lett.} {\bf B137} (1984) 245.
\bibitem{donoghueetal:86} 
{ J.F. Donoghue, E. Golowich, B.R. Holstein and J. Trampeti{\'c},}
{ Phys. Lett.} {\bf B179} (1986) 361. 
\bibitem{burasgerard:87}
{ A.J. Buras} and { J.-M. G{\'e}rard},
{ Phys. Lett.} {\bf B192} (1987) 156; 
 { Phys. Lett.} {\bf B203} (1988) 272.
 \bibitem{lusignoli:89}
H.-Y. Cheng, { Phys. Lett.} {\bf B201} (1988) 155;
{ M. Lusignoli,} { Nucl. Phys.} {\bf B325} (1989) 33. 
\bibitem{bardeen:87}
{ W.A. Bardeen}, { A.J. Buras} and { J.-M. G{\'e}rard},
 { Phys. Lett.} {\bf B180} (1986) 133;
{ Nucl. Phys.} {\bf B293} (1987) 787;
{ Phys. Lett.} {\bf B192} (1987) 138.
\bibitem{flynn:89}
{ J.M. Flynn} and { L. Randall},
{ Phys. Lett.} {\bf B224} (1989) 221; erratum ibid.\ { Phys.
  Lett.} {\bf B235} (1990) 412.
\bibitem{buchallaetal:90}
{ G.~Buchalla}, { A.J. Buras}, and { M.K. Harlander},
{ Nucl. Phys.} {\bf B337} (1990) 313.
\bibitem{PW91}
{ E.A. Paschos and Y.L. Wu,} { Mod. Phys. Lett.} {\bf A6} (1991) 93;
{ M. Lusignoli, L. Maiani, G. Martinelli and L. Reina,} 
{ Nucl. Phys.} {\bf B369} (1992) 139.
\bibitem{ciuchini:95}
{ M.~Ciuchini}, { E.~Franco}, { G.~Martinelli}, {L.~Reina
 and   L.~Silvestrini},
 { Z. Phys.} {\bf C68} (1995) 239.
\bibitem{BJL96a}
{ A.J. Buras}, { M.~Jamin}, and { M.E. Lautenbacher},
{ Phys. Lett.} {\bf B389} (1996) 749.
\bibitem{paschos:96}
{ J.~Heinrich}, { E.A. Paschos}, { J.-M. Schwarz} and { Y.L. Wu},
{ Phys. Lett.} {\bf B279} (1992) 140; Y.L. Wu, Int. J. Mod. Phys. 
{\bf A7} (1992) 2863.
\bibitem{BERT98}
S. Bertolini, M. Fabbrichesi and J.O. Eeg, 
{ Rev. Mod. Phys} {\bf 72} (2000) 65;
hep-ph/0002234; M. Fabbrichesi, { Phys. Rev.} {\bf D62} (2000) 097902. 
\bibitem{BUJA}
A.J. Buras, hep-ph/9908395; M. Jamin, hep-ph/9911390; 
U. Nierste, Nucl. Phys. Proc. Suppl. {\bf 86} (2000) 329; 
 S. Bertolini, hep-ph/0101212.
\bibitem{CM00}
M. Ciuchini and G. Martinelli, hep-ph/0006056.
\bibitem{REVOTH}
T. Hambye and P.H. Soldan, hep-ph/0009073;
J.O. Eeg, hep-ph/0010042; A. Pich, hep-ph/0010181.
\bibitem{EP99}
S. Bosch, A.J. Buras, M. Gorbahn, S. J{\"a}ger, M. Jamin, 
M.E. Lautenbacher and L. Silvestrini, 
{ Nucl. Phys.} {\bf B 565} (2000) 3.
\bibitem{buraslauten:93}
{ A.J. Buras} and { M.E. Lautenbacher},
{ Phys. Lett.} {\bf B318} (1993) 212.
\bibitem{Lubicz}
V. Lubicz, hep-lat/0012003.
\bibitem{NARM}
S. Narison, hep-ph/9911454.
\bibitem{JAM}
M. Jamin, Nucl. Phys. B. Proc. Suppl. {\bf 64} (1998) 250.
\bibitem{MAL}
K. Maltman, { Phys. Lett.} {\bf B462} (1999) 195.
\bibitem{PiPr}
J. Prades and A. Pich,  JHEP {\bf 9910} (1999) 004.
\bibitem{GUPTA01}
R. Gupta and K. Maltman, hep-ph/0101132.
\bibitem{GUPTA98}
R. Gupta, hep-ph/9801412.
\bibitem{QCDS}
{ M.~Jamin} and { M.~M{\"u}nz},
 { Z. Phys.} {\bf C66} (1995) 633;
{ S.~Narison},
{ Phys. Lett.} {\bf B358} (1995) 113;
K.G. Chetyrkin, D.~Pirjol, and 
  K.~Schilcher,
{ Phys. Lett.} {\bf B404} (1997) 337;
P. Colangelo, F. De Fazio, G. Nardulli, and N. Paver,
{ Phys. Lett.} {\bf B408} (1997) 340.
\bibitem{MSBOUND}
L. Lellouch, E. de Rafael, and J. Taron, 
{ Phys. Lett.} {\bf B414} (1997) 195;
F.J. Yndurain, { Nucl. Phys.} {\bf B 517} (1998) 324;
H.G. Dosch and S. Narison, { Phys. Lett.} {\bf B417} (1998) 173.
\bibitem{Nierste}
Y.-Y. Keum, U. Nierste and A.I. Sanda, 
{ Phys. Lett.} {\bf B457} (1999) 157.
\bibitem{DORT98}
T. Hambye, G.O. K\"ohler, E.A. Paschos, P.H. Soldan and W.A. Bardeen,
{ Phys. Rev.} {\bf D58} (1998) 014017.
\bibitem{GKS}
G. Kilcup, R. Gupta and S.R. Sharpe, 
{ Phys. Rev.} {\bf D57} (1998) 1654.
\bibitem{G67}
R. Gupta, T. Bhattacharaya, and S.R. Sharpe, 
{ Phys. Rev.} {\bf D55} (1997) 4036.
\bibitem{APE}
L. Conti, A. Donini, V. Gimenez, G. Martinelli, M. Talevi and
A. Vladikas, { Phys. Lett.} {\bf B421} (1998) 273.
\bibitem{Golowich}
J.F. Donoghue and E. Golowich, { Phys. Lett.} {\bf B478} (2000) 172.
\bibitem{Prades}
J. Bijnens and J. Prades, J. High Energy Phys. {\bf 0001} (2000) 002;
{\bf 0001} (2000) 023; {\bf 0006} (2000) 035; hep-ph/0009156;
hep-ph/0010008.
\bibitem{Narison}
S. Narison, { Nucl. Phys.} {\bf B 593} (2001) 3.
\bibitem{Bardeen}
W.A. Bardeen, private communication.
\bibitem{kilcup:91}
{ G.W. Kilcup},
 { Nucl. Phys. (Proc. Suppl.)} {\bf B20} (1991) 417.
\bibitem{sharpe:91}
{ S.R. Sharpe},
 { Nucl. Phys. (Proc. Suppl.)} {\bf B20} (1991) 429.
\bibitem{kilcup:99}
D. Pekurovsky and G. Kilcup, hep-lat/9812019.
\bibitem{Knecht}
M. Knecht, S. Peris and E. de Rafael, 
{ Phys. Lett.} {\bf B457} (1999) 227.
\bibitem{ECKER99}
G. Ecker, G. M\"uller, H. Neufeld and A. Pich,
{ Phys. Lett.} {\bf B 477} (2000) 88.
\bibitem{Dortmund} T. Hambye, G.O. K\"ohler, E.A. Paschos and
P.H. Soldan, { Nucl. Phys.} {\bf B 564} (2000) 391,
hep-ph/0001088.
\bibitem{ROMA99}
M. Ciuchini, E. Franco, L. Giusti, V. Lubicz and G. Martinelli,
hep-ph/9910237. 
\bibitem{BEL}
A.A. Belkov, G. Bohm, A.V. Lanyov and A.A. Moshkin, hep-ph/9907335;
hep-ph/0010142.
\bibitem{Tajpei}
H.-Y. Cheng, hep-ph/9911202.
\bibitem{Beijing}
Y.-L. Wu, hep-ph/0012371.
\bibitem{PAPI99}
E. Pallante and A. Pich, { Phys. Rev. Lett.} {\bf 84} (2000) 2568;
{ Nucl. Phys.} {\bf B 592} (2000) 294.
\bibitem{PA99}
E.A. Paschos, hep-ph/9912230.
\bibitem{Kambor}
J. Kambor, J. Missimer and D. Wyler, { Phys. Lett.} {\bf B 261}
 (1991) 496.
\bibitem{DI12}
{ W.A. Bardeen}, { A.J. Buras} and { J.-M. G{\'e}rard},
{ Phys. Lett.} {\bf B192} (1987) 138;
{ A. Pich and E. de Rafael}, { Nucl. Phys.} {\bf B358} (1991) 311;
{ M. Neubert and B. Stech}, { Phys. Rev.} {\bf D 44} (1991) 775;
{ M. Jamin and A. Pich}, { Nucl. Phys.} {\bf B425} (1994) 15; 
{ J. Kambor, J. Missimer and D. Wyler},
{ Nucl. Phys.} {\bf B346} (1990) 17;
{ Phys. Lett.} {\bf B261} (1991) 496;
 S. Bertolini, J.O. Eeg, M. Fabrichesi and E.I. Lashin,
{ Nucl. Phys.} {\bf B514} (1998) 63;
J. Bijnens and J. Prades, JHEP {\bf 9901} (1999) 023;
J. Bijnens et al.,  Nucl. Phys. {\bf B521} (1998) 305.
T. Hambye, G.O. K\"ohler and P.H. Soldan, Eur. Phys. J. {\bf C10}
(1999) 271.
\bibitem{GW00}
J.-M. G{\'e}rard and J. Weyers, hep-ph/0011391. Here the assumption
of trace anomaly dominance in weak K decays successfully reproduces
the $\Delta I=1/2$ rule in $K_S\to \pi\pi$, $K_L\to\pi\pi\pi$,
$K_S\to\gamma\gamma$ and $K_L\to\pi^0\gamma\gamma$.
\bibitem{Truong}
T.N. Truong, { Phys. Lett.} {\bf B 207} (1988) 495.
\bibitem{AITAL}
A.J. Buras, M. Ciuchini, E. Franco, G. Isidori, G. Martinelli and
L. Silvestrini, { Phys. Lett.} {\bf B 480} (2000) 80.
\bibitem{ISI00}
G. Isidori, hep-ph/0011017.
\bibitem{LLs}
L. Lellouch and M. L\"uscher, hep-lat/0003023.
\bibitem{Gardner}
S. Gardner and G. Valencia,
{ Phys. Lett.} {\bf B466} (1999) 355.
\bibitem{Wolfe}
C.E. Wolfe and K. Maltman,
{ Phys. Lett.} {\bf B482} (2000) 77.
\bibitem{CIRI1}
V. Cirigliano, J.F. Donoghue and E. Golowich,
{ Phys. Lett.} {\bf B450} (1999) 241.
\bibitem{CIRI2}
V. Cirigliano, J.F. Donoghue and E. Golowich,
{ Phys. Rev.} {\bf D61} (2000) 093001; ibid
093002. 
\bibitem{CIRI3}
V. Cirigliano, J.F. Donoghue and E. Golowich,
Eur. Phys. J. {\bf C18} (2000) 83.
\bibitem{Ecker00}
G. Ecker, G. Isidori, G. M\"uller, H. Neufeld and A. Pich,
{ Nucl. Phys.} {\bf B 591} (2000) 419.
\bibitem{Gardner2}
S. Gardner and G. Valencia, { Phys. Rev.} {\bf D62} (2000) 094024.
\bibitem{MM99}
A. Masiero and H. Murayama, { Phys. Rev. Lett.} {\bf 83} (1999) 907.
\bibitem{BS99}
A.J. Buras, G. Colangelo, G. Isidori, A. Romanino and
L. Silvestrini,  { Nucl. Phys.} {\bf B 566} (2000) 3; 
L. Silvestrini, hep-ph/0009284; G. Isidori, hep-ph/0101121.
\bibitem{KN}
A. Kagan and M. Neubert, { Phys. Rev. Lett.} {\bf 83} (1999) 4929.
\bibitem{SIL99}
G. Eyal, A. Masiero, Y. Nir and L. Silvestrini, JHEP 9911 (1999) 032;
R. Barbieri, R. Contino and A. Strumia, 
{ Nucl. Phys.} {\bf B 578} (2000) 153.
\bibitem{HE}
X-G. He and B.H.J. McKellar, { Phys. Rev.} {\bf D51} (1995) 6484; 
X-G. He, { Phys. Lett.} {\bf B460} (1999) 405.
\bibitem{Huang}
C-S. Huang, W-J. Huo and Y-L. Wu, hep-ph/0005227.
\bibitem{Jang}
J.-h. Jang, K.Y. Lee, S.Ch. Park and H.S. Song, hep-ph/0010107.
\bibitem{Frampton}
J. Agrawal and P. Frampton,  { Nucl. Phys.} {\bf B 419} (1994) 254.
\bibitem{TERR}
F. Terranova, hep-ph/0005188.
\bibitem{BS98}
A.J. Buras and L. Silvestrini,
{ Nucl. Phys.} {\bf B 546} (1999) 299.
\bibitem{GMS}
E. Gabrielli, A. Masiero and L. Silvestrini, 
{ Phys. Lett.} {\bf B374} (1996) 80;
F. Gabbiani, E. Gabrielli, A. Masiero and L. Silvestrini,
{ Nucl. Phys.} {\bf B 477} (1996) 321.
\bibitem{ISI}
G. Colangelo and G. Isidori, JHEP 09 (1998) 009.
\bibitem{CHANOWITZ}
M.S. Chanowitz, hep-ph/9905478(v2).
\bibitem{RS}
{ D. Rein and L.M. Sehgal,} { Phys. Rev.} {\bf D39} (1989) 3325;
{ J.S. Hagelin and L.S. Littenberg,} { Prog. Part. Nucl. Phys.}
{\bf 23} (1989) 1;
{ M. Lu and M.B. Wise,} { Phys. Lett.} {\bf B324} (1994) 461;
{ S. Fajfer}, [hep-ph/9602322]; { C.Q. Geng, I.J. Hsu and Y.C. Lin},
{ Phys. Rev.} {\bf D54} (1996) 877.
\bibitem{GBGI}
G. Buchalla and G. Isidori, { Phys. Lett.} {\bf B440} (1998) 170.
\bibitem{FalkLP}
A.F. Falk, A. Lewandowski and A.A. Petrov, hep-ph/0012099.
\bibitem{BB4}
{ G. Buchalla and A.J. Buras}, 
{ Phys. Lett.} {\bf B333} (1994) 221.
\bibitem{novikovetal:77}
{ V.A. Novikov, A.I. Vainshtein, V.I. Zakharov and M.A. Shifman,}
Phys. Rev. {\bf D16}, (1977) 223;
{ J. Ellis and J.S. Hagelin,} { Nucl.~Phys.} {\bf B217} (1983) 189;
{ C.O. Dib, I. Dunietz and F.J. Gilman,} { Mod. Phys. Lett.}
{\bf A6} (1991) 3573.
\bibitem{MP}
{ W. Marciano and Z. Parsa}, Phys. Rev. {\bf D53}, R1 (1996).
\bibitem{Adler97}
S. Adler et al., { Phys. Rev. Lett.} {\bf 79}, (1997) 2204.
\bibitem{Adler00}
S. Adler et al., { Phys. Rev. Lett.} {\bf 84}, (2000) 3768.
\bibitem{LITT00}
{ L. Littenberg,} hep-ex/0010048.
\bibitem{AGS2}
B. Bassalleck et al., E949 Proposal, BNL 67247, TRI-PP-00-06, 1999.
\bibitem{Coleman}
 R. Coleman et al., (CKM collaboration), 
Charged Kaons at the Main Injector, Fermilab-P-0905, 1998.
\bibitem{littenberg:89}
{ L. Littenberg,} { Phys. Rev.} {\bf D39} (1989) 3322.
\bibitem{NIR96}
{ Y. Grossman, Y. Nir and R. Rattazzi}, hep-ph/9701231,
in \cite{BULIND}, page 755;  
Y. Nir, hep-ph/9904271.
\bibitem{BUCH96}
{ G. Buchalla}, hep-ph/9612307.
\bibitem{BB96}
{ G. Buchalla} and { A.J. Buras},
 { Phys. Rev.} {\bf D54} (1996) 6782.
\bibitem{KTeV00X}
A. Alavi-Harati et al., { Phys. Rev.} {\bf D61} (2000) 072006.
\bibitem{KEKKL}
{ T. Inagaki, et al.,} KEK Internal 96-13, November 1996.
\bibitem{FNALKL}
E. Chen et al., hep-ex/9709026.
\bibitem{KOPIO}
I.-H. Chiang et al., AGS Experiment Proposal 926 (1996).
\bibitem{Perez}
S. Bergmann and G. Perez, JHEP {\bf 0008} (2000) 034.
\bibitem{KLBSM}
Y. Grossman and Y. Nir, { Phys. Lett.} {\bf B398} (1997) 163;
C.E. Carlson, G.D. Dorada and M. Sher,
{ Phys. Rev.} {\bf D54} (1996) 4393; 
G. Burdman, { Phys.\ Lett.} {\bf B409} (1997) 443;
A. Berera, T.W. Kephart and M. Sher, 
{ Phys. Rev.} {\bf D56} (1997) 7457;
Gi-Chol Cho, hep-ph/9804327;
T. Hattori, T. Hasuike and S. Wakaizumi, hep-ph/9804412.
\bibitem{KLMT}
C. Dib, I. Dunietz and F.J. Gilman, 
{Phys. Rev.} {\bf D39} (1989) 2639; J. Flynn and L. Randall,
 { Nucl. Phys.} {\bf B326} (1989) 31, erratum, ibid.
{\bf B334} (1990) 580.
\bibitem{KL}
{ G. Ecker, A.~Pich} and { E.~de~Rafael},
{ Nucl. Phys.} {\bf B291} (1987) 692,
{ Nucl. Phys.} {\bf B303} (1988) 665,
{ Phys. Lett.} {\bf B237} (1990) 481;
{ A.G. Cohen, G.~Ecker,} and { A.~Pich},
{ Phys. Lett.} {\bf B304} (1993) 347;
{ L.M.~Seghal,}
{ Phys. Rev.} {\bf D38} (1988) 808;
{ P. Heiliger} and { L.M.~Seghal,}
{ Phys. Rev.} {\bf D47} (1993) 4920;
{ C. Bruno and J.~Prades},
{ Z. Phys.} {\bf C57} (1993) 585;
{ J. F. Donoghue} and { F.~Gabbiani},
{ Phys. Rev.} {\bf D51} (1995) 2187;
G. D'Ambrosio and J. Portol{\'e}s, 
{ Nucl. Phys.} {\bf B492} (1997) 417.
\bibitem{AEIP}
G. D'Ambrosio, G. Ecker, G. Isidori and J. Portol{\'e}s, 
JHEP {\bf 08} (1998) 004.
\bibitem{KGG}
A. Alavi-Harati et al. (KTeV Collaboration), 
{ Phys. Rev. Lett.} {\bf 83}, (1999) 917;
V.D. Kekelidze (NA48 Collaboration), talk presented at ICHEP2000,
Osaka, Japan.
\bibitem{ke}
A. Alavi-Harati et al. (KTeV Collaboration), hep-ex/0009030.
\bibitem{mpmm}
D. Ambrose et al. (E871), { Phys. Rev. Lett.} {\bf 84}, (2000) 1389.
\bibitem{dambrosio}
G. D'Ambrosio, G. Isidori and J. Portol{\'e}s, 
{ Phys. Lett.} {\bf B423} (1998) 385.
\bibitem{pich}
D. Gomez Dumm and A. Pich, Nucl. Phys. Proc. Suppl. {\bf 74} (1999)
186.
\bibitem{GVAL}
G. Valencia, hep-ph/9711377; M. Knecht, S. Peris, M. Perrottet and
E. de Rafael, { Phys. Rev. Lett.} {\bf 83}, (1999) 5230.
\bibitem{CPRARE}
{ L. Littenberg and G. Valencia,}
{ Ann. Rev. Nucl. Part. Sci.} {\bf 43} (1993) 729;
{ J.L. Ritchie and S.G. Wojcicki,} { Rev. Mod. Phys.} {\bf 65} (1993)
1149; A. Pich, hep-ph/9610243; G. D'Ambrosio and G. Isidori,
hep-ph/9611284; P. Buchholz and B. Renk, Prog. Part. Nucl. Phys. 
{\bf 39} (1997) 253; R. Peccei, hep-ph/9909236; 
R. Belusevic, KEK 97-264 (March 1998); G. Buchalla, hep-ph/9912369;
G. D'Ambrosio, hep-ph/0002254; A.R. Barker and S.H. Kettel,
hep-ex/0009024.
\bibitem{GREUB00}
C. Greub, hep-ph/9911348.
\bibitem{Mikolaj}
M. Misiak, hep-ph/0002007; hep-ph/0009033.
\bibitem{AH1}
A. Ali and G. Hiller, Eur. Phys. J. {\bf C8} (1999) 619.
\bibitem{AH2}
A. Ali, P. Ball, L.T. Handoko and G. Hiller, 
{ Phys. Rev.} {\bf D61} (2000) 074024
\bibitem{BHI}
G. Buchalla, G. Hiller and G. Isidori, 
{ Phys. Rev.} {\bf D63} (2001) 014015;
G. Isidori, hep-ph/0009024.
\bibitem{Lunghi}
E. Lunghi, A. Masiero, I. Scimemi and L. Silvestrini,
{ Nucl.~Phys.} {\bf B568} (2000) 120;
E. Lunghi and I. Scimemi,
 { Nucl.~Phys.} {\bf B574} (2000) 43.
\bibitem{KrLu}
F. Kr\"uger and E. Lunghi, { Phys. Rev.} {\bf D63} (2001) 014013.
\bibitem{Hewett}
J.L. Hewett and D. Wells, { Phys. Rev.} {\bf D55} (1997) 5549; 
Q. Yan, C. Huang, W. Liao and S. Zhu, 
{ Phys. Rev.} {\bf D62} (2000) 094023.
\bibitem{AGMU}
A. Ali and C. Greub, Z. Phys. {\bf C60} (1993) 433;
A.J. Buras, M. Misiak, M. M\"unz and S. Pokorski,
{ Nucl.~Phys.} {\bf B424} (1994) 374.
\bibitem{AH3}
A. Ali,  G. Hiller, L.T. Handoko and T. Morozumi, 
{ Phys. Rev.} {\bf D55} (1997) 4105.
\bibitem{CLEOBS}
S. Ahmed et al., (CLEO), hep-ex/9908022.
\bibitem{ALEPHBS}
R. Barate et al., (ALEPH),  Phys.~Lett. {\bf B429} (1998) 169.
\bibitem{BELLEBS}
A. Abashian et al., BELLE-CONF-0003.
\bibitem{DGG00}
G. Degrassi, P. Gambino and G.F. Giudice, hep-ph/0009337.
\bibitem{CGNW00}
M. Carena, D. Garcia, U. Nierste and C.E.M. Wagner,
hep-ph/0010003.
\bibitem{CBB}
T.E. Coan et al., (CLEO), { Phys. Rev. Lett.} {\bf 84}, (2000) 5283;
C. Jessop et al., (BABAR), hep-ex/0011054; A. Abashian et al., 
BELLE-CONF-00-03.
\bibitem{Pirjol}
D. Pirjol, hep-ph/0101045; A. Ali, hep-ph/0101154.
\bibitem{Burdman}
G. Burdman,
{ Phys. Rev.} {\bf D57} (1998) 4254.
\bibitem{Charles}
J. Charles et. al., { Phys. Rev.} {\bf D60} (1999) 014001;
Phys.~Lett. {\bf B451} (1999) 187.
\bibitem{Overmb}
G. Buchalla and G. Isidori,
{ Nucl.~Phys.} {\bf B525} (1998) 333. A.F. Falk, M. Luke and M.J. Savage,
{ Phys. Rev.} {\bf D49} (1994) 3367.
\bibitem{BUC97}
G. Buchalla, G. Isidori and S.-J. Rey, 
{ Nucl.~Phys.} {\bf B511} (1998) 594. 
\bibitem{CRS}
J.-W. Chen, G. Rupak and M.J. Savage, { Phys.~Lett.} {\bf B410} (1997) 285.
\bibitem{KSMETHOD}
F. Kr\"uger and L.M. Sehgal,
{ Phys.~Lett.} {\bf B380} (1996) 199.
\bibitem{Glenn}
S. Glenn et al. (CLEO), { Phys. Rev. Lett.} {\bf 80}, (1998) 2289.
\bibitem{Affolder1}
T. Affolder et al. (CDF), { Phys. Rev. Lett.} {\bf 83}, (1999) 3378.
\bibitem{CM78} 
{ N. Cabibbo and L. Maiani}, 
{ Phys.~Lett.} {\bf B79} (1978) 109.
\bibitem{KIMM}
{ C.S. Kim and A.D. Martin},
{ Phys.~Lett.} {\bf B225} (1989) 186.
\bibitem{Aleph96}
H. Kroha, (ALEPH Collaboration), in Proceedings of 
the 28th International Conference
on High Energy Physics, July 1996, Warsaw, Poland,
eds. Z. Ajduk and A.K. Wroblewski, World Scientific,
Singapore (1997).
\bibitem{NUNUK}
Y. Grossman, Z. Ligeti and E. Nardi, 
{ Nucl. Phys.} {\bf B465} (1996) 369; D. Melikhov, N. Nikitin and 
S. Simula, { Phys.~Lett.} {\bf B428} (1998) 171; T.M. Aliev and 
C.S. Kim, { Phys. Rev.} {\bf D58} (1998) 013003; C.S. Kim, Y.G. Kim
and T. Morozumi, { Phys. Rev.} {\bf D60} (1999) 094007; T.M. Aliev, 
A. \"Ozpinezi and M. Savci, hep-ph/0101066.
\bibitem{CDFMU}
F. Abe et al. (CDF), { Phys. Rev.} {\bf D57} (1998) R3811.
\bibitem{CLEOMU}
T. Bergfeld et al. (CLEO), { Phys. Rev.} {\bf D62} (2000) 091102.
\bibitem{LARGET}
K. Babu and C. Kolda,  { Phys. Rev. Lett.} {\bf 84}, (2000) 228;
H.E. Logan and U. Nierste, { Nucl. Phys.} {\bf B577} (2000) 88;
C.S. Huang, W. Liao, Q.-S. Yan and S.-H. Zhu, hep-ph/0006250;
P.H. Chankowski and L. Slawianowska, hep-ph/0008046.
\bibitem{GLN96}
J. Kalinowski and W. Skiba, { Nucl. Phys.} {\bf B404} (1993) 3.
Y. Grossman, Z. Ligeti and E. Nardi,
{ Phys. Rev.} {\bf D55} (1997) 2768.
\bibitem{NQ}
{Y. Nir and H.R. Quinn}
{ Ann. Rev. Nucl. Part. Sci.} {\bf 42}
(1992) 211 and
 in `` B Decays ", ed S. Stone
 (World Scientific, 1994),
p. 520; {I. Dunietz,} ibid p.550 and refs. therein;
H.R. Quinn and A.I. Sanda, Eur. Phys. J. {\bf C15} (2000) 626.
\bibitem{NIR08}
Y. Nir, hep-ph/0008226.
\bibitem{RF97}
{ R. Fleischer}, { Int. J. of Mod. Phys.}
 {\bf A12} (1997) 2459.
\bibitem{B95}
{ A.J. Buras,} { Nucl. Instr. Meth.} {\bf A368} (1995) 1.
\bibitem{CLEO99}
D. Cronin-Hennessy et al., (CLEO),   { Phys. Rev. Lett.}
 {\bf 85} (2000) 515 and 525. 
\bibitem{ITAL}
M. Ciuchini, E. Franco, G. Martinelli and  L. Silvestrini,
{ Nucl. Phys.} {\bf B501} (1997) 271;
M. Ciuchini, R. Contino, E. Franco, G. Martinelli and  L. Silvestrini,
{ Nucl. Phys.} {\bf B512} (1998) 3.
\bibitem{CPASYM}
{ M. Gronau and D. London,} { Phys. Rev. Lett.}
 {\bf 65} (1990) 3381.
\bibitem{SNYD}
A. Snyder and H.R. Quinn, { Phys. Rev.} {\bf D48} (1993) 2139;
{ A.J. Buras and R. Fleischer,}
{ Phys. Lett.} {\bf B360} (1995) 138;
J.P. Silva and L. Wolfenstein, 
{ Phys. Rev.} {\bf D49} (1995) R1151; 
{ A.S. Dighe, M. Gronau and J. Rosner}, 
{ Phys. Rev.} {\bf D54} (1996) 3309; 
R. Fleischer and T. Mannel, 
{ Phys. Lett.} {\bf B397} (1997) 269;
C.S. Kim, D. London and T. Yoshikawa, 
{ Phys. Rev.} {\bf D57} (1998) 4010. 
\bibitem{HRQ}
Y. Grossman and H.R. Quinn, { Phys. Rev.} {\bf D56} (1997) 7259;
H.R. Quinn and J.P. Silva,  { Phys. Rev.} {\bf D62} (2000) 054002.
\bibitem{BrBaBar}
B. Aubert et al., (BaBar), hep-ex/0008057.
\bibitem{Bellepi}
K. Trabelsi, the Belle Collaboration.
\bibitem{NEUB}
M. Neubert, hep-ph/0011064.
\bibitem{BSANDA}
A.B. Carter and A.I. Sanda, { Phys. Rev. Lett.} {\bf 45} (1980) 952;
{ Phys. Rev.} {\bf D23} (1981) 1567.
{ I.I. Bigi and A.I. Sanda,}
{ Nucl. Phys.} {\bf B193} (1981) 85.
\bibitem{PHI}
D. London and A. Soni, { Phys. Lett.} {\bf B407} (1997) 61;
Y. Grossman and M.P. Worah, { Phys. Lett.} {\bf B395} (1997) 241;
M. Ciuchini et al., { Phys. Rev. Lett.} {\bf 79} (1997) 978;
R. Barbieri  and A. Strumia, { Nucl. Phys.} {\bf B508} (1997) 3.
\bibitem{adk}
{ R. Aleksan, I. Dunietz and B. Kayser,}
 { Z. Phys.} {\bf C54} (1992) 653;\\
R. Fleischer and I. Dunietz, { Phys. Lett.} {\bf B387} (1996) 361.
\bibitem{Wyler}
{ M. Gronau and D. Wyler,} { Phys. Lett.} {\bf B265} (1991) 172.
\bibitem{DUN2}
M. Gronau and D. London, { Phys. Lett.} {\bf B253} (1991) 483;
{ I. Dunietz}, { Phys. Lett.} {\bf B270} (1991) 75.
\bibitem{V97}
D. Atwood, I. Dunietz and A. Soni, 
{ Phys. Rev. Lett.} {\bf B78} (1997) 3257.
\bibitem{FW01}
R. Fleischer and D. Wyler, { Phys. Rev.} {\bf D62} (2000) 057503.
\bibitem{BrBelle}
Abashian et al., (Belle), BELLE-CONF-0005 and 006. 
\bibitem{FSI}
L. Wolfenstein, {Phys.\ Rev.} {\bf D52} (1995) 537; 
J. Donoghue, E. Golowich, A.~Petrov and J. Soares, 
{ Phys. Rev. Lett.} {\bf 77} (1996) 2178; 
B. Blok and I. Halperin, { Phys. Lett.} {\bf B385} (1996) 324; 
B. Blok, M. Gronau and J.L.~Rosner,
{ Phys.\ Rev.\ Lett.} {\bf 78} (1997) 3999;
A.J. Buras, R. Fleischer and T. Mannel, { Nucl. Phys.} {\bf B533} (1998) 3;
J.-M. G{\'e}rard and J. Weyers, Eur. Phys. J. {\bf C7} (1999) 1;
D. Delepine, J.-M. G{\'e}rard, J. Pestieau and J. Weyers,
{ Phys. Lett.} {\bf B429} (1998) 106; 
M. Neubert, { Phys. Lett.} {\bf B424} (1998) 152; 
A.F. Falk, A.L. Kagan, Y. Nir and A.A. Petrov, 
{Phys.\ Rev.} {\bf D57} (1998) 4290;
D. Atwood and A. Soni (1997), {Phys.\ Rev.} {\bf D58} (1998) 036005; 
R. Fleischer, { Phys. Lett.} {\bf B435} (1998) 221;
M. Gronau and J.L. Rosner, {Phys.\ Rev.} {\bf D58} (1998) 113005.
\bibitem{FM}
R. Fleischer, { Phys. Lett.} {\bf B365} (1996) 399; 
R. Fleischer and T. Mannel, {Phys.\ Rev.} {\bf D57} (1998) 2752.
\bibitem{GRRO}
M. Gronau and J.L. Rosner, {Phys.\ Rev.} {\bf D57} (1998) 6843.
\bibitem{GPAR1}
R. Fleischer, Eur. Phys. J. {\bf C6} (1999) 451.
\bibitem{GPAR3}
A.J. Buras and R. Fleischer, Eur. Phys. J. {\bf C11} (1999) 93; 
Eur. Phys. J. {\bf C16} (2000) 97; hep-ph/0008298. 
\bibitem{GPAR2}
M. Neubert, JHEP 9902 (1998) 014.
\bibitem{NRBOUND}
M. Neubert and J.L. Rosner, { Phys. Lett.} {\bf B441} (1998) 403; 
{ Phys. Rev. Lett.} {\bf 81} (1998) 5076.
\bibitem{RFA}
R. Fleischer, hep-ph/9904313.
\bibitem{NEU99}
M. Neubert, hep-ph/9904321.
\bibitem{HHY}
X.-G. He, W.-S. Hou and K-Ch. Yang, hep-ph/9902256;
W.-S. Hou and K.-Ch. Yang, { Phys. Rev.} {\bf D61} (2000) 073014;
W.-S. Hou, J.G. Smith and F. W\"urthwein, hep-ex/9910014.
\bibitem{BBNS2}
M. Beneke, G. Buchalla, M. Neubert and C.T. Sachrajda,
hep-ph/0007256.
\bibitem{Muta2}
D. Du, D. Yang and G. Zhu, hep-ph/0005006; T. Muta, A. Sugamoto, 
M.Z. Yang and Y.D. Yang, hep-ph/0006022.
\bibitem{RF99}
R. Fleischer, { Phys.\ Lett.} {\bf B459} (1999) 306,
 Eur. Phys. J. {\bf C10} (1999) 299.
\bibitem{RF991}
R. Fleischer, { Phys. Lett.} {\bf B459} (1999) 306.
\bibitem{GRCW}
M. Gronau and J.L. Rosner,   { Phys.\ Lett.} {\bf B482} (2000) 71;
C.W. Chiang and L. Wolfenstein, { Phys.\ Lett.} {\bf B493} (2000) 73.
\bibitem{G00}
M. Gronau, hep-ph/0008292.
\bibitem{CDGG}
M. Ciuchini, G. Degrassi, P. Gambino and G.F. Giudice,
{ Nucl. Phys.} {\bf B 534} (1998) 3.
\bibitem{UUT}
A.J. Buras, P. Gambino, M. Gorbahn, S. J\"ager and L. Silvestrini,
hep-ph/0007085.
\bibitem{BNEW}
M. Gronau and D. London, {Phys.\ Rev.} {\bf D55}
(1997) 2845;
Y. Grossman, Y. Nir and M.P. Worah, 
{ Phys.\ Lett.} {\bf B407} (1997) 307; Y. Grossman, hep-ph/0012216;
I. Dunietz, R. Fleischer and U. Nierste, hep-ph/0012219.
\bibitem{MPR}
M. Misiak, S. Pokorski and J. Rosiek, hep-ph/9703442, 
in \cite{BULIND}, page 795.
\bibitem{BRS}
A.J. Buras, A. Romanino and L. Silvestrini, 
{ Nucl. Phys.} {\bf B520} (1998) 3.
\bibitem{GGMS} 
 F. Gabbiani, E. Gabrielli, A. Masiero and L. Silvestrini,
 { Nucl. Phys.} {\bf B477} (1996) 321. 
\bibitem{AJB94}
{ A.J. Buras}, { Phys. Lett.} {\bf B333} (1994) 476.
\bibitem{HKR} 
 L.J. Hall, V.A. Kostelecky and S. Rabi,  
 { Nucl. Phys.} {\bf B267} (1986) 415.
\bibitem{GN1}
Y. Nir and M.P. Worah, { Phys.\ Lett.} {\bf B423} (1998) 319.
\bibitem{SUSY}
R. Barbieri and A. Strumia, { Nucl. Phys.} {\bf B508} (1997) 3; 
R. Barbieri, L. Hall, A. Stocchi and N. Weiner, 
{ Phys.\ Lett.} {\bf B425} (1998) 119.
D.A. Demir, A. Masiero and O. Vives, {Phys.\ Rev.} {\bf D61}
(2000) 075009, { Phys. Lett.} {\bf B479} (2000) 230;
A. Masiero and O. Vives, hep-ph/0001298, hep-ph/007320; 
G. Barenboim, F.J. Botella 
and O. Vives, hep-ph/0012197; S. Baek, J.-H. Jang, P. Ko and 
J.H. Park, hep-ph/0009105; M. Dine, E. Kramer, Y. Nir and Y. Shadmi,
hep-ph/0101092.
\bibitem{POKOR}
There is a vast literature on this subject. A collection of
useful references can be found in K. Choi, E.J. Chun and 
K. Hwang, hep-ph/0004101. See in particular [3] and [4] in
this paper.
\bibitem{SCPV}
G. Barenboim, { Nucl. Phys.} {\bf B534} (1998) 318;
G. Barenboim, J. Bernabeu, J. Matias and M. Raidal,
{ Phys. Rev.} {\bf D60} (1999) 016003; G. Barenboim and M. Raidal,
{ Phys.\ Lett.} {\bf B457} (1999) 109; 
P. Ball, J.M. Frere and J. Matias, { Nucl. Phys.} {\bf B572} (2000) 3;
P. Ball and R. Fleischer, { Phys.\ Lett.} {\bf B475} (2000) 111;
G.C. Branco, F. Kr\"uger, 
J.C. Rom$\tilde{a}$o and A. Teixeira, hep-ph/0012318. 
\bibitem{Oliver}
J. Charles and L. Oliver, hep-ph/0010304.
\bibitem{Hyperon}
CP violation in hyperon decays is reviewed in S. Pakvasa, 
hep-ph/0002210; X.-G. He, hep-ph/0005299; G. Valencia,
 hep-ph/0006178.
\bibitem{CLEOD}
R. Godang et al., (CLEO), { Phys. Rev. Lett.} {\bf 84} (2000) 5038;
hep-ex/0001060.
\bibitem{FOCUSD}
J.M. Link et al., (FOCUS),  { Phys. Lett.} {\bf B485} (2000) 62; 
hep-ex/0004034.
\bibitem{D0D0}
Theoretical discussion of the CLEO and FOCUS data is given in
S. Bergmann, Y. Grossman, Z. Ligeti, Y. Nir and A.A. Petrov,
hep-ph/0005181; I.I. Bigi, hep-ph/0012161.
\bibitem{Higgs1}
W. Bernreuther, T. Schr\"oder and T.N. Pham, 
{ Phys. Lett.} {\bf B279} (1992) 389; 
B. Grzadkowski and J.F. Gunion,{ Phys. Lett.} {\bf B294} (1992) 361;
W. Bernreuther and A. Brandenburg, {Phys.\ Rev.} {\bf D49} (1994) 4481;
J.F. Gunion, B. Grzadkowski and X.-G. Xe, 
{ Phys. Rev. Lett.} {\bf 77} (1996) 5172.  
W. Bernreuther and
P. Overmann, Z. Phys. {\bf C72} (1996) 461;
B. Grzadkowski and Z. Hioki, Phys. Lett. {\bf B476} (2000) 87.
\bibitem{Higgs2}
A. Pilaftsis and C.E. Wagner, { Nucl. Phys.} {\bf B553} (1999) 3;
M. Carena, J. Ellis, 
A. Pilaftsis and C.E. Wagner, { Nucl. Phys.} {\bf B586} (2000) 92;
hep-ph/0009212; S.Y. Choi, M. Drees and J.S. Lee, 
{ Phys. Lett.} {\bf B481} (2000) 57.
\bibitem{LIND}
See for instance K. Dick, M. Freund, M.Lindner and A. Romanino,
{ Nucl. Phys.} {\bf B562} (1999) 29.
\bibitem{WBUCH}
See recent review by W. Buchm\"uller, hep-ph/0101102 and 
references therein.
\bibitem{Feng}
Most recent review is given in J.L. Feng, hep-ph/0101122.
\end{thebibliography}
\end{document}